\documentclass[letterpaper,11pt,fleqn]{article}
\usepackage{jheppub}

\usepackage[utf8]{inputenc}
\usepackage[T1]{fontenc}
\usepackage{extarrows}
\usepackage{bm,amsmath,amssymb}
\usepackage[mathscr]{eucal}
\usepackage{graphicx}
\usepackage{graphbox}
\usepackage{cancel}
\usepackage{subcaption}
\usepackage{dsfont}
\usepackage{enumerate}

\long\def\comment#1{ }

\newcommand{\eqn}[1]{Eq.~\eqref{#1}}
\newcommand{\beq}{\begin{equation}}
\newcommand{\eeq}{\end{equation}}
\newcommand{\bal}{\begin{align}}
\newcommand{\eal}{\end{align}}

\newcommand{\order}[1]{\mathcal{O}{\left(#1\right)}}

\newcommand{\nn}{\nonumber\\}
\newcommand{\rmd}{{\rm d}}
\newcommand{\dif}{{\rm d}}

\newcommand{\bk}{\bm{k}}

\newcommand{\bx}{\bm{x}}
\newcommand{\by}{\bm{y}}

\newcommand{\bz}{\bm{z}}

\newcommand{\br}{\bm{r}}
\newcommand{\bb}{\bm{b}}
\newcommand{\xbj}{x_{\rm \scriptscriptstyle Bj}}
\newcommand{\mcal}{\mathcal}
\newcommand{\bK}{\bm{K}}
\newcommand{\bP}{\bm{P}}
\newcommand{\bR}{\bm{R}}
\newcommand{\QP}{\mathcal{Q}_{\mathbb P}}
\newcommand{\YP}{Y_{\mathbb P}}

\title{TMD factorisation for diffractive jets in photon-nucleus interactions}

\author[a]{S.~Hauksson,}
\author[a]{E.~Iancu,}
\author[b]{A.H.~Mueller,}
\author[c]{D.N.~Triantafyllopoulos}
\author[d]{and S.Y.~Wei}

\affiliation[a]{Universit\'{e} Paris-Saclay, CNRS, CEA, Institut de physique th\'{e}orique, F-91191, Gif-sur-Yvette, France}
\affiliation[b]{Department of Physics, Columbia University, New York, NY 10027, USA}
\affiliation[c]{European Centre for Theoretical Studies in Nuclear Physics and Related Areas (ECT*)\\and Fondazione Bruno Kessler, Strada delle Tabarelle 286, I-38123 Villazzano (TN), Italy}
\affiliation[d]{Key Laboratory of Particle Physics and Particle Irradiation (MOE), Institute of frontier and interdisciplinary science, Shandong University, Qingdao, Shandong 266237, China}

\emailAdd{sigtryggur.hauksson@ipht.fr}
\emailAdd{edmond.iancu@ipht.fr}
\emailAdd{ahm4@columbia.edu}
\emailAdd{trianta@ectstar.eu}
\emailAdd{shuyi@sdu.edu.cn}

\usepackage[normalem]{ulem}

\abstract{
Using the colour dipole picture and the colour glass condensate effective theory, we  study the
diffractive  production of two or three jets via coherent photon-nucleus interactions at high energy.
We consider the hard regime where the photon virtuality and/or the transverse momenta of the produced 
jets are much larger than the saturation momentum $Q_s$ of the nuclear target. We show that, despite this 
hardness, the leading-twist contributions are controlled by relatively large parton configurations, 
with transverse sizes  $R\sim 1/Q_s$,  which undergo strong scattering and probe gluon saturation.
For {\it exclusive} dijets, this implies that both final jets have {\it semi-hard} transverse
momenta ($P_\perp\sim Q_s$) and that one of them is aligned with the photon.
The dominant contributions to the diffractive production of {\it hard} dijets ($P_\perp\gg Q_s$) rather come from 
three-jet final states, which are very asymmetric and will be referred to as 2+1 jets: two of the jets are hard, 
while the third one is semi-hard. We demonstrate that the leading-twist contributions to
both exclusive dijets and the diffractive production of 2+1 jets admit
transverse-momentum dependent (TMD) factorisation, in terms of quark and gluon diffractive TMD distribution
functions, for which we obtain explicit expressions from first principles. 
We show that the contribution of 2+1 jets to diffractive
SIDIS (semi-inclusive  deep inelastic scattering)
 takes the form of one step in the DGLAP evolution of the quark diffractive PDF. 
}

\begin{document}
\maketitle

\section{Introduction}
\label{sec:intro}

Multiparticle production in QCD collisions at very high energies is often a multi-scale problem,
that can naturally be addressed within the colour glass condensate (CGC) effective picture
\cite{Iancu:2002xk,Iancu:2003xm,Gelis:2010nm,Kovchegov:2012mbw}. For processes which also
involve hard scales, the CGC predictions must agree with the respective predictions of 
the collinear factorisation (specialised to high energies) \cite{Sterman:1993hfp,Collins:2011zzd}. 

The CGC framework is best suited
to deal with the ``semi-hard'' aspects of the collision, by which we mean the high gluon density
 in the target and its physical consequences in terms of gluon saturation and multiple scattering for
the partons in the projectile. As implicit in the above, we have in mind a ``dilute--dense collision'',
where the target is a dense system of partons, like a large nucleus, which is subject to saturation,
while the projectile (proton, virtual photon...) is comparatively dilute. And the ``semi-hard'' scale is,
of course, the gluon saturation momentum $Q_s$ in the target. Yet, such a problem may involve
additional transverse momentum scales, which are much harder than $Q_s$. Consider the example of
electron-nucleus ($eA$) deep inelastic scattering (DIS), which will be our general set-up throughout 
this work: in that case, the hard scales could be the virtuality $Q^2$ of the exchanged photon and/or
the transverse momenta (with a typical value $P_\perp$) of the particles (hadrons, jets) produced 
in the final state. Despite the presence of these hard scales, such a process can still be sensitive to 
the semi-hard physics of saturation, via correlations among the produced particles.

Several instructive examples of $eA$ and $pA$ processes leading to multiparticle production have been 
discussed in the seminal paper \cite{Dominguez:2011wm}, where the relation between the CGC approach 
and the collinear factorisation has been systematically studied (at leading order) for the first time.
The prototype example is the inclusive production of a pair of hard jets in the ``correlation limit'': 
their individual transverse momenta, of order $P_\perp=|\bk_1-\bk_2|/2$, are much larger than both 
$Q_s$ and their momentum imbalance $K_\perp=|\bk_1+\bk_2|$, hence the jets propagate nearly
back to back in the transverse plane. For such a process, Ref.~\cite{Dominguez:2011wm} demonstrated
that the cross-section computed in the CGC approach exhibits transverse-momentum dependent (TMD)
factorisation, a more exclusive form of collinear factorisation in which the momentum imbalance is
measured as well. (See also Refs.~\cite{Metz:2011wb,Dominguez:2011br,Iancu:2013dta,Dumitru:2015gaa,Kotko:2015ura,Marquet:2016cgx,vanHameren:2016ftb,Marquet:2017xwy,Albacete:2018ruq,Dumitru:2018kuw,Boussarie:2021ybe,Kotko:2017oxg}
for related studies and \cite{Iancu:2020mos,Caucal:2021ent,Taels:2022tza,Bergabo:2022tcu,Iancu:2022gpw,Caucal:2023nci,Caucal:2023fsf} for extensions to next-to-leading order (NLO) --- all within the CGC framework.)
Namely, the cross-section can be written as the product of a ``hard factor'' describing the formation
of the hard pair and its distribution in $P_\perp$, and a ``gluon TMD'' (transverse-momentum dependent
gluon distribution, a.k.a. ``unintegrated gluon distribution'') describing the scattering between that
hard pair and the target, and the resulting  $K_\perp$--dependence. The emergence of this form
of factorisation from the CGC effective theory not only establishes a bridge between this approach and
the more traditional approach of collinear factorisation, but also extends the latter, in that the gluon TMD is 
also predicted for semi-hard values of $K_\perp$, where saturation effects are important. 
Remarkably, this factorisation has been recently proven to hold after adding the NLO corrections \cite{Taels:2022tza,Caucal:2023nci,Caucal:2023fsf}. 


That said, the restriction to inclusive processes with hard scales strongly reduces the sensitivity
of the observables to the physics of saturation. For hard dijets in the correlation limit, the effects of
saturation could be visible in the deviation of the final dijets from the back-to-back configuration, 
as measured via azimuthal correlations \cite{Marquet:2007vb,Albacete:2010pg,Stasto:2011ru,Lappi:2012nh,Zheng:2014vka}.
In practice though, this is complicated by the fact that
the $K_\perp$--distribution exhibits a slowly decaying tail $\propto 1/K_\perp^2$ at large 
momenta $K_\perp\gg Q_s$,  as produced via hard scattering off the dilute part of the target gluon distribution.
Accordingly, the {\it typical} dijet events are characterised by a large momentum imbalance,
 which is insensitive to saturation.
The problem is further amplified by the ``Sudakov effect''  \cite{Mueller:2013wwa,Taels:2022tza,Caucal:2023nci}, 
i.e. by the recoil due to the final-state radiation, which brings a new, and comparatively large,
contribution to the dijet imbalance \cite{Zheng:2014vka}.

One can however enhance the sensitivity to saturation by looking at even more exclusive processes,
with diffraction being an obvious candidate. Indeed, it has been realised long time ago that the
leading twist contribution to the diffractive structure function $F_2^D$ at high $Q^2$ is controlled by saturation
\cite{Wusthoff:1997fz,GolecBiernat:1999qd,Hebecker:1997gp,Buchmuller:1998jv,Hautmann:1998xn,Hautmann:1999ui,
Hautmann:2000pw,Golec-Biernat:2001gyl}. 
This is best seen in the framework of the {\it colour dipole picture} for DIS at high energy --- the factorisation scheme
underlying the applications of the CGC effective theory to DIS,  which is known by now to next-to-leading
order accuracy \cite{Balitsky:2010ze,Balitsky:2012bs,Beuf:2016wdz,Beuf:2017bpd,Beuf:2021qqa,Beuf:2021srj,Beuf:2022ndu,
Beuf:2022kyp,Beuf:2024msh}. This picture is formulated in a Lorentz
frame in which the virtual photon is ultrarelativistic, so it is naturally a picture of {\it projectile evolution}:
all the partons partaking in the collision are first generated as fluctuations of the virtual photon.
At leading order, the photon splits into a quark-antiquark
($q\bar q$)  colour dipole, which then scatters off the gluons in the target. The  dipole 
scattering is strong --- meaning, 
saturation effects become important --- provided the dipole transverse size is large enough: $R\gtrsim 1/Q_s$. 
This size $R$ is controlled by the effective virtuality 
 $\bar Q^2\equiv \vartheta(1-\vartheta) Q^2$ of the $q\bar q$ fluctuation,  $R\sim 1/\bar Q$, so it
 also depends upon the longitudinal momentum fractions $\vartheta$ and $1-\vartheta$ of the quark and, 
 respectively, the antiquark. This simple argument shows that saturation effects in DIS may be important
 even at high $Q^2\gg Q_s^2$, provided the dipole fluctuation of the virtual photon is extremely asymmetric 
 in its longitudinal sharing: $\vartheta(1-\vartheta)\lesssim Q_s^2/ Q^2\ll 1$.  
 
%
 We are now prepared to explain the fundamental difference between diffractive and inclusive DIS at large
 $Q^2\gg Q_s^2$.  Diffraction proceeds via {\it elastic} scattering, so the cross-section is
 proportional to the {\it square} of the dipole scattering amplitude. This  favours
 the  asymmetric $q\bar q$  configurations with $ \vartheta(1-\vartheta) \sim Q_s^2/Q^2$, which are relatively large
 ($R\sim 1/Q_s$) and therefore suffer strong scattering. These configurations 
 yield the dominant  contribution to $F_2^D$ at high $Q^2$. They lead to asymmetric quark-antiquark
 final states,  in which one of the  quarks is very soft, 
while the other one (the ``aligned jet'') carries the quasi-totality of the photon longitudinal momentum.
Both quarks have semi-hard transverse momenta\footnote{One should not confuse transverse momenta
and virtualities: for the soft member of the pair, the virtuality is indeed small, of order $K_\perp^2$;
on the other hand, the aligned jet has a large virtuality of order $Q^2$, so it will likely materialise
like a genuine jet in the final state.} $K_\perp\sim Q_s$, as fixed by the 
Fourier transform from $R$ to $K_\perp$.   One sees that, 
despite the hardness of the photon, this elastic  process is fully controlled by saturation.


Inclusive DIS, on the other hand, is controlled by {\it inelastic}
collisions, so it is less sensitive to the strength of the scattering --- by the optical theorem, the cross-section
is proportional to the imaginary part of the dipole scattering amplitude ---, but more sensitive to the phase-space 
available to the $q\bar q$  fluctuations. Accordingly, the leading twist contribution to  the inclusive structure function  $F_2$
comes from relatively small dipoles ($R\ll 1/Q_s$), which scatter only weakly. These small dipoles materialise
as {\it hard} jets or hadrons ($P_\perp\gg Q_s$) in the final state.

The differences between the final states produced via elastic and respectively inelastic scattering also explain 
the different types of factorisations found for these processes at high $Q^2$ and/or high $P_\perp^2$. 
For hard {\it inclusive} dijets, TMD factorisation has been identified  ``at the dipole level''  \cite{Dominguez:2011wm}: 
the $q\bar q $ dipole is still viewed
as a fluctuation of the virtual photon, but the gluon exchanged in the $t$--channel is interpreted as a small--$x$
constituent of the target, with a transverse momentum  distribution given by the Weiszäcker-Williams (WW) unintegrated
gluon distribution.  



In the case of {\it elastic} scattering,  the ``aligned jet'' $q\bar q$
configurations produced at leading order could hardly qualify as ``dijets'':  both quarks
are semi-hard and one of them is also soft, with longitudinal momentum fraction
$\vartheta_1\sim Q_s^2/Q^2\ll 1$, and therefore very difficult to measure in practice. 
A better suited observable for such an asymmetric final state is {\it semi-inclusive DIS} (SIDIS),
in which one measures just the aligned jet. (This can
be observed as a jet or a leading hadron at forward rapidities  \cite{Iancu:2020jch}.) For this observable,
in Sect.~\ref{sec:qqbar} we shall establish TMD factorisation ``at the photon level'': instead of the virtual
photon decaying into a $q\bar q$ pair, we shall describe this process as the absorption of the 
photon by a quark constituent of the target. This looks like the  standard partonic picture of DIS that is
expected in the  leading twist (LT) approximation at high $Q^2$. As we shall see, the LT approximation
is indeed implicit in our treatment of asymmetric jets: it is equivalent to keeping just the first term in the expansion
in the small parameter $\vartheta_1\sim Q_s^2/Q^2\ll 1$. This softness condition $\vartheta_1\ll 1$ is also
what allows us to transfer this quark from the wavefunction 
of the virtual photon to that of the target.



The key ingredient of the TMD factorisation for diffractive SIDIS at leading order
 is the {\it quark diffractive TMD} (DTMD) \cite{Hatta:2022lzj}.
Physically, this describes the transverse momentum distribution of a quark constituent of the  Pomeron 
(the colourless fluctuation of the hadronic target that mediates elastic scattering). 
Within the framework of collinear factorisation\footnote{As a matter of fact, the quark and gluon diffractive TMD
were not explicitly mentioned in the literature on collinear factorisation. They have been first introduced in the
recent works~\cite{Iancu:2021rup,Hatta:2022lzj,Iancu:2022lcw}, following progress with understanding the 
TMD factorisation for diffractive jets in the colour dipole picture \cite{Iancu:2021rup}.}
 \cite{Trentadue:1993ka,Berera:1995fj,Collins:1997sr},  this would be treated
as a non-perturbative  quantity, but here it is explicitly determined by the dipole picture, and more precisely
by the physics of saturation: its calculation is controlled by large dipole sizes $R\sim 1/Q_s$ and
 would be ill defined in the absence of saturation. 
Our result for the quark DTMD  is not new --- it has recently been presented in Ref.~\cite{Hatta:2022lzj} ---, 
but the original discussion in~\cite{Hatta:2022lzj}  is  rather succinct and its relation to the colour
dipole picture is only sketchily explained. Our accent  in Sect.~\ref{sec:qqbar} will be on
pedagogy: diffractive SIDIS is the simplest set-up for demonstrating the emergence of diffractive 
TMD factorisation from the colour dipole picture. We shall use this set-up to demonstrate that 
the aligned jet configurations are selected by the condition of strong scattering and that they 
control the cross-section in the leading twist approximation.
The discussion of exclusive dijets  in Sect.~\ref{sec:qqbar}  will greatly simplify our subsequent treatment
of diffractive (2+1)--jets, which is technically more involved and also physically more interesting,
in that it gives one access to {\it hard diffractive dijets}.

Indeed, the dominant contribution to the diffractive production of {\it hard} dijets does not come 
from  the exclusive $q\bar q$ pairs, despite the fact that this process occurs at leading order in pQCD.
As already mentioned, the {\it typical} $q\bar q$ pairs produced via elastic scattering are {\it semi-hard},
with transverse momenta $P_\perp\sim Q_s$. Another way of looking at that, is by computing the contribution
of this exclusive process to  the cross-section for producing a pair of hard jets with $P_\perp\gg Q_s$: 
one finds a higher-twist contribution, which decreases like the power  $1/P_\perp^6$ 
\cite{Salazar:2019ncp,Iancu:2022lcw}
(to be compared with the power $1/P_\perp^4$ for {\it inclusive} dijets  \cite{Dominguez:2011wm}).
This strong suppression is of course related to the colour transparency 
of small colour dipoles ($R\sim 1/P_\perp\ll 1/Q_s$).
As observed in  Ref.~\cite{Iancu:2021rup,Iancu:2022lcw}, there exists a natural mechanism which avoids this suppression
and generates a leading-twist ($\propto 1/P_\perp^4$) contribution to the cross-section for hard diffractive dijets:
the simultaneous production of {\it 2+1 jets}. Namely, the hard dijets that we are primarily
interested in should be accompanied by (at least) one additional parton
radiated at a relatively large transverse distance $R\sim 1/Q_s$ from the other partons. 
This pattern ensures that the colour flow irrigates  a large transverse area, thus allowing for strong 
elastic scattering. 
For this to be possible, 
the third parton must be sufficiently soft, with longitudinal momentum fraction $\vartheta_1\sim Q_s^2/Q^2\ll 1$,
and also semi-hard, with transverse momentum $K_\perp\sim Q_s$. Clearly, 
this is very similar to the aligned jet configurations, except that now, two of the jets can be hard.

The contribution from 2+1 jets to the hard dijet cross-section is suppressed by a factor of $\alpha_s$
(formally, it is a part of the NLO corrections as computed in \cite{Boussarie:2014lxa,Boussarie:2016ogo,Fucilla:2022wcg}),
yet it represents the dominant pQCD contribution at sufficiently large transverse momenta $P_\perp\gg Q_s$, where
it dominates over the exclusive dijet production by a large factor $P_\perp^2/Q_s^2$. In fact, this is the
contribution that is implicitly accounted for in the collinear factorisation approach to diffractive jets (a leading twist
formalism) \cite{Hautmann:2002ff,Britzger:2018zvv,H1:2007jtx,Guzey:2016tek}.
Indeed, that approach is based on diffractive parton distribution functions (DPDFs), which describe
the distribution of the partons from the target that are involved in the diffractive scattering, while being inclusive
in the structure of the final state; hence, this approach implicitly allows for the production of additional partons, 
which are not measured.

Being of leading-twist order, the cross-section for diffractive 2+1 jets admits TMD factorisation:
it can be written as the product of a hard factor describing the hard dijet times a quark or gluon diffractive TMD,
representing unintegrated parton distributions of the Pomeron.  We have previously demonstrated this factorisation
for the case where the hard dijet is made with a quark and an antiquark, while the soft ``jet''
is a gluon~\cite{Iancu:2021rup,Iancu:2022lcw}.  In this paper, we shall extend this construction
to the case where the soft parton is a quark, or an antiquark. 
The general physics argument is very similar --- once again, we need to transfer  the soft parton from
the wavefunction of the virtual photon to that of the target --- but its mathematical implementation is quite different, 
notably because of the lack of symmetry of the final state (the hard dijet is now made with a quark and a gluon).
The two amplitudes contributing to this process (gluon emission by the quark and, respectively, by the antiquark)
describe different projectile evolutions and  have very different mathematical properties
(see Sect.~\ref{sec:2plus1} for details). So, it is a non-trivial
consistency check of our calculation that we manage to demonstrate TMD factorisation for all the three
contributions to the cross-section (direct gluon emissions by the quark and by the antiquark, and their interference).
Importantly, all these contributions involve the same expression for the quark TMD, which moreover coincides
with that previously identified  in Sect.~\ref{sec:qqbar}, in relation with diffractive SIDIS
(see Sect.~\ref{sec:tmd} and notably  \eqn{crosstotal} which exhibits the complete result).
This is the first explicit proof of the {\it universality} of the new
diffractive TMDs introduced in Refs.~\cite{Hatta:2022lzj,Iancu:2021rup,Iancu:2022lcw}.  

A further, equally non-trivial, test of our calculations comes from evaluating the contribution of 2+1 jets to diffractive 
SIDIS. By integrating out the kinematics of the soft jet and of one of two hard jets, one gets
a contribution to the cross-section for measuring the other hard jet.
 A priori, this is a part of the NLO corrections to diffractive SIDIS \cite{Fucilla:2023mkl}, 
yet, due to its kinematics, this contribution is quite special: it demonstrates the emergence of
the DGLAP evolution \cite{Gribov:1972ri,Altarelli:1977zs,Dokshitzer:1977sg} for the diffractive PDFs from the dipole picture
 (see Sect.~\ref{sec:SIDIS} for details).
Namely, this is the dominant leading-twist contribution to diffractive SIDIS for the case where the transverse momentum 
$P_\perp$ of the measured jet is {\it moderately} hard: $Q^2\gg P_\perp^2\gg Q_s^2$. 
Within the standard parton picture of the target,
one would expect such a contribution to be generated via a hard, DGLAP--like, splitting of a semi-hard parton 
constituent of the Pomeron. This is indeed the result that we obtain from ``integrating out''  the 2+1 jets. What is truly
remarkable about this result, is the way how the DGLAP splitting functions get reconstructed from gluon emissions in
dipole picture: we generate one step in the DGLAP evolution in the $t$--channel by summing over all possible
gluon emissions in the dipole picture in the $s$--channel. This construction strongly 
suggests that the compatibility between the CGG picture and the collinear factorisation (for such
multi-scale problems) persists beyond the leading-order approximation: the CGC approach appears to be
able to also accommodate the DGLAP evolution of the target (diffractive) PDFs.

Our analysis confirms that the colour dipole picture for the wavefunction of the virtual photon together with the CGC description
of the hadronic target provide a powerful framework for studies of diffractive photon-hadron interactions
from first principles. Besides global quantities like the diffractive structure function $F_2^D$, that has recently
been computed in this framework to NLO \cite{Beuf:2024msh}, this approach also gives us access to more exclusive
final states, like the diffractive production of single jets (or hadrons) \cite{Iancu:2020jch,Hatta:2022lzj}
 and that of dijets (or dihadrons) \cite{Iancu:2021rup,Iancu:2022lcw,Iancu:2023lel}  --- both currently known
to NLO \cite{Fucilla:2022wcg,Fucilla:2023mkl}.  The essential common denominator of these diffractive observables
is the fact that they are controlled by strong scattering in the vicinity of the black disk limit, hence they are sensitive
to unitarity corrections and {\it a fortiori} to gluon saturation. In fact, their calculation within the CGC effective theory
becomes possible only when the gluon distribution in the hadronic target is dense enough for the associated saturation momentum to be semi-hard, $Q_s^2\gg\Lambda^2$. This condition is easier to realise for a large nuclear target with 
mass number $A\gg 1$. And in any case it requires sufficiently high energies, or, for the specific 
case of diffraction, sufficiently large values for the rapidity gap $\YP$. 

The sensitivity of the diffractive observables to saturation
extends to large values of the transverse momentum scales in the problem (photon virtuality, dijet transverse momentum ...),
where the CGC formalism overlaps with the traditional collinear factorisation  and it is expected  to be consistent with it.
This opens the way for first principle calculations of the diffractive TMDs and the diffractive PDFs --- at least,
up to genuinely non-perturbative aspects, like the effects of confinement on the transverse inhomogeneity of the target.
It is furthermore promising in view of applications to phenomenology, notably to diffractive jet production in
DIS at the Electron Ion Collider  \cite{Accardi:2012qut,Aschenauer:2017jsk,AbdulKhalek:2022hcn} 
and in ultra-peripheral nucleus-nucleus collisions (UPCs) at the LHC  \cite{ATLAS:2017kwa,ATLAS:2022cbd,CMS:2020ekd,CMS:2022lbi}. 

This paper is organised as follows. Sect.~\ref{sec:qqbar} is devoted to exclusive quark-antiquark production and
its contribution to diffractive SIDIS. Using the light-cone wavefunction formalism, we successively construct the 
$q\bar q$ fluctuation of the virtual photon, the cross-section for elastic scattering, and the cross-section 
 for diffractive SIDIS. We show that the leading-twist contribution to diffractive SIDIS 
 comes from aligned jet configurations and that it exhibits TMD factorisation. This allows us to isolate
 the leading-order contribution to the quark diffractive TMD and relate it to the scattering amplitude of a colour
 dipole. The discussion in Sects.~\ref{sec:LTA} and \ref{sec:TMDSIDIS}  illustrates the importance of properly choosing the 
longitudinal momentum variables so that the TMD factorisation (and its physical interpretation as target evolution) becomes manifest. In Sect.~\ref{sec:MS} we demonstrate that the
quark DTMD is controlled by multiple scattering and gluon saturation. 

Starting with Sect.~\ref{sec:2plus1},
we address the main physical problem of interest for us here: the diffractive production of 2+1 jets, with the focus
on 2+1 configurations where the soft  jet is a quark. In Sect.~\ref{sec:2plus1} we construct the 
light-cone wavefunctions describing the $q\bar q g$ fluctuations of the virtual photon for the two channels
which contribute to the final state. We then deduce the cross-section, including the
interference term. In  Sect.~\ref{sec:tmd}, we show that the  (2+1)--jet cross-section previously constructed
admits TMD factorisation, with the same expression for the quark TMD as found in Sect.~\ref{sec:qqbar} 
for diffractive SIDIS.  In Sect.~\ref{sec:SIDIS}, we study the contributions from 2+1 jets with either
a soft quark, or a soft gluon, to diffractive SIDIS. This allows us reconstruct the DGLAP splitting functions for
the evolution of the quark DPDF from gluon emissions in the dipole picture. 

Sect.~\ref{sec:qPomeron} presents
a systematic, analytic and numerical, study of the structure and properties of the (quark and gluon)
diffractive TMDs and of the respective PDFs, as determined by the colour dipole picture supplemented with 
the DGLAP evolution.
The leading-order (or ``tree-level'') results for the DTMDs are computed in terms of the dipole
scattering amplitude, itself obtained from numerical solutions to the Balitsky-Kovchegov equation \cite{Balitsky:1995ub,Kovchegov:1999yj} with collinear improvement \cite{Iancu:2015vea,Ducloue:2019ezk}. 
Following the analysis in Sect.~\ref{sec:SIDIS}, 
the matching between the DTMDs and the DPDFs is performed  by treating the 
tree-level estimates for the former as {\it source terms} in the DGLAP equations for the 
latter (see Sect.~\ref{sec:DGLAP} for details). As a simple application, we use the DGLAP solutions
for the quark DPDF to estimate the diffractive structure function $F_2^{D(3)}$ (see Sect.~\ref{sec:f2d3}).

Sect.~\ref{sec:conc} contains a summary of our results together with open questions and perspectives.
All of our developments in the main text
focus on the case where the virtual photon has transverse polarisation, as this is the only one to contribute
to diffractive SIDIS at leading twist and also to diffractive 2+1 jets in the limit where the photon virtuality
is negligible, as in UPCs. Yet,  the case of a longitudinal photon is interesting as well (notably,
for the behaviour of $F_2^{D(3)}$ near $\beta=1$) and it is addressed
in some detail in App.~\ref{sec:long}, from which we quote the relevant results in the main text.
Other appendices present more technical details on the calculations.

\section{From the dipole picture to TMD factorisation for exclusive dijets}
\label{sec:qqbar}

As explained in the Introduction, our main purpose is to demonstrate the emergence of transverse-momentum dependent (TMD) factorisation from the colour dipole picture for coherent diffraction in deep inelastic scattering (DIS) at high energy.
In this section, we consider the simplest diffractive process: the exclusive production of a quark-antiquark ($q\bar q$) pair
by a virtual photon with relatively large virtuality, $Q^2\gg Q^2_s$, where $Q_s$ denotes the target saturation momentum.

In the dipole picture, this $q\bar q$  pair first appears as a colourless fluctuation
of the virtual photon, which is put on-shell via {\it elastic} scattering off the hadronic target (see below for details).
In what follows, we would like to show that the typical $q\bar q$ pairs produced in this way are {\it semi-hard},
i.e.~the final quarks carry transverse momenta of the order of $Q_s$, 
and also {\it very asymmetric}, in the sense that one of the two fermions ---  the ``aligned jet'' --- carries the quasi-totality
of the longitudinal momentum of the virtual photon. In turn, the prominence of the ``aligned jet'' configurations is a 
consequence of the elastic nature of the process, which favours strong scattering in the vicinity of the black disk limit.

These configurations are accurately described by the leading-twist approximation at high $Q^2$, which opens the
way for a partonic interpretation in terms of the {\it target} wavefunction: when the process is viewed in the target
infinite momentum frame (and in a suitable gauge, see below), the ``aligned jet'' is the quark from the target
which absorbs the photon. As we shall see,  this partonic interpretation naturally emerges from the dipole picture,
which leads to a {\it factorised} expression for the diffractive cross-section in the leading-twist approximation.
The target factor appearing in this expression is the quark diffractive,  transverse-momentum dependent,
distribution function --- the ``quark DTMD''.  This factorisation, together with an independent calculation for 
the quark DTMD, have recently been presented in \cite{Hatta:2022lzj}, but similar ideas have emerged long time
ago \cite{Wusthoff:1997fz,GolecBiernat:1999qd,Hebecker:1997gp,Buchmuller:1998jv,Hautmann:1998xn,Hautmann:1999ui,
Hautmann:2000pw,Golec-Biernat:2001gyl}.

\subsection{General discussion and kinematics}
\label{sec:kin}


The colour dipole picture is formulated in a specific frame --- the {\it dipole frame} --- where the virtual photon is 
ultrarelativistic and develops long-lived partonic fluctuations, which interact with the hadronic target.
To be specific, we choose the photon to be a right-mover with 4-momentum   $q^{\mu} = (q^+, -Q^2/2q^+,\bm{0}_{\perp})$ 
and $q^+\gg \sqrt{Q^2}$, while the nucleus is a left-mover  with 4-momentum $P_N^{\mu} = (0,P_N^-,\bm{0}_{\perp})$ 
per nucleon (we neglect the nucleon mass).  In this frame, a  {\it diffractive} process consists in the
elastic scattering between the partons in the photon wavefunction and the nuclear target. In particular, the
process is {\it coherent} when the target emerges unbroken from the collision. 

The elastic scattering involves the exchange of a colourless partonic fluctuation 
between the target and the  ``diffractive system'' (the ensemble of particles produced by the collision), 
usually referred to as the {\it Pomeron}.  To lowest order in pQCD, the Pomeron is just a two-gluon exchange,
but in general its structure can be more complicated. For instance, when the Pomeron has a relatively large 
extent $Y_{\mathbb P}$ in rapidity, such that $\alpha_s Y_{\mathbb P}\gtrsim 1$, then the gluons inside the
Pomeron can multiply via the high-energy evolution, possibly leading to a dense gluon system. The high-density
effects are expected to be enhanced when the target is a large nucleus with mass number $A\gg 1$.
The distinguished experimental signature of a diffractive process is the existence of a pseudo-rapidity gap in the final state
(between the diffractive system and the outgoing target), that is roughly equal with $Y_{\mathbb P}$.

Another important feature of coherent diffraction is the fact  
that the transverse momentum  $\Delta_\perp$ transferred by the target to the diffractive system is very small and
can be neglected for most purposes. Indeed, this is solely determined by the target inhomogeneity in the
transverse plane, so for a large nucleus it can be estimated as
 $\Delta_\perp\sim 1/R_A$, with $R_A \approx A^{1/3} R_{N}$  and $R_N\simeq 1.1$~fm (the radius of a nucleon).
 With $A\approx 200$ (Pb or Au), one finds $\Delta_\perp\approx 30$~MeV, which is indeed 
much smaller than the typical transverse momenta of the produced particles (see below).

When discussing diffraction, it is customary to introduce a few kinematical invariants, which apply for a generic final state
and physically represent fractions of the target longitudinal momentum $P_N^-$ transferred to the diffractive system
($x_{\mathbb{P}}$), or to the struck quark ($\xbj$). These variables are defined as
\begin{align}\label{xP}
x_{\mathbb{P}}\equiv \,\frac{Q^2+M_{X}^2+\Delta_\perp^2}{2q\cdot P_N}\,,\qquad 
\beta\equiv \,\frac{Q^2}{Q^2+M_{X}^2+\Delta_\perp^2},\,\qquad
\xbj=\, \frac{Q^2}{2q\cdot P_N}\,=\beta x_{\mathbb{P}} \,,
\end{align}
where $M_X$ denotes the invariant mass of the diffractive system.
The rapidity gap is then computed as  $Y_{\mathbb P}=\ln(1/x_{\mathbb{P}})$. 
The complimentary rapidity interval $\ln(1/\beta) = Y-Y_{\mathbb P}$, with $Y\equiv \ln(1/\xbj)$,
is occupied by the partonic fluctuation of the photon.
The relevant value of the target saturation momentum is $Q_s(Y_{\mathbb P})$ and
includes the high-energy evolution of the Pomeron. As mentioned, in what follows we shall neglect
$\Delta_\perp$ in these (and all the subsequent) expressions. This is formally equivalent to assuming the
target to be homogeneous in the transverse plane.

\begin{figure}
	\begin{center}
	\includegraphics[align=c,width=0.4\textwidth]{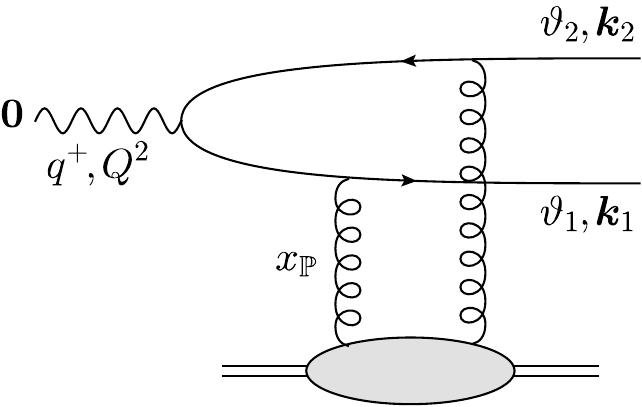}
	\hspace*{0.08\textwidth}
		\includegraphics[align=c,width=0.4\textwidth]{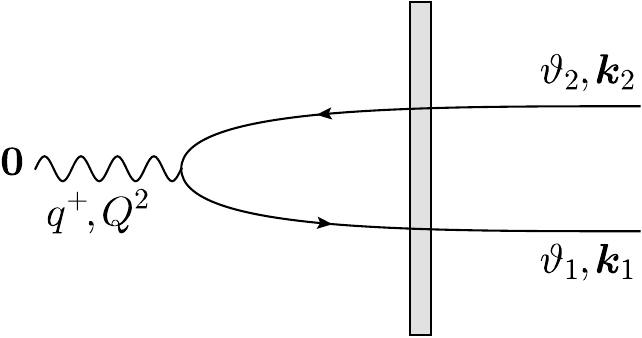}
	\end{center}
	\caption{\small Exclusive dijet production in DIS off a nucleus, as described in the dipole picture. Left panel: one of the diagrams in the two-gluon exchange approximation. Right panel: the scattering off the nucleus is represented by a shockwave. It implicitly contains all diagrams with an arbitrary number of gluon exchanges, which can be hooked to the projectile system in all possible ways. In the present work it is assumed that the overall exchange, independently of its complexity, is always colourless.}
\label{fig:2jets_2g_SW}
\end{figure}

 
As announced, we start our analysis with the simplest diffractive process in DIS:
the exclusive production of a quark-antiquark pair (``dijet'') and its contribution to the total diffractive cross-section.
The amplitude and the kinematics for this process are illustrated in Fig.~\ref{fig:2jets_2g_SW}, where the elastic scattering off the nucleus is either drawn in the two-gluon exchange approximation, or depicted by a shockwave.
 (This shockwave picture is indeed appropriate in the dipole frame, 
  since the lifetime of the projectile system is much larger than the longitudinal
  extent of the Lorentz contracted nucleus.) In order to establish contact with the collinear factorisation,
we shall consider the case of a highly virtual photon, with $Q^2\gg Q_s^2(Y_{\mathbb P})$. As
we shall see, the {\it typical} values of the transverse momenta $k_{1\perp}$ and $k_{2\perp}$ of
the produced ``jets'' --- those which control the diffractive structure function --- are of order $Q_s$;
that is, they are considerably lower than the photon virtuality, but at the same time much larger
than the transverse momentum $\Delta_\perp$ transferred by the target (and which controls
the dijet imbalance, as we have seen: $\Delta_\perp=|\bk_1+\bk_2|$).
Hence, as anticipated, we shall ignore the dijet imbalance in what follows; that is, we shall
assume the final jets to propagate back-to-back in the transverse plane: $\bk_2 = -\bk_1\equiv \bK$.

%
%

The contribution of this process to the total diffractive cross section is obtained after integrating over the kinematics of the two jets
 for a fixed value of the rapidity gap $Y_{\mathbb P}$. Within collinear factorisation, this is conveniently written as
  \begin{align}
  \label{F2D}
  \frac{\rmd \sigma_D^{\gamma A
  \rightarrow  X A}}
  {\rmd Y_{\mathbb P}} 
  = \frac{4\pi^2 \alpha_{em}}{Q^2}\, 
 x_{\mathbb{P}} F_2^{D(3)}(\xbj, x_{\mathbb{P}}, Q^2)\,,
\end{align} 
where $X$ represents the ``diffractive system'' (for exclusive dijets, $X=q\bar q $), the factor
$(4\pi^2 \alpha_{em})/{Q^2}$ is the cross-section for photon absorption by a point-like particle
with unit electric charge, 
and $F_2^{D(3)}$ is the diffractive structure function integrated over the momentum $t\equiv \Delta_\perp^2$ 
transferred by the target\footnote{In the approximations of interest, where we neglect $\Delta_\perp$
in the transverse momentum imbalance, the integrand of \eqn{F2D3def} is proportional to $\delta(t)$.}:
\beq\label{F2D3def}
 x_{\mathbb{P}} F_2^{D(3)}(\xbj, x_{\mathbb{P}}, Q^2)\,\equiv\int\rmd t
 \,  \frac{\rmd F_2^{D}}
  {\rmd Y_{\mathbb P}\,\rmd t} (\xbj, x_{\mathbb{P}}, Q^2, t)\,.\eeq
In the leading twist (LT) approximation at large $Q^2$ --- the main approximation
underlying collinear factorisation --- $F_2^{D(3)}$  is only weakly dependent upon $Q^2$ (via the DGLAP evolution)
and can be expressed as a linear combination of diffractive parton distribution functions (DPDFs), 
with coefficient functions that can be computed within perturbative QCD.
(These  coefficient functions are the same as for inclusive DIS; see e.g.~\cite{Golec-Biernat:2001gyl} and references therein.)
To leading order (LO) in pQCD, this combination involves the quark and antiquark DPDFs weighted by their electric charge
squared and summed over the active flavours\footnote{Notice that our conventions are slightly different than those used
in \cite{Golec-Biernat:2001gyl}: our definitions for the DPDFs include an additional factor of 
$x_{\mathbb{P}}$ compared to those in \cite{Golec-Biernat:2001gyl}. This becomes clear e.g.~by comparing our 
\eqref{xqD} to Eq.~(24) in  \cite{Golec-Biernat:2001gyl}.}:
\begin{align}
\label{xqD}
x_{\mathbb{P}} F_2^{D(3)}(\xbj, x_{\mathbb{P}}, Q^2)=
\sum_f e_f^2\big[\beta q^D_f(\beta, x_{\mathbb{P}}, Q^2)+\beta \bar q^D_f(\beta, x_{\mathbb{P}}, Q^2)\big]
\qquad\mbox{(LT \& LO)}.
\end{align} 
The diffractive DPDFs can be interpreted as the number of partons inside the Pomeron. 
And indeed, the quark and antiquark DPDFs appearing in the r.h.s.~of \eqn{xqD} are 
expressed in terms of the longitudinal momentum fraction $\beta$
of the measured (anti)quark {\it with respect to the Pomeron}.

Within collinear factorisation, the diffractive PDFs are treated as non-perturbative quantities, 
that can be parameterised and fitted from the data. Within the dipole picture on the other hand,
they can be explicitly computed, via mostly perturbative calculations
\cite{Wusthoff:1997fz,GolecBiernat:1999qd,Hebecker:1997gp,Buchmuller:1998jv,Hautmann:1998xn,Hautmann:1999ui,
Hautmann:2000pw,Golec-Biernat:2001gyl}. 
For instance, the leading order contribution to $ F_2^{D(3)}$, hence to the quark DPDF, 
emerges from the dipole picture calculation of exclusive dijet production (see below).
%


Our main purpose in what follows is to demonstrate that results similar to \eqref{F2D}--\eqref{xqD} can be derived
within the dipole picture already before (fully) integrating over the kinematics of the final state. In particular,
for the $q\bar q$ final state to be discussed in this section, we shall show that one can write an ``unintegrated'' version of 
Eqs.~\eqref{F2D}--\eqref{xqD},
in which the (common) transverse momentum $\bK$ of the two final jets is measured as well:
\begin{align}
  \label{qqTMDfact}
  \frac{\rmd \sigma_D^{\gamma A
  \rightarrow  q\bar q A}}
  {\rmd \ln(1/\beta)\,\rmd^{2}\bm{K}} 
  = \frac{4\pi^2 \alpha_{em}}{Q^2}\ 2\sum_f e_f^2\,
  \frac{\rmd \, \beta q^D_f
	(\beta, x_{\mathbb{P}}, K_{\perp}^2)}{\rmd^2 \bK}\,\qquad\mbox{(LT \& LO)}.
\end{align} 
Such a process, in which one measures a single jet (or hadron) in the final state of DIS diffraction, is known as
diffractive SIDIS (semi-inclusive deep inelastic scattering). For the leading-order process at hand, where the final
quarks have equal but opposite transverse momenta, the difference between dijet production and SIDIS is
rather trivial: it amounts to a $\delta$-function $\delta^{(2)}(\bk_1+\bk_2)$ that has been integrated over
in obtaining \eqn{qqTMDfact}. The second factor in the r.h.s.~of this equation plays the role of a {\it quark diffractive TMD}
(transverse-momentum dependent  distribution function). Our subsequent 
derivation of \eqn{qqTMDfact} from the dipole picture will support its interpretation as the unintegrated quark
distribution of the Pomeron (see the right panel in Fig.~\ref{fig:2jets_TMD} for a pictorial interpretation). 
As compared to \eqn{xqD}, in \eqref{qqTMDfact} we have included just the quark DTMD and multiplied
the result by a factor of two. Indeed, either the quark, or the antiquark, can be measured in the final state, and the 
respective contributions are identical, since the Pomeron is invariant under charge conjugation.

Unlike the more familiar relations  \eqref{F2D}--\eqref{xqD} between the diffractive cross-section and the quark DPDF, 
its ``unintegrated''  version in \eqref{qqTMDfact} has never been discussed to our knowledge in the context of collinear factorisation. Yet, results which are suggestive of \eqref{qqTMDfact} have emerged in early discussions of DIS 
diffraction in the dipole picture \cite{Wusthoff:1997fz,GolecBiernat:1999qd,Hebecker:1997gp,Buchmuller:1998jv,Hautmann:1998xn,Hautmann:1999ui,
Hautmann:2000pw,Golec-Biernat:2001gyl}, although the connection to TMD factorisation has not been mentioned
at that time. Very recently, this connection has been established in 
Ref.~\cite{Hatta:2022lzj} (see especially Appendix B there),  which is closer in spirit to our present approach.

As we shall see, the leading-twist approximation to diffraction is quite subtle, in that it includes a
whole series of (formally, power-suppressed) corrections associated with multiple scattering and which are often
referred to in the literature as ``higher twist effects''.  To avoid any confusion on this point, it is important to distinguish 
between two types of higher-twist corrections, which appear when computing diffraction within the dipole picture:
the genuine ``DIS twists'', which are proportional to powers of $K_\perp^2/Q^2$, and  the ``saturation twists'', which 
scale like powers of $Q_s^2/K_\perp^2$.  The standard ``leading-twist approximation'' of the collinear factorisation 
is the lowest order term in the series of DIS twists, so its validity requires $K_\perp^2\ll Q^2$.
Physically, it describes a situation 
where  the virtual photon is absorbed by a single quark in the target wavefunction\footnote{Of course,
this physical interpretation becomes manifest only when working in the ``target picture'', that is, in a
frame and gauge where the usual partonic picture for the target makes sense.}. Yet, this truncation
of the DIS twists series imposes no  constraint on the  saturation twists, which describe 
multiple scattering between the $q\bar q$ dipole and the target gluon distribution. 
As we shall see, the leading-twist contribution  to diffractive SIDIS 
 includes ``saturation twists'' to all orders  and is even dominated by strong scattering.
 In more physical terms, it is controlled by the physics of gluon saturation.

\subsection{The cross-section for exclusive dijets}
\label{sec:hard2jets}

To compute diffraction in the colour dipole picture, we shall employ the light-cone wavefunction (LCWF) formalism,
where the particles produced in the final state start as partonic fluctuations of the virtual photon: these fluctuations
develop already long before the collision and are eventually put on-shell by their scattering with the nuclear target.
This formalism is developed in the dipole frame, where the virtual photon (the ``projectile'') and its partonic 
fluctuations are relativistic right movers, and in the projectile light-cone gauge $A^+=0$. This particular gauge
choice is beneficial for at least two reasons: it permits a partonic (Fock--state) description for the wavefunction of the projectile and it allows for the eikonal coupling between the partons in the projectile and the ``large'' component $A^-_a$
of the colour field generated by the left-moving target. We are using the CGC 
description for the target wavefunction, in which the information about the gluon distribution is encoded in the 
correlations 
of the classical, random, fields $A^-_a$ \cite{Iancu:2002xk,Iancu:2003xm,Gelis:2010nm,Kovchegov:2012mbw}.

The general strategy for  obtaining dijet cross sections in the CGC framework is well-known, see e.g.~\cite{Dominguez:2011wm} for a variety of processes computed at leading order. For definiteness, we shall 
follow the approach in \cite{Iancu:2022gpw}, which considers both dijet and trijet 
production in DIS off a large nucleus. This approach has been 
originally developed for inclusive production, but its extension to diffractive processes is straightforward
(see also  \cite{Iancu:2021rup,Iancu:2022lcw,Beuf:2022kyp} for related work).
Our discussion will be rather succinct: we will merely sketch the crucial steps and the relevant results, while
deferring the details to the literature.

\begin{figure}
	\begin{center}
		\includegraphics[align=c,width=0.42\textwidth]{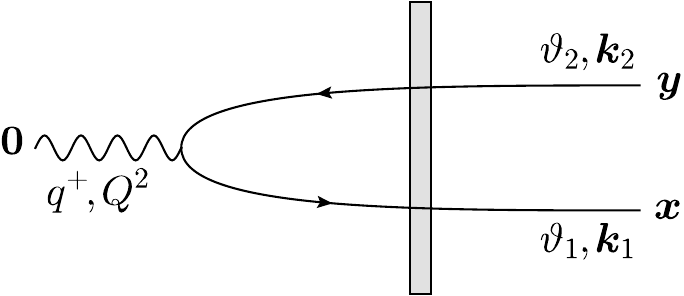}
		\hspace*{0.08\textwidth}
		\includegraphics[align=c,width=0.42\textwidth]{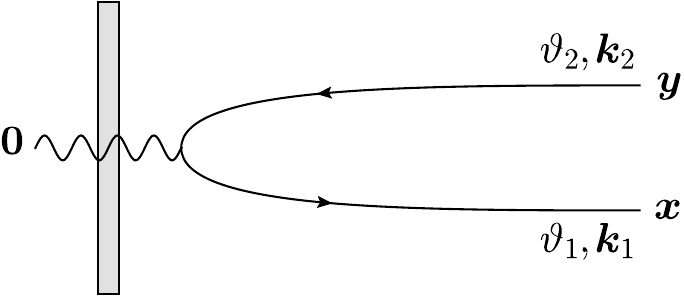}
	\end{center}
	\caption{\small The probability amplitude for forward dijet production in DIS off a nuclear shockwave. Left panel: The $q\bar{q}$ pair interacts with the nucleus. Right panel: The no-scattering limit.}
\label{fig:2jets}
\end{figure}

In this section, we consider exclusive dijet production, so we focus on the $q\bar{q}$ component of the LCWF 
of the virtual photon,  as depicted in Fig.~\ref{fig:2jets}. We take the quarks to be massless, so in practice 
we limit ourselves to the 3 light flavours $u,\,d,$ and $s$ (the sum over flavours will often be implicit in what follows). 
We present explicit calculations for the case of a virtual photon with transverse polarisation, but we shall later add
the corresponding  results for a longitudinal photon.

The general strategy can be summarised as follows  \cite{Iancu:2018hwa,Iancu:2020mos,Iancu:2022gpw}.
We first compute the $q\bar q$ component in the absence of scattering and in momentum space,
by using the Feynman rules of the light-cone perturbation theory (see e.g.~\cite{Beuf:2016wdz,Beuf:2017bpd,Iancu:2018hwa}).
Then we make a Fourier transform to the transverse coordinate representation and introduce the effects of
the collision. The coordinate representation is better suited for that purpose since the transverse coordinates 
of the partons from the projectile are not modified by their high-energy scattering in the eikonal approximation.
Finally, we return to momentum space, in order to specify the (longitudinal and transverse) momenta of the final
quark and antiquark, assumed to be on-shell. 

The $q\bar q$ component  of the  LCWF reads (in momentum space and in the absence of any collision)
\begin{align}
	\label{qqmom}
	\big|\gamma_{\scriptscriptstyle T}^{i}\big\rangle_{q\bar{q}} 
	=\, \delta_{\alpha\beta}
	\int_{0}^{1} & \rmd
	\vartheta_1 \rmd \vartheta_2 \,
	\delta(1 - \vartheta_1 - \vartheta_2)
	\int \rmd^{2}\bm{k}_1  
	\rmd^{2}\bm{k}_{2}\,
	\delta^{(2)}(\bm{k}_1 +\bm{k}_2)
	\nn 
	& \times
	\psi^i_{\lambda_{1}\lambda_{2}}(\vartheta_1,\bk_1)\,
	\big|{q}_{\lambda_{1}}^{\alpha}(\vartheta_1, \bk_1)\,
	\bar{q}_{\lambda_{2}}^{\beta}(\vartheta_2, \bk_2)
	\big\rangle,
	\end{align}
with $i=1,2$ characterizing the photon polarization and  $\alpha, \beta$ being color indices in the fundamental representation. $\lambda_1$, $\bk_1$ and $\vartheta_1\equiv k_1^+/q^+$ 
are the helicity, transverse momentum and longitudinal fraction (w.r.t.~the virtual photon) of the quark, and likewise for the antiquark. It is understood that there is a summation over the repeated indices $\alpha, \beta$ and $\lambda_1, \lambda_2$.
The $\delta$--function in colour space,  $\delta_{\alpha\beta}$, shows that the $q\bar q$ pair forms a colour dipole.
 The $q\bar{q}$ amplitude appearing in Eq.~\eqref{qqmom} is given by
\begin{align}
	\label{psiqq}
	\psi^i_{\lambda_1 \lambda_{2}} (\vartheta_1, \bk_1) =
	\frac{\delta_{\lambda_1\lambda_2}}{2\sqrt{{2 q^+}}}
	\frac{ee_{f}}{(2\pi)^3}\,
	\frac{1}{\vartheta_1(1-\vartheta_1)}\,
	\frac{\varphi^{ij}(\vartheta_1,\lambda_1)\,k_1^{j}}
	{E_{q}+E_{\bar{q}}-E_{\gamma}},
\end{align}
where $e$ is the QED coupling, $e_f$ the fractional electric charge of the quark with flavor $f$. The function
\begin{align}
	\label{phidef}
	\varphi^{ij}(\vartheta,\lambda)
	\equiv
	(2\vartheta-1)\delta^{ij}
	+ 2 i \lambda \varepsilon^{ij}
\end{align}
with $\varepsilon^{ij}$ the Levi-Civita symbol in two dimensions, encodes the helicity structure of the photon splitting vertex, which is also proportional to the relative transverse velocity of the two fermions, namely (omitting an overall factor $1/q^+$)
\begin{align}
	\label{vrelqqbar}
	\frac{\bk_1}{\vartheta_1}- 
	\frac{-\bk_1}{1-\vartheta_1}
	=
	\frac{\bk_1}{\vartheta_1(1-\vartheta_1)}.
\end{align}
The light-cone (LC) energies and their combination appearing in the denominator read
\begin{align}
	\label{EDsqq}
	E_q= \frac{k_{1\perp}^2}{2 \vartheta_1 q^+},
	\quad
	E_{\bar{q}}= \frac{k_{1\perp}^2}{2 (1-\vartheta_1) q^+},
	\quad
	E_\gamma= -\frac{Q^2}{2 q^+},
	\quad
	E_{q}+E_{\bar{q}}-E_{\gamma} = \frac{k_{1\perp}^2 +\bar{Q}^2}{2 \vartheta_1 (1-\vartheta_1) q^+},
\end{align}
where we have defined $\bar{Q}^2 \equiv \vartheta_1 (1-\vartheta_1) Q^2$. Substituting the above energy denominator into Eq.~\eqref{psiqq} we trivially find
\begin{align}
	\label{psiqqnew}
	\psi^i_{\lambda_1 \lambda_{2}} (\vartheta_1, \bk_1) =
	\delta_{\lambda_1\lambda_2}\,
	\sqrt{\frac{q^+}{2}}\,
	\frac{ee_{f}}{(2\pi)^3}\,
	\frac{\varphi^{ij}(\vartheta_1,\lambda_1)\,k_1^{j}}
	{k_{1\perp}^2 +\bar{Q}^2}.
\end{align}
This probability amplitude can be transformed to coordinate space via the replacement
\begin{align}
	\label{qqFT}
	\frac{k_1^j}{k_{1\perp}^2 +\bar{Q}^2}
	\to
	\int \frac{\dif^2\bk_1}{(2\pi)^2}\,
	e^{i \bk_1 \cdot \bR}\,
	\frac{k_1^j}{k_{1\perp}^2 +\bar{Q}^2} = 
	\frac{i}{2 \pi}\, \frac{R^j}{R}\,
	\bar{Q} K_1(\bar{Q} R),
\end{align}
where $\bR \equiv \bx- \by$, with $\bx$ and $\by$ the transverse coordinates of the quark and the antiquark respectively. 
The modified Bessel function  $K_1$ 
 vanishes exponentially for dipoles sizes $R \equiv |\bR|$ much larger than $1/\bar{Q}$. This shows that
 the typical value for the dipole size is $R\sim 1/\bar Q$. Importantly, this depends not only upon the virtuality
of the incoming photon, but also upon the longitudinal momentum sharing by the two fermions.

As already explained, the excursion through the coordinate space is useful for adding the effects of
the interaction with the nuclear target (a Lorentz contracted shockwave): as a result of this interaction,
each partonic component of the projectile acquires a colour rotation implemented by a suitable Wilson line. 
For the $q\bar q$ component, we can write
\beq
\hat S	\big|{q}_{\lambda_{1}}^{\alpha}(\vartheta_1, \bx)\,
	\bar{q}_{\lambda_{2}}^{\beta}(\vartheta_2, \by)
	\big\rangle\,=\,
\left(V(\bm{x})\,V^{\dagger}(\bm{y})\right)_{\alpha\beta}\,
\big|{q}_{\lambda_{1}}^{\alpha}(\vartheta_1, \bx)\,
	\bar{q}_{\lambda_{2}}^{\beta}(\vartheta_2, \by)
	\big\rangle\eeq
where $V(\bm{x})$ (for the quark) and $V^{\dagger}(\bm{y})$ (for the antiquark)
 are Wilson lines in the fundamental representation of the colour group, e.g.
 \begin{align}
	\label{wilson}
	V(\bm{x})={\rm T}
	\exp\left[i g \int \rmd x^{+} 
	t^{a} A^{-}_a(x^{+},\bm{x})\right],
\end{align}
where the  symbol T stands for LC time ($x^+$) ordering. Hence, after changing to the coordinate representation
and adding the effects of the collision, the colour structure of the LCWF \eqref{qqmom} gets modified as follows:
\begin{align}
	\label{scat}	
	\delta_{\alpha \beta}
	\rightarrow
	\big[V(\bx) V^\dagger(\by) - \mathds{1} \big]_{\alpha \beta},	
\end{align}
where we have also subtracted the no-scattering limit, which corresponds to the diagram shown in
the right panel of Fig.~\ref{fig:2jets}. 
More precisely, the replacement in \eqn{scat} leads to the {\it inclusive} production of a dijet. In order to have a
{\it diffractive} process to the order of accuracy, we must ensure that the colour dipole scatter elastically 
off the nuclear target, i.e.~it remains a colour singlet in the final state. This is achieved by the following projection
\begin{align}
	\label{scatdiff}
	\big[V(\bx) V^\dagger(\by) - \mathds{1} \big]_{\alpha \beta}
	\rightarrow	
	\left\{ \frac{1}{N_c}\,
	{\rm tr} \big[V(\bx) V^\dagger(\by)\big] - 1 \right\} 
	\delta_{\alpha \beta}
	\equiv [S(\bx,\by) - 1]\, \delta_{\alpha \beta}
	\equiv -T(\bx,\by)\, \delta_{\alpha \beta},
\end{align} 
where $S(\bx,\by)$ is the dipole $S$-matrix for elastic scattering and $T(\bx,\by)$ is the corresponding $T$-matrix. 

When calculating the cross section, one encounters the product $T(\bx,\by) T(\bar{\by},\bar{\bx})$, with $(\bar{\bx},\bar{\by})$ the dipole coordinates in the complex conjugate amplitude (CCA). In the CGC framework this product must be averaged over all possible target field configurations, with a weight function which takes into account the BK/JIMWLK evolution
\cite{Balitsky:1995ub,Kovchegov:1999yj,JalilianMarian:1997jx,JalilianMarian:1997gr,Kovner:2000pt,Weigert:2000gi,Iancu:2000hn,Iancu:2001ad,Ferreiro:2001qy}
of the gluon distribution up to the rapidity scale of interest (see below). Still for the purpose of the elastic scattering,
one must keep only the factorised piece in the average of the product, that is
\begin{align}
	\label{factorize}
	\left \langle 
	T(\bm{x},\bm{y}) T(\bar{\bm{y}},\bar{\bm{x}}) 
	\right \rangle
	\to
	\left \langle 
	T(\bm{x},\bm{y}) 
	\right \rangle 
	\left \langle
	T(\bar{\bm{y}},\bar{\bm{x}}) 
	\right \rangle.
\end{align}
Indeed, the colour flow on the target side must close separately in the direct ampltiude (DA) and in the CCA,
 as a necessary condition to have a coherent process\footnote{The non-factorised piece  $\left \langle 
	T(\bm{x},\bm{y}) T(\bar{\bm{y}},\bar{\bm{x}}) 
	\right \rangle - \left \langle 	T(\bm{x},\bm{y}) 	\right \rangle \left \langle
	T(\bar{\bm{y}},\bar{\bm{x}}) \right \rangle$ is suppressed at large $N_c$ (at least, within the present framework)
	and it is generally assumed to describe ``incoherent diffraction'', i.e.~a process where the scattering is elastic
	on the projectile side, but inelastic on the target side \cite{Marquet:2010cf,Mantysaari:2019hkq,Rodriguez-Aguilar:2023ihz}.}.
	For simplicity,  we will also assume that the nucleus is homogeneous and therefore any target average can depend only on the differences between the various transverse coordinates. Thus, the {\it average}
	dipole scattering amplitude takes the form $\left \langle T(\bm{x},\bm{y}) \right \rangle = \mcal{T}(\bR)$, with
	$\bR=\bx-\by$.
	(Note that we shall systematically use calligraphic notations for the averaged quantities.)
	All in all, when calculating coherent diffraction with a homogeneous nucleus, the eikonal scattering is taken into account by simply letting
\begin{align}
	\label{deltat}
	\delta_{\alpha\beta} \to 
	-\mcal{T}(\bR)\,
	\delta_{\alpha\beta},
\end{align}
within the coordinate-space version of the LCWF for the $q\bar q$ component. After this step,
Eq.~\eqref{qqFT} gets replaced by 
\begin{align}
	\label{qqFTinv}
	\frac{i}{2 \pi}\, \frac{R^j}{R}\,
	\bar{Q} K_1(\bar{Q} R) 
	&\to 
	-\frac{i}{2 \pi}\, \frac{R^j}{R}\,
	\bar{Q} K_1(\bar{Q} R)
	\mcal{T}(\bR)
	\to 
	-\frac{i}{2\pi}
	\int \dif^2 \bR\, e^{ - i \bk_1 \cdot \bR}\,
	\frac{R^j}{R}\,
	\bar{Q} K_1(\bar{Q} R)
	\mcal{T}(\bR) 
	\nn
	& = -\frac{k_1^j}{k_{1\perp}}
	\int \dif R\, R\, J_1(k_{1\perp} R)\,
	\bar{Q} K_1(\bar{Q} R)\,
	\mcal{T}(R),
\end{align}
where we have also performed the inverse Fourier transform to transverse momentum space and, in doing that,  
we assumed target isotropy, for simplicity: $ \mcal{T}(\bR)= \mcal{T}(R)$. 

To summarise, the generalisation of the probability amplitude in Eq.~\eqref{psiqqnew} which includes the
effects of the collision with the nuclear target reads as follows
\begin{align}
	\label{psiqqdiff}
	\psi^{i,D}_{\lambda_1 \lambda_{2}} (\vartheta_1, \bk_1) =
	- \delta_{\lambda_1\lambda_2}\,
	\sqrt{\frac{q^+}{2}}\,
	\frac{ee_{f}}{(2\pi)^3}\,
	\varphi^{ij}(\vartheta_1,\lambda_1)\,
	\frac{k_1^{j}} {k_{1\perp}}\,
	\frac{\mcal{Q}_{T}(\bar{Q},k_{1\perp})}{\bar{Q}},
\end{align}
where we have introduced the dimensionless, scalar function
\begin{align}
\label{QP}
	\mcal{Q}_{T}(\bar{Q},K_{\perp}, Y_{\mathbb P} ) \equiv
	\bar{Q}^2 \int \dif R\, R\, J_1(K_{\perp} R)\,
	K_1(\bar{Q} R)\,
	\mcal{T}(R, Y_{\mathbb P}).
\end{align} 
Strictly speaking, the momentum $\bk_1$ of the quark after the scattering, as appearing in Eq.~\eqref{psiqqdiff}, is 
different from the one before the scattering in Eq.~\eqref{psiqqnew}, even though we have used the same notation.
But it is important to notice that the $\delta$--function imposing $\bk_1+\bk_2=0$ still holds, like in \eqn{qqmom}.
This is due to the elastic nature of the scattering (which enforced the factorisation of the average scattering amplitude,
cf.~Eq.~\eqref{factorize}) and also to the assumed homogeneity of the target (which is tantamount to our original
assumption that $\Delta_\perp\sim 1/R_A\sim\ 30$~MeV is truly negligible compared to $k_{1\perp}$
and  $k_{2\perp}$).

The cross section for exclusive dijet production is finally determined by calculating the number of quarks and antiquarks in the final state with the desired longitudinal and transverse momenta. For the precise normalization of the respective number operators, as well as of the states appearing in Eq.~\eqref{qqmom}, we follow \cite{Iancu:2018hwa} to obtain
\begin{align}
	\label{crossqq}
\frac{\rmd\sigma_{\scriptscriptstyle}^{\gamma_{\scriptscriptstyle T}^* A\rightarrow q\bar q A}}
	{\rmd \vartheta_1 \rmd \vartheta_2\, 
	\rmd^{2}\bm{k}_{1}
  \rmd^{2}\bm{k}_{2}}
   = \frac{(2\pi)^3 S_{\perp} N_c}{2q^+}\,
   \delta(1-\vartheta_1-\vartheta_2)\,
   \delta^{(2)}(\bk_1+\bk_2) 
   \sum_{i,\lambda_1,\lambda_2}
   \big| \psi_{\lambda_1\lambda_2}^{i, D} \big|^2,
\end{align}
where $S_{\perp}$ is the transverse are of the nucleus. The above is of course proportional to the square of the probability amplitude in Eq.~\eqref{psiqqdiff} which is easily calculated. By further doing the trivial integration over the antiquark using the $\delta$-functions and summing over the photon transverse polarisations, over the quark helicities, and over all active flavors, we arrive at  	
\begin{align}
	\label{crossqqnew}
	\frac{\rmd\sigma_{\scriptscriptstyle}^{\gamma_{\scriptscriptstyle T}^* A\rightarrow q\bar q A}}
	{\rmd \vartheta_1\,
	\rmd^{2}\bK} = 
	\frac{S_{\perp} \alpha_{\rm em} N_c}{2\pi^2} \left( \sum e_f^2\right)
	\left[\vartheta_1^2 +(1-\vartheta_1)^2 \right] 
	\frac{[\mcal{Q}_{T}(\bar{Q},K_{\perp}, Y_{\mathbb P})]^2}{\bar{Q}^2},
\end{align}
where $\alpha_{\rm em} = e^2/4\pi$ and we defined $\bK \equiv -\bk_1 = \bk_2$ for convenience. 
The $\gamma\to q\bar q$ splitting function visible in the r.h.s.~has been generated via the identity:
\beq
\varphi^{ij}(\vartheta,\lambda)\,
\varphi^{il\,*}(\vartheta,\lambda)\,
=\,2\delta^{jl}\left[\vartheta^2+(1-\vartheta)^2\right].
\eeq

As indicated by our above notations, see e.g.~\eqn{QP}, the quantity
$\mcal{Q}_T$ also depends upon $Y_{\mathbb P}$, via the corresponding dependence of the dipole amplitude
$\mcal{T}(R, Y_{\mathbb P})$. We recall that $Y_{\mathbb P}$ is the rapidity separation between the diffractive
system (here, the $q\bar q$ pair) and the target. Importantly, this refers to {\it target} rapidity, i.e.~to the fraction
$ x_{\mathbb P}$ of the longitudinal momentum $P_N^-$ of a nucleon from the target which is transferred
to the diffractive system via the collision. We more precisely have $Y_{\mathbb P} = \ln 1 / x_{\mathbb P}$, where the 
value of $x_{\mathbb P}$ is fixed by the condition that the final $q\bar q$ pair be on-shell:
\begin{align}
\label{xPqqdef}
	x_{\mathbb P} P_N^- = 
	\frac{1}{2 q^+} \left( \frac{k_{1\perp}^2}{\vartheta_1} + 
	\frac{k_{2\perp}^2}{\vartheta_2} + Q^2\right)
	\,\Longrightarrow\,
	 x_{\mathbb P}= \frac{1}{\hat{s}}
	 \left[Q^2 + \frac{K_{\perp}^2}{\vartheta_1 (1-\vartheta_1)} \right],
\end{align}
with $\hat{s} = 2 q^+ P_N^-$. This has the structure anticipated in \eqn{xP}, with the expected
value for the diffractive mass of the $q\bar q$ final state\footnote{Indeed, one can successively write
$M_{q\bar q}^2   \equiv (k_1+k_2)^2 =2k_1\cdot k_2= {(\vartheta_2\bk_1-\vartheta_1\bk_2)^2}/{\vartheta_1\vartheta_2} 
={K_{\perp}^2}/{\vartheta_1 (1-\vartheta_1)}$, where in the last equality we have used $\vartheta_1 +\vartheta_2=1$ and $\bk_2 = -\bk_1\equiv \bK$.}, that is, 
$M_X^2={K_{\perp}^2}/\vartheta_1 \vartheta_2$. For this particular process, the 
diffractive variable $\beta$ introduced too in \eqn{xP} takes the form:
\begin{align}
	\label{beta}
	\beta \equiv \frac{x_{\rm \scriptscriptstyle Bj}}{x_{\mathbb P}} = 
	\frac{\bar{Q}^2}{\bar{Q}^2 + K_{\perp}^2}.
\end{align}

\comment{
\subsection{Exclusive production of hard dijets}
\label{sec:exclusive}

As a first application of \eqn{crossqqnew}, we consider the production of a pair of jets which are relatively  hard,
$K_\perp\gg Q_s$, and nearly symmetric: $ \vartheta_1\sim \vartheta_2\sim 1/2$. Even though this is not
the {\it typical} configuration, as we shall see, it is nevertheless the most interesting situation for the 
experimental measurements of dijets (since hard and relatively symmetric dijets are easier to measure).

Since the transverse momenta of the produced jets are hard, the process remains perturbative
for any value $Q^2$ of the photon virtuality. So, in this section, we shall take $Q^2$ to be arbitrary and 
we shall demonstrate that the exclusive production of hard dijets is strongly suppressed, 
as a large inverse power of the hardest scale in the problem (either $K_\perp^2$, or $Q^2$).
This suppression is ultimately related to the elastic nature of the scattering, i.e., to the fact that
the cross-section \eqref{crossqqnew} is proportional to the {\it square} of the
dipole amplitude $\mcal{T}$, hence it is strongly suppressed when the scattering is weak ($\mcal{T}\ll 1$).


Since $K_\perp\gg Q_s$, the integral over $R$ in \eqn{QP} is restricted to relatively small values $R\ll 1/Q_s$ 
by the Bessel function $ J_1(K_{\perp} R)$ (indeed, this function is rapidly oscillating when its argument is large).
A small colour dipole scatters only weakly, by colour transparency, so in this regime one can rely on the single
scattering approximation $\mcal{T}(R)\sim R^2 Q_s^2$. For definiteness,
we shall use the expression of $\mcal{T}(R)$ given by the McLerran-Venugopalan (MV)
model \cite{McLerran:1993ni,McLerran:1994vd}, which assumes that the dipole projectile scatters off the 
valence quarks in the nuclear target. This model is meant to apply for a large nucleus
($A\gg 1$) at moderate energies ($x_{\mathbb P}\gtrsim 10^{-2}$) and provides a physically
motivated initial condition for the high-energy (BK-JIMWLK) evolution with increasing $Y_{\mathbb P}$.
In this model,
\begin{align}
	\label{SMV}
	\mcal{T}(R) = 1- \exp
	\left(- \frac{R^2 Q_A^2}{4} \, \ln \frac{4}{R^2 \Lambda^2} \right)
	\simeq \frac{R^2 Q_A^2}{4} \, \ln \frac{4}{R^2 \Lambda^2}\,,
\end{align}
where the second, approximate, equality holds for $R^2 Q_s^2 \ll 1$. The quantity $Q_A^2$ is proportional to the density of the color 
charge squared of the valence quarks in the transverse plane. The saturation momentum $Q_s$ is the scale which defines
the onset of multiple scattering. It is conventionally defined as the value of $2/R$ for which the
exponent becomes of order one:
\begin{align}
	\label{QsMV}
	 Q_s^2=Q_A^2 \ln \frac{Q_s^2}{\Lambda^2}\,.
\end{align}
Using the single scattering approximation in \eqn{SMV}, we have been able to compute the integral in 
 \eqn{QP} exactly (see Appendix~\ref{sec:single} for details). The results are conveniently shown for the ratio
$ [\mcal{Q}_{\mathbb P}(\bar{Q},K_{\perp})]^2/{\bar{Q}^2}$, since this is the quantity which determines
the strength of the cross-section  \eqref{crossqqnew}. We recall that $\bar Q^2= \vartheta_1\vartheta_2 Q^2$,
hence  $\bar Q^2 \sim Q^2$ for the relatively symmetric dijets with $ \vartheta_1\sim  \vartheta_2\sim 1/2$.

  There are two interesting kinematical regimes:

\begin{enumerate}[(i)]
	\item When $\bar Q^2$ is comparable to, or larger than, $K_{\perp}^2$, one finds
	\begin{align}
	\label{QPoverQ-largeQ}
	\frac{\big[\mcal{Q}_{T}(\bar{Q},K_{\perp})\big]^2}{\bar{Q}^2}\, 
	\simeq \,\frac{4 Q_A^4 K_{\perp}^2\bar{Q}^4}{(K_{\perp}^2+\bar{Q}^2)^6}\,
	\ln^2\frac{(K_{\perp}^2+\bar{Q}^2)^2}{\bar{Q}^2\Lambda^2}
    \qquad {\rm for} \qquad  \bar Q\,\gtrsim \,
    K_\perp\, \gg \,Q_s.
    \end{align}
    The argument of the logarithm is large and the above result holds to leading logarithmic accuracy,
    but the coefficient under the log is also known (see \eqn{app:QPtot}).

 \item In the photo-production limit $\bar Q^2\ll K_\perp^2$ (as relevant e.g.~for ultra-peripheral
 nucleus-nucleus collisions \cite{Iancu:2023lel}), \eqn{app:QPKlarge} yields
 \begin{align}
	\label{QPoverQ-lowQ}
	\frac{\big[\mcal{Q}_{T}(\bar{Q},K_{\perp})\big]^2}{\bar{Q}^2}\, 
	\simeq \,\frac{Q_A^4}{K_{\perp}^6} \qquad {\rm for} \qquad 
    K_\perp\, \gg \,\bar Q,\, Q_s.
    \end{align}

\end{enumerate}

To discuss the physical consequences of these results, we shall use $\bar Q^2 \sim Q^2$ to reduce the number
of scales. Then the above equations show that,
when  $K_\perp^2\gtrsim Q^2\gg Q_s^2$, the exclusive dijet cross-section behaves like $Q_s^4/K_\perp^6$ (up to logarithmic
factors), whereas for larger virtualities $Q^2\gg K_\perp^2\gg Q_s^2$ it decays even faster, like $Q_s^4 K_\perp^2/Q^8$.
In both cases, this is a very small cross-section,  as it can be appreciated by 
recalling the case of {\it inclusive}
dijet production in DIS: at large $K_\perp\gg Q_s$, the respective cross-section decays only 
like $Q_s^2/K_\perp^4$ \cite{Dominguez:2011wm,Dominguez:2011br}. 

This strong suppression of hard exclusive dijets motivates our quest for alternative mechanisms for 
the diffractive production of hard dijets. The processes that we propose in that sense --- in our earlier publications
 \cite{Iancu:2021rup,Iancu:2022lcw,Iancu:2023lel} and also in Sect.~\ref{sec:2plus1} of this paper --- 
include a third, semi-hard,  jet in the final state. They are formally suppressed by a power of $\alpha_s$, yet they give the dominant contributions at large $K_\perp\gg Q_s$, 
 where the associated cross-sections decrease only like  $1/K_\perp^4$. But before we move to 
 more complicated final states, let us stay with exclusive dijets for the remaining part of this section and
 consider the {\it typical} such configurations: those where the final dijets are {\it semi-hard} ($K_\perp\sim Q_s$).
}

\subsection{Diffractive SIDIS and its leading-twist approximation}
\label{sec:LTA}

In this section, we shall use \eqn{crossqqnew} to construct the cross-section  \eqref{qqTMDfact} for
diffractive SIDIS and then we shall isolate its leading-twist contribution and eventually recover the factorised structure 
anticipated in the r.h.s.~of \eqn{qqTMDfact}. 

Clearly, in order to go from \eqn{crossqqnew} 
to \eqn{qqTMDfact} one just needs to perform a change of longitudinal variables, from $\vartheta_1$ to $\beta$,
which can be easily done with the help of  \eqref{beta}. Although mathematically trivial, this change of variables
is truly important if one is interested in building a physical interpretation for this process in terms of the evolution
of the target. Indeed,  $\beta$ represents the longitudinal momentum fraction w.r.t.~the Pomeron
of the quark from the target that is struck by the virtual photon. This interpretation implicitly assumes that the 
photon was absorbed by a {\it single} quark, which is indeed the case in the {\it leading-twist} (LT) approximation.
In this analysis, we shall discover that the LT contribution to the cross-section corresponds to {\it aligned jet} configurations
in the dipole picture: $\vartheta_1(1- \vartheta_1)\ll 1$. This is important too for building a target picture: 
in order for a parton (here, one of the two final quarks) to be transferred from the wavefunction of the projectile to that of the target, that parton must have a negligible longitudinal momentum along the direction of motion of the
projectile. 

One can perform the aforementioned change of variables by integrating over $\vartheta_1$ with a 
$\delta$--function enforcing the condition \eqref{beta}:
\begin{align}
  \label{2jetSIDIS0}
  \frac{\rmd \sigma^{\gamma_T^* A
  \rightarrow q\bar q A}}
  {\rmd \ln(1/\beta)\,\rmd^{2}\bm{K}} = \int_0^1 \rmd \vartheta_1 \,\beta\delta\left(\beta-
  \frac{\bar{Q}^2}{\bar{Q}^2 + K_{\perp}^2}\right)
  \frac{\rmd\sigma_{\scriptscriptstyle}^{\gamma_{\scriptscriptstyle T}^* A\rightarrow q\bar q A}}
	{\rmd \vartheta_1\,
	\rmd^{2}\bK} \,.
\end{align}
The $\delta$--function is conveniently rewritten as (recall that $\bar Q^2= \vartheta_1(1-\vartheta_1)Q^2$)
  \begin{align}  \label{delta}
  \hspace*{-.8cm}
  \beta\delta\left(\beta-\frac{\bar Q^2}{\bar Q^2+ K_\perp^2}\right)&\,=\frac{\vartheta_1(1-\vartheta_1)}{1-\beta}\,
  \delta\left(\vartheta_1(1-\vartheta_1)-\frac{\beta}{1-\beta}\frac{K_\perp^2}{Q^2}\right)\\*[0.2cm]
  &\, =\, \frac{\vartheta_1(1-\vartheta_1)}{1-\beta}\,\frac{
    \delta\left(\vartheta_1-\vartheta_*\right) + \delta\left(\vartheta_1-1+\vartheta_*\right)}{1-2\vartheta_*}
    \ \longrightarrow\  \frac{2\vartheta_*(1-\vartheta_*)}{(1-2\vartheta_*)(1-\beta)}\,
        \delta\left(\vartheta_1-\vartheta_*\right).\nonumber
  \end{align}
Here $\vartheta_*$ and $1-\vartheta_*$ are the two values\footnote{These two values correspond to the
quark and the antiquark contributions, which are equal to each other.}
 of $\vartheta_1$ selected by the $\delta$--function and
we conveniently chose $\vartheta_*$ to be smaller than 1/2:
\beq
	\label{K2max}
	\vartheta_*\,=\,\frac{1}{2}\left(1 -\sqrt{1-\frac{4\beta}{1-\beta}\frac{K_\perp^2}{Q^2}}\right)
	\qquad
	\mbox{with}\quad K_\perp^2\,\le\,\frac{1-\beta}{4\beta}\,Q^2\,.
\eeq
The integrand in \eqn{2jetSIDIS0} is symmetric under the exchange $\vartheta_1\to 1-\vartheta_1$, hence it is sufficient
to keep this smaller solution $\vartheta_*$ and multiply the result by 2, as indicated in the second line of 
\eqn{delta}. In terms of these new variables $K_\perp$ and $\beta$, the value of $\bar Q$ is computed as
\begin{align}
	\label{Q2M2}
	\bar{Q}^2 \,=\, \frac{\beta}{1-\beta}\,K_{\perp}^2 \equiv \mcal{M}^2,
\end{align}
hence it is independent of $Q^2$. A simple calculation yields
\begin{align}
  \label{2jetSIDIS}
  \frac{\rmd \sigma^{\gamma_T^* A
  \rightarrow q\bar q A}}
  {\rmd \ln(1/\beta)\,\rmd^{2}\bm{K}} = 
  \frac{S_{\perp} \alpha_{\rm em} N_c}{\pi^2} \left( \sum e_f^2\right)\,\frac{1}{Q^2}\,\frac{1-\frac{2\mcal{M}^2}{Q^2}}
 {\sqrt{1-\frac{4\mcal{M}^2}{Q^2}}}\,
	\frac{[\mcal{Q}_{T}(\mcal{M},K_{\perp}, Y_{\mathbb P})]^2}{1-\beta}\,.
  \end{align}
  When $ \mcal{M}^2\ll Q^2$, one can expand the expression under the square root, thus generating
  the twist-expansion of \eqn{2jetSIDIS}: an infinite series of power corrections which scale
like powers of  $\mcal{M}^2/Q^2$.
For generic values of $\beta$, which are not very small, nor very close to one, one has $\mcal{M}^2\sim 
K_{\perp}^2$, so these corrections can be recognised  as the ``DIS higher-twists'' discussed 
at the end of Sect.~\ref{sec:kin}. To make contact with  \eqn{qqTMDfact}, we need to isolate the 
leading-twist piece in this series, that is, the dominant contribution in the limit 
$Q^2\gg \mcal{M}^2$. (The relevance of this limit for the physical problem at hand will be clarified
in the next sections.) One easily finds
 \begin{align}
  \label{2jetLT}
  \frac{\rmd \sigma^{\gamma_T^* A
  \rightarrow q\bar q A}}
  {\rmd \ln(1/\beta)\,\rmd^{2}\bm{K}} = 
  \frac{S_{\perp} \alpha_{\rm em} N_c}{\pi^2} \left( \sum e_f^2\right)\,\frac{1}{Q^2}\,	
  \frac{[\mcal{Q}_T(\mcal{M},K_{\perp}, Y_{\mathbb P})]^2}{1-\beta}\,\quad\mbox{when}\quad
  Q^2\gg \mcal{M}^2\,.
  \end{align}
 As anticipated, this result has the factorised structure anticipated in  \eqn{qqTMDfact}.
  The whole $Q^2$--dependence is now isolated  in the overall power-law $1/Q^2$, which is the hallmark
  of the elementary cross-section for photon absorption by a point-like quark form the target.

Recalling \eqn{K2max}, one sees that the condition $\mcal{M}^2/Q^2\ll 1$ for the validity of \eqn{2jetLT}
automatically selects the asymmetric $q\bar q$ pairs with $\vartheta_*\simeq\mcal{M}^2/Q^2\ll 1$.  
This strong correlation between the LT approximation and the aligned jet configurations  can be
 understood as follows: when $\vartheta_*\ll 1$, one of the two fermions  of the dipole picture --- that with longitudinal
fraction $1-\vartheta_*\simeq 1$ --- carries most of the longitudinal momentum $q^+$ of the incoming photon and hence
can be unambiguously identified as the quark from the target that has absorbed the photon. 

This simple analysis  illustrates some of the main steps that we shall encounter in more complicated situations, in the
process of shifting from the colour dipole picture to the target picture: a change in longitudinal variables (from
projectile-oriented to target-oriented longitudinal momentum fractions), the leading-twist approximation at high $Q^2$
(or for large jet transverse momenta), 
and the focus on asymmetric configurations. In practice, the last two steps often come together: when working in the
target-oriented variables, the restriction to asymmetric dijets automatically implements the relevant LT approximation.


\subsection{TMD factorisation for diffractive SIDIS}
\label{sec:TMDSIDIS}

The leading-twist approximation \eqref{2jetLT} to the cross-section for diffractive SIDIS exhibits
 a factorised form, which is consistent with  TMD factorisation. 
Specifically,
 \begin{align}
  \label{2jetTMD}
  \frac{\rmd \sigma^{\gamma_T^* A
  \rightarrow q\bar q A}}
  {\rmd \ln(1/\beta)\,\rmd^{2}\bm{K}} 
  = \frac{4\pi^2 \alpha_{em}}{Q^2}\big(\sum e_f^2\big)\,2\,
  \frac{\rmd xq_{\mathbb{P}}
	(x, x_{\mathbb{P}}, K_{\perp}^2)}{\rmd^2 \bK}\bigg |_{x=\beta}.	
\end{align} 
with the following expression for the quark (or antiquark) diffractive TMD\footnote{
\eqn{qDTMD} refers to a single quark flavour, but we shall generally omit the flavour ($f$) index, since
the diffractive TMDs are identical for all massless flavours.}:
\begin{align}
  \label{qDTMD}
  \frac{\rmd xq_{\mathbb{P}}
  (x, x_{\mathbb{P}}, K_{\perp}^2)}{\rmd^2 \bK}\equiv
  \frac{S_{\perp} N_c}{4 \pi^3}\,	
	\frac{[\mcal Q_{\mathbb{P}}(x, x_{\mathbb{P}}, K_{\perp}^2)]^2}
	{2\pi(1-x)}\,\equiv\, S_{\perp} \,\frac{2N_c}{(2 \pi)^3}\,	\Psi_{\mathbb{P}}
  (x, x_{\mathbb{P}}, K_{\perp}^2)\,,
\end{align}
where $x$ is the longitudinal momentum fraction\footnote{For the SIDIS problem at hand and to LT accuracy,
the variables $x$ and $\beta$ coincide with each other, as indicated in the r.h.s.~of \eqn{2jetTMD}. But in general
they are different. For a generic diffractive final state, the variable $\beta$ is defined as in \eqn{xP},
while $x$ always denotes the fraction of a parton from the Pomeron w.r.t.~$x_{\mathbb{P}}$.}
 of the measured (anti)quark w.r.t.~the Pomeron and
 $\QP(x, x_{\mathbb{P}}, K_{\perp}^2)$ is the same as the quantity
 $\mcal{Q}_T(\bar Q,K_{\perp}, Y_{\mathbb P})$  previously 
introduced in Eq.~\eqref{QP}, but which is now viewed as a function of only target kinematical variables,
like the TMD:
\begin{align}
\label{QPpm}
	\mcal Q_{\mathbb{P}}(x, x_{\mathbb{P}}, K_{\perp}^2)
	\equiv
	\mcal{M}^2 \int_0^{\infty} \rmd R\, R\, J_1(K_\perp R)\,
        K_1(\mcal{M} R)\, \mcal{T}(R,Y_{\mathbb P})
\end{align}
It is understood that $\mcal{M}^2=\frac{x}{1-x}K_\perp^2$ (cf.~\eqn{Q2M2} with $\beta\to x$).
The interpretation of the quantity $\Psi_{\mathbb{P}}  (x, x_{\mathbb{P}}, K_{\perp}^2)$ will be shortly explained.

In what follows, we would like to argue that this quark TMD can naturally be interpreted as the
{\it unintegrated quark distribution of the Pomeron} --- that is, the number of quarks (or antiquarks) of a given flavour
that can be found in the light-cone wavefunction of the Pomeron
with longitudinal momentum fraction $x$  and with transverse momentum $\bK$:
\begin{align}
  \label{qDTMDdef}
  \frac{\rmd xq_{\mathbb{P}}
  (x, x_{\mathbb{P}}, K_{\perp}^2)}{\rmd^2 \bK}\,=\,x\,\frac{\rmd N_{q\in\mathbb{P}}}{\rmd x\rmd^2\bm{K}} \,.
\end{align}
With this interpretation, the quantity $\Psi_{\mathbb{P}}  (x, x_{\mathbb{P}}, K_{\perp}^2)$ introduced in the last equality 
\eqn{qDTMD} has the meaning of the {\it  quark occupation number}.

\begin{figure}
	\begin{center}
		\includegraphics[align=c,width=0.60\textwidth]{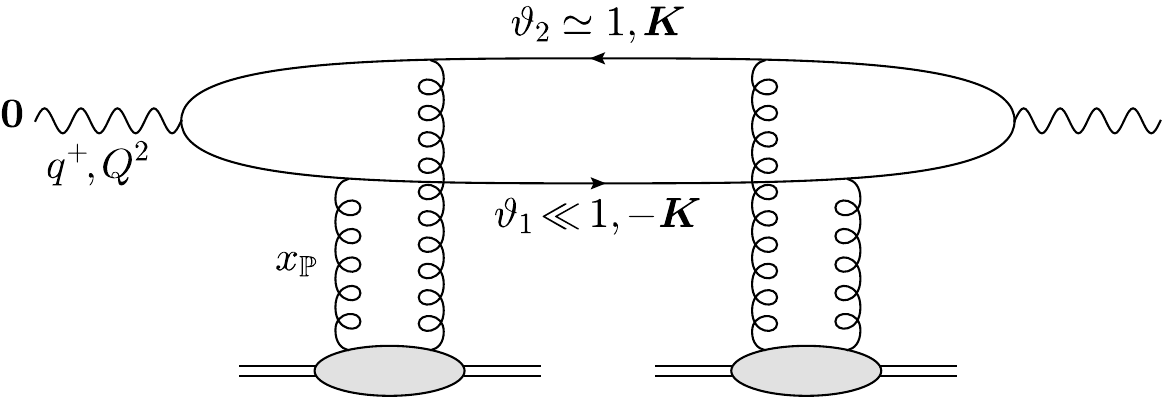}
		\hspace*{0.04\textwidth}
		\includegraphics[align=c,width=0.34\textwidth]{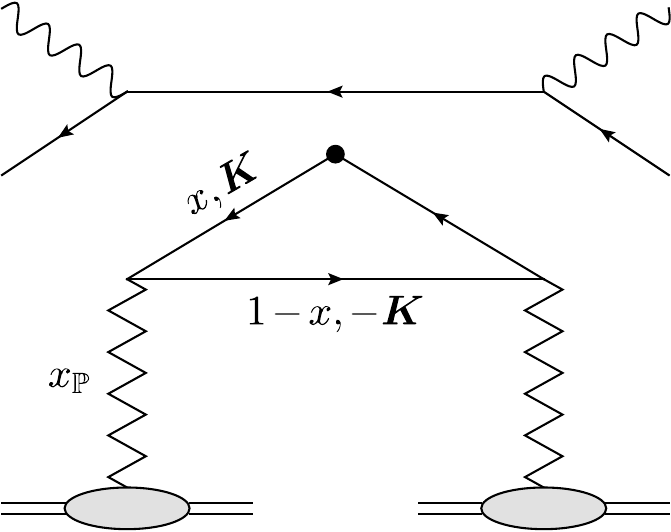}
	\end{center}
	\caption{\small Cross section for the exclusive dijet production with a soft quark in DIS off a nucleus. Left panel: the viewpoint of the colour dipole picture. Right panel: the pictorial illustration of TMD factorisation. The soft quark, with longitudinal fraction $1-x$, is now radiated from the Pomeron and thus belongs to the target wavefunction. The virtual photon probes the (anti)quark DTMD evaluated at  fraction $x$ and transverse momentum $\bK$.}
\label{fig:2jets_TMD}
\end{figure}

This partonic interpretation of the quark DTMD could be unambiguously established only by working in the target infinite momentum (or Bjorken) frame\footnote{This frame
is obtained from the original dipole frame by boosting the longitudinal momenta according to $q^+\to q^+/\gamma$
and $q^-\to \gamma q^-$, where $\gamma\sim q^+/Q\gg 1$. Also, the time-variable in this new frame is $x^-$ rather than $x^+$.} 
and in the target light-cone gauge $A^-=0$: indeed, these are the frame and gauge where the construction of the light-cone
wavefunction makes sense for a relativistic left-mover. That said, the cross-section is boost and gauge invariant, hence the
partonic interpretation can also be {\it a posteriori} recognised when working in the dipole picture, provided one focuses
on the leading-twist approximation at large $Q^2\gg Q_s^2(Y_{\mathbb{P}})$ 
and one uses the target longitudinal momentum fraction
$x$  instead of the projectile momentum fraction $\vartheta_1$. 

Specifically, we would like to argue that \eqn{2jetTMD} is consistent with the following physical picture in the Bjorken 
frame, as illustrated in the right panel in Fig.~\ref{fig:2jets_TMD}: the virtual photon, which in this new frame is relatively slow
($q^+\sim Q$), is absorbed by a quark-antiquark pair which represents the final evolution of a colourless partonic 
fluctuation of the target --- the ``Pomeron''.  This Pomeron carries a tiny (net) transverse momentum
$\Delta_\perp \sim 1/R_A$, which is negligible for our purposes, and a small fraction $ x_{\mathbb{P}}\ll 1$ of the
target longitudinal momentum $P_N^-$, that is transmitted to the $q\bar q$ pair. One of the fermions in this pair,
say the quark with splitting fraction $1-x$ and transverse momentum $-\bm{K}$, is already on mass-shell
and appears in the final state. This quark does not interact with the virtual photon, rather it plays the role of a spectator.
The other fermion --- a $t$--channel antiquark with splitting fraction $x$ and transverse momentum $\bm{K}$ ---
is virtual and scatters with the hard photon in order to produce the final, on-shell, antiquark. 
From the viewpoint of the color dipole picture, cf.~the left panel in Fig.~\ref{fig:2jets_TMD}, the virtual antiquark is the intermediate quark in between the photon decay vertex and the scattering with the target. This fermion looks like a right-moving quark in the dipole frame, but like a left-moving antiquark after boosting to the Bjorken frame.

The spectator ($s$--channel) quark carries a fraction $(1-x)x_{\mathbb P}$ of the target momentum 
and a fraction $\vartheta_1$ of the longitudinal momentum $q^+$ of the virtual photon, hence the condition to be on-shell reads
\begin{align}
	\label{onshellq}
	2 \vartheta_1 
	(1-x)x_{\mathbb P} 
	q^+P_N^- =
	K_\perp^2 
	\,\Longrightarrow\, 
	(1-x)x_{\mathbb P}=\frac{K_\perp^2}{\vartheta_1 \hat s}.
\end{align} 
The $t$--channel antiquark has $k^+=-\vartheta_1 q^+$ and $k^-=xx_{\mathbb P}P_N^-$,
hence a {\it space-like} virtuality
\beq\label{tchannel}
k^2 = 2k^+k^--K_\perp^2 = - \vartheta_1 xx_{\mathbb P}\hat s-K_\perp^2 = -\frac{K_\perp^2}{1-x}\,.\eeq
Taking the ratio of Eqs.~\eqref{xPqqdef} and \eqref{onshellq} and solving either for $\bar{Q}$ or for $x$ we find
\begin{align}
	\label{Q2andx}
	\bar{Q}^2 = \frac{x -\vartheta_1}{1-x}\,K_{\perp}^2
	\,\Longleftrightarrow\, 
	x = \frac{\bar{Q}^2 + \vartheta_1 K_{\perp}^2}
	{\bar{Q}^2 + K_{\perp}^2}.
\end{align}
This equation shows that, in general,
$\bar Q^2$ cannot be expressed uniquely in terms of the two variables $x$ and $K_\perp^2$  
which characterise the production of the $q\bar q$ pair by the Pomeron --- it also depends upon the
projectile variable $\vartheta_1$.  Moreover, $x$ is not the same as $\beta$ (compare the second 
equation above to Eq.~\eqref{beta}).
Yet, when the spectator quark is soft w.r.t.~the virtual photon, i.e.~when $\vartheta_1 \ll 1$, the first equation in \eqref{Q2andx} reduces to
\begin{align}
	\label{Q2M2-1}
	\bar{Q}^2 \simeq \vartheta_1 Q^2 \simeq \frac{x}{1-x}\,K_{\perp}^2, 
\end{align}
which is now a function of only $x$ and $K_{\perp}^2$, as required by factorisation. 
Still for $\vartheta_1 \ll 1$, the second equation in \eqref{Q2andx} becomes 
\begin{align}
	\label{xbeta}
	x \simeq 
	\frac{\bar{Q}^2}
	{\bar{Q}^2 + K_{\perp}^2} = \beta,
\end{align}
and then Eqs.~\eqref{Q2M2-1} and \eqref{Q2M2} are indeed consistent with each other. 

Clearly, if within the original expression
\eqref{crossqqnew} for the cross-section one makes a change of variables from $\vartheta_1$ to $x$ and one uses $\vartheta_1 \ll 1$,  together with \eqref{Q2M2-1} and \eqref{xbeta}, then one finds the leading-twist approximation
in \eqn{2jetLT} with $\beta=x$. We thus recover the correlation between 
aligned jets and the LTA, as discussed at the end of Sect.~\ref{sec:LTA}. This correlation is easy to understand on
the basis of \eqn{Q2M2-1} which shows that, for fixed and generic values of $x$, the variable  $\vartheta_1$ scales
like  $\vartheta_1\sim K_{\perp}^2/Q^2$. Hence, by working at leading order in the small parameter
$\vartheta_1$, one automatically selects the dominant contributions in the twist expansion at high $Q^2$.

To summarise, both the mathematical derivation of the TMD factorisation for diffractive SIDIS from the dipole picture, and its
physical interpretation in terms of parton evolution in the target, crucially rely on the fact that the produced $q\bar q$ 
pair is very asymmetric: $\vartheta_1\vartheta_2 \ll 1$. In the next subsection, we shall argue that this condition is very
well satisfied for the {\it typical} diffractive processes --- those which give the dominant contribution to $F_2^D$.
%
%
\subsection{Multiple scattering and parton saturation in the Pomeron}
\label{sec:MS}

In this section, we will study in some detail the quark DTMD defined in \eqn{qDTMD},  with the purpose of 
demonstrating (via parametric estimates) 
two important features: \texttt{(i)} this distribution {\it saturates} at relatively low momenta,
$K_\perp\lesssim Q_s$, but \texttt{(ii)}  it abruptly falls, like the power $Q_s^4/K_\perp^4$, at much larger
momenta. This shows that the {\it typical} quark-antiquark pairs produced by the Pomeron are {\it semi-hard}, i.e.~they carry transverse momenta $K_\perp\sim Q_s$.
(In \cite{Iancu:2021rup,Iancu:2022lcw}, similar conclusions have been reached for the {\it gluon} diffractive TMD.)
These results  can be translated to the original, dipole, picture with the help of 
 the relation \eqref{Q2M2-1} between projectile and target variables: for $K_\perp\sim Q_s$ and generic values of $x$,
 this relation implies
\begin{align}
	\label{thetaQs}
	\vartheta_1 \simeq \frac{x}{1-x}\,\frac{K_{\perp}^2}{Q^2}\,\sim\,\frac{Q_s^2}{Q^2}\,\ll 1\,.
	\end{align}
This shows that the $q\bar q$ fluctuations of the virtual photon which are predominantly produced via elastic
scattering are very asymmetric, with a low effective virtuality $\bar Q^2\simeq \vartheta_1 Q^2\sim Q_s^2$
and hence a large transverse size $R\sim 1/Q_s$. Such large colour dipoles undergo {\it strong scattering},
$\mcal{T}(R)\sim 1$, and thus are strongly favoured by the elastic nature of the scattering, i.e.~by the fact that
the SIDIS cross-section is proportional to the square $\mcal{T}^2$ of the dipole scattering amplitude.

These conclusions heavily rely on the interplay between the structure of the virtual photon LCWF and the
scattering of the $q\bar q$ dipole. In the target viewpoint  underlying the TMD factorisation \eqref{2jetTMD},
these two fundamental aspects are both encoded in the function
$\mcal Q_{\mathbb{P}}(x, x_{\mathbb P}, K_{\perp}^2)$ defined in  \eqn{QPpm}. In particular, the quantity
$\mcal{M}^2= \frac{x}{1-x}\,K_{\perp}^2$ appearing in this equation is nothing but the effective virtuality
$\bar Q^2$ of the dipole picture expressed in terms of target variables, cf.~\eqn{Q2M2-1}. 

For the subsequent
discussion, we also need an explicit form for the dipole scattering amplitude $\mcal{T}(R)$, that we shall take
from  the McLerran-Venugopalan (MV) model \cite{McLerran:1993ni,McLerran:1994vd}.
This model is meant to apply for a large nucleus
($A\gg 1$) at moderate energies ($x_{\mathbb P}\gtrsim 10^{-2}$) and provides a physically
motivated initial condition for the high-energy (BK-JIMWLK) evolution with increasing $Y_{\mathbb P}$.
It assumes that the dipole projectile scatters off a Gaussian  colour charge distribution representing the
valence quarks. One finds
\begin{align}
	\label{SMV}
	\mcal{T}(R) = 1- \exp
	\left(- \frac{R^2 Q_A^2}{4} \, \ln \frac{4}{R^2 \Lambda^2} \right),
\end{align}
where $Q_A^2$ is essentially the density of the colour 
charge squared of the valence quarks in the transverse plane. By expanding the exponential in the
r.h.s.~one generates the multiple scattering series (the ``saturation twists'' alluded to at the end of Sect.~\ref{sec:kin}).
The first term in this expansion, which coincides with the exponent in \eqn{SMV}, represents the amplitude
for a single elastic scattering via two gluon exchange. For a sufficiently large dipole though, the scattering
becomes strong (the exponent is of order one, or larger), and then one needs to include multiple scattering
to all orders. The saturation momentum $Q_s$ is the scale which defines
the onset of multiple scattering. It is conventionally defined as the value of $2/R$ for which the
exponent becomes of order one:
\begin{align}
	\label{QsMV}
	 Q_s^2=Q_A^2 \ln \frac{Q_s^2}{\Lambda^2}\,.
\end{align}

We are now in a position to analyse the behaviour of  \eqn{QPpm} in various limits.
The integral over $R$  in  \eqn{QPpm} is controlled by the interplay between three functions:
 the two Bessel functions $K_1( \mcal{M} R)$ and  $J_1(K_{\perp} R)$, and the dipole scattering amplitude
$\mcal{T}(R,Y_{\mathbb P})$. The Bessel functions are rapidly decreasing (or oscillating)
for large values of their respective arguments, hence they restrict the maximal value of $R$. On the contrary,
the dipole amplitude favours large dipole sizes $R\gtrsim 1/Q_s(Y_{\mathbb P})$, for which the
scattering is as strong as possible. 

 For parametric estimates, it suffices to use a piecewise approximation to the dipole amplitude in \eqn{SMV}
 (more sophisticated approximations, which take into account the full structure of the MV model amplitude and
also include  its  BK evolution with increasing $Y_{\mathbb P}$,   will be considered in Sect.~\ref{sec:qDTMD}):
\begin{align}\label{TGBW}
\mcal{T}(R)\,\simeq 
    \begin{cases}    \displaystyle{\ R^2 Q_s^2}
         &
        \text{for \ $R\!\ll 1/Q_s$,}
                  \\*[0.2cm]
               \displaystyle{\ 1} &
        \text{for \ $R\!\gtrsim 1/Q_s$}\,.
         \end{cases}
\end{align}
The first line expresses the single scattering approximation, as appropriate for a small dipole. 
The second line represents the black disk limit, as obtained by
resumming multiple scattering to all orders.

To the same level of accuracy, the two Bessel functions in  \eqn{QPpm} can be approximated as:
\beq\label{Bessel}
K_1( \mcal{M} R)\,\sim\,\frac{\Theta(1- \mcal{M} R)}{ \mcal{M} R}\,,\qquad J_1(K_{\perp} R)
\,\sim\,\Theta(1- K_{\perp} R)\frac{K_{\perp} R}{2}
\eeq
Recalling that $\mcal{M}^2= \frac{x}{1-x}\,K_{\perp}^2$, it is clear that the dipole size
is limited  to  $R\lesssim R_{\rm max}$ with
\begin{align}
	\label{Rmax}
	R^2_{\max} \,\equiv\, \frac{1-x}{K^2_{\perp}}\,.
\end{align}  
The result of the integral over $R$ depends upon the competition between $R_{\rm max}$ and $1/Q_s$, or,
equivalently, between $K_\perp$ and the {\it effective} saturation momentum $\tilde{Q}_s(x,Y_{\mathbb{P}})$ defined
as 
\begin{align}
	\label{Qsx}
	\tilde{Q}_s^2(x,\YP)
	\equiv
	(1-x) Q_s^2(\YP).
	\end{align}
There are two interesting, limiting, regimes:

\texttt{(i)} When $K_\perp\gg \tilde Q_s(x)$, we have $R_{\rm max}\ll 1/Q_s$ and the scattering
is weak. Using the first line of \eqref{TGBW} together with \eqref{Bessel} within \eqn{QPpm}, one easily finds
\begin{align}
	\label{QPlargemom}
	\mcal Q_{\mathbb{P}}&\,\sim \,\mcal{M}^2\!\int \rmd R \,R\,\frac{\Theta(R_{\rm max}- R)}{ \mcal{M} R}\,(K_{\perp} R)\,
	(R^2 Q_s^2)\,\sim\, \mcal{M} K_\perp Q_s^2\, R_{\rm max}^4
	\nonumber\\*[0.2cm]
	&\, \sim\,\sqrt{x(1-x)^3}\ \frac{Q_s^2}{K_\perp^2}
	\, \sim\,\sqrt{x(1-x)}\ \frac{\tilde Q_s^2(x)}{K_\perp^2}
	\qquad
	\mbox{for}\quad K_\perp\gg \tilde Q_s(x).
\end{align}

\texttt{(ii)} When $K_\perp\ll \tilde Q_s(x)$, 	we have $R_{\rm max}\gg 1/Q_s$ and 
then the integration is dominated by large dipole sizes, within the interval $1/Q_s \lesssim R\le R_{\rm max}$, 
for which $\mcal{T}(R)=1$. The dominant contribution comes from the upper limit $R= R_{\rm max}$ and reads
\begin{align}
\label{QPsmallmom}
	\mcal Q_{\mathbb{P}}&\,\sim \, \mcal{M}^2\!\int \rmd R R\,\frac{\Theta(R_{\rm max}- R)}{ \mcal{M} R}\,(K_{\perp} R)\,
	\Theta(R Q_s-1) \sim \mcal{M} K_\perp R_{\rm max}^2\nonumber\\*[0.2cm]
	&\,  \sim\,\sqrt{x(1-x)}\,\qquad
	\mbox{for}\quad K_\perp\ll \tilde Q_s(x).
\end{align}

These parametric estimates do not allow us to control the transition region around $K_\perp\sim \tilde Q_s(x)$,
so it is reassuring to observe that the above results are consistent with each other in that region.
It is also interesting to notice that,  when $\mcal{T}=1$, the integral in  \eqn{QPpm} can be {\it exactly} computed:
\begin{align}
	\label{QPsmallK}
	\mcal{M}^2 \int_0^{\infty} \rmd R\, R\, J_1(K_\perp R)\,
        K_1(\mcal{M} R)\, =\, \sqrt{x(1-x)}\,.
\end{align}

We can combine these results in a single piecewise formula for the quark occupation number
(cf.~Eq.~\eqref{qDTMD}):
\begin{align}
	\label{qDTMDpw}
	\Psi_{\mathbb{P}}  (x, x_{\mathbb{P}}, K_{\perp}^2)\,\simeq 
	\frac{x}{2\pi}
	\begin{cases}    
	\displaystyle{\,1}
    &\quad {\rm for} \quad
    K_\perp \ll \tilde{Q}_s(x,Y_{\mathbb{P}})
    \\*[0.2cm]
    \displaystyle{\,
    \frac{\tilde Q_s^4(x,Y_{\mathbb{P}})}{K_\perp^4}}
    &\quad {\rm for} \quad
    K_\perp \gg \tilde{Q}_s(x,Y_{\mathbb{P}})\,.
    \end{cases}
\end{align}
The overall  normalisation in the r.h.s.~has been chosen to reproduce the exact
result \eqref{QPsmallK} in the black disk limit.
Albeit quite crude, the piecewise approximation in \eqn{qDTMDpw} exhibits some important features that can be checked
via more advanced, analytic and numerical, calculations (see  Sect.~\ref{sec:qDTMD}):

\texttt{(i)} For relatively low momenta $K_\perp\lesssim  \tilde{Q}_s(x,Y_{\mathbb{P}})$, the quark occupation number of the Pomeron
{\it saturates}, i.e.~it reaches a {\it universal} value, which is independent of $K_\perp$, of the QCD coupling $\alpha_s$, and also upon
the nature of the hadronic target (e.g.~it does not depend upon the nuclear mass number $A$).
From the viewpoint of the original dipole picture, this saturation corresponds
to the black disk limit, $\mcal{T}(R)= 1$ when $R\gg 1/Q_s(Y_{\mathbb{P}})$, which is universal as well.

%

\texttt{(ii)} For larger transverse momenta $K_\perp\gg  \tilde{Q}_s(x,Y_{\mathbb{P}})$, the quark occupation number decreases
very fast, as $1/K_\perp^4$. This power, which in the dipole picture reflects the elastic nature of the scattering
(the cross-section is proportional to the elastic amplitude squared, hence to the {\it square} of the
dipole scattering amplitude), should correspond
in the target picture to the fact that the $q\bar q$ pair is created via a colourless fluctuation of the parton distribution.

These observations imply that the bulk of the quark distribution of the Pomeron lies at low momenta, in the saturation domain
at $K_\perp\lesssim  \tilde{Q}_s(x,Y_{\mathbb{P}})$. The rapidly decaying 
tail at higher  momenta $K_\perp\gg  \tilde{Q}_s(x,Y_{\mathbb{P}})$ 
describes rare events, which give a negligible contribution to the diffractive structure function
(see \eqn{xqP} below). This discussion confirms that our previous approximations to diffractive SIDIS
--- the LT approximation  in \eqn{2jetLT} and its TMD factorisation
in Eqs.~\eqref{2jetTMD}--\eqref{qDTMD} --- are indeed well justified when
 $Q^2\gg   {Q}_s^2(Y_{\mathbb{P}})$. It furthermore shows that, in order to compute the bulk of the cross-section 
at $K_\perp\lesssim  \tilde{Q}_s(x,Y_{\mathbb{P}})$, one has to resum the multiple scattering series 
to all orders: the leading-twist result  shown in \eqn{2jetLT}
includes an infinite series of  ``saturation twists'', in the form of unitarity corrections to the dipole
scattering amplitude. 


The above discussion also suggests a deep correspondence between parton saturation in the Pomeron 
wavefunction and the unitarity limit for the elastic dipole-hadron scattering:  these two seemingly different phenomena
are in fact  {\it dual descriptions} of a same phenomenon, but
which is viewed in different Lorentz frames and in different gauges.
The black disk limit is the natural picture in the dipole frame and the dipole LC gauge $A^+=0$, a framework well adapted
to the description of scattering. The quark saturation, on the other hand, emerges in the Bjorken
frame and in the target LC gauge $A^-=0$, where the parton picture of the target wavefunction makes sense. 




\comment{To better appreciate how non-trivial this picture is, one should recall that the case of the Weiszäcker-Williams 
(WW) gluon TMD (the
unintegrated gluon distribution of the target),  which  enters the TMD factorisation for inclusive dijet production 
in the correlation limit \cite{Dominguez:2011wm,Dominguez:2011br}). The WW distribution too saturates at low 
$K_\perp\ll Q_s$, albeit at a value of order $1/\alpha_s$ 
\cite{Iancu:2002xk,Iancu:2003xm,Gelis:2010nm,Kovchegov:2012mbw}. Once again, the saturation picture appears to
be dual to the unitarity limit for dipole scattering. However, at larger momenta $K_\perp\gg Q_s$, the WW gluon TMD 
decreases only like $1/K_\perp^2$, since generated via single {\it inelastic} scattering. Accordingly, the distribution
of inclusive dijet exhibits a slowly decaying tail at large $K_\perp$ and the TMD factorisation (which again requires 
$Q^2\gg K_\perp^2$) is truly justified only to leading logarithmic accuracy.}

The contribution of exclusive dijets to the diffractive structure function is obtained by integrating 
\eqn{2jetTMD} over $K_\perp^2$ up to the maximally allowed value of order $\sim Q^2$, 
cf.~\eqn{K2max}:
\begin{align}
  \label{DF2}
\frac{\rmd \sigma^{\gamma_T^* A
  \rightarrow q\bar q A}}
 {\rmd \ln(1/\beta)} 
  = \frac{4\pi^2 \alpha_{em}}{Q^2}\ 2\left(\sum e_{f}^{2}\right) \, x q_{\mathbb{P}}
  \left(x, x_{\mathbb{P}}, \frac{1-x}{4x}Q^2\right)\bigg |_{x=\beta}. \end{align}
The r.h.s.~features the quark diffractive PDF $xq_{\mathbb{P}}(x, x_{\mathbb{P}}, Q^2)$, 
as obtained from the color dipole picture after isolating the leading-twist contribution:
\begin{align}
	\label{xqP}	
	xq_{\mathbb{P}}(x, x_{\mathbb{P}}, Q^2)	= \int
	\dif^2\bK\,
	\frac{\rmd xq_{\mathbb{P}}(x, x_{\mathbb{P}}, K_{\perp}^2)}
	{\rmd^2 \bK} 
	= 
	\frac{S_\perp N_c}{4\pi^2} 
	\int_0^{Q^2}\!
	\dif K_{\perp}^2\,
	\Psi_{\mathbb P}(x,x_{\mathbb P},K_{\perp}^2).
 \end{align}
It is easy to verify that the structure of the integrand, that can be read from Eqs.~\eqref{qDTMD} and \eqref{QPpm},
is fully consistent with the respective results in \cite{Wusthoff:1997fz,GolecBiernat:1999qd,Hebecker:1997gp,Buchmuller:1998jv,Hautmann:1998xn,Hautmann:1999ui,
Hautmann:2000pw,Golec-Biernat:2001gyl} (see e.g.~Eq.~(25) in \cite{Golec-Biernat:2001gyl}). Yet, the essential fact that
the integrand is strongly peaked at $K_\perp \sim \tilde Q_s(x)$, and the profound relation between diffraction and
parton saturation in the Pomeron wavefunction, have been overlooked by those early studies. These features
has been properly recognised only in the recent studies \cite{Iancu:2021rup,Hatta:2022lzj,Iancu:2022lcw,Iancu:2023lel}.

Since the above integral is controlled by $K_\perp\sim \tilde Q_s(x)$, one can easily deduce the following parametric estimate
for the quark DPDF (see  Sect.~\ref{sec:qPomeron} for more precise, numerical, results, including the effects of
the DGLAP evolution with increasing $Q^2$)
\begin{align}
\label{xqPsat}
 xq_{\mathbb{P}}
	(x, x_{\mathbb{P}}, Q^2)\,\simeq\,\frac{S_{\perp} N_c }{4\pi^3}\, \kappa_q \,{x(1-x)} Q_s^2(Y_{\mathbb{P}}),
\end{align}	
with $\kappa_q$ a slowly varying function that can be treated as a number of $\order{1}$.
This result linearly vanishes at the endpoints of the phase-space in $x$, i.e.~when either $x\to 0$, or $x\to 1$, in agreement with the previous calculation of the quark DPDF by Hatta et al,
Ref.~\cite{Hatta:2022lzj}. 

\begin{figure}
	\begin{center}
		\includegraphics[width=0.9\textwidth]{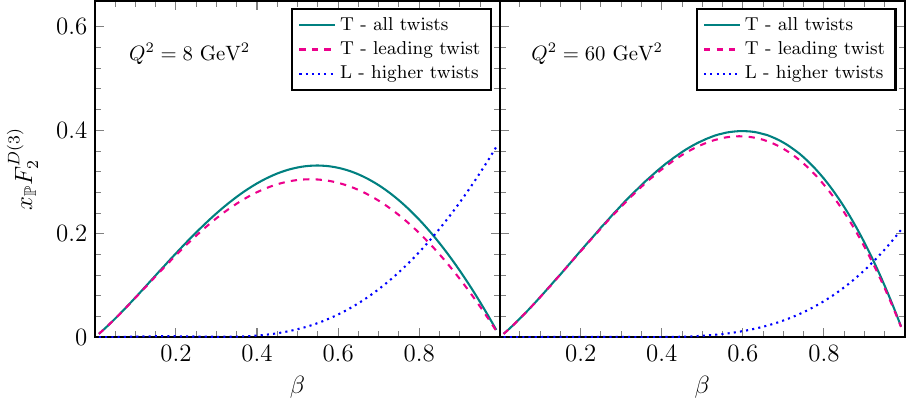}
	\end{center}
	\caption{\small Transverse (all twist and its leading twist approximation) and longitudinal contributions to the diffractive structure function $x_{\mathbb{P}} F_2^{D(3)}$ as functions of $\beta$ for two representative values of $Q^2$. The scattering amplitude is given by the MV model with a saturation scale $Q_{s}^2=0.88\,\mathrm{GeV}^2$ (in the fundamental representation). A factor $S_{\perp}/4 \pi^3$ has been omitted.}
\label{fig:three-int}
\end{figure}

To verify the quality of the leading twist approximation,  we compare in Fig.~\ref{fig:three-int} the numerical results 
for the diffractive structure function $x_{\mathbb{P}} F_2^{D(3)}$ 
 as obtained by integrating over $K_{\perp}^2$ the full (all-twist) result in  Eq.~\eqref{2jetSIDIS} and,
respectively, its LTA in \eqn{2jetLT}. We more precisely display the contribution of a single flavour of massless quarks to the
the diffractive structure function $x_{\mathbb{P}} F_2^{D(3)}$, cf.~\eqn{xqD} 
(and excluding an irrelevant prefactor $S_{\perp}/4\pi^3$).  
The full and LT results are seen to agree very well with each other and their (small) discrepancy decreases with increasing $Q^2$.  In the same figure, we also show the results for the corresponding contribution to $x_{\mathbb{P}} F_2^{D(3)}$ from the longitudinal sector. This is reviewed and studied in detail in Appendix \ref{sec:long} and in particular Eq.~\eqref{2jetSIDISL} shows that it is of higher twist order. As visible in Fig.~\ref{fig:three-int}, this contribution is indeed suppressed, but only 
for small or moderate values of $\beta$. However it dominates when  $1-\beta$ gets small, since it approaches a finite value when $\beta\to 1$, cf.~Eq.~\eqref{DF2L},  unlike the transverse contribution, which vanishes
 in that limit, cf.~\eqn{xqPsat}. 
We shall discuss more thoroughly the diffractive structure $x_{\mathbb{P}} F_2^{D(3)}$ in Sect.~\ref{sec:f2d3},
notably by including the DGLAP evolution of the (quark and gluon) DPDFs.

\comment{

\begin{figure}[t]\centerline{\includegraphics[width=0.5\textwidth]{TMD-3jets-softQ-ampl.pdf}\hspace*{0.8cm}
\includegraphics[width=0.36\textwidth]{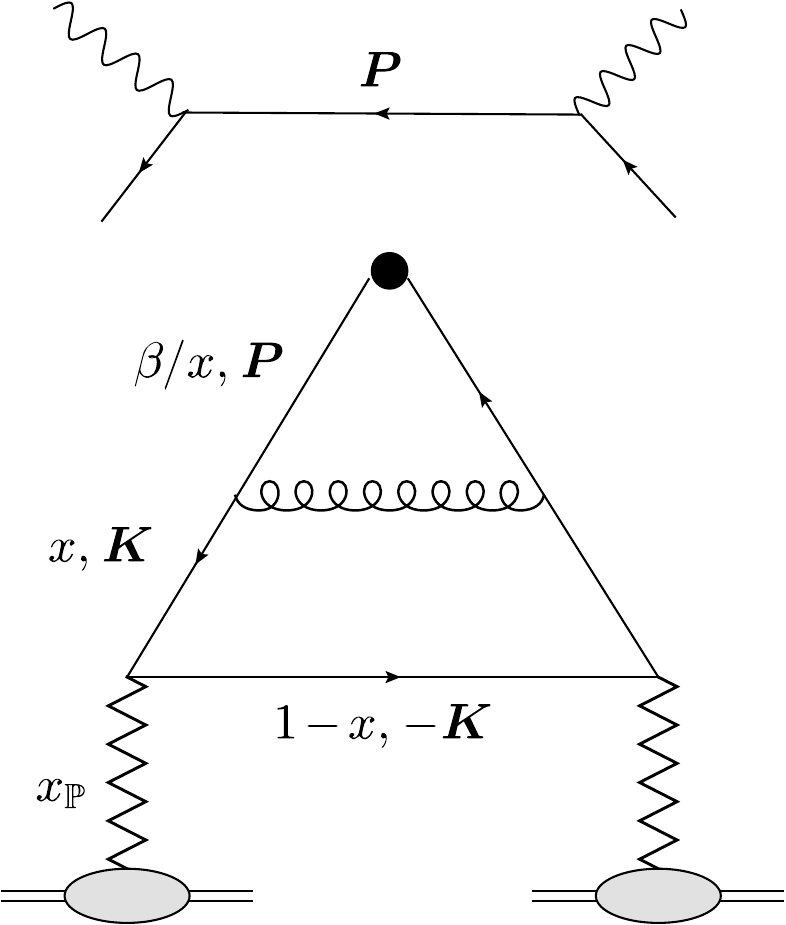}}
\caption{\small (a) Diffractive 2+1 jet production with a soft quark in the dipole picture. The gluon and the antiquark
are hard, $k_{3\perp}\simeq k_{2\perp}\simeq P_\perp$, and carry most of the photon longitudinal
momentum: $\vartheta_3+\vartheta_2\simeq 1$. The quark is semi-hard, $K_\perp\sim Q_s\ll P_\perp$,
and soft: $\vartheta_1\sim K_\perp^2/P_\perp^2 \ll 1$.
(b) TMD factorisation for the associated SIDIS process, in the regime where $Q^2\gg P_\perp^2$
and the hard dijets are asymmetric.}
 \label{fig:sidis-3jets-softQ}
\end{figure}

\subsection{Towards 2+1 diffractive jets}
\label{sec:disc}

We conclude this section with a few remarks on the structure of the diffractive final state, which also prepare the transition to
the topics to be addressed in the next sections. The previous discussion suggests that the hadron, or jet, that is typically
measured in diffractive SIDIS is necessarily semi-hard, $K_\perp\sim Q_s$, since produced by 
the Pomeron via elastic scattering. This is correct, but only so long as we stay at the level of the leading-order
approximation, that is, of exclusive $q\bar q$ production. However, this conclusion changes if one allows for next-to-leading order
(NLO) corrections, i.e.~for additional parton branchings. 

Consider e.g.~the process illustrated in Fig.~\ref{fig:sidis-3jets-softQ}: the $t$-channel
quark originally produced by the Pomeron with semi-hard $K_\perp\sim Q_s$ is now allowed to undergo
a $q\to qg$ branching, which can be hard --- as standard in the DGLAP evolution ---, thus producing a hard quark-gluon pair,
with relative transverse momentum $P_\perp\gg K_\perp$. The distribution of this pair in $P_\perp$ follows
the standard bremsstrahlung spectrum $\sim 1/P_\perp^2$. The hard quark from this pair can then be put on-shell
by the scattering with the photon and thus appear as a hard jet/hadron in the final state. The contribution of this channel
to the SIDIS cross-section is suppressed by a power of $\alpha_s$ compared to exclusive dijets, yet
it dominates over the latter at large transverse momenta $P_\perp\gg Q_s$, where it falls only like $1/P_\perp^2$.
By also integrating out the hard pair, one obtains a contribution to the diffractive structure function which is
enhanced by the large logarithm $\int\rmd P_\perp^2/P_\perp^2=\ln(Q^2/Q_s^2)$. This contribution is recognised
as one step in the DGLAP evolution of the leading-order result in \eqn{DF2}.
  
\begin{figure}[t]\centerline{\includegraphics[width=0.5\textwidth]{TMD-3jets-ampl.pdf}\hspace*{0.8cm}
\includegraphics[width=0.36\textwidth]{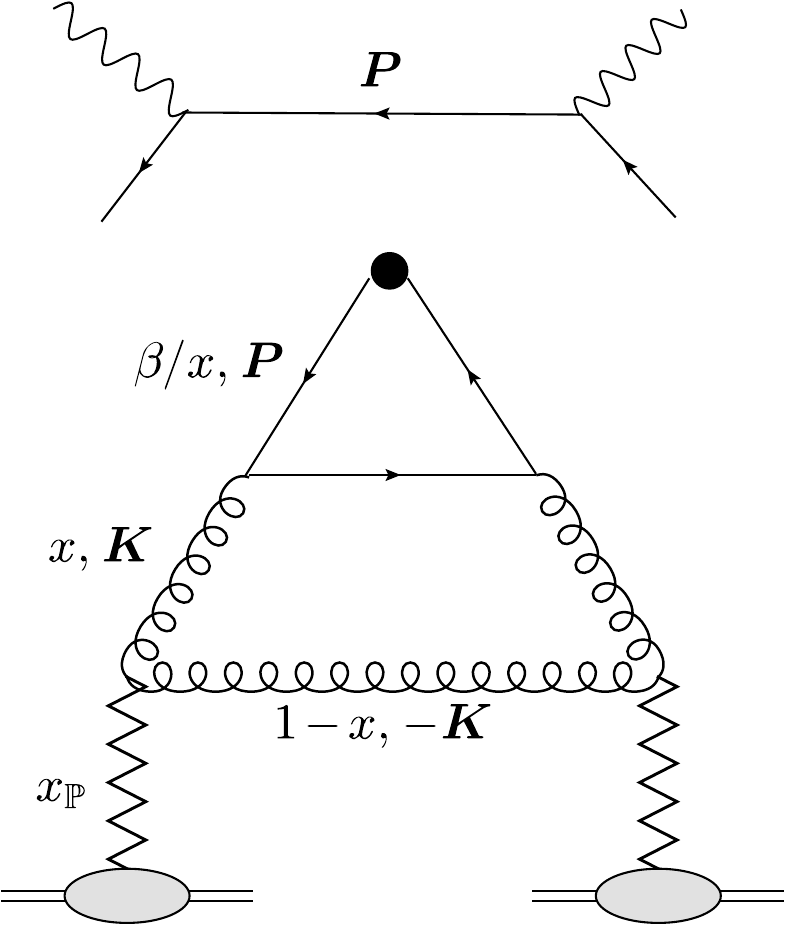}}
\caption{\small (a) Diffractive 2+1 jet production with a soft gluon in the dipole picture. The $q\bar q$ pair is hard,
$k_{1\perp}\simeq k_{2\perp}\simeq P_\perp$, and carries most of the photon longitudinal
momentum: $\vartheta_1+\vartheta_2\simeq 1$. The gluon is semi-hard, $K_\perp\sim Q_s\ll P_\perp$,
and soft: $\vartheta_3\sim K_\perp^2/P_\perp^2 \ll 1$.
(b) TMD factorisation for the associated SIDIS process, in the regime where
$Q^2\gg P_\perp^2$ and the hard dijets are asymmetric.}
 \label{fig:sidis-3jets}
\end{figure}

A similar discussion applies to the process illustrated in Fig.~\ref{fig:sidis-3jets}, where the Pomeron creates a semi-hard
gluon-gluon dipole and one of the gluons in that pair splits into a hard $q\bar q$ pair, which absorbs the virtual photon.
Once again, the hard $q\bar q$ pair has a  bremsstrahlung spectrum $1/P_\perp^2$, and the integral over $P_\perp$ 
yields a logarithmically-enhanced contribution to $F_2^D$ --- here generated via the DGLAP evolution of the gluon 
distribution of the Pomeron (the gluon DPDF).

These considerations, which corroborate the discussion of hard exclusive dijets in Sect.~\ref{sec:exclusive},
 illustrate the importance of higher-order diffractive processes, which can dominate over
the (formally) leading-order ones in some of the most interesting regions in phase-space. In the remaining part
of this paper, we shall be interested in diffractive processes like those shown in Fig.~\ref{fig:sidis-3jets-softQ} 
and Fig.~\ref{fig:sidis-3jets}, whose final state involves two hard jets accompanied by a semi-hard one
(``2+1 jets'').
Using the colour dipole picture, we shall compute the respective contributions to hard dijet production. 
By subsequently integrating our dijet cross-sections over the
kinematics of the hard dijets, one can deduce (large) contributions to diffractive SIDIS and to $F_2^D$.
}

\section{Diffractive (2+1)--jets with a soft quark}
\label{sec:2plus1}

The discussion in the previous section shows that the typical dijets produced via elastic scattering are {\it semi-hard},
with transverse momenta $K_\perp\sim Q_s(\YP)$, even when the virtual photon is much harder, $Q^2\gg Q_s^2(\YP)$.
The cross-section for producing much harder jets, with\footnote{For more clarity, from now on we shall reserve
the notation $K_\perp$ for transverse momenta which are comparable to $Q_s$, while $P_\perp$ will denote transverse
momenta which are much larger: $P_\perp\gg K_\perp\sim Q_s$.} $P_\perp\gg Q_s$,
decreases very fast, like $1/P_\perp^6$ (see  \eqn{crossqqhard} in Appendix~\ref{sec:single} 
for a succinct discussion and also App. A in Ref.~\cite{Iancu:2022lcw} for more details).  
For dijet production, this $1/P_\perp^6$ spectrum should be viewed as a higher-twist effect; for comparison, the inclusive
production of a pair of jets in DIS leads to the harder spectrum $1/P_\perp^4$  \cite{Dominguez:2011wm}.
The ultimate reason for the strong suppression of hard exclusive dijets  is the colour transparency of small dipoles: 
a hard $q\bar q$ pair has a small transverse size, hence it interacts only weakly.


However, all that is strictly true only so long as one stays at leading order in QCD perturbation theory, that is, if 
one limits oneself to the {\it exclusive} production of a quark-antiquark pair.  As originally observed in  \cite{Iancu:2021rup}, 
a pair of hard jets with relative transverse momentum $P_\perp\gg Q_s$ can be efficiently produced via coherent diffraction 
if one allows for the formation of a third, {\it semi-hard}, jet, with transverse momentum  $K_\perp\sim Q_s$. 
Such a ``(2+1)--jet'' process starts with a $q\bar q g$ fluctuation of the virtual photon, 
hence the respective cross-section is suppressed by a power of $\alpha_s$. 
Yet, the ensuing spectrum for the hard dijets decays only like $1/P_\perp^4$ and thus
dominates over the exclusive $q\bar q$ production  at large $P_\perp$. The emergence of this harder
spectrum is closely connected to the presence of the third jet in the final state, with $K_\perp\sim Q_s$:
because of the latter, the overall transverse size $R$ of the $q\bar q g$
fluctuation is quite large, $R\sim 1/Q_s$, despite the fact that two of the jets are hard. This allows for strong scattering
in the black disk limit and thus avoids the suppression due to colour transparency.

\begin{figure}
	\begin{center}
	\includegraphics[align=c,width=0.45\textwidth]{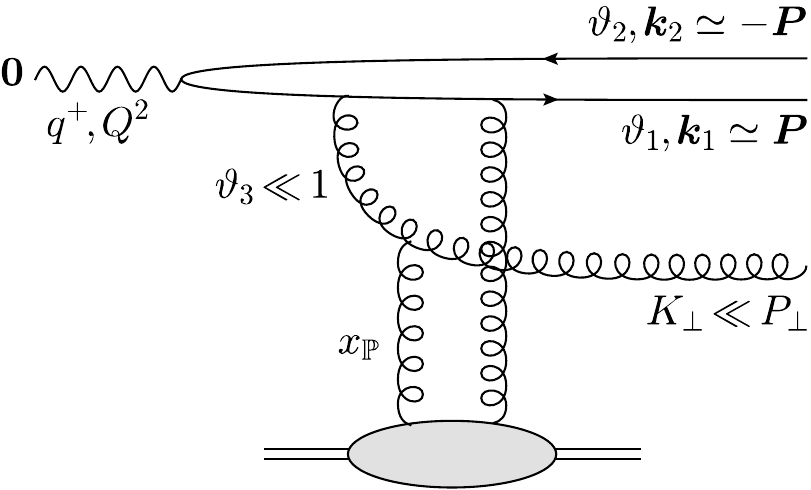}
	\hspace*{0.08\textwidth}
	\includegraphics[align=c,width=0.45\textwidth]{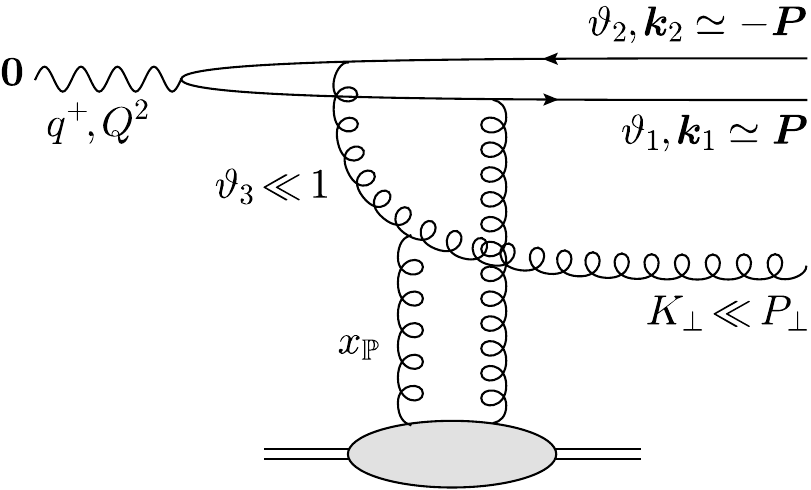}
	\end{center}
	\caption{\small Diffractive scattering with a final state composed of a hard $q\bar{q}$ dijet and a semi-hard gluon. The gluon is emitted either from the quark (left panel) or from the antiquark (right panel) and at a large
	transverse distance $R\sim 1/K_\perp$  from the small $q\bar{q}$ pair (with transverse size $r\sim 1/P_\perp$).}
\label{fig:3jets_gen}
\end{figure}

\begin{figure}
	\begin{center}
	\includegraphics[align=c,width=0.45\textwidth]{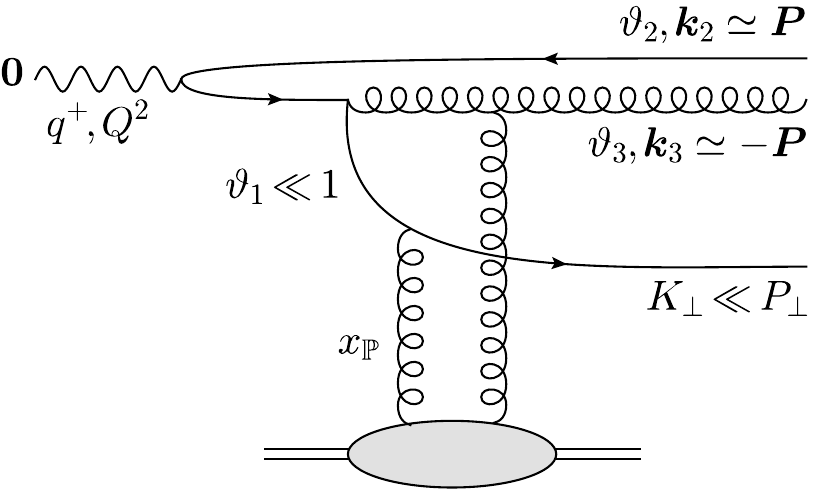}
	\hspace*{0.08\textwidth}
	\includegraphics[align=c,width=0.45\textwidth]{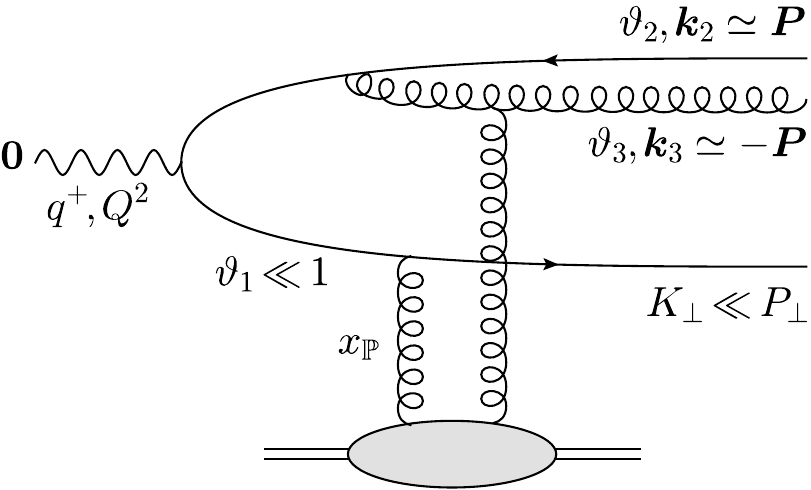}
	\end{center}
	\caption{\small Diffractive scattering with a final state composed of a hard $\bar{q}g$ dijet and a semi-hard quark. The gluon is emitted either from the quark (left panel) or from the antiquark (right panel). The quark is always far from the small $\bar{q}g$ dijet.}
\label{fig:3jets_gen_softq}
\end{figure}

The Feynman graphs in Figs.~\ref{fig:3jets_gen} and \ref{fig:3jets_gen_softq} illustrate the leading-order amplitudes for different final states with 2+1 ``jets'' (actually, partons). In Fig.~\ref{fig:3jets_gen} the final gluon is semi-hard, while the hard dijet is made with the quark and the antiquark ($q\bar q$). The total amplitude for this process is the sum of the two contributions, since the gluon can be emitted by the quark (left panel) or the antiquark (right panel). For the two other processes, as shown in Fig.~\ref{fig:3jets_gen_softq}, the final state includes a hard  antiquark-gluon ($\bar q g$) dijet together with a semi-hard quark. (There are of course similar processes where the semi-hard final ``jet'' is the antiquark.) Once again, the gluon can be emitted by either the quark, or the antiquark, but since the two fermions have widely different kinematics --- in the final state, one is hard, while the other one is semi-hard ---, the respective amplitudes have different properties 
and need be separately computed. These two amplitudes interfere with each other (since they lead to the same final state), hence they must be added before computing the cross-section.

Our drawing of the three amplitudes in Figs.~\ref{fig:3jets_gen} and \ref{fig:3jets_gen_softq} is meant to emphasise their space-time structure and, especially, their distribution in the transverse coordinate space, which in turn reflects the kinematics of the final state. In all these cases, one observes a large transverse separation $R\sim 1/K_\perp$ with, typically, $K_\perp\sim Q_s(\YP)$, between the semi-hard parton and the hard dijet, which is relatively compact, with transverse size $r\sim 1/P_\perp\ll 1/Q_s$. Since $1/Q_s$ is also the typical scale for transverse variations in the gluon distribution of the target, it should be clear that, in so far as the scattering is concerned, the small dijet can be effectively treated as a point-like particle in the appropriate representation of the colour group SU($N_c$). For instance, in the case of a semi-hard gluon, cf.~Fig.~\ref{fig:3jets_gen}, the hard $q\bar q$ pair scatters like an effective gluon and the amplitude is proportional to the elastic scattering amplitude $\mcal{T}_{g}(R, \YP)$ of a gluon-gluon dipole with large size $R\sim 1/Q_s$ (which is of order one, as anticipated). Similarly, for the diagrams with a soft quark in Fig.~\ref{fig:3jets_gen_softq}, one encounters an effective quark-antiquark dipole, with elastic scattering amplitude $\mcal{T}_{q\bar q}(R, \YP)\sim\order{1}$.

Notice that the photon virtuality $Q^2$ plays no special role in these arguments: so long as $P_\perp\gg Q_s$,
the dijet is hard irrespective of the photon virtuality and our subsequent analysis holds for arbitrary values of $Q^2$
(including  the photo-production limit $Q^2\to 0$). Yet, for the sake of the discussion, we shall often treat 
$Q^2$ as a hard scale which is comparable to $P_\perp^2$.

The (2+1)--jet process with a semi-hard gluon has been discussed in detail in \cite{Iancu:2021rup,Iancu:2022lcw} (see also Refs.
\cite{Wusthoff:1997fz,GolecBiernat:1999qd,Hebecker:1997gp,Buchmuller:1998jv,Hautmann:1998xn,Hautmann:1999ui,
Hautmann:2000pw,Golec-Biernat:2001gyl} for earlier related work), with results that will be briefly summarised in the next subsection.
After that, we will present the corresponding analysis for the processes involving a semi-hard quark.
Before we proceed, let us specify our notations and conventions.
We shall use the subscripts 1, 2, and 3 to label the momenta of the quark, antiquark, and gluon
respectively. The ``plus'' components of their 4-momenta will be written as $k_i^+=\vartheta_i q^+$, with
$\vartheta_1+\vartheta_2+\vartheta_3=1$. For coherent diffraction, we can neglect the transverse momentum 
transferred by the target, which implies $\bk_1+\bk_2+\bk_3=0$. Two of these jets are hard, while the third
one is semi-hard. By momentum conservation,  the hard dijets propagate back-to-back in the transverse plane: their 
relative transverse momentum $P_\perp$ is much larger than their momentum imbalance $K_\perp$,
which is fixed by the third jet. For instance, for the process with
a semi-hard gluon in Fig.~\ref{fig:3jets_gen}, we have $P_\perp\gg K_\perp$, where
$P_\perp\simeq k_{1\perp}\simeq k_{2\perp}$ and  $K_\perp \equiv |\bk_1+\bk_2|=k_{3\perp}$.

 A final point, which is important for what
follows: besides being ``semi-hard'', in the sense of having a transverse momentum $K_\perp\sim Q_s(\YP)$, 
the third jet is also {\it soft}, in the sense of carrying a small fraction $\vartheta \sim K_\perp^2/P_\perp^2\ll 1$ of the longitudinal 
momentum $q^+$ of the incoming photon. This property follows from formation time constraints, and more precisely
from the condition to have a large effective dipole at the time of scattering
(see below for details). Because of this, it becomes possible to factorise out this semi-hard
parton from the hard dijet and to reinterpret it as a part of the target wavefunction ---  very much like the factorisation
of the soft quark from the $q\bar q$ pair in the calculation of SIDIS in the previous section.
This factorisation has been demonstrated in Refs.~\cite{Iancu:2021rup,Iancu:2022lcw} for (2+1)--jets with
a soft gluon and will be extended below to the corresponding processes with a soft fermion. 
To summarise, the third jet is both semi-hard and soft, and will often be referred
to as  ``the soft jet'' (or ``the semi-hard jet''), for brevity.

%

\subsection{2+1 jets with a soft gluon: a brief review}
\label{sec:softg}

This case is illustrated by the amplitude in Fig.~\ref{fig:3jets_gen}: the quark-antiquark pair is hard and nearly back-to-back, $k_{1\perp}\simeq k_{2\perp} \gg Q_s(\YP)$, whereas the gluon is semi-hard, $k_{3\perp}\sim  Q_s(\YP)$, and also soft: $\vartheta_3\ll 1$. As already mentioned, this particular case has been studied in great detail in \cite{Iancu:2021rup,Iancu:2022lcw},
from which we quote here the final results. We recall that there
are only two  independent momentum variables, conveniently chosen as
\begin{align}
	\label{PandK}
	\bm{P} \equiv
	\frac{\vartheta_1 \bm{k}_2 - \vartheta_2 \bm{k}_1}
	{\vartheta_1 +\vartheta_2}
	\simeq 
	\vartheta_1 \bm{k}_2 -
	\vartheta_2 \bm{k}_1
	\,,\qquad
	\bm{K} \equiv \bm{k}_1 + \bm{k}_2,
\end{align}
where we have also used $\vartheta_3\ll 1$, hence $\vartheta_1 +\vartheta_2\simeq 1$. 
Physically,   $\bP$ is the relative  momentum  of the hard $q\bar q$ pair, $\bK$ is
their imbalance, and $\bk_3=-\bK$.

To start with, let us clarify the physical origin of the softness condition
$\vartheta_3\ll 1$. In order for the scattering to be strong, the gluon must be emitted before
the collision with the nuclear target. That is, its formation time $\tau_g\simeq 2\vartheta_3q^+/k_{3\perp}^2$ must
be smaller than the photon coherence time $ \tau_{\gamma}=2q^+/Q^2$ 
(the lifetime of the $q\bar q$ fluctuation in the absence of scattering). This is satisfied provided 
$\vartheta_3\lesssim K_\perp^2/Q^2$, which is indeed small for the interesting values
$Q^2\gg K_\perp^2\sim Q_s^2$.  A similar condition holds when the photon is real ($Q^2\to 0$):
 in that case the lifetime of the $q\bar q$ pair is estimated as
 $\tau_{q\bar q}\simeq 2\vartheta_1\vartheta_2 q^+/P_\perp^2$ and the softness condition 
becomes $\vartheta_3\lesssim  K_\perp^2/P_\perp^2$. (We generally treat
$\vartheta_1$ and $\vartheta_2$ as comparable quantities of order 1/2, but asymmetric configurations
with $\vartheta_1\vartheta_2\ll 1$ will be interesting in special cases.)

 Yet, when computing the amplitude, it is not always possible to work in the lowest order 
approximation in $\vartheta_3$  (e.g.~one cannot treat the gluon emission vertex in the eikonal approximation): 
the subleading corrections become important for the largest allowed values  $\vartheta_3\sim K_\perp^2/P_\perp^2$,
which are the most interesting in practice \cite{Iancu:2022lcw}. This is related to the fact that the would-be 
leading-order contributions in the double limit  $\vartheta_3\to 0$ and $K_\perp\to 0$ precisely
cancel between the two amplitudes representing gluon emission from the quark and, respectively, the antiquark.
So, one needs to keep terms of next-to-leading order in either  $\vartheta_3$, or $K_\perp/P_\perp$,
and the respective contributions are comparable to each other for the interesting values
$\vartheta_3\sim K_\perp^2/P_\perp^2$ (see Sect.~4 in \cite{Iancu:2022lcw} for details).

This argument also has important consequences for the new analysis in this paper: it explains why
the diffractive production of (2+1)--jets with a soft quark is not suppressed compared to the respective
process with a soft gluon, despite the fact that the latter is {\it a priori} favoured by the soft
singularity of bremsstrahlung: this enhancement is in practice compensated by the dipolar nature
of the soft gluon radiation, i.e.~by the fact that the gluon can be emitted by either the quark, or the
antiquark, and these two emitters partially screen each other when seen from large distances.
These compensation is fully effective only when $\vartheta_3\sim K_\perp^2/P_\perp^2$.
The (2+1)--processes which involve much
softer gluons with $\vartheta_3\ll K_\perp^2/P_\perp^2$ are still dominant, but they are less interesting
since they have lower rapidity gaps. To understand this, let us examine the rapidity gap
$\YP=\ln (1/x_{\mathbb P})$ for the $q\bar q g$ final state: one can write\footnote{Notice
the identity $ M_{q\bar q g}^2   \equiv (k_1+k_2+k_3)^2 =\frac{k^2_{1\perp}}{\vartheta_1} +\frac{k^2_{2\perp}}{\vartheta_2} 
 +\frac{k^2_{3\perp}}{\vartheta_3} -(\bk_1+\bk_2+\bk_3)^2$.}
 (cf.~\eqn{xP})
 \beq\label{xgdef}
 x_{\mathbb{P}}=\frac{1}{2q^+ P_N^-}\left(\frac{k^2_{1\perp}}{\vartheta_1} +\frac{k^2_{2\perp}}{\vartheta_2} 
 +\frac{k^2_{3\perp}}{\vartheta_3} +Q^2\right)
 \ge x_{q\bar q} ,\eeq
where  $x_{q\bar q} P_N^-$ is the light-cone energy needed to put the hard $q\bar q$ pair on-shell:
\beq\label{xqqdef}
 x_{q\bar q} =\frac{1}{2q^+P_N^-}\left(\frac{k^2_{1\perp}}{\vartheta_1} +\frac{k^2_{2\perp}}{\vartheta_2} 
  +Q^2\right)\,\simeq\, \frac{1}{\hat s}\left(\frac{P_\perp^2}{\vartheta_1\vartheta_2}+ Q^2\right)\,.
\eeq
The quantity $Y_{q\bar q} \equiv \ln(1/x_{q\bar q})$ --- that is, the rapidity separation between the 
$q\bar q$ dijet and the target --- provides an upper limit on the rapidity gap
$\YP$. So long as $\vartheta_3\sim K_\perp^2/P_\perp^2$, $x_{\mathbb P}$ and  $x_{q\bar q}$
are comparable with each other, hence $\YP$ is close to its maximal value  $Y_{q\bar q}$. But for much
softer gluons with  $\vartheta_3\ll K_\perp^2/P_\perp^2$, the  rapidity gap $\YP$ is considerably smaller
than $Y_{q\bar q}$, and then the process is less interesting (e.g.~it is less favourable to gluon saturation).
In fact, such very soft gluons should more naturally be seen as part of the high-energy evolution
of the projectile \cite{Kovchegov:1999ji,Hatta:2006hs}.

\begin{figure}
	\begin{center}
		\includegraphics[align=c,width=0.59\textwidth]{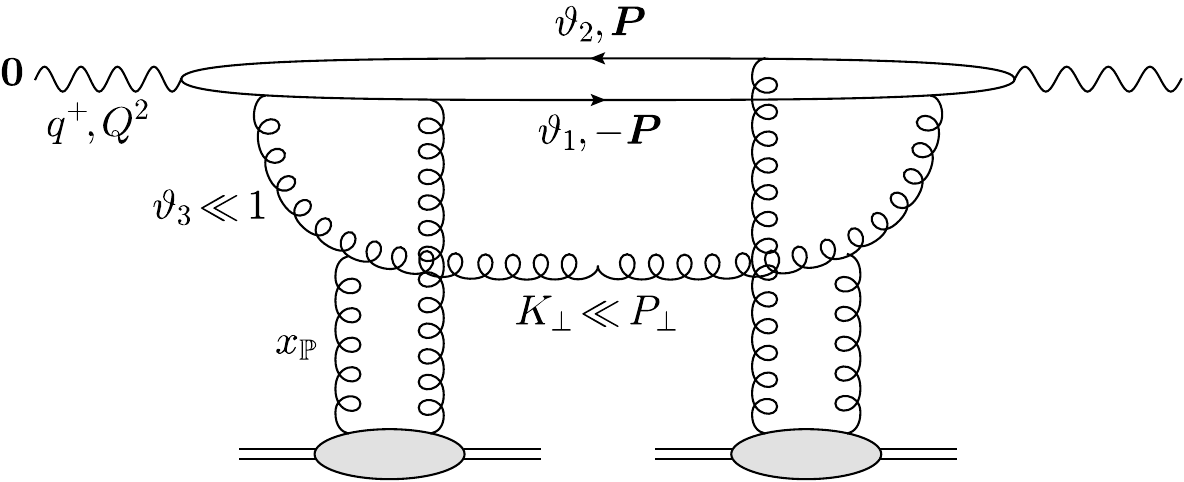}
		\hspace*{0.04\textwidth}
		\includegraphics[align=c,width=0.35\textwidth]{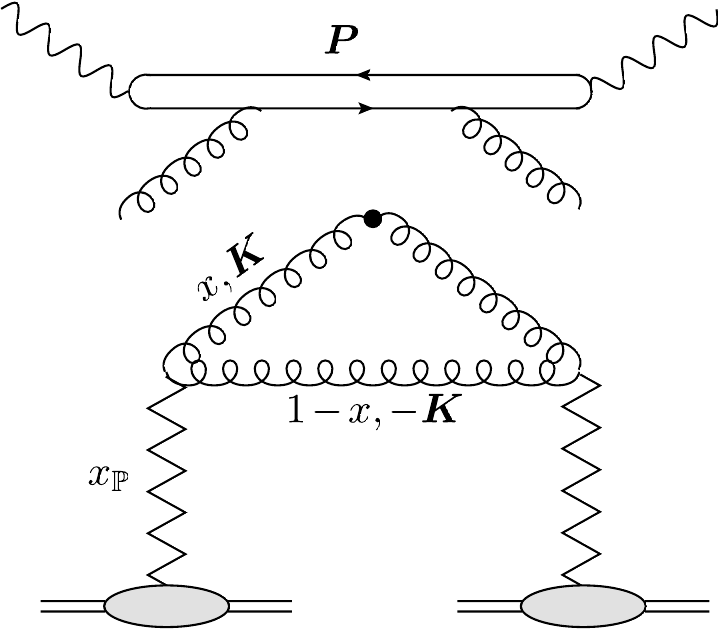}
		\end{center}
	\caption{\small Diagrams representing the cross-section for the diffractive production of (2+1)--jets with a soft gluon. Left panel: the viewpoint of the colour dipole picture. Right panel: the TMD factorisation of the soft gluon from the hard $q\bar q$ pair.}
\label{fig:TMD-softg}
\end{figure}

The strong separation of scales, both transverse and longitudinal, between the semi-hard gluon and 
the hard $q\bar q$ pair 
leads to important simplifications which allows one to write the cross-section
for diffractive (2+1)--jet  production in the TMD factorised form \cite{Iancu:2022lcw}
\begin{align}
	\label{3jetsD1}
	\frac{\rmd \sigma
	^{\gamma_{T,L}^* A
	\rightarrow q\bar q (g) A}}
  	{\rmd \vartheta_1
  	\rmd \vartheta_2
  	\rmd^{2}\!\bm{P}
  	\rmd^{2}\!\bm{K}
  	\rmd Y_{\mathbb P}} = 
  	 \tilde{H}_{T,L}(\vartheta_1,\vartheta_2, {Q}^2, P_{\perp}^2)\,
  	 \frac{\dif xG_{\mathbb{P}}(x, x_{\mathbb{P}}, K_\perp^2)}
  	 {\dif^2\bm{K}},
 \end{align}
 which is further illustrated in the right panel in Fig.~\ref{fig:TMD-softg}. The final state is indicated as $ q\bar q (g) A$ with the gluon label shown in parentheses, to emphasise that this is the soft parton in the final state. 
 The ``hard factors'' have been defined as (with $\bar{Q}^2=\vartheta_1\vartheta_2{Q}^2$)
 \begin{align}
 	\label{HardT}
	\tilde{H}_T(\vartheta_1,\vartheta_2, {Q}^2, P_{\perp}^2)
	\equiv 
	{\alpha_{em}\alpha_s}
	\Big(\sum e_{f}^{2}\Big) \,
	\delta(1-\vartheta_1-\vartheta_2)  
	\left(\vartheta_1^{2} + 
	\vartheta_2^{2}\right)
	\frac{P_{\perp}^4 + \bar{Q}^4}
	{(P_{\perp}^2 + \bar{Q}^2)^4}\,
 \end{align}
 for a photon with transverse polarisation and,  respectively,
\begin{align}
	\label{HardL}
	\tilde{H}_L(\vartheta_1,\vartheta_2,{Q}^2, P_{\perp}^2)
	\equiv 
	{\alpha_{em}\alpha_s}\Big(\sum e_{f}^{2}\Big)\,
	\delta(1-\vartheta_1-\vartheta_2)  
	\vartheta_1 \vartheta_2 \,
	\frac{8P_{\perp}^2 \bar{Q}^2}{(P_{\perp}^2 + \bar{Q}^2)^4}\,
 \end{align}
 for the case of a longitudinal photon.  If $P_{\perp}^2$ and $\bar{Q}^2$ are of the same order, then,
 clearly, both $\tilde{H}_T$ and $\tilde{H}_L$ decrease like $1/P_\perp^4$, as anticipated. On the other hand,
 if $P_\perp^2\gg \bar Q^2$, then $\tilde{H}_T$ is still proportional to $1/P_\perp^4$, but $\tilde{H}_L$ decreases faster,
 like $\bar Q^2/P_\perp^6$, and thus becomes negligible. In particular, the longitudinal contribution vanishes
 for a real photon ($Q^2\to 0$), unlike $\tilde{H}_T$, which has a finite value in that limit.

 Furthermore, the ``semi-hard factor'' in \eqn{3jetsD1} is the gluon diffractive TMD,  
 defined as (compare to \eqn{qDTMD})
 \begin{align}
	\label{pomugddef}	
	\frac{\rmd xG_{\mathbb{P}}
	(x, x_{\mathbb{P}}, K_{\perp}^2)}
	{\rmd^2 \bK} \equiv 
	\frac{S_\perp (N_c^2-1)}{4\pi^3}\,  \frac{
	[\mcal{G}_{\mathbb{P}}(x,x_{\mathbb P}, K_{\perp}^2)]^2}{2\pi(1-x)} 
	\equiv \frac{S_\perp (N_c^2-1)}{4\pi^3}\,  
	\Phi_{\mathbb P}(x,x_{\mathbb P},K_{\perp}^2)
 \end{align}
with (below,  $\mcal{M}^2=\frac{x}{1-x}K_\perp^2$)
\begin{align}
	\label{Gscalarnew}
	\mcal{G}_{\mathbb{P}}(x,x_{\mathbb P}, K_{\perp}^2)=
	\mcal{M}^2 \int_0^\infty 
	\rmd R\, R\, 
	J_2(K_{\perp} R) 
	K_2(\mcal{M} R) \mcal{T}_g(R,Y_{\mathbb P}).
\end{align}

The hard factors describe the formation of the $q\bar q$ pair and its coupling to a $t$-channel gluon
produced by the Pomeron. The gluon DTMD represents the  unintegrated gluon distribution of the Pomeron,
i.e.~the number of gluons in the Pomeron wavefunction which have a transverse momentum $K_\perp$ and
a ``minus'' longitudinal momentum fraction $x$ w.r.t.~the Pomeron. In particular, $\Phi_{\mathbb P}$ denotes
the gluon occupation number in the Pomeron. 

When computing the cross-section \eqref{3jetsD1}, $x$ is not an independent variable. Rather, its
value is fixed by the diffractive gap together with the kinematics of the hard dijet, via
(recall \eqn{xqqdef})
     \beq
 x= \frac{x_{q\bar q}}{x_{\mathbb P}}\,=\,\beta\, \frac{x_{q\bar q}}{\xbj}
 \,\simeq\,\beta\, \frac{\bar Q^2+P_\perp^2}{\bar Q^2}\,.\eeq
  
 To get more insight into  the gluon distribution of the Pomeron, it is useful to have a piecewise approximation
 to the gluon occupation number, like  \eqn{qDTMDpw} for the quark occupation number. 
 Via manipulations similar to those in Sect.~\ref{sec:MS},
 one finds  \cite{Iancu:2021rup,Iancu:2022lcw}
\begin{align}\label{Philimits}
\Phi_{\mathbb{P}}(x, x_{\mathbb P}, K_\perp^2)\,\simeq \frac{1-x}{2\pi}
    \begin{cases}    \displaystyle{\ 1}
         &
        \text{for \ $K_\perp\!\ll \tilde Q_{s,g}(x, \YP)$}
                  \\*[0.4cm]
               \displaystyle{\ \frac{\tilde Q_{s,g}^4(x, \YP)}{K_\perp^4}} &
        \text{for \ $K_\perp\!\gg \tilde Q_{s,g}(x, \YP)$}\,.
         \end{cases}
\end{align}
with $\tilde Q_{s,g}^2(x, \YP)=(1-x) Q_{s,g}^2(\YP)$, where $Q_{s,g}(\YP)$ is the value of the saturation momentum
probed by a gluon-gluon dipole. It is approximately related to the corresponding value for a $q\bar q$
dipole via the ratio of the respective Casimir factors: $Q_{s,g}^2(\YP) \simeq (N_c/C_F)Q_{s}^2(\YP)\simeq 2Q_{s}^2(\YP)$. 

\eqn{Philimits} is similar to \eqn{qDTMDpw} in the sense that they both show saturation
at low momenta and a rapidly decaying tail $\propto 1/K_\perp^4$ at high momenta. Note however the different
dependence upon $x$: the gluon occupation number is non-zero when $x\to 0$, but it vanishes faster when 
$x\to 1$.

If one is not interested in measuring the transverse momentum imbalance of the hard dijets (hence, neither
the momentum $K_\perp$ of the gluon jet), then one can integrate the cross-section \eqref{3jetsD1} over $K_\perp$
up to an upper limit of order $P_\perp$. Clearly, this operation produces the gluon distribution of the Pomeron
(a.k.a. the gluon diffractive PDF) on the resolution scale $\sim P_\perp^2$. One finds
\begin{align}
	\label{gluondipColl}
	\frac{\rmd \sigma
	^{\gamma_{T,L}^* A
	\rightarrow q\bar q (g) A}}
  	{\rmd \vartheta_1
  	\rmd \vartheta_2
  	\rmd^{2}\!\bm{P}
  	\rmd Y_{\mathbb P}} = 
  	 \tilde{H}_{T,L}(\vartheta_1, \vartheta_2, {Q}^2, P_{\perp}^2)\,
  	 xG_{\mathbb{P}}(x, x_{\mathbb{P}}, (1-x)P_\perp^2),
 \end{align}
with the following approximation for the gluon DPDF:
\begin{align}
	\label{xGP}	
	xG_{\mathbb{P}}(x, x_{\mathbb{P}}, Q^2)
	\equiv \int
	\dif^2\bK\,
	\frac{\rmd xG_{\mathbb{P}}(x, x_{\mathbb{P}}, K_{\perp}^2)}
	{\rmd^2 \bK} 
	= 
	\frac{S_\perp (N_c^2-1)}{4\pi^2} 
	\int_0^{Q^2}\!
	\dif K_\perp^2\, 
	\Phi_{\mathbb P}(x,x_{\mathbb P},K_{\perp}^2).
 \end{align}
 The precise value of the resolution scale which appears in \eqn{gluondipColl},
 namely $(1-x)P_\perp^2$,  will be justified  later (see the discussion after \eqn{crosstotalint}), 
 but it is not very important at this point: given the rapid decay of the integrand in \eqref{xGP}	
  at large momenta $K_\perp \gg \tilde Q_{s,g}(x, \YP)$, cf.~\eqn{Philimits},  it is clear that
 the result of the integral is quasi-independent of its upper limit $Q^2$ so long as 
 $Q^2\gg \tilde Q_{s,g}^2(x, \YP)$. 
  
 Using \eqref{Philimits}, one finds a simple
approximation (compare to \eqn{xqPsat}),
\begin{align}
	\label{xGPhigh}
	xG_{\mathbb{P}}(x, x_{\mathbb{P}}, Q^2)
	=
	\frac{S_\perp (N_c^2-1)}{4\pi^3}\,
	\kappa_g
	(1-x)^2{Q}_{s,g}^2(Y_ {\mathbb P}),
\end{align}
with $\kappa_g\sim \order{1}$,  which can be useful for parametric estimates.  

When the resolution scale $Q^2$ in \eqn{xGP}	is so large 
that $\alpha_s\ln(Q^2/Q_{s,g}^2(\YP))\sim 1$, one expects the present approximation
to be modified by the effects of the DGLAP evolution. 
We shall return to this point after  computing the cross-sections
for the diffractive (2+1)--jet processes with a soft quark.   

%

\subsection{2+1 jets with a soft quark: gluon emission from the antiquark}
\label{sec:qbarsource}

As mentioned at the beginning of this section, we consider the configurations illustrated  
in Fig.~\ref{fig:3jets_gen_softq}, where the final state includes a hard 
antiquark-gluon ($\bar q g$) dijet together with a semi-hard quark ($q$). We shall separately compute
the amplitudes corresponding to these two configurations, starting with the case
where the gluon is emitted by the antiquark. As suggested by the graphical representation of the respective amplitude in the right panel in Fig.~\ref{fig:3jets_gen_softq}, this is
the case where the $q\bar q$ dipole produced by the decay of the virtual photon is relatively large,
with a typical transverse size $R\sim 1/Q_s$.

Before specialising to that case, we briefly
review the calculation of the three-parton ($q\bar q g$) component of the light-cone wavefunction 
 (LCWF) of the virtual photon.
The LCWF is measured at LC time $x^+ \to \infty$, whereas the scattering with the nuclear shockwave 
(SW) is localised near $x^+=0$. There are three possible time orderings: 
\texttt{(1)} both the photon decay and the gluon emission occur at negative $x^+$ (before the scattering with the SW), 
\texttt{(2)} the photon decays before crossing the SW ($x^+<0$) , but the gluon is emitted only after ($x^+>0$), and
 \texttt{(3)} the photon decays after crossing the SW ($x^+>0$). 
In case \texttt{(1)}, all the three partons can interact with the target. 
In case \texttt{(2)}, it is only the initial $q\bar q$ fluctuation which can interact.
Finally, in  case \texttt{(3)}, there is no interaction at all. 

In the absence of any scattering (that is, if there is no SW at $x^+=0$), the three
contributions to the LCWF must add to zero: indeed, a space-like (or even light-like) photon cannot decay into an on-shell partonic
system, by energy-momentum conservation. Thus, when computing the full LCWF (including the scattering with the SW),
it suffices to consider the first two time-orderings above; the third case can be simply accounted for by 
subtracting the no-scattering limits from the two other cases. We would like to point out that in addition to the aforementioned possibilities, according to the rules of LCPT one should also take into account diagrams with instantaneous interactions. In general, the latter contribute to the LCWF under consideration, however they are suppressed in the particular kinematics we are interested in, as we show in Appendix \ref{sec:inst}.

 As anticipated, we are interested in three-parton configurations which are quite special:
 a hard antiquark-gluon pair accompanied by a quark which is both soft ($\vartheta_1 \ll 1$) and semi-hard ($k_{1\perp}\sim Q_s$).
 This special kinematics allows for important simplifications, that are most conveniently implemented by working in the momentum space
 representation. On the other hand, the (multiple) scattering with the nuclear target is easier to include in the transverse coordinate representation, where the eikonal approximation takes a simple form. Accordingly, 
 our strategy in this section will be as follows. We shall first compute the momentum-space $q\bar q g$ components 
 of the photon LCWF corresponding to the first two time-orderings above and in the absence of scattering. 
 We shall then identify and implement the relevant kinematical approximations, as permitted
 by the special structure of the final state. Finally, we shall change representation (from  transverse momenta to transverse coordinates)
 via a Fourier transform and include the effects of the scattering in the eikonal approximation.
 
The general structure of the $q\bar{q}g$ component of the virtual photon LCWF
 in the absence of scattering and in momentum space reads as follows (compare to the $q\bar q$ component 
 in \eqn{qqmom})
\begin{align}
	\label{qqgmom}
	\big|\gamma_{\scriptscriptstyle T}^{i}\big\rangle_{q\bar{q}g} 
	= \,t^a_{\alpha\beta}
	\int_{0}^{1} & \rmd
	\vartheta_1 \rmd \vartheta_2 \, \rmd \vartheta_3 \,
	\delta(1 - \vartheta_1 - \vartheta_2 -\vartheta_3)
	\int \rmd^{2}\bm{k}_1\,  
	\rmd^{2}\bm{k}_{2}\,
	\rmd^{2}\bm{k}_{3}\,
	\delta^{(2)}(\bm{k}_1 +\bm{k}_2+\bk_3)
	\nn 
	& \times
	\Psi^{im}_{\lambda_{1}\lambda_{2}}(\vartheta_1,\bk_1,
	\vartheta_2,\bk_2,
	\vartheta_3,\bk_3)\,
	\big|{q}_{\lambda_{1}}^{\alpha}(\vartheta_1, \bk_1)\,
	\bar{q}_{\lambda_{2}}^{\beta}(\vartheta_2, \bk_2)\,
	g_m^a(\vartheta_3, \bk_3)
	\big\rangle.
	\end{align}
The notation related to the two fermions is the same as the one used in the previous section starting from Eq.~\eqref{qqmom}. In addition, $\bk_3$, $\vartheta_3$, $a$ and $m=1,2$ stand respectively for the transverse momentum, longitudinal fraction (w.r.t.~the virtual photon), color index and polarisation state of the transverse gluon. The gluon can be emitted by either the antiquark, or the quark, and its emission
can occur either before, or after, the scattering with the SW.  Accordingly, the probability amplitude $\Psi^{im}_{\lambda_{1}\lambda_{2}}$ receives several contributions, that will be computed in what follows. 
Similar calculations can be found in the literature, so in practice we shall simply adapt the results of
Ref.~\cite{Iancu:2022gpw} to the kinematical conditions of interest.

\begin{figure}
	\begin{center}
		\includegraphics[align=c,width=0.44\textwidth]{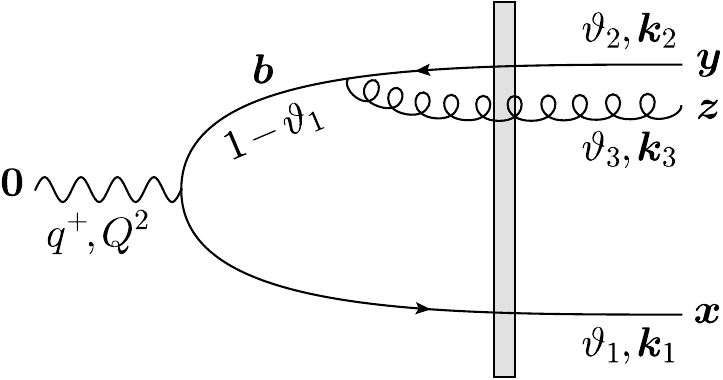}
		\hspace*{0.08\textwidth}
		\includegraphics[align=c,width=0.44\textwidth]{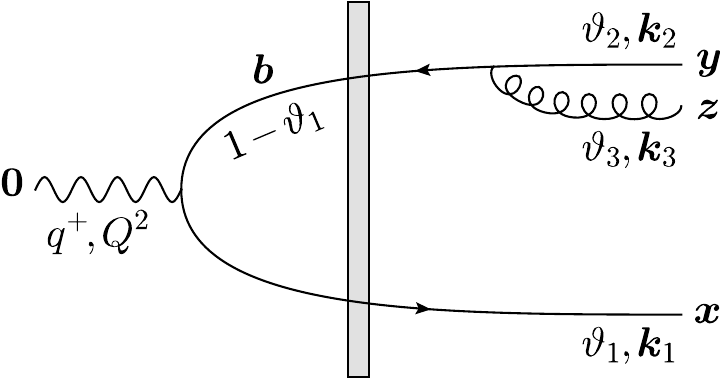}
	\end{center}
	\caption{\small The amplitude for (2+1)-jet production in DIS off a nuclear shockwave via the channel $\bar{q}'\to \bar{q}g$. The large pair is formed already at the photon vertex. Left panel: The gluon is radiated before the scattering off the shockwave. Right panel: The gluon is radiated after the scattering off the shockwave. }
\label{fig:fromqbar}
\end{figure}

In this section we consider the case where the gluon is emitted by the antiquark, so we shall insert a  lower label ``($\bar{q}$)''
on the respective components of the amplitude. There are two possible time-orderings (with time $t\equiv x^+$),
see Fig.~\ref{fig:fromqbar}:
\texttt{(1)} the photon decays at some time $ t_1 < 0$ and the gluon is then emitted at a time $t_2$ which
is negative as well, $t_1 < t_2 < 0$, and \texttt{(2)} the photon decays at time $t_1 < 0$, but the gluon 
is emitted at some positive time $t_2 > 0$. The integrals over $t_1$ and $t_2$ generate the
energy denominators, which are different in the two cases  \cite{Iancu:2022gpw}.
  
When the gluon is emitted prior to the scattering ($t_1 < t_2 < 0$), the amplitude in momentum space and in absence of
the scattering takes the following form   \cite{Iancu:2022gpw}.
\begin{align}
	\label{Psi1} 
	\Psi^{im\,(1)}_{\lambda_1\lambda_2\,(\bar{q})}=
	\delta _{\lambda_{1}\lambda_2}\,
	\frac{1}{8q^+}\,
	\frac{e e_f g }{(2\pi)^6}\,
	\frac{1}{\vartheta_1(1-\vartheta_1)\vartheta_2\vartheta_3}\,
	\frac{1}{\sqrt{\vartheta_3}}\,
	\frac{\varphi^{ij}(\vartheta_1,\lambda_1)\,k_1^j }
	{E_q+E_{\bar q'}-E_\gamma}\,
	\frac{\tau^{mn}(\vartheta_1,\vartheta_2,\lambda_1)P^n}
	{E_q+E_{\bar q}+E_g -E_\gamma}.
\end{align}
The LC energies in the denominators in Eq.~\eqref{Psi1} read 
\begin{align}
	\label{LCEs1}
	E_q=\frac{k_{1\perp}^2}{2\vartheta_1 q^+},
	\quad
	E_{\bar q}=\frac{k_{2\perp}^2}{2\vartheta_2 q^+},
	\quad
	E_g=\frac{k_{3\perp}^2}{2\vartheta_3 q^+},
	\quad  	
	E_{\bar q'} = \frac{k_{1\perp}^2}{2(1-\vartheta_1) q^+},
	\quad 
	E_\gamma=-\frac{Q^2}{2q^+},
\end{align}
where we have denoted with a prime the antiquark in the intermediate state (before the gluon emission).
The first energy denominator is generated by the photon decay,
while the second one corresponds to the gluon emission.
The function $\varphi^{ij}(\vartheta,\lambda)$ has been defined in Eq.~\eqref{phidef}, while the new function
\begin{align}
	\label{taudef}
	\tau^{mn}(\vartheta_1,\vartheta_2,\lambda)
	\equiv
	(1-\vartheta_1+\vartheta_2)
	\delta^{mn}
	+2i \lambda\vartheta_3
	\varepsilon^{mn}
\end{align}
with $\vartheta_3 = 1 \!-\! \vartheta_1 \!-\! \vartheta_2$, arises from the polarization structure of the $\bar{q}\to\bar{q}g$ vertex. The latter is also proportional to the relative transverse velocity of the two daughter partons, that is
\begin{align}
	\label{vrelqg}
	\frac{1}{\vartheta_2 +\vartheta_3}
	\left(\frac{\bk_2}{\vartheta_2}-
	\frac{\bk_3}{\vartheta_3}\right) =\,
	\frac{\bP}{\vartheta_2\vartheta_3}
	\,\Longleftrightarrow\,	
	\bm{P} \equiv 
	\frac{\vartheta_3 \bm{k}_2 -\vartheta_2 \bm{k}_3}
	{\vartheta_2 +\vartheta_3},
\end{align}
where $\bP$ is the relative momentum of the $\bar{q}g$ pair, which also appears in \eqn{Psi1}. For what follows,
it is convenient to choose $\bP$ and $\bk_1$ as independent momenta and express $\bk_2$ and $\bk_3$
in terms of them (by also using the condition of transverse momentum conservation: $\bk_1+\bk_2+\bk_3 = \bm{0}$):
\begin{align}
	\label{k2k3}
	\bm{k}_2 = \bm{P} - \frac{\vartheta_2\bm{k}_1}{\vartheta_2 +\vartheta_3},
	\qquad  
	\bm{k}_3 =- \bm{P} - \frac{\vartheta_3\bm{k}_1}{\vartheta_2 +\vartheta_3}.
\end{align}


When the gluon emission occurs at positive times ($t_1<0<t_2$), 
the amplitude differs from the one in Eq.~\eqref{Psi1} only in the second energy denominator, which now involves the LC energy difference between the intermediate $q\bar{q}$ state and the final $q\bar{q}g$ one, that is  \cite{Iancu:2022gpw},
\begin{align}
	\label{Psi2} 
	\Psi^{im\,(2)}_{\lambda_1\lambda_2\,(\bar{q})}=
	\delta _{\lambda_{1}\lambda_2}\,
	\frac{1}{8q^+}\,
	\frac{e e_f g }{(2\pi)^6}\,
	\frac{1}{\vartheta_1(1-\vartheta_1)\vartheta_2\vartheta_3}\,
	\frac{1}{\sqrt{\vartheta_3}}\,
	\frac{\varphi^{ij}(\vartheta_1,\lambda_1)\,k_1^j }
	{E_q+E_{\bar q'}-E_\gamma}\,
	\frac{\tau^{mn}(\vartheta_1,\vartheta_2,\lambda_1)P^n}
	{E_{\bar{q}'} - E_{\bar q} -E_g}.
\end{align}

At this level, one would now need to perform a Fourier transform to coordinate space, then insert the proper scattering $S$-matrices which are generally different for each of the two configurations and, only after that, sum the resulting probability amplitudes. However, this procedure can be considerably simplified for the problem at hand, as we now explain.

Recall that we are interested in special configurations where the antiquark and the gluon are hard, $P_\perp\gg Q_s(\YP)$,
whereas the quark is semi-hard, $k_{1\perp}\sim Q_s(\YP)$. For such configurations, the overall size of the $q\bar q g$
system $R\sim 1/Q_s$ is much larger than the transverse separation $r\sim 1/P_\perp$ between the gluon
and the antiquark.  In particular, the $\bar q g$ pair is so small --- its transverse size $r$ is
much smaller than the correlation length $1/Q_s$ of the colour 
fields in the target --- that it scatters in the same way as its parent antiquark $\bar q'$.
Hence, in so far as the scattering is concerned, there is no difference between the two time-orderings:
in both cases, the partonic system scatters as a $q\bar q$ dipole with size $R$.

This allows us to add the respective amplitudes already in the {\it  absence of scattering}, as shown (in momentum-space) in 
Eqs.~\eqref{Psi1} and \eqref{Psi2}. Their sum involves the following linear combination of energy denominators
\begin{align}
	\label{sumEDs}
	\hspace*{-0.85cm}
	\frac{1}{E_q + E_{\bar q'} - E_\gamma}\!
	\left(\frac{1}{E_q + E_{\bar q} + E_g - E_\gamma} + 
	\frac{1}{E_{\bar q'} - E_{\bar q} - E_g}\right)=
	\frac{1}{E_{\bar q'} - E_{\bar q} - E_g}
	\frac{1}{E_q + E_{\bar q} + E_g - E_\gamma},
\end{align}
whose result is formally the same as the product of energy denominators in the amplitude
describing {\it final-state} evolution (the situation where both the photon decay and the gluon emission occur 
at positive times, $0< t_1< t_2$, after the photon crosses the shockwave). 
Making use of Eqs.~\eqref{k2k3} and \eqref{LCEs1} we immediately get
\begin{align}
	\label{ED1}
	2q^+\big(E_{\bar q'}-E_{\bar q}-E_g\big)
	=\frac{k_{1\perp}^2}{1-\vartheta_1}-
	\frac{k_{2\perp}^2}{\vartheta_2}-
	\frac{k_{3\perp}^3}{\vartheta_3} 
	= -
	\frac{1-\vartheta_1}{\vartheta_2\vartheta_3}\,P_\perp^2\,
	\simeq
	- \frac{P_\perp^2}{\vartheta_2\vartheta_3},
\end{align}
and
\begin{align}
	\label{ED2}
	2q^+\big(E_q + E_{\bar q} + E_g -E_\gamma\big)&\,=
	\frac{k_{1\perp}^2}{\vartheta_1}+
	\frac{k_{2\perp}^2}{\vartheta_2}+
	\frac{k_{3\perp}^2}{\vartheta_3}+
	Q^2 
	=
	\frac{k_{1\perp}^2}{\vartheta_1(1-\vartheta_1)}+
	\frac{1-\vartheta_1}{\vartheta_2\vartheta_3}\,{P_\perp^2}+
	Q^2\,
	\nn
	&\,
	\simeq
	\frac{k_{1\perp}^2+\mcal{M}^2}{\vartheta_1},
\end{align}
where we have defined\footnote{We shall see later on, cf.~Eq.~\eqref{M2target}, that there are good reasons to use the same notation $\mcal{M}^2$ as in Eq.~\eqref{Q2M2}.}
\begin{align}
	\label{Mdef1}
	\mcal{M}^2\,\equiv\,
	\vartheta_1\left(\frac{P_\perp^2}{\vartheta_2\vartheta_3}
	+Q^2\right).
\end{align} 
In writing the last approximate equalities in Eqs.~\eqref{ED1} and \eqref{ED2} we have assumed the quark to be
soft, that is, $\vartheta_1 \ll 1$. This condition is  needed to ensure that the quark transverse momentum remains relatively
small, $k_{1\perp}\sim Q_s\ll P_\perp$,  as required by strong scattering. Indeed, the typical values of $k_{1\perp}^2$
are fixed by the  effective ``mass-term'' $\mcal{M}^2$ in the energy denominator. Hence, in order for $k_{1\perp}$ to be
semi-hard, we need $\mcal{M}\sim Q_s$, which together with \eqn{Mdef1} implies $\vartheta_1 \sim Q_s^2/P_\perp^2\ll 1$. 
In more physical terms, this condition on $\vartheta_1$ means that the lifetime
$\tau_{q\bar q}\sim 2\vartheta_1 q^+/k_{1\perp}^2$ of the intermediate $q\bar q$
state and the formation time $\tau_{q\bar q g}\sim 2q^+/P^2_\perp$ of the $q\bar q g$ system must be comparable to
each other, so that they have comparable chances to scatter off the nuclear shockwave.

Notice that the r.h.s.~of \eqn{Mdef1} involves a linear combination of $P_\perp^2$ and $Q^2$, which is hard 
provided at least one of these two scales is much larger than $Q_s^2$. So, all the results to be obtained
in this section remain true in the photo-production limit $Q^2\to 0$, so long as  $P_\perp^2$ is hard enough, of course.

 Putting everything together, we find the total amplitude $\Psi^{im}_{\lambda_1\lambda_2\,(\bar{q})} = \Psi^{im\,(1)}_{\lambda_1\lambda_2\,(\bar{q})}+ \Psi^{im\,(2)}_{\lambda_1\lambda_2\,(\bar{q})}$ in the absence of scattering:
 \begin{align}
	\label{Psitot}
	\Psi^{im}_{\lambda_1\lambda_2\,(\bar{q})}
	\simeq 
	-\,\delta _{\lambda_{1}\lambda_2}\,
	\frac{e e_f g q^+ }{2(2\pi)^6}\,
	\frac{1}{\sqrt{\vartheta_3}}\,
	\frac{\varphi^{ij}(0,\lambda_1)k_1^j }
	{k_{1\perp}^2+\mcal{M}^2}\,
	\frac{\tau^{mn}(0,\vartheta_2,\lambda_1)P^n}{P_\perp^2}.
\end{align}
 As anticipated, this depends only upon two transverse momentum variables: $\bk_1$ and $\bP$.
 
We are now in a position to add the effects of the scattering. To that aim, it is advantageous to change from 
the transverse momentum to the transverse coordinate representation. Let $\bx$, $\by$ and $\bz$ denote
the transverse positions of the (final) quark, antiquark and gluon, respectively. It is convenient to change these variables
 to $\br$, $\bR$ and $\bb$, which are the transverse separation of the legs in the $\bar{q}g$ pair, the transverse separation of the legs in the $q\bar{q}'$ pair (before the gluon emission), and the transverse position of the intermediate antiquark, respectively. We have
\begin{align}
	\label{rRbdef}
	\br \equiv \by-\bz,
	\qquad 
	\bm{R}\equiv \bx-\bm{b},
	\qquad
	\bm{b}\equiv
	\frac{\vartheta_2\by + \vartheta_3\bz}
	{\vartheta_2 +\vartheta_3},
\end{align} 
while the inverse transformation reads
\begin{align}
	\label{xyz}
	\bx = \bb + \bR,
	\qquad
	\by = \bb + \frac{\vartheta_3}{\vartheta_2 + \vartheta_3}\, \br,
	\qquad 
	\bz = \bb - \frac{\vartheta_2}{\vartheta_2 + \vartheta_3}\, \br, 
\end{align}
where, to the order of accuracy, one is allowed to let $\vartheta_2 + \vartheta_3 \simeq 1$. In terms of the new variables in Eqs.~\eqref{vrelqg} and \eqref{xyz}, the exponential phase which appears when performing the Fourier transform from momentum to coordinate space can be expressed as
\begin{align}
	\label{phaseqbar}
	\bk_1 \cdot \bx + \bk_2 \cdot \by + \bk_3 \cdot \bz
	\,=\, 
	\bP \cdot \br + \bk_1 \cdot \bR,
\end{align}
where we have made used of the fact that the total transverse momentum in the projectile wavefunction vanishes.
(For coherent diffraction in which the target is homogeneous, this property remains true after adding the effects
of the collision; recall e.g.~the discussion in Sect.~\ref{sec:qqbar}.) 

Eq.~\eqref{phaseqbar} indicates that $\br$ is the dual variable to the hard momentum $\bP$ and similarly $\bR$ is dual to the semi-hard momentum $\bk_1$, so we have that $r \ll R$. Moreover, from Eq.~\eqref{xyz} we readily find that the recoil 
$|\by - \bb|$ of the antiquark (due to the gluon emission) is of the order of $r$. We conclude that the gluon and the antiquark (both before and after the emission) are effectively at the same transverse position when ``seen'' by the quark,
which is separated  from the other partons by a large distance $R$ (see Fig.~\ref{fig:fromqbar}):
\begin{align}
	\label{zybapprox}
	\bz \simeq \by \simeq \bb = \bx-\bR
\end{align}
Using this, one can check that the effect of the collision is indeed the same for both time orderings.
Indeed, consider case \texttt{(1)}  in which the gluon is emitted at $x^+<0$, hence it participates in the scattering.
Before doing any diffractive projection, one can include the effect of the collision by replacing
(analogously to Eq.~\eqref{scat})
\begin{align}
	\label{scatqqg}
	t^a_{\alpha \beta}
	\to \big[U^{ab}(\bz) V(\bx)t^b V^{\dagger}(\by) - t^a \big]_{\alpha\beta}
	\simeq \big[V(\bx) V^{\dagger}(\bx-\bR) t^a - t^a \big]_{\alpha\beta}, 
\end{align} 
in the coordinate-space version of Eq.~\eqref{qqgmom}. Here, $U(\bz)$ denotes a Wilson line in the adjoint 
representation and we have also subtracted the no-scattering limit. (As previously explained, this subtraction is
introduced by the third time ordering, where the photon decays at $x^+>0$.)
To arrive at the approximate equality above we have used Eq.~\eqref{zybapprox} and standard Fierz identities. The
final result describes the scattering of a $q\bar q$ dipole with its two legs separated by $\bR$. As anticipated, 
this is the same as the scattering amplitude in case \texttt{(2)}.

To determine the precise form of the diffractive scattering, we proceed as in the exclusive dijet case developed in the Sect.~\ref{sec:qqbar}. We start by requiring elastic scattering on the projectile side, then we average over the target wavefunction already at the amplitude level as a necessary condition to have a coherent process and finally we assume that the nucleus is homogeneous. These three steps lead to the following replacements in the scattering matrix in \eqref{scatqqg}
\begin{align}
	\label{scatR}
	\big[V(\bx) V^{\dagger}(\bx-\bR) t^a - t^a \big]_{\alpha\beta}
	\to
	-T(\bx, \bx-\bR)\, t^a_{\alpha\beta}
	\to
	-\mcal{T}(\bx, \bx-\bR)\, t^a_{\alpha\beta} 
	\to 
	-\mcal{T}(\bR)\, t^a_{\alpha\beta}.
\end{align}  
To summarise, the effect of coherent diffraction for the special $q\bar q g$ configuration at hand amounts to
the replacement
\begin{align}
	\label{tatoT}
	t^a_{\alpha\beta}
	\to 
	-\mcal{T}(\bR)\, t^a_{\alpha\beta},
\end{align}
in the coordinate-space version of Eq.~\eqref{qqgmom}.
This is very similar to \eqref{deltat} for the exclusive dijets. In particular, it depends upon the large size $\bR$
of the intermediate $q\bar q$ dipole, but not also upon the small size $\br$ of the final $\bar q g$ pair. This in turn
implies that the scattering cannot change the hard momentum $\bP$ (the momentum variable dual to $\br$),
which we recall is the {\it relative} momentum of the $\bar q g$ dijet: it only modifies their imbalance $\bk_2+\bk_3$,
or, equivalently, the momentum $\bk_1$ of the semi-hard quark.

This discussion shows that there is no need to explicitly perform the Fourier transform from $\bP$ to $\br$, and
back: that would be redundant, since the $\bP$ dependence of the no--scattering amplitude in \eqn{Psitot} 
remains unchanged after adding the effects of the collision. Rather, it suffices to perform the Fourier transform from $\bk_1$ to $\bm{R}$, then multiply by $-\mcal{T}(\bR)$ as required by Eq.~\eqref{tatoT}, and finally perform the inverse Fourier transform from $\bm{R}$ to $-\bK$. Here $\bK \equiv \bk_2+\bk_3$ is imbalance of the two hard jets in the final state
and coincides with minus the {\it final} transverse momentum of the semi-hard quark.  This gives (analogously to
Eqs.~\eqref{qqFT} and \eqref{qqFTinv})
\begin{align}
	\label{qqgFT}
	\frac{k_1^j}{k_{1\perp}^2 +\mcal{M}^2}
	& \to
	-\frac{i}{2\pi}
	\int \dif^2 \bR\, e^{ i \bK \cdot \bR}\,
	\frac{R^j}{R}\,
	\mcal{M} K_1(\mcal{M} R)
	\mcal{T}(\bR) 
	\nn
	 & =\,  \frac{K^j}{K_{\perp}}
	\int \dif R\, R\, J_1(K_{\perp} R)\,
	\mcal{M} K_1(\mcal{M} R)\,
	\mcal{T}(R)
	= \frac{K^j}{K_{\perp}}
	\frac{\mcal{Q}_{T}(\mcal{M}, K_{\perp})}{\mcal{M}},
\end{align}
where in the second line we also assumed target isotropy and we have expressed the final 
result in terms of the function $\mcal{Q}_{T}(\mcal{M}, K_{\perp})$ defined in Eq.~\eqref{QP} (with $\bar Q\to 
\mcal{M}$). From the discussion in Sect.~\ref{sec:MS}, we know that this function is strongly suppressed
except for semi-hard values $\mcal{M}\sim K_{\perp}\sim Q_s$, which allow for strong scattering.
 In view of \eqref{Mdef1}, this implies $\vartheta_1\sim Q_s^2/P_\perp^2\ll 1$.

To summarise, the diffractive amplitude for (2+1)-jet production via the channel $\bar{q}'\to \bar{q}g$ and such that the quark jet is semi-hard (in transverse momentum) and soft (in longitudinal momentum) reads 
\begin{align}
\label{Psidiff}
	\Psi^{im,D}_{\lambda_1\lambda_2 (\bar{q})}
	\simeq \,\delta _{\lambda_{1}\lambda_2}\,
	\frac{e e_f g q^+ }{2(2\pi)^6}\,
	\frac{\Phi^{ijmn}_{(\bar{q})}(\vartheta_2,\lambda_1)}{\sqrt{\vartheta_3}}\,
	\frac{P^n}{P_\perp^2}\,\frac{K^j}{K_{\perp}}\,
	\frac{\mcal{Q}_{T}(\mcal{M}, K_{\perp})}{\mcal{M}}
\end{align}
where it is understood that $\vartheta_2+\vartheta_3\simeq 1$. The spinorial and polarisation structure is encoded in (notice that we absorb a minus sign into the definition)
\begin{align}
	\label{Phi}
	\Phi^{ijmn}_{(\bar{q})}(\vartheta,\lambda)\,
	\equiv\,
	-\,\varphi^{ij}(0,\lambda)
	\tau^{mn}(0,\vartheta,\lambda)
	= 
	\big[\delta^{ij} -
	2i\lambda\varepsilon^{ij}\big]
	\big[(1+\vartheta)\delta^{mn}
	+2i\lambda (1-\vartheta)\varepsilon^{mn}\big].
\end{align}
\smallskip

\subsection{2+1 jets with a soft quark: gluon emission from the quark}
\label{sec:qsource}

Now we consider the case in which the hard gluon is emitted by the quark.  The respective amplitude is
depicted in the left panel in Fig.~\ref{fig:3jets_gen_softq}, which anticipates the fact that, for the interesting kinematics, 
the $q\bar q$ dipole produced by the photon decay is relatively small,
with transverse size $r\sim 1/P_\perp\ll 1/Q_s$. Like before,  $\bP$ is the relative transverse
momentum of the hard $\bar q g$ dijet, as defined in Eq.~\eqref{vrelqg}.

\begin{figure}
	\begin{center}
		\includegraphics[align=c,width=0.44\textwidth]{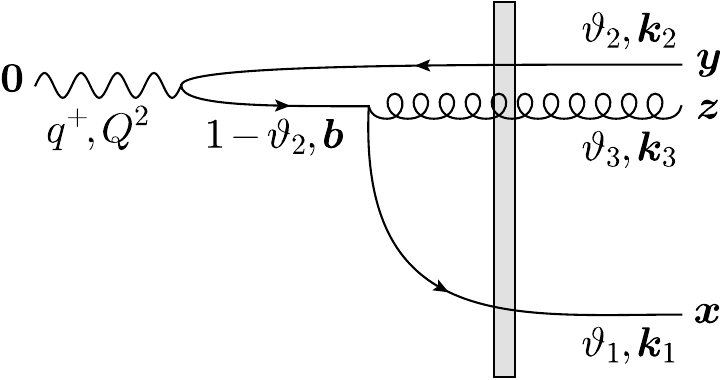}
		\hspace*{0.08\textwidth}
		\includegraphics[align=c,width=0.44\textwidth]{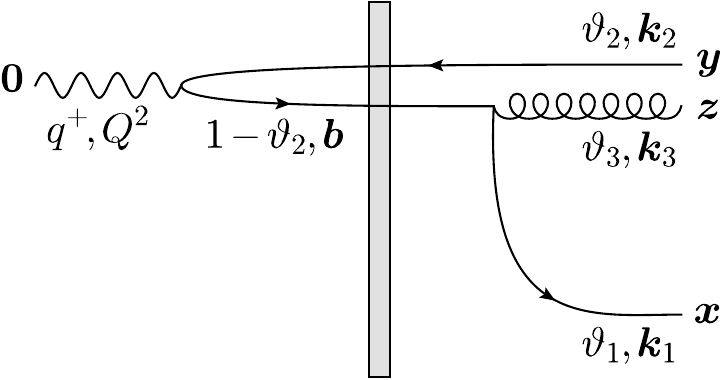}
	\end{center}
	\caption{\small The probability amplitude for (2+1)-jet production in DIS off a nuclear shockwave via the channel $q'\to q$g. The large pair is formed at the gluon vertex. Left panel: The gluon is radiated before the scattering off the shockwave. Right panel: The gluon is radiated after the scattering off the shockwave. }
\label{fig:fromq}
\end{figure}

We have again two possibilities, since the gluon can be emitted either before, or after, the scattering, as shown in Fig.~\ref{fig:fromq}, and the respective amplitudes differ in one energy denominator. 
Following the notation of the previous subsection we have \cite{Iancu:2022gpw}
\begin{align}
	\label{Psi1q} 
	\Psi^{im\,(1)}_{\lambda_1\lambda_2(q)}=
	-\,\delta _{\lambda_{1}\lambda_2}\,
	\frac{1}{8q^+}\,
	\frac{e e_f g }{(2\pi)^6}\,
	\frac{1}{\vartheta_1 \vartheta_2\vartheta_3 (1-\vartheta_2)} \,
	\frac{1}{\sqrt{\vartheta_3}}\,
	\frac{\varphi^{ij*}(\vartheta_2,\lambda_1)\,k_2^j }
	{E_{\bar{q}}+E_{q'}-E_\gamma}\,
	\frac{\tau^{mn*}(\vartheta_2,\vartheta_1,\lambda_1)\bar{K}^n}
	{E_q+E_{\bar q}+E_g -E_\gamma},
\end{align}
for the emission before the scattering. Notice that this can be obtained from Eq.~\eqref{Psi1} by changing the overall sign, taking the complex conjugate, letting $\vartheta_1 \leftrightarrow \vartheta_2$ and $\bk_1 \leftrightarrow \bk_2$ and finally interchanging $q$ and $\bar{q}$ in the energy denominators. In particular the momentum $\bar{\bK}$ associated with the $q'\to qg$ decay, with $q'$ the intermediate quark state, reads
\begin{align}
	\label{kbar}
	\bar{\bm{K}} =
	\frac{\vartheta_3 \bm{k}_1 -\vartheta_1 \bm{k}_3}
	{\vartheta_1 +\vartheta_3}.
\end{align}
Using the above together with the conservation of transverse momentum allows us to express the momenta $\bk_1$ and $\bk_3$ in terms of $\bk_2$ and $\bar{\bK}$ as
\begin{align}
	\label{k1k3}
	\bm{k}_1 = \bar{\bK} - \frac{\vartheta_1\bm{k}_2}{\vartheta_1 +\vartheta_3},
	\qquad  
	\bm{k}_3 =- \bar{\bK} - \frac{\vartheta_3\bm{k}_2}{\vartheta_1 +\vartheta_3}
\end{align}
and the LC energies which appear in the energy denominators in Eq.~\eqref{Psi1q} are given by
\begin{align}
	\label{LCEs1q}
	E_q=\frac{k_{1\perp}^2}{2\vartheta_1 q^+},
	\quad
	E_{\bar q}=\frac{k_{2\perp}^2}{2\vartheta_2 q^+},
	\quad
	E_g=\frac{k_{3\perp}^2}{2\vartheta_3 q^+},
	\quad  	
	E_{q'} = \frac{k_{2\perp}^2}{2(1-\vartheta_2) q^+},
	\quad 
	E_\gamma=-\frac{Q^2}{2q^+}.
\end{align}
Regarding the configuration in which the gluon is emitted after the shockwave we just need to modify the second denominator in Eq.~\eqref{Psi1q}, namely
\begin{align}
	\label{Psi12} 
	\Psi^{im\,(1)}_{\lambda_1\lambda_2(q)}= -\,
	\delta _{\lambda_{1}\lambda_2}\,
	\frac{1}{8q^+}\,
	\frac{e e_f g }{(2\pi)^6}\,
	\frac{1}{\vartheta_1 \vartheta_2\vartheta_3 (1-\vartheta_2)} \,
	\frac{1}{\sqrt{\vartheta_3}}\,
	\frac{\varphi^{ij*}(\vartheta_2,\lambda_1)\,k_2^j }
	{E_{\bar{q}}+E_{q'}-E_\gamma}\,
	\frac{\tau^{mn*}(\vartheta_2,\vartheta_1,\lambda_1)\bar{K}^n}
	{E_{q'} - E_q - E_g}.
\end{align}

At this point, it is useful to remember that we are interested in special configurations
which are such that the final antiquark and gluon are hard and nearly back-to-back, $k_{2\perp}\simeq k_{3\perp}
\simeq P_\perp\gg Q_s(\YP)$, whereas the final quark is semi-hard, $k_{1\perp}\sim Q_s(\YP)$, and also soft,
$\vartheta_1 \ll 1$. Since the hard gluon is now emitted by the quark,  
this means that the intermediate quark $q'$ must be hard as well:
its transverse momentum reads $\bk_1+\bk_3\simeq\bk_3$. Hence the partonic system starts as a hard quark-antiquark
dipole $q'\bar q$, with small transverse size $r\sim 1/P_\perp$, which then evolves by  emitting a soft quark 
at some large distance $R\sim 1/k_{1\perp}$ from the other 
partons\footnote{We implicitly assume here that the transverse position of the hard emitter is 
not significantly affected by the emission of the soft quark; that is, the transverse size $r$ of the final $\bar q g$ pair is comparable to that of the intermediate $q'\bar q$ dipole; this will be verified in a moment.}.
This is the evolution illustrated in the left panel in Fig.~\ref{fig:3jets_gen_softq}.  In so far as the distribution of colour is concerned, 
the intermediate  $q'\bar q$ pair behaves like a small dipole of size $r$, whereas the final $q\bar qg$ system behaves
like a large quark-antiquark dipole with size $R$.
Since the scattering of a small dipole, with size $r\ll 1/Q_s$, is power suppressed, this immediately implies that
one can ignore the  second configuration, where the gluon is emitted after the nuclear shockwave.
Hence, we shall consider in detail only the first configuration (gluon emission before the scattering). 

Returning to Eq.~\eqref{Psi1q}, we write the two energy denominators as
\begin{align}
	\label{ED1q}
	2q^+\big(E_{\bar{q}} + E_{\bar{q}'}-E_\gamma \big)
	=\frac{k_{2\perp}^2}{\vartheta_2} +
	\frac{k_{2\perp}^2}{1-\vartheta_2}
	+ Q^2
	\simeq
	\frac{P_\perp^2}{\vartheta_2\vartheta_3} + Q^2
\end{align}
and
\begin{align}
	\label{ED2q}
	2q^+\big(E_q + E_{\bar q} + E_g -E_\gamma\big)&\,=
	\frac{k_{1\perp}^2}{\vartheta_1}+
	\frac{k_{2\perp}^2}{\vartheta_2}+
	\frac{k_{3\perp}^2}{\vartheta_3}+
	Q^2 
	\simeq
	\frac{k_{1\perp}^2+\mcal{M}^2}{\vartheta_1},
\end{align}
where $\mcal{M}^2$ is the scale already defined in Eq.~\eqref{Mdef1}. In arriving at the last approximate equalities in Eqs.~\eqref{ED1q} and \eqref{ED2q} we have used the fact that $k_{2\perp}\simeq k_{3\perp}
\simeq P_\perp$ and we have assumed  $\vartheta_1  \ll 1$, hence  $\vartheta_2 + \vartheta_3\simeq 1$.
The softness of $\vartheta_1$ follows from the same formation-time argument as presented in Sect.~\ref{sec:softg}
for  (2+1)--jets with a soft gluon: the formation time $\tau_q\sim 2\vartheta_1 q^+/k_{1\perp}^2$
of the soft quark must be shorter than the lifetime $\tau_{q'\bar q}\sim 2q^+/P_\perp^2$ of the intermediate
dipole before its scattering with the target. This implies $\vartheta_1 \lesssim k_{1\perp}^2/P_\perp^2\ll 1$,
which in turn shows that the relative momentum $\bar{\bK}$ of the quark-gluon pair is {\it semi-hard},
$\bar{\bK} \simeq \bk_1$, despite the fact that the final gluon is hard.

Inserting the above approximations into Eq.~\eqref{Psi1q} we obtain the following expression for
the amplitude corresponding to the first configuration (below, $\tilde{Q}^2 \equiv \vartheta_2\vartheta_3 Q^2$):
\begin{align}
	\label{Psiqtot}
	\Psi^{im}_{\lambda_1\lambda_2(q)}
	\simeq  -\,
	\delta _{\lambda_{1}\lambda_2}\,
	\frac{e e_f g q^+ }{2(2\pi)^6}\,
	\frac{1}{\vartheta_3^{3/2}}\,
	\frac{\varphi^{ij*}(\vartheta_2,\lambda_1)P^j }
	{P_{\perp}^2+ \tilde{Q}^2}\,
	\frac{\tau^{mn*}(\vartheta_2,0,\lambda_1)k_{1}^n}
	{k_{1\perp}^2+ \mcal{M}^2},
\end{align}
Like in the previous subsection, we expect the system of the three partons to effectively scatter like a large quark-antiquark dipole. To see that, let us introduce appropriate variables in coordinate space:
\begin{align}
	\label{rRbdefq}
	\br \equiv \by-\bb,
	\qquad 
	\bm{R}\equiv \bx-\bm{z},
	\qquad
	\bm{b}\equiv
	\frac{\vartheta_1\bx + \vartheta_3\bz}
	{\vartheta_1 +\vartheta_3}.
\end{align} 
Here, $\br$ is the separation between the antiquark and the intermediate quark, $\bR$ is the one between the final quark and the gluon and $\bb$ is the position of the intermediate quark. The inverse transformation reads
\begin{align}
	\label{xyzq}
	\bx = \bb + \frac{\vartheta_3}{\vartheta_1 + \vartheta_3}\, \bR,
	\qquad 
	\by = \bb + \br,
	\qquad
	\bz = \bb - \frac{\vartheta_1}{\vartheta_1 + \vartheta_3}\, \bR,
\end{align}
so that the phase involved in the Fourier transform from momentum to coordinate space becomes
\begin{align}
	\label{phaseq}
	\bk_1 \cdot \bx + \bk_2 \cdot \by + \bk_3 \cdot \bz
	\,=\, \bk_2 \cdot \br + \bar{\bK} \cdot \bR 
	\,\simeq\, 
	\bP \cdot \br + \bk_1 \cdot \bR.
\end{align}
The final result is like in Eq.~\eqref{phaseqbar}, so the comments following the latter 
apply also here. In particular, we have $r \ll R$, that is, $q'$ and $\bar{q}$ are very close to each other. 
From Eq.~\eqref{xyzq} we also find that when $\vartheta_1 \ll k_{1\perp}/P_{\perp}\ll 1$ the emitted gluon has the same transverse position as its source. 
Eventually the soft final quark $q$ ``sees'' the two other final partons at the same transverse position:
$\bz \simeq \by \simeq \bb$.
The general structure of the scattering matrix corresponding to this configuration 
is again given by \eqn{scatqqg} and its projection on elastic scattering is obtained by replacing  $t^a_{\alpha\beta} \to 
	-\mcal{T}(\bR)\, t^a_{\alpha\beta}$, cf.~\eqref{tatoT}, in the LCWF \eqref{qqgmom}. We can further  
convince ourselves that the second configuration, where the gluon is emitted at $x^+>0$, is proportional to $\mcal{T}(\br)$, which confirms our expectation that it is power-suppressed. 

The functional dependence on $\bk_1$ of the total wavefunction $\Psi_{(q)}$ in Eq.~\eqref{Psiqtot} is the same as the one we encountered when the gluon source was the antiquark, cf.~\eqref{Psitot}, so we just need to do the replacement dictated by \eqref{qqgFT} to properly include the scattering described by $\mcal{T}(\bR)$. We finally arrive at the diffractive amplitude for (2+1)-jet production via the channel $q'\to qg$ and such that the quark jet is semi-hard (in transverse momentum) and soft (in longitudinal momentum)
\begin{align}
	\label{Psiqdiff}
	\Psi^{im,D}_{\lambda_1\lambda_2(q)}
	\simeq 
	\delta _{\lambda_{1}\lambda_2}\,
	\frac{e e_f g q^+ }{2(2\pi)^6}\,
	\frac{\Phi^{ijmn}_{(q)}(\vartheta_2,\lambda_1)}{\vartheta_3^{3/2}}\,
	\frac{P^j}
	{P_{\perp}^2+\tilde{Q}^2}\,
	\frac{K_{\perp}^n}
	{K_{\perp}}\,
	\frac{\mcal{Q}_{T}(\mcal{M}, K_{\perp})}{\mcal{M}},
\end{align}
with $\vartheta_2+\vartheta_3 \simeq 1$ and where we recall that $\bK=\bk_2+\bk_3$ is the transverse momentum imbalance in the hard $\bar{q}g$ dijet. The spinorial and polarisation structure is contained in
\begin{align}
	\label{Phiq}
	\Phi^{ijmn}_{(q)}(\vartheta,\lambda)\,
	\equiv\,
	-\,\varphi^{ij*}(\vartheta,\lambda)
	\tau^{mn*}(\vartheta,0,\lambda)
	=
	(1-\vartheta)\big[(1- 2\vartheta ) \delta^{ij} +
	2i\lambda\varepsilon^{ij}\big]
	\big[\delta^{mn}
	-2i \lambda \varepsilon^{mn}\big].
\end{align}
\smallskip

\subsection{The cross section for diffractive 2+1 jets with a soft quark}
\label{sec:cross-section}

The cross section for the diffractive (2+1)-jet production is determined similarly to the one for exclusive dijets in Eq.~\eqref{crossqqnew}, i.e.~by counting the number of partons in the final state with the desired longitudinal and transverse momenta. Following once more the normalizations in \cite{Iancu:2018hwa} we find
\begin{align}
	\label{crossqqg}
	\hspace{-0.7cm}
\frac{\rmd\sigma
^{\gamma_{\scriptscriptstyle T}^* A\rightarrow (q)\bar q g A}}
	{\rmd \vartheta_1 
	\rmd \vartheta_2
	\rmd \vartheta_3 
	\rmd^{2}\bm{k}_{1}
  	\rmd^{2}\bm{k}_{2} 
  	\rmd^{2}\bm{k}_{3}}
   = \frac{(2\pi)^6 S_{\perp} C_F N_c}{2(q^+)^2}\,
   \delta(1\!-\!\vartheta_1 \!-\!
   \vartheta_2 \!-\!
   \vartheta_3)\,
   \delta^{(2)}(\bk_1+\bk_2+\bk_3) 
   \!\sum_{\substack {i,m\\ \lambda_1,\lambda_2}}\!
   \big| \Psi_{\lambda_1\lambda_2}^{im,\, D} \big|^2,
\end{align}
where the overall colour factor emerged as ${\rm tr}[t^a t^a] = C_F N_c$. In writing the final state, 
the quark label is  shown in parentheses, ``$(q)$'', to emphasise that this is the soft parton from the (2+1)--jet.
The wavefunction $\Psi_{\lambda_1\lambda_2}^{im, D}$ in the above is simply the sum of the expressions in Eqs.~\eqref{Psidiff} and Eqs.~\eqref{Psiqdiff}, which were obtained for gluon emission from the antiquark and the quark respectively. By using the $\delta$-function to perform one of the two-dimensional integrations and switching to the $\bP$ and $\bK$ variables we trivially get
\begin{align}
	\label{crossqqgnew}
	\hspace*{-1cm}
\frac{\rmd\sigma
^{\gamma_{\scriptscriptstyle T}^* A\rightarrow (q)\bar q g A}}
	{\rmd \vartheta_1 
	\rmd \vartheta_2
	\rmd \vartheta_3
	\rmd^{2}\bP
  	\rmd^{2}\bK}
   = \frac{(2\pi)^6 S_{\perp} C_F N_c}{2(q^+)^2}\,
   \delta_{\vartheta}
   \!\sum_{\substack {i,m\\ \lambda_1,\lambda_2}}\!
   \left\{
   \big| \Psi_{\lambda_1\lambda_2(\bar{q})}^{im, D} \big|^2
   \!+\! 
   \big| \Psi_{\lambda_1\lambda_2(q)}^{im, D} \big|^2
   \!+\!
   2 \Re\! \left[\Psi_{\lambda_1\lambda_2(\bar{q})}^{im, D}
   \Psi_{\lambda_1\lambda_2(q)}^{im, D*}\right]  
   \right\}\!,
\end{align}
where
for the economy of the notation we defined $\delta_{\vartheta} \equiv \delta(1-\vartheta_1-\vartheta_2-\vartheta_3) \simeq \delta(1-\vartheta_2-\vartheta_3)$. In what follows, we shall separately compute the three contributions to the cross section.

\paragraph{(i) Direct term from gluon emission by the antiquark.}
Let us consider the first term in Eq.~\eqref{crossqqgnew}. Starting with Eq.~\eqref{Phi}, it is an easy exercise to find (recall that $\lambda^2=1/4$)
\begin{align}
	\label{Phisquare}
	\sum_{i,m,\lambda}
	\Phi^{ijmn}_{(\bar{q})}(\vartheta,\lambda) 
	\Phi^{ikmr*}_{(\bar{q})}(\vartheta,\lambda) = 
	8(1+\vartheta^2) \delta^{jk} \delta^{nr} + 
	8(1-\vartheta^2) \varepsilon^{jk} \varepsilon^{nr}.
\end{align}
Notice that when we calculate $\big| \Psi_{(\bar{q})}^{D} \big|^2$, we encounter the momenta tensorial structure $K^j K^k P^n P^r$ and the second, antisymmetric, term in Eq.~\eqref{Phisquare} does not contribute. Substituting into Eq.~\eqref{crossqqgnew} we obtain for this part of the cross section in which the gluon is emitted by the antiquark both in the DA and in the CCA 
\begin{align}
	\label{cross1}
	\frac{\rmd\sigma_{(\bar{q}\bar{q})}^{\gamma_{\scriptscriptstyle T}^* A\rightarrow (q)\bar q g A}}
	{\rmd \vartheta_1 
	\rmd \vartheta_2
	\rmd \vartheta_3 \,
	\rmd^{2}\bP
  	\rmd^{2}\bK}
   = 
   \frac{S_{\perp} \alpha_{\rm em} N_c}{2\pi^4} 
   \left( \sum e_f^2\right)
   \delta_{\vartheta}\,
   \frac{\alpha_s C_F}{P_{\perp}^2}
   \frac{1+(1-\vartheta_3)^2}{2\vartheta_3}\,
   \frac{\big|\mcal{Q}_{T}(\mcal{M}, K_{\perp},Y_{\mathbb P})\big|^2}{\mcal{M}^2},
   \end{align}
with $\alpha_s = g^2/4\pi$ the strong coupling constant and $\mcal{M}^2$ as defined in \eqn{Mdef1}. One recognises
here the DGLAP splitting function 
\begin{align}
	\label{Pgq}
	P_{gq}(\vartheta_3) = C_F \frac{1+(1-\vartheta_3)^2}{\vartheta_3}\,
\end{align}
and the characteristic transverse momentum 
spectrum $1/P_\perp^2$ for a hard splitting $\bar q'\to \bar q g$ occurring in the final state.

\paragraph{(ii) Direct term from gluon emission by the quark.} Now we move on to the second term in Eq.~\eqref{crossqqgnew}. Starting with Eq.~\eqref{Phiq}, we get
\begin{align}
	\label{Phiqsquare}
	\sum_{i,m,\lambda}
	\Phi^{ijmn}_{(q)}(\vartheta,\lambda) 
	\Phi^{ikmr*}_{(q)}(\vartheta,\lambda) = 
	8(1-\vartheta)^2 
	\left\{ \left[ 
	\vartheta^2 +(1-\vartheta)^2 
	\right] 
	\delta^{jk} \delta^{nr} + 
	(1-2 \vartheta) \varepsilon^{jk} \varepsilon^{nr} \right\}.
\end{align}
When we calculate $ \big| \Psi_{(q)}^{D} \big|^2$, we encounter the structure $P^j P^k K^n K^r$, so that again the antisymmetric term in Eq.~\eqref{Phiqsquare} does not contribute. Substituting into Eq.~\eqref{crossqqgnew} we obtain for this part of the cross section in which the gluon is emitted by the quark in  both the DA and the CCA 
(with $\tilde{Q}^2 \equiv \vartheta_2\vartheta_3 Q^2$)
\begin{align}
	\label{cross2}
	\hspace*{-0.6cm}
	\frac{\rmd\sigma_{(qq)}^
	{\gamma_{\scriptscriptstyle T}^* A \rightarrow (q)\bar q g A}}
	{\rmd \vartheta_1 
	\rmd \vartheta_2
	\rmd \vartheta_3 \,
	\rmd^{2}\bP
  	\rmd^{2}\bK}
   = 
   \frac{S_{\perp} \alpha_{\rm em} N_c}{2\pi^4} 
   \left( \sum e_f^2\right)
   \delta_{\vartheta}\,
   \frac{\alpha_s C_F P_{\perp}^2}{(P_{\perp}^2 + \tilde{Q}^2)^2}
   \frac{\vartheta_2^2 +(1-\vartheta_2)^2}{2\vartheta_3}\,
   \frac{\big|\mcal{Q}_{T}(\mcal{M}, K_{\perp},Y_{\mathbb P})\big|^2}{\mcal{M}^2}.
   \end{align}

\paragraph{(iii) Interference term.} Finally we go to the last, interference, term in Eq.~\eqref{crossqqgnew}. Using Eqs.~\eqref{Phi} and \eqref{Phiq}, we obtain
\begin{align}
	\label{PhiPhiq}
	\sum_{i,m,\lambda}
	\Phi^{ijmn}_{(\bar{q})}(\vartheta,\lambda) 
	\Phi^{ikmr*}_{(q)}(\vartheta,\lambda) = 
	- 8 \vartheta^2  (1-\vartheta) 
	 (\delta^{jk} \delta^{nr} - 
	\varepsilon^{jk} \varepsilon^{nr}).
\end{align}
In the wavefunction $\Psi_{(\bar{q})}^{D}$ the semi-hard momentum $\bK$ is attached to the photon vertex and the hard momentum $\bP$ in the gluon vertex, while in $\Psi_{(q)}^{D}$ the opposite happens. This is just a consequence of the fact that the large dipole is created directly at the photon vertex in the former case and in the gluon vertex in the latter. Hence, when we calculate the cross product $\Psi_{(\bar{q})}^{D} \Psi_{(q)}^{D*}$, we meet a substantially different tensorial structure in the momentum sector, namely $K^j P^k P^n K^r$. As a result the term involving the antisymmetric symbols is now relevant and we have
\begin{align}
	\label{ddeepk}
	(\delta^{jk} \delta^{nr} - 
	\varepsilon^{jk} \varepsilon^{nr}) 
	K^j P^k P^n K^r = 
	(\bP \cdot \bK)^2 +
	(\bP \times \bK)^2 = 
	P_{\perp}^2  
	K_{\perp}^2, 
\end{align}
so that the dependence on the two momenta factorizes. Remarkably, the particular combination of the two terms in Eq.~\eqref{ddeepk} eliminates any potential angular correlation between the hard and semi-hard momenta.

Putting everything together and substituting into Eq.~\eqref{crossqqgnew} we obtain the interference contribution 
to the cross section in the form
\begin{align}
	\label{cross3}
	\frac{\rmd\sigma_{(\bar{q}q)}^
	{\gamma_{\scriptscriptstyle T}^* A\rightarrow (q)\bar q g A}}
	{\rmd \vartheta_1 
	\rmd \vartheta_2
	\rmd \vartheta_3 \,
	\rmd^{2}\bP
  	\rmd^{2}\bK}
   = -
   \frac{S_{\perp} \alpha_{\rm em} N_c}{2\pi^4} 
   \left( \sum e_f^2\right)
   \delta_{\vartheta}\,
   \frac{\alpha_s C_F}{P_{\perp}^2 + \tilde{Q}^2}\,
   \frac{\vartheta_2^2}{\vartheta_3}\,
   \frac{\big|\mcal{Q}_{T}(\mcal{M}, K_{\perp},Y_{\mathbb P})\big|^2}{\mcal{M}^2}.
   \end{align}
This is negative, as typically expected for an interference term.

\section{TMD factorisation for diffractive (2+1)--jets with a soft quark}
\label{sec:tmd}

So far, the cross-section for diffractive (2+1)--jet production with a soft quark has been
constructed  by working in the colour dipole picture, where the parton evolution is exclusively  associated
with the {\it projectile}: the three partons appearing in the final state  have all been generated via 
radiation in the wavefunction of the virtual photon. In what follows, we would like to show that, after a suitable 
change of variables, the result can be reinterpreted in terms of target evolution:
the soft quark can alternatively be viewed as a constituent of the Pomeron. The partonic
picture of the Pomeron is exactly the same as discussed  in Sect.~\ref{sec:qqbar}, in relation
with diffractive SIDIS (i.e.~exclusive dijets in the aligned jet configuration).
The Pomeron fluctuates into a semi-hard quark-antiquark pair: an on-shell
quark in the $s$--channel (which appears in the final state) and a space-like antiquark in the $t$--channel
(which interacts with the projectile). The differences between exclusive dijets and  (2+1)--jets
with a soft quark refer to {\it final state interactions}, as we now explain.

\begin{figure}
	\begin{center}
		\includegraphics[align=c,width=0.41\textwidth]{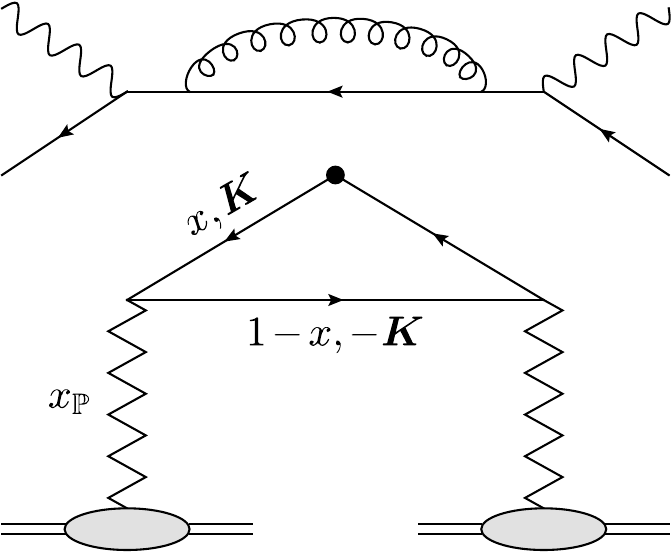}
		\hspace*{0.08\textwidth}
		\includegraphics[align=c,width=0.44\textwidth]{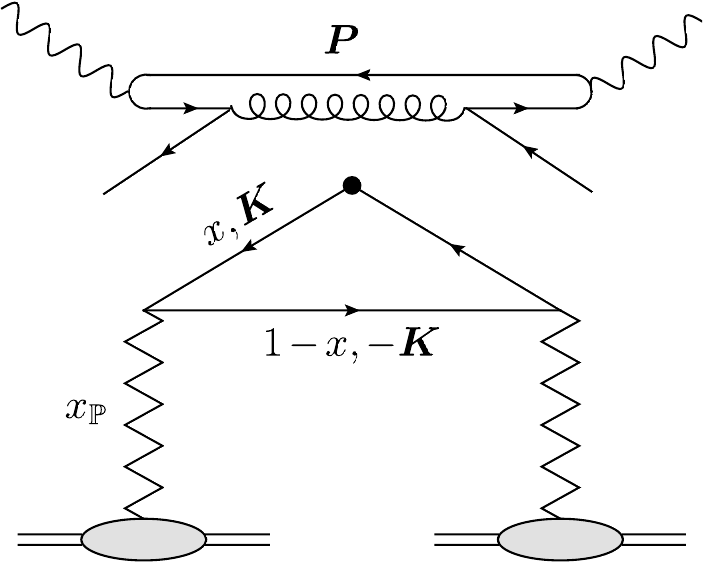}\\
		\vspace*{1.5cm}
		\includegraphics[align=c,,width=0.44\textwidth]{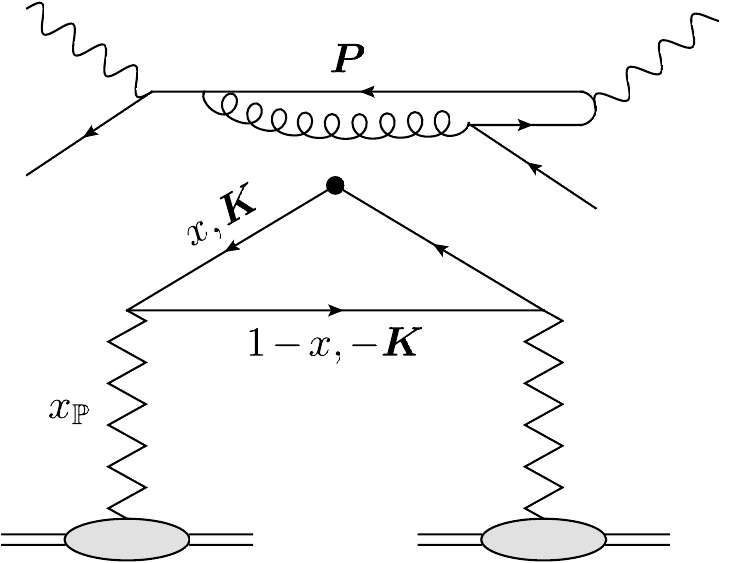}
	\end{center}
	\caption{\small TMD factorization in forward dijet production in DIS off a nucleus. The soft quark from the dipole picture is now 
	viewed as a constituent of the target wavefunction. The Pomeron splits  into a quark-antiquark dipole: an
	 $s$--channel quark	with longitudinal fraction $1-x$, which appears in the final state, and a $t$--channel antiquark
	 with longitudinal fraction $x$, which participates in the scattering with the virtual photon.
	 Left panel: The $t$--channel antiquark first absorbs the photon and then splits into a hard antiquark-gluon pair.
          Right panel:  The $t$--channel antiquark annihilates against the quark from a hard quark-antiquark fluctuation of the photon.
          Bottom panel: An interference contribution.}
\label{fig:3jets_TMD}
\end{figure}

From Sect.~\ref{sec:qqbar}, we recall that in diffractive SIDIS the  
$t$--channel antiquark gets on-shell after absorbing the virtual photon
and appears in the final state (see the right panel in Fig.~\ref{fig:2jets_TMD}). 
For the (2+1)--jet process depicted in the right panel in Fig.~\ref{fig:3jets_gen_softq} (gluon emission by the antiquark), 
the antiquark remains virtual after absorbing the photon and eventually
decays into a hard $\bar q g$ pair, which appears in the final state.
This interpretation is illustrated in the left diagram in Fig.~\ref{fig:3jets_TMD}, which looks like the SIDIS
graph in the right panel in Fig.~\ref{fig:2jets_TMD}, but with an additional gluon emission in the final state.
For the other (2+1)--jet process, where the gluon is emitted by the quark, cf.~the left panel in Fig.~\ref{fig:3jets_gen_softq},
the $t$--channel antiquark does not directly couple to the virtual photon,  rather it
annihilates against the ($s$--channel) quark from the hard $q\bar q$ dipole generated by the photon decay.
This annihilation process, which gives rise to a hard gluon in the final state, is illustrated 
in the right diagram in Fig.~\ref{fig:3jets_TMD}. By combining the two pictures above, one deduces a
graphical representation for the interference contribution to the (2+1)--jet cross-section, shown 
as the bottom diagram in Fig.~\ref{fig:3jets_TMD}.

The physical picture that we just described is suggestive of TMD factorisation, with the same quark diffractive TMD
as introduced in Sect.~\ref{sec:qqbar}. This is furthermore supported by the fact that all three contributions to the 
cross-section --- the direct gluon emissions in Eqs.~\eqref{cross1} and \eqref{cross2},
 and the interference contribution in \eqn{cross3} --- are proportional to  $|\mcal{Q}_{T}|^2$, 
which is the main ingredient of the quark DTMD (recall Eqs.~\eqref{2jetLT}--\eqref{qDTMD}).
But in order to make the TMD  factorisation manifest, one still needs to perform a change of 
longitudinal variables, from the projectile-oriented variable 
$\vartheta_1$ (the ``plus'' longitudinal momentum fraction of the final quark
w.r.t.~the virtual photon) to the target-oriented variable $x$ (the  ``minus'' longitudinal fraction of the 
$t$--channel antiquark w.r.t.~the Pomeron).  To that aim, one can use the same argument as in  
Sect.~\ref{sec:TMDSIDIS}, that is, the condition that  the $s$--channel quark with ``minus'' 
longitudinal fraction $1-x$ be on-shell, cf.~Eq.~\eqref{onshellq}.

Alternatively, and perhaps more instructive for the present purposes, one can
use the conservation of the minus longitudinal momentum: the momentum $x x_{\mathbb P} P_N^-$
carried by the $t$--channel virtual quark gets transmitted to the ``hard'' part of the amplitude, 
hence it must compensate the (negative) LC energy of the virtual photon,
$q^-=-Q^2/2q^+$, and put the final $\bar q g$ pair on shell. This argument implies  $x x_{\mathbb P} =
x_{\bar q g}$, with
\begin{align}
\label{xbarqg}
\hspace{-0.25cm}
x_{\bar q g}\equiv \frac{1}{2q^+P_N^-}\left(\frac{k^2_{2\perp}}{\vartheta_2} +\frac{k^2_{3\perp}}{\vartheta_3} 
 +Q^2\right)= \frac{1}{\hat s}\left( \frac{\vartheta_2 +\vartheta_3}
	{\vartheta_2 \vartheta_3}\, 
	P_\perp^2 +	
	\frac{K_{\perp}^2}{\vartheta_2 +\vartheta_3} + Q^2\right)
 \simeq \frac{1}{\hat s}\left(\frac{P_\perp^2}{\vartheta_2\vartheta_3}+ Q^2\right)\,,
\end{align}
where we have also used the conditions $\vartheta_1\ll 1$ and $K_\perp^2\ll P_\perp^2, \,Q^2$ to
simplify the result. Similarly,
\begin{align}
\label{xPqqg}
	x_{\mathbb P} = 
	\frac{1}{2 q^+ P_N^-} \left( \frac{k_{1\perp}^2}{\vartheta_1} + 
	\frac{k_{2\perp}^2}{\vartheta_2} +
	\frac{k_{3\perp}^2}{\vartheta_3}+  
	Q^2\right)  \simeq \frac{1}{\hat s}\left(\frac{P_\perp^2}{\vartheta_2\vartheta_3}+ Q^2+
	 \frac{K_{\perp}^2}{\vartheta_1}\right)\,.
\end{align}
Using these results together with $x=x_{\bar q g}/ x_{\mathbb P}$, one finds the following relation
between $\vartheta_1$ and $x$,
\begin{align}\label{theta1x}
		\vartheta_1 = \frac{x}{1-x} \,
		 \frac{K_\perp^2}{\frac{P_{\perp}^2}{\vartheta_2 \vartheta_3} + {Q}^2},
\end{align} 
which we recall is only valid so long as the  $\vartheta_1\ll 1$ (the produced quark must be soft w.r.t.~the projectile in order to be reinterpreted as a constituent of the target). The r.h.s.~of \eqn{theta1x}
is indeed much smaller than one in the transverse kinematics of interest ($K_\perp\ll P_\perp$)
and for generic values of $x$. Vice-versa, our approximations fail when $1-x$ is so small that
the r.h.s.~of  \eqn{theta1x} becomes of order one. 

Using  \eqn{theta1x}, one can now finalise our
change of variables and thus demonstrate TMD factorisation. Recalling the definition \eqref{Mdef1}
of the quantity $\mcal{M}^2$, one can use  \eqref{theta1x} to express this quantity in terms
of the target variables $x$ and $K_\perp$, as required by factorisation:
\begin{align}
	\label{M2target}
	\mcal{M}^2 = \frac{x}{1-x} \, K_{\perp}^2 \,.
\end{align}
One also needs to change the measure in the differential cross-sections; this is done by using
\begin{align}
	\label{xmeasure}
	\frac{\dif \vartheta_1}{\vartheta_1} = 
	\frac{1}{1-x}\, \frac{\dif x}{x}\,.
\end{align} 
Omitting unnecessary variables, it is easy to see that the differential cross-section per unit $x$
can be obtained as
\begin{align}	\label{dsigma}
x\frac{\rmd\sigma}{\rmd x}\,=\,\frac{\vartheta_1}{1-x}\,
\frac{\rmd\sigma}{\rmd \vartheta_1 }\,=\,\frac{x}{(1-x)^2}\,\frac{\vartheta_2\vartheta_3 K_\perp^2}
{P_{\perp}^2 + \tilde{Q}^2}\,\frac{\rmd\sigma}{\rmd \vartheta_1 }\,,
\end{align} 
where we recall the notation $\tilde{Q}^2\equiv \vartheta_2\vartheta_3 Q^2$. 

To summarise, the new independent variables 
are the kinematical variables $\vartheta_2$, $\vartheta_3$ and $\bP$ of the hard  dijets 
together with  the target variables  $x$ and $\bK$ (besides the general DIS variables $Q^2$ and $\xbj$). 
We recall that $\bK$ has several meanings (by virtue of transverse momentum
conservation): it represents the
transverse momentum of the virtual antiquark exchanged in the  $t$--channel, but also
 minus the transverse momentum of the final quark ($\bk_1=-\bK$)
 and, finally, the transverse momentum imbalance between the hard dijets ($\bK =\bk_2+\bk_3$).
 Notice also that the more standard diffractive variables $x_{\mathbb{P}}$ and $\beta$ are not
 independent variables, rather they can be expressed as 
 \begin{align}
	\label{xbetaqqg}
	x_{\mathbb{P}}=\frac{x_{\bar q g}}{x}=\frac{\xbj}{x}\, \frac{P_{\perp}^2 + \tilde{Q}^2}{\tilde{Q}^2}
	\,,\qquad
	 \beta=\frac{\xbj}{x_{\mathbb{P}}} =x\, \frac{\tilde{Q}^2}{P_{\perp}^2 + \tilde{Q}^2}\,.
\end{align}
In practice, one can use any of the three variables $x$, $x_{\mathbb{P}}$ and $\beta$ as the
independent ``target longitudinal momentum variable'' and then use \eqn{xbetaqqg} to express
the two others. In this section, we shall write the cross-section per unit of $\ln(1/x)$, but in
 Sect.~\ref{sec:softg} we have chosen $\YP=\ln(1/x_{\mathbb{P}})$ as the independent variable
(recall  \eqn{3jetsD1}) and in the next section we shall find it more convenient to work with $\ln(1/\beta)$.
 
We are now prepared to rewrite the cross-section for diffractive (2+1)--jets
with a soft quark in terms of the new variables and thus  demonstrate
TMD factorisation.   We shall separately consider the
three contributions to the cross-section isolated in Sect.~\ref{sec:cross-section}.

\paragraph{(i) Direct term from gluon emission by the antiquark.} Making the aforementioned change of variables, cf.~Eqs.~\eqref{theta1x}--\eqref{dsigma}, we can put Eq.~\eqref{cross1} in the manifestly factorised form
\begin{align}
	\label{cross1new}
	\frac{\rmd\sigma_{(\bar{q}\bar{q})}^
	{\gamma_{\scriptscriptstyle T}^* A\rightarrow (q)\bar q g A}}
	{\rmd \vartheta_2
	\rmd \vartheta_3
	\rmd^{2}\bP
  	\rmd^{2}\bK\,
  	 \rmd \ln(1/x)}
   = 
   \frac{4 \pi^2 \alpha_{\rm em}}{Q^2}\,  \left( \sum e_f^2\right)\,
   \mcal{H}_T^{(\bar{q}\bar{q})}
   (\vartheta_2,\vartheta_3,P_{\perp}^2,\tilde{Q}^2)\,
   \frac{\rmd xq_{\mathbb{P}}
  (x, x_{\mathbb{P}}, K_{\perp}^2)}
  {\rmd^2 \bK}.
\end{align}
This expression exhibits a {\it double} factorisation: it is the product of two hard factors and a semi-hard one
(see the left panel in Fig.~\ref{fig:3jets_TMD}).
The first hard factor describes the absorption of the virtual photon by a semi-hard antiquark from the target;
this process is ``hard'' when $Q^2$ is large. The second hard factor, defined as
(from now on, it is understood that $\delta_{\vartheta} = \delta(1-\vartheta_2-\vartheta_3)$)
\begin{align}
	\label{hqqg}
	\mcal{H}^{(\bar{q}\bar{q})}_T
 	(\vartheta_2,\vartheta_3,P_{\perp}^2,\tilde{Q}^2)
	\equiv 
	\delta_{\vartheta}\,
	\frac{\alpha_s}{2\pi^2}\,
	P_{gq}(\vartheta_3)\,
	\frac{\tilde{Q}^2}
	{P^2_\perp(P_{\perp}^2 + \tilde{Q}^2)},
\end{align}   
describes the decay of the intermediate antiquark $\bar{q}'$ into a hard antiquark-gluon pair $\bar{q}g$
(indeed, it features the DGLAP splitting function $P_{gq}(\vartheta_3)$ which was introduced in Eq.~\eqref{Pgq}).
Being hard, this decay fully factorises from the remaining part of the process, which describes
the formation of the initial $q\bar q$ pair and admits TMD factorisation for the reasons explained 
in Sect.~\ref{sec:qqbar}.  And indeed, the last (``semi-hard'') factor in  \eqn{cross1new} is the quark TMD 
already introduced in Eq.~\eqref{qDTMD}.

Although natural at high $Q^2$, the double factorisation  in \eqn{cross1new} somehow obscures that fact that
 this cross-section applies for any value of $Q^2$ and in particular it remains finite in the photo-production limit $Q^2\to 0$
  (as relevant e.g.~for UPCs). Indeed, the inverse power $1/Q^2$ from the first hard factor cancels against the overall
  power $\tilde{Q}^2= \vartheta_2\vartheta_3 Q^2$ from the second hard factor.

\paragraph{(ii) Direct term from gluon emission by the quark.} For this process too, one can rewrite the cross-section
\eqref{cross2} in a form which looks similar to \eqn{cross1new}, namely 
\begin{align}
	\label{cross2new}
	\hspace{-0.3cm}
	\frac{\rmd\sigma_{(qq)}^
	{\gamma_{\scriptscriptstyle T}^* A\rightarrow (q)\bar q g A}}
	{\rmd \vartheta_2
	\rmd \vartheta_3
	\rmd^{2}\bP
  	\rmd^{2}\bK\,
  	\rmd \ln(1/x)}
  	=
   \frac{4 \pi^2 \alpha_{\rm em}}{Q^2}\left( \sum e_f^2\right)
      \mcal{H}_T^{(qq)}
   (\vartheta_2,\vartheta_3,P_{\perp}^2,\tilde{Q}^2)\,
   \frac{\rmd xq_{\mathbb{P}}
  (x, x_{\mathbb{P}}, K_{\perp}^2)}
  {\rmd^2 \bK},
   \end{align}
    where the second hard factor reads
\begin{align}
\hspace{-0.5cm}
	\label{hqqg2}\hspace*{-0.5cm}
	\mcal{H}_T^{(qq)}
   (\vartheta_2,\vartheta_3,P_{\perp}^2,\tilde{Q}^2)
	\equiv 
	\delta_{\vartheta}\,
	\frac{\alpha_s C_F}{\pi^2}\,P_{q\gamma}(\vartheta_2) \,
	\frac{1}{\vartheta_3}\,
	\frac{\tilde{Q}^2 P^2_{\perp}}
	{(P_{\perp}^2 + \tilde{Q}^2)^3}\quad \mathrm{with} 
	\quad P_{q\gamma}(\vartheta_2) =\frac{1}{2}\left[ \vartheta_2^2 +(1-\vartheta_2)^2\right].
\end{align} 
In this case, however, the writing in \eqn{cross2new} 
must be taken with a grain of salt: despite this writing,  this process  
does {\it not} admit a genuine ``double factorisation''. 
Its hard part  includes two elementary processes --- the photon decay into a  hard $q\bar q$ pair and the
 annihilation of the ($s$--channel) quark from this pair against the ($t$--channel) antiquark from the Pomeron --- 
 which strictly speaking do {\it not} factorise from each other. Rather, they are together described by the product of
 the two ``hard factors'' in the r.h.s.~of \eqn{cross2new}, which physically cannot be separated from each other.
 For instance, the second factor  $\mcal{H}_T^{(qq)}$ describes both the photon decay $\gamma \to q \bar{q}$
(as obvious from the presence of the electromagnetic splitting function $P_{q\gamma}(\vartheta_2)$) 
and the subsequent gluon emission $q'\to qg$ (the ``annihilation process'').
The overall process is illustrated by the second diagram  in Fig.~\ref{fig:3jets_TMD}. 

It is perhaps useful to recall the origin of the factor $1/\vartheta_3$ in \eqn{hqqg2}: 
unlike in \eqn{hqqg}, this factor has not been generated by the gluon
emission. Indeed, the splitting fraction of the gluon at the $q'\to qg$ vertex is $\vartheta_3/(\vartheta_1 + \vartheta_3) \simeq 1$, so the 
respective splitting function becomes trivial. Rather, by inspection of the calculation in Sect.~\ref{sec:qsource}, one sees that this
factor has been generated as $1/\vartheta_1'\simeq 1/\vartheta_3$, where  $\vartheta_1'\equiv \vartheta_1+ \vartheta_3$
is the longitudinal fraction of the intermediate quark state, prior to the gluon emission. In turn, this factor $1/\vartheta_1'$
comes from the normalisation of the respective single-particle state.

\paragraph{(iii) Interference term.} Finally, the contribution to the cross section from the interference term,
cf.~Eq.~\eqref{cross3}, becomes
\begin{align}
	\label{cross3new}
	\hspace{-0.2cm}
	\frac{\rmd\sigma_{(\bar{q}q)}^
	{\gamma_{\scriptscriptstyle T}^* A\rightarrow (q)\bar q g A}}
	{\rmd \vartheta_2
	\rmd \vartheta_3
	\rmd^{2}\bP
  	\rmd^{2}\bK\,
  	\rmd \ln(1/x)}    = 	
    -\frac{4 \pi^2 \alpha_{\rm em}}{Q^2} \left( \sum e_f^2\right)
   	\delta_{\vartheta}\,
	\frac{\alpha_s C_F}{\pi^2}\,
	\frac{\vartheta_2^2}{\vartheta_3}\,
	\frac{\tilde{Q}^2}
	{(P_{\perp}^2 + \tilde{Q}^2)^2}\,
   	\frac{\rmd xq_{\mathbb{P}}
   	(x, x_{\mathbb{P}}, K_{\perp}^2)}
  	{\rmd^2 \bK},
\end{align}
and is illustrated by the lower graph in Fig.~\ref{fig:3jets_TMD}.

\paragraph{(iv) Total cross-section in the transverse sector.} 
Putting together the three contributions in Eqs.~\eqref{cross1new}, \eqref{cross2new} and \eqref{cross3new} we arrive at the total cross section for (2+1)-jet diffractive production for the case of an initial photon with transverse polarisation. It reads
\begin{align}
	\label{crosstotal}
	\frac{\rmd\sigma
	^{\gamma_{\scriptscriptstyle T}^* A\rightarrow (q)\bar q g A}}
	{\rmd \vartheta_2
	\rmd \vartheta_3
	\rmd^{2}\bP
  	\rmd^{2}\bK\,
  	\rmd \ln(1/x)}
  	=\frac{4 \pi^2 \alpha_{\rm em}}{Q^2} \left( \sum e_f^2\right)
  	\mcal{H}_T(\vartheta_2,\vartheta_3,P_{\perp}^2,\tilde{Q}^2)\,
  	\frac{\rmd xq_{\mathbb{P}}
   	(x, x_{\mathbb{P}}, K_{\perp}^2)}
  	{\rmd^2 \bK},
\end{align}
where the combined hard factor acquires the relatively simple form
\begin{align}
	\label{hardtotal}
	\mcal{H}_T(\vartheta_2,\vartheta_3,P_{\perp}^2,\tilde{Q}^2) =
		\delta_{\vartheta}\,
	\frac{\alpha_s C_F}{\pi^2}\,\frac{1}{2\vartheta_3}\,
	\frac{\tilde{Q}^2\big[(P_{\perp}^2 + \tilde{Q}^2)^2 + 
	\vartheta_2^2 \tilde{Q}^4+
	\vartheta_3^2 P_{\perp}^4 \big]}
	{P_{\perp}^2 (P_{\perp}^2 + \tilde{Q}^2)^3}.
\end{align}
This cross section remains finite in the photo-production limit $Q^2\to 0$, hence
the above results are also relevant for applications to UPCs. At  large  $P_\perp$
(by which we mean either $P_\perp^2\gg \tilde{Q}^2, \, Q_s^2$, or 
$P_\perp^2\sim \tilde{Q}^2\gg Q_s^2$),  the cross-section decreases like $1/P_{\perp}^4$, which is
much slower than the  $1/P_\perp^6$ decay of the cross-section for exclusive dijets
 (cf.~the discussion in Appendix \ref{sec:single}). This confirms our main argument,
that the diffractive production of  hard dijets is controlled by (formally, next-to-leading order)
processes involving 2+1 jets.

%

\paragraph{(v) The longitudinal sector.} For the case of a virtual photon with longitudinal polarisation, the calculations of
the $q\bar q g$ component of the photon LCWF and of the  cross-section for the diffractive (2+1)-jets 
with a soft quark are briefly presented in Appendix~\ref{sec:long}. The main difference refers to the splitting function for
the decay $\gamma^*_L\to q\bar q$ of the longitudinal photon, which reads
\beq\label{PgammaL}
P_{q\gamma}^L(\vartheta) \,=\,4 \vartheta (1-\vartheta).
\eeq
This is suppressed for aligned-jet configurations and, vice-versa, it favours relatively symmetric splittings with
$\vartheta (1-\vartheta)\sim 1/4$.
Clearly, this has important consequences for the systematics of diffractive jet production and for its  twist expansion.

Among the two amplitudes contributing to (2+1)-jet production with a soft quark ($\vartheta_1\ll 1$),
 it should be clear that it is only one of them ---
  that where the initial $q\bar q$ dipole is hard and the gluon is emitted by the quark, cf.~Fig.~\ref{fig:fromq} --- which contributes at leading twist. The direct cross-section associated to this amplitude
 (the right diagram  in Fig.~\ref{fig:3jets_TMD}) involves
the photon splitting function $P_{q\gamma}^L(\vartheta_2)\simeq 4\vartheta_2\vartheta_3$, 
which is of order one when the hard final dijets are quasi-symmetric.
The other amplitude, which starts with a semi-hard $q\bar q$ dipole, cf.~Fig.~\ref{fig:fromqbar}, leads to a cross-section
involving the splitting function $P_{q\gamma}^L(\vartheta_1)$, which is suppressed  
by a factor $\vartheta_1\sim K_\perp^2/Q^2\sim Q_s^2/Q^2\ll 1$ w.r.t.~to the respective contribution \eqref{cross1new}
of the transverse sector (see \eqn{cross1L} for an explicit result). Clearly, this implies a corresponding suppression for
the interference term. 

To summarise, the dominant contribution to the cross-section for (2+1)--jets in this case comes 
from the second diagram in Fig.~\ref{fig:3jets_TMD} (gluon emission by the quark) and reads (cf.~\eqn{cross2L})
\begin{align}
	\label{cross2newL}
	\frac{\rmd\sigma
	^{\gamma_{\scriptscriptstyle L}^* A\rightarrow (q)\bar q g A}}
	{\rmd \vartheta_2
	\rmd \vartheta_3
	\rmd^{2}\bP
  	\rmd^{2}\bK\,
  	\rmd \ln(1/x)}
  	=
   \frac{4 \pi^2 \alpha_{\rm em}}{Q^2}\,
    \left( \sum e_f^2\right)
      \mcal{H}_L
   (\vartheta_2,\vartheta_3,P_{\perp}^2,\tilde{Q}^2)\,
   \frac{\rmd xq_{\mathbb{P}}
  (x, x_{\mathbb{P}}, K_{\perp}^2)}
  {\rmd^2 \bK},
   \end{align} 
%
with the following hard factor (compare to \eqn{hqqg2})
\begin{align}
	\label{hqqg2L}
	\mcal{H}_L
   (\vartheta_2,\vartheta_3,P_{\perp}^2,\tilde{Q}^2)
	\equiv 
	\delta_{\vartheta}\,	\frac{\alpha_s C_F}{\pi^2}\,
	P^{L}_{q\gamma}(\vartheta_2)\,
	\frac{1}{2\vartheta_3}\,
	\frac{\tilde{Q}^4}
	{(P_{\perp}^2 + \tilde{Q}^2)^3}\,,
\end{align} 
which describes the decay $\gamma^*_L\to q\bar q$ of the longitudinal photon followed by the annihilation of the
quark against the antiquark coming from the Pomeron. So long as $\vartheta_2 (1-\vartheta_2)\sim 1/4$
and $Q^2$ is sufficiently hard, $Q^2\sim P_\perp^2$,
this contribution is comparable to that from the transverse photon in \eqn{cross2new}.
On the other hand, the longitudinal contribution vanishes, as expected, in the limit $Q^2\to 0$
where the photon becomes quasi-real.

\paragraph{The quark diffractive PDF: towards the DGLAP evolution.}
So far, we have focused on exclusive final states which, besides the hard dijet (a gluon and an antiquark),
also involves a semi-hard third jet (a quark). The kinematics of the three final partons was fully specified.
Although formally correct to leading order accuracy in pQCD, such a precise characterisation of the final 
state introduces strong constraints on the higher order corrections, which can amplify the strength of the latter.
Indeed, when $P_\perp\gg k_{3\perp}\sim Q_s(\YP)$, there is a large probability 
$\sim\alpha_s\ln(P_\perp^2/Q_s^2)$  for additional radiation, either in the $t$--channel exchange between the
hard dijets and the Pomeron, or from the two hard jets produced in the final state. 
This radiation naturally leads to a broadening of the dijet distribution in their imbalance $K_\perp$, allowing for 
$K_\perp$--values much larger than $Q_s$ (logarithmically distributed within the interval
$Q_s^2 \ll K_\perp^2 \ll P_\perp^2$).   Vice-versa, the fact of imposing a relatively small value 
$K_\perp \sim Q_s$ for this imbalance, introduces a veto on the additional radiation, 
which penalises the cross-section via the ``Sudakov effect''
--- a strong reduction in the cross-section due to the mismatch between ``real'' and ``virtual'' 
radiative corrections (see e.g.~the discussion in \cite{Mueller:2013wwa}). 
Although formally of higher order in $\alpha_s$, the Sudakov
corrections are numerically large, since enhanced by the double logarithm $\ln^2(P_\perp^2/Q_s^2)$, and could 
significantly modify our leading-order estimates.

Such complications can be avoided by giving up the measurement of the hard dijet imbalance $K_\perp$, 
hence of  the detailed structure of the final state. This allows for more general (less constrained) final states, 
which, besides the hard dijets, can
involve additional radiation (one semi-hard jet, at least) with transverse momenta $k_\perp\ll P_\perp$.
The cross-section for such a  more inclusive process is obtained by integrating the previous results in
this section over  $K_\perp$ up to a hard scale of order $P_\perp$. This amounts to replacing the quark diffractive TMD 
with the corresponding PDF in all the previous formulae. E.g.~the cross-section for the diffractive production
of two hard jets accompanied by {\it at least} one soft quark is computed as (cf.~\eqn{crosstotal})
\begin{align}
	\label{crosstotalint}
	\frac{\rmd\sigma
	^{\gamma_{\scriptscriptstyle T}^* A\rightarrow (q)\bar q g A}}
	{\rmd \vartheta_2
	\rmd \vartheta_3
	\rmd^{2}\bP
  	\rmd \ln(1/x)}
  	=\frac{4 \pi^2 \alpha_{\rm em}}{Q^2} \left( \sum e_f^2\right)
  	\mcal{H}_T(\vartheta_2,\vartheta_3,P_{\perp}^2,\tilde{Q}^2)\,
  	xq_{\mathbb{P}}\left(x, x_{\mathbb{P}}, (1-x)P_\perp^2\right),	
	\end{align}
with the quark DPDF as defined in \eqn{xqP}.
This is the analog of \eqn{gluondipColl} for diffractive dijets accompanied by (at least) one soft gluon.

The resolution scale $(1-x)P_\perp^2$  for the quark DPDF has been introduced by the upper limit on the integral 
over $K_\perp^2$ and can be understood in several ways:
 \texttt{(i)} as the condition that the size $r\sim 1/P_\perp$ of the hard dijet be much smaller than the (maximal)
 size $R_{\rm max}\sim \sqrt{1-x}/K_\perp$ of the effective $q\bar q$ dipole (cf.~\eqn{Rmax}), or 
  \texttt{(ii)} as the condition that the virtuality \eqref{tchannel} of the $t$-channel quark exchanged between
  the Pomeron and the hard dijets be much smaller than the transverse momentum of the latter.

From the experience with the collinear factorisation, we know that the integral over $K_\perp$ ensures 
the cancellation of the Sudakov double logs between real and virtual corrections, but it leaves behind 
the single logarithmic corrections describing the DGLAP evolution of the quark diffractive PDF. 
Clearly, it would be instructive to explicitly observe the emergence of the DGLAP evolution 
also in the framework of the colour dipole picture.  This would represent an important additional argument
in favour of the mutual consistency between the two approaches. This will be discussed  in the next section.

\section{(2+1)--jet contributions to diffractive SIDIS: the emergence of DGLAP}
\label{sec:SIDIS}

To unveil the DGLAP evolution from the colour dipole picture for (2+1)--jet production, it is convenient
to consider their contributions to more inclusive quantities,
like diffractive SIDIS and the diffractive structure function $ F_2^{D(3)}(\xbj, x_{\mathbb{P}}, Q^2)$.
By ``diffractive SIDIS'' in  the context of (2+1)-jets, we mean that only one of the two hard jets 
--- a priori, either one of them --- is measured in the final state. Without loss of generality, we can denote the 
transverse momentum of the measured jet by $\bP$; then the other jet will have a momentum
$-\bP+\bK$ and the SIDIS cross-section is obtained by integrating over all the kinematical variables
that are not measured. For instance, for (2+1)-jets with a soft quark and for a virtual photon with
transverse polarisation, the original cross-section \eqref{crosstotal} must be integrated over 
$\vartheta_2,\,\vartheta_3$, $\bK$, and $x$ at a fixed value of the rapidity gap $\YP$ --- or,
equivalently, of the diffractive variable $\beta$. In what follows, we choose to work with $\beta$,
that is, we shall construct the analog of \eqn{2jetSIDIS0} for a diffractive process involving 2+1 jets.
The integration over $\bK$ trivially replaces the quark diffractive TMD by the corresponding
PDF, cf.~\eqn{crosstotalint}. The remaining, non-trivial, integrations are
\begin{align}
	\label{qqgDIFFR}
	\hspace*{-0.5cm}
	\frac{\rmd \sigma^{\gamma_{\scriptscriptstyle T}^* A\rightarrow (q)\bar q g A}}{ \rmd^{2}\bm{P} \,\rmd \ln(1/\beta)} 
	&\,=\int \rmd \vartheta_2
  	\rmd \vartheta_3 \int \frac{\rmd x}{x}
 \,\beta\delta\left(\beta-
 x \frac{\tilde{Q}^2}{\tilde{Q}^2 + P_{\perp}^2}\right)
  		\frac{\rmd\sigma^
	{\gamma_{\scriptscriptstyle T}^* A\rightarrow (q)\bar q g A}}
	{\rmd \vartheta_2
	\rmd \vartheta_3
	\rmd^{2}\bP
  	\rmd \ln(1/x)}
\\*[0.2cm]
       &	=
       \int \rmd \vartheta_2\rmd \vartheta_3
  	  \int_\beta^1\rmd x\,\delta\left(x-\beta\, \frac{P_\perp^2+\tilde{Q}^2}{\tilde{Q}^2}\right)
  	{H}_T(\vartheta_2,\vartheta_3,P_{\perp}^2,\tilde{Q}^2)\,
  	xq_{\mathbb{P}}\left(x, x_{\mathbb{P}}, (1-x)P_\perp^2\right),	  \nonumber
	\end{align}
 where  $H_T$ is a succinct notation  for the complete hard factor in \eqn{crosstotalint}.
   
 Like in Sect.~\ref{sec:qqbar}, we shall assume that the photon virtuality is the largest scale
 in the problem,  $Q^2\gg P_\perp^2$, and we shall demonstrate that the 
 leading-twist contribution to  \eqn{qqgDIFFR} --- that is, the leading term in its expansion in powers of 
 $P_\perp^2/Q^2$ ---- admits TMD factorisation. The outcome of this analysis will be a new contribution to the quark 
 DTMD which decreases like $Q^2_s/P_\perp^2$  at large transverse momenta\footnote{A small remark
 on the notation: so far, the transverse momentum argument of the quark DTMD has been systematically 
 denoted as $K_\perp$ and it was typically semi-hard, $K_\perp\sim Q_s(\YP)$. However, for the new
 contribution to the quark DTMD emerging from  \eqn{qqgDIFFR}, the natural argument is $P_\perp$
 and is always hard, $ P_\perp^2\gg Q_s^2(Y_{\mathbb{P}})$.}
$ P_\perp\gg Q_s(Y_{\mathbb{P}})$, where it dominates over the leading-order (LO) result \eqref{qDTMD}
(which decays  much faster, like $Q_s^4/P_\perp^4$, cf.~the second line in \eqn{qDTMDpw}).
This discussion also explains why, in writing  \eqn{qqgDIFFR},  we have not included the respective contribution of a
 longitudinal photon, cf.~\eqn{cross2newL}:  as we shall check in Appendix~\ref{sec:3SIDIS}, that
contribution is of higher twist.

The twist-expansion of \eqn{qqgDIFFR} turns out to be quite subtle, because of the singular behaviour of
the integrand near the lower limit $x=\beta$. The detailed analysis presented  in Appendix~\ref{sec:3SIDIS}
shows that \eqn{qqgDIFFR} encompasses two types of  leading-twist (LT) contributions:
one which comes from hard but {\it asymmetric} antiquark-gluon configurations, with 
$\vartheta_3\sim P_\perp^2/Q^2\ll \vartheta_2\simeq 1$,  
and one which corresponds to relatively {\it symmetric} $\bar q g$ pairs, with $\vartheta_2\vartheta_3\sim 1/4$. 
The first LT contribution turns out to be more interesting for our present purposes (besides being
numerically larger): indeed, as we shall see, it can be recognised as the first step in the DGLAP
evolution of the LO quark TMD \eqref{qDTMD}. Accordingly, in this section, we shall present a simplified 
analysis of \eqn{qqgDIFFR}  which aims at isolating only the first LT contribution --- that of the DGLAP type. 
The other contribution (associated with symmetric $\bar q g$ configurations)  
will be computed in Appendix~\ref{sec:3SIDIS}, but will be physically discussed in this section.

 To perform the integrals over $\vartheta_2$ and $\vartheta_3$ in \eqref{qqgDIFFR},  we first use the
 $\delta$--function $\delta_{\vartheta} = \delta(1-\vartheta_2-\vartheta_3)$ which is implicit in the
 structure of $H_T$ to replace $\vartheta_2=1-\vartheta_3$ and then observe that the 
 second $\delta$-function, which is explicit in  \eqref{qqgDIFFR}, implies (recall that
 $\tilde{Q}^2= \vartheta_2\vartheta_3 Q^2$)
\beq\label{thetabeta}
\vartheta_3(1-\vartheta_3)\,=\,\frac{\beta}{x-\beta}\,\frac{P_\perp^2}{Q^2}\,\quad \Longrightarrow\quad\,
x-\beta\,\ge\,4\beta\,\frac{P_\perp^2}{Q^2}\,.
\eeq
The last inequality follows from $(1-\vartheta_3)\vartheta_3\le 1/4$ and introduces a lower limit 
$x_{\rm min} =\beta\big(1+4P_\perp^2/Q^2\big)$ on the integral over $x$. This limit is larger than the original
lower limit $\beta$ in \eqn{qqgDIFFR} and this has important consequences, as we shall see.
 The condition $x_{\rm min} < 1$ (for the integral to exist)
 implies an upper limit on $P_\perp$ similar to that encountered for exclusive dijets (cf.~\eqn{K2max})
 \beq\label{Pmax}
 x_{\rm min} < 1 \quad\Longrightarrow\quad P_\perp^2 < \,\frac{1-\beta}{4\beta}\,Q^2\,.\eeq
 The lower limit $x_{\rm min}$ is indeed important since,
as explained in Appendix~\ref{sec:3SIDIS} (see also below), the integrand has singularities at $x=\beta$. Since moreover
the difference $x_{\rm min} -\beta$ is small and of order ${P_\perp^2}/{Q^2}$, it should be clear that
this singular behaviour mixes with the twist expansion. Specifically, the analysis in  Appendix~\ref{sec:3SIDIS}
shows that the integrand for the final integral over $x$ includes two types of terms:

\texttt{(i)} Terms which are at most logarithmically sensitive to the lower limit on $x-\beta$. By integrating
over these terms, one generates LT contributions which are controlled by values of $x$ in the bulk, 
such that $(x-\beta)/\beta\sim 1$. In view of \eqref{thetabeta}, these contributions are associated with
asymmetric $\bar q g$ configurations, for which $\vartheta_2\vartheta_3\sim P_\perp^2/Q^2\ll 1$.

\texttt{(ii)} Terms which exhibit power-like divergences at $x=\beta$. Their LT contributions are therefore
controlled by the lower limit $x_{\rm min}$, i.e.~by values of $x$ satisfying $(x-\beta)/\beta\sim
 P_\perp^2/Q^2\ll 1$. Via  \eqref{thetabeta}, they correspond to symmetric $\bar q g$ configurations
 with $\vartheta_2\vartheta_3\sim 1/4$. 

 As already announced, in this section we shall consider only the first type of contributions, for which 
 $\vartheta_2\vartheta_3\ll 1$. This condition allows for 
two possibilities: either $\vartheta_3\ll 1$ and therefore $\vartheta_2=1-\vartheta_3\simeq 1$, 
or $\vartheta_2\ll 1$ and $\vartheta_3\simeq 1$. 
Recalling the factor $1/\vartheta_3$ in the structure of the hard factor \eqref{hardtotal}, 
it is clear that the first situation (small $\vartheta_3$)
should give a larger result. So, we will focus on this case. Then \eqn{thetabeta} is easily solved as
\beq\label{theta*}
\vartheta_3\,\simeq\,\frac{\beta}{x-\beta}\,\frac{P_\perp^2}{Q^2}\equiv \vartheta_*\,,\eeq
whereas the $\delta$--function can be rewritten as 
\beq\label{delta2}
\delta\left(x-\beta\, \frac{P_\perp^2+\tilde{Q}^2}{\tilde{Q}^2}\right)
=\frac{\vartheta_2\vartheta_3}{x-\beta}\,\delta\left(\vartheta_2\vartheta_3-\frac{\beta}{x-\beta}\,\frac{P_\perp^2}{Q^2}\right)
\simeq\,\frac{\vartheta_3}{x-\beta}\,\delta\left(\vartheta_3-\vartheta_*\right)\,.
\eeq
\eqn{theta*} should be compared to the softness condition \eqref{theta1x} on $\vartheta_1$,
where we recall that the typical value for the dijet imbalance is $K_\perp\sim Q_s$.
One sees that  the contribution to SIDIS that we are about to compute is characterised by the 
following hierarchy among the ``plus''  longitudinal momenta of the three jets:
\beq\label{theta-hierarchy}
\vartheta_2\,\simeq\,1\,\gg\, \vartheta_3\,\sim \,\frac{P_\perp^2}{Q^2}\,\gg\,\vartheta_1\,\sim \,\frac{Q_s^2}{Q^2}\,.\eeq
We recover the condition that a parton must be soft compared to the projectile so that it is be possible
to transfer that parton from the wavefunction of the virtual photon to that of the target. 
By first assuming the quark to be soft, $\vartheta_1\sim Q_s^2/Q^2\ll 1$, for  generic values of
$\vartheta_2$ and $\vartheta_3$, we were able to reinterpret that quark as a constituent of the Pomeron.
By also assuming the gluon to be soft, $\vartheta_3\sim P_\perp^2/Q^2\ll 1$, we will be able
to interpret its emission as one step in the DGLAP evolution of the Pomeron.
The hierarchy $\vartheta_3\gg \vartheta_1$ is also important for that purpose: it reflects the fact, within
the context of the DGLAP evolution, the ``plus'' longitudinal momenta of the emitted partons are
strongly increasing, or equivalently their (target) light-cone lifetimes are strongly decreasing,
 when moving from the target towards the projectile.

Under these assumptions, $P_\perp^2$ and  $\tilde{Q}^2\simeq \vartheta_3 Q^2$ are of the same order,
while the last term $\vartheta_3^2 P_{\perp}^4$ in the numerator of  \eqref{hardtotal} is truly negligible.
Then the non-trivial piece of the hard factor becomes
\beq\label{Hnontr}
\frac{\tilde{Q}^2\big[(P_{\perp}^2 + \tilde{Q}^2)^2 + 
	\vartheta_2^2 \tilde{Q}^4\big]}
	{P_{\perp}^2 (P_{\perp}^2 + \tilde{Q}^2)^3} \simeq\,\frac{1}{P_\perp^2}\,\frac{\beta}{x-\beta}\left(\frac{x-\beta}{x}\right)^3
	\left[\left(\frac{x}{x-\beta}\right)^2+\left(\frac{\beta}{x-\beta}\right)^2\right]=\frac{1}{P_\perp^2}\,\frac{\beta}{x^3}(x^2+\beta^2).
	\eeq
This must be multiplied by a factor $1/(2\vartheta_3)$ from the structure of  \eqref{hardtotal} and by the Jacobian ${\vartheta_3}/(x-\beta)$ from \eqn{delta2}; putting all the factors together, we can write
\beq\label{Pemergence}
 \mcal{H}_T\,\frac{\vartheta_3}{x-\beta}\,\to\, \frac{\alpha_s C_F}{\pi^2}\,\frac{1}{P_\perp^2}\,\frac{\beta}{x^3}\,\frac{x^2+\beta^2}{2(x-\beta)}\,
  =\, \frac{\alpha_s}{2\pi^2}\,\frac{1}{P_\perp^2}\,\frac{\beta}{x^2}\ P_{qq}\left(\frac{\beta}{x}\right)
\eeq
where we have identified the DGLAP splitting function 
\begin{align}
	\label{Pqq}
	P_{qq}(z)=C_F\frac{1+z^2}{1-z}	
\end{align}
with $z=\beta/x$. After inserting this result in \eqn{qqgDIFFR}, one obtains the following contribution to the cross-section
for diffractive SIDIS,
\begin{align}
	\label{qqgSIDIS}
	\frac{\rmd \sigma^{\gamma_{\scriptscriptstyle T}^* A\rightarrow (q)\bar q g A}}{ \rmd^{2}\bm{P} \,\rmd \ln(1/\beta)} 
	&\,=  \frac{4 \pi^2 \alpha_{\rm em}}{Q^2} \left( \sum e_f^2\right) \, \frac{\alpha_s }{2\pi^2}\,\frac{1}{P_\perp^2}\,
\int_{x_{\rm min} }^1\frac{\rmd x}{x}\,\frac{\beta}{x}\ P_{qq}\left(\frac{\beta}{x}\right)
xq_{\mathbb{P}}\left(x, x_{\mathbb{P}}, (1-x)P_\perp^2\right).
 \end{align}	
This is manifestly a leading-twist quantity (so, it describes a process in which the virtual photon is absorbed
by a single quark from the target) and it exhibits the bremsstrahlung spectrum $1/P^2_\perp$
for the measured jet.  To the accuracy of interest, one can approximate the 
 resolution scale of the quark distribution as $(1-x)P_\perp^2\simeq P_\perp^2$:  
indeed, the quark DPDF vanishes as $x\to 1$, cf.~\eqn{xqPsat},
so the integral in  \eqn{qqgSIDIS} has little sensitivity to its upper limit.

Together, the splitting function $P_{qq}(\beta/x)$ and the bremsstrahlung spectrum 
describe a hard $\bar q\to \bar q g$ splitting occurring in
the target wavefunction (see Fig.~\ref{fig:sidis-3jets} for a pictorial interpretation): 
the parent parton is the $t$--channel antiquark produced by the Pomeron with ``minus''
longitudinal momentum fraction $x$;  the daughter partons are a $t$--channel antiquark and 
an $s$--channel gluon, with splitting fractions $\beta/x$ and respectively $1-\beta/x$. 
This splitting is hard in the sense that the transverse
momenta of the daughter partons, of order $P_\perp$, 
are much larger than the respective momentum $K_\perp\sim Q_s(\YP)$
of the parent antiquark. The precise value of $K_\perp$ is unimportant and indeed the cross-section 
\eqref{qqgSIDIS} involves only the total number $xq_{\mathbb{P}}\left(x, x_{\mathbb{P}}, P_\perp^2\right)$
of  parent (anti)quarks with momenta $K_\perp\ll P_\perp$.
The daughter antiquark has a longitudinal momentum fraction 
 $\beta$ w.r.t.~the Pomeron, meaning a ``minus'' fraction $\beta x_{\mathbb{P}} =\xbj$ w.r.t.~the target.
So, clearly,  this hard antiquark is the parton which absorbs the virtual photon before emerging in the final 
state\footnote{
This is also consistent with the fact that the final
antiquark carries the quasi-totality of the ``plus'' longitudinal momentum of the photon: $\vartheta_2\simeq 1$.}.
Thus, the LT cross-section \eqref{qqgSIDIS} is truly a 
measurement of the number of hard antiquarks in the target wavefunction. We shall return to this
interpretation towards the end of this section.

\begin{figure}
\begin{center}
	\includegraphics[align=c,width=0.40\textwidth]{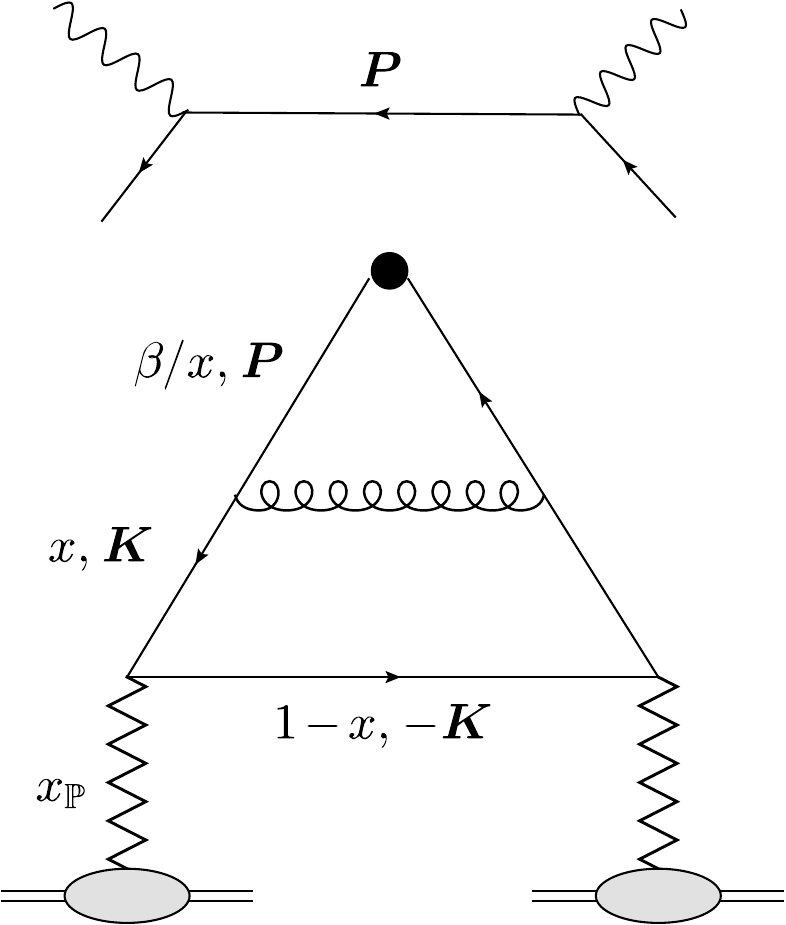}
	\hspace*{0.12\textwidth}
	\includegraphics[align=c,width=0.40\textwidth]{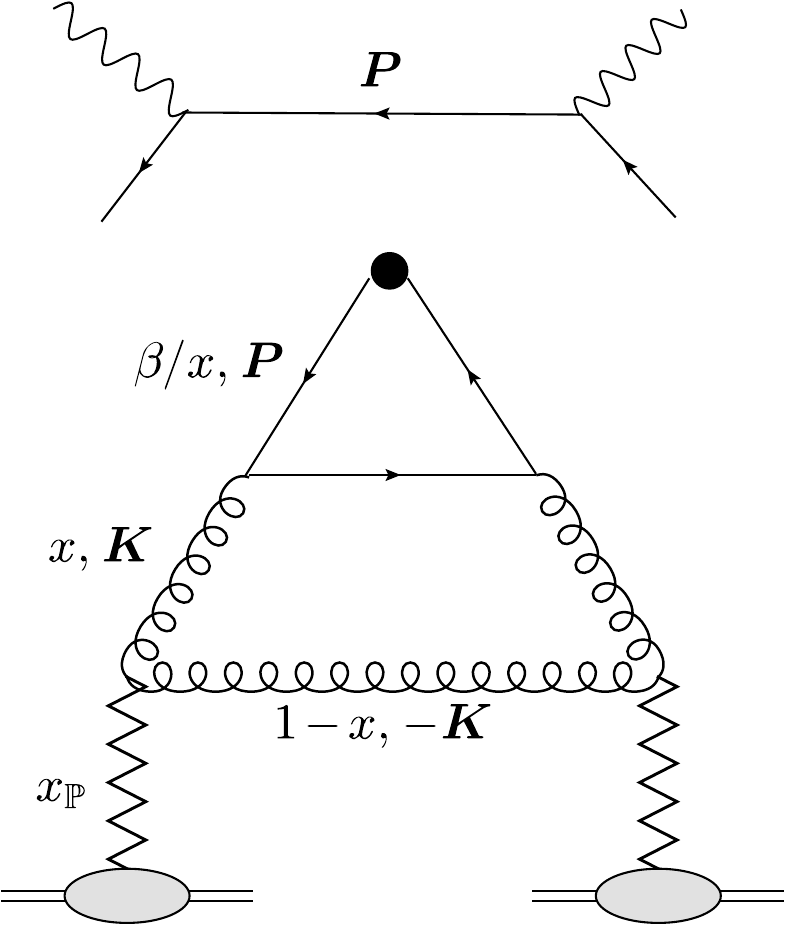}
\end{center}
\caption{\small TMD factorisation for the (2+1)--jet contributions to diffractive SIDIS. Left panel: The soft quark case. Right panel: The soft gluon case.}
 \label{fig:sidis-3jets}
\end{figure}

Whereas the emergence of the DGLAP splitting function $P_{qq}\left({\beta}/{x}\right)$ may look natural
(in view of the kinematics),  what is truly remarkable is the way how this gets reconstructed  
from the Feynman graphs of the colour dipole picture. As visible in
\eqn{Hnontr} (see also Appendix~\ref{sec:3SIDIS} for more details), this
receives contributions from all the three graphs in Fig.~\ref{fig:3jets_TMD}: the direct 
gluon emissions by the antiquark and by the quark, and the  interference piece.
This is remarkable because the relative time-ordering of the two branching vertices
--- the photon vertex  and the gluon emission vertex  --- appears
to change when going from the projectile picture (as illustrated by the amplitude
graphs in Sects.~\ref{sec:2plus1} and also by the cross-section graphs in Fig.~\ref{fig:3jets_TMD})
to the target picture,  cf.~the left panel of Fig.~\ref{fig:sidis-3jets}.
One should however keep in mind that the time variables are different in the two pictures: the role of ``time'' 
is played by $x^+$ in the colour dipole and by $x^-$ in the target picture.

The splitting function $P_{qq}(\beta/x)$ exhibits a pole at $x=\beta$, which in the target picture has an obvious
physical interpretation: it corresponds to the emission of  a soft gluon in the $s$--channel, see the left panel in
Fig.~\ref{fig:sidis-3jets}.  It is interesting to clarify the origin of this pole in the colour dipole picture.
The  analysis  in  Appendix~\ref{sec:3SIDIS} shows that the singular piece of $P_{qq}(\beta/x)$
is exclusively generated by the first 
process in Fig.~\ref{fig:3jets_TMD} --- a gluon emission  by the antiquark in the final state.
It is well known that such final-state emissions develop collinear singularities from 
integrating over large  formation times. In this case,  the relevant formation time 
reads $\tau_f \simeq 2 \vartheta_2\vartheta_3 q^+/P_\perp^2$ and becomes arbitrarily
large in the collinear limit $P_\perp\to 0$. Using \eqn{thetabeta} for $\vartheta_2\vartheta_3 $,
one finds  $\tau_f \sim  \tau_\gamma/(x-\beta)$ (with 
$ \tau_{\gamma}=2q^+/Q^2$ the coherence time of the virtual photon), which confirms
that the singularity at $x=\beta$ corresponds to a large formation time. Hence, the soft singularity
of the hard splitting in the target picture is inherited from a collinear singularity in the projectile picture.

In the standard formulation of the DGLAP evolution, the poles of the splitting functions associated
with soft gluon emissions are screened by the plus prescription, as introduced
by the virtual corrections. Here, however, the virtual corrections are missing since we measure the transverse
momentum $P_\perp$ of the daughter antiquark. (The compensation of soft divergences
between real and virtual emissions should only hold  in more inclusive processes, in which 
$P_\perp$ is integrated as well; see the discussion at the end of this section.) 
Instead, the would-be logarithmic divergence of the integral over $x$
is cut off by the explicit lower limit $x_{\rm min}$ in \eqn{qqgSIDIS}, which follows from
\eqn{thetabeta}.  This generates
a logarithmically enhanced contribution 
$\ln [1/(x_{\rm min}-\beta)] \simeq \ln(Q^2/P_\perp^2)$ which is divergent when $P_\perp\to 0$, 
thus reflecting the original collinear singularity in the projectile picture. 

The  LT contribution of 2+1 diffractive jets with a soft gluon to diffractive SIDIS can be similarly computed
(see Appendix~\ref{sec:3SIDIS} for details) and the result looks indeed similar to \eqn{qqgSIDIS}, namely
 \begin{align}
	\label{qqgSIDISsoftg}
	\frac{\rmd \sigma^{\gamma_{\scriptscriptstyle T}^* A\rightarrow q\bar q (g) A}}{ \rmd^{2}\bm{P} \,\rmd \ln(1/\beta)} 
	&\,=  \frac{4 \pi^2 \alpha_{\rm em}}{Q^2} \left( \sum e_f^2\right) \, \frac{\alpha_s}{\pi^2}\,\frac{1}{P_\perp^2}\,
\int_\beta^1\frac{\rmd x}{x}\,\frac{\beta}{x}\ P_{qg}\left(\frac{\beta}{x}\right)
xG_{\mathbb{P}}\left(x, x_{\mathbb{P}}, (1-x)P_\perp^2\right),	
 \end{align}	
where 
\begin{align}
	\label{Pqg}
	P_{qg}(z) = \frac{z^2 +(1-z)^2}{2} 	
 \end{align}
is the DGLAP splitting function for the process  $g\to q\bar q$. Once again, this result corresponds to asymmetric hard dijets, that is, to 
$q\bar q$ pairs with $ \vartheta_1\vartheta_2\ll 1$.
For instance, if the measured jet is the quark, the ``plus'' longitudinal momenta of the final partons obey the strong 
ordering condition  (compare to \eqn{theta-hierarchy})
 \beq\label{theta-hierarchy-sg}
\vartheta_1\,\simeq\,1\,\gg\, \vartheta_2\,\sim \,\frac{P_\perp^2}{Q^2}\,\gg\,\vartheta_3\,\sim \,\frac{Q_s^2}{Q^2}\,.\eeq
For a  measured antiquark,  the variables  $\vartheta_1$ and $\vartheta_2$ get exchanged with each other.

 \eqn{qqgSIDISsoftg} has a natural interpretation as target evolution, shown in the right panel of  Fig.~\ref{fig:sidis-3jets}:
 a gluon from the Pomeron wavefunction with longitudinal momentum fraction $x$ undergoes a hard splitting
 into a $q\bar q$ pair. Either the quark, or the antiquark, from this pair absorbs the virtual photon and
emerges in the final state with ``minus'' longitudinal fraction $\xbj$ and ``plus'' 
 longitudinal fraction close to unity. This splitting function $P_{qg}({\beta}/{x})$  has no singularity at $x\to \beta$, 
which explains why the lower limit of the above integral has been extended down to $\beta$.  This was to be expected:
the original process in the dipole picture, cf.~Fig.~\ref{fig:TMD-softg}, involves no final-state emission, hence there
is no associated collinear divergence. This argument also explains why this particular process
  brings no additional LT contribution from symmetric $q\bar q$ pairs ($ \vartheta_1\vartheta_2
  \sim 1/4$), as will be explicitly verified in  Appendix~\ref{sec:3SIDIS}.

At this level it should be clear that the LT contributions to diffractive SIDIS shown in Eqs.~\eqref{qqgSIDIS}--\eqref{qqgSIDISsoftg} describe diffractive measurements of the (anti)quark distribution in the target and hence
can be interpreted as {\it contributions to  the quark diffractive TMD}. Formally, these are next-to-leading order
corrections, as they are suppressed by a factor $\alpha_s$ compared to the respective contribution from 
exclusive dijets, cf.~Eqs.~\eqref{2jetTMD}--\eqref{qDTMD}. Yet, as already emphasised,
they give the dominant contribution
at large momenta $P_\perp\gg Q_s(\YP)$, where the LO result is strongly suppressed. As shown by the
previous discussion, they should be viewed as the result of one step in the DGLAP evolution of the LO
distribution \eqref{qDTMD}. It is in fact natural to add together these various contributions and write
\begin{align}
	\label{SIDISqqg}
	\frac{\rmd \sigma^{\gamma_{\scriptscriptstyle T}^* A\rightarrow (q \,{\rm or}\,\bar q) XA}}{ \rmd^{2}\bm{P} \,\rmd \ln(1/\beta)} 
	&\,=\,\frac{\rmd \sigma^{\gamma_{\scriptscriptstyle T}^* A\rightarrow q\bar q A}}{ \rmd^{2}\bm{P} \,\rmd \ln(1/\beta)} 
	+ 2\,\frac{\rmd \sigma^{\gamma_{\scriptscriptstyle T}^* A\rightarrow (q)\bar q g A}}{ \rmd^{2}\bm{P} \,\rmd \ln(1/\beta)} 
	+\frac{\rmd \sigma^{\gamma_{\scriptscriptstyle T}^* A\rightarrow q\bar q (g) A}}{ \rmd^{2}\bm{P} \,\rmd \ln(1/\beta)} 
	\nonumber\\*[0.2cm]
	&\,\equiv\,  \frac{4 \pi^2 \alpha_{\rm em}}{Q^2} \left( \sum e_f^2\right) \, 2\,\frac{\rmd xq_{\mathbb{P}}
  (x, x_{\mathbb{P}}, P_{\perp}^2)}{\rmd^2 \bP}\bigg |_{x=\beta}
	 \end{align}
with the following approximation for the quark diffractive TMD:
\begin{align}
  \label{qDTMDNLO}
  \frac{\rmd xq_{\mathbb{P}}
  (x, x_{\mathbb{P}}, P_{\perp}^2)}{\rmd^2 \bP}\,=\, \frac{\rmd xq_{\mathbb{P}}^{q\bar q}
  (x, x_{\mathbb{P}}, P_{\perp}^2)}{\rmd^2 \bP}\,+\,\frac{\rmd xq_{\mathbb{P}}^{(q)\bar q g}
  (x, x_{\mathbb{P}}, P_{\perp}^2)}{\rmd^2 \bP}\,+\,\frac{\rmd xq_{\mathbb{P}}^{q\bar q (g)}
  (x, x_{\mathbb{P}}, P_{\perp}^2)}{\rmd^2 \bP}.
  \end{align}
The first term in the r.h.s.~is the exclusive $q\bar q$ pair contribution from 
  \eqn{qDTMD}, the second term comes from 2+1 jets with a soft quark\footnote{This term is multiplied by a factor
  of 2 to account for the (identical) contribution of  (2+1)--jets with a soft antiquark.}
   and can be read off \eqn{qqgSIDIS}, 
while the third term refers to 2+1 jets with a soft gluon and can be inferred from \eqn{qqgSIDISsoftg}. Specifically,
    \begin{align}
  \label{qDTMDhard}
  \frac{\rmd xq_{\mathbb{P}}^{(q)\bar q g}
  (x, x_{\mathbb{P}}, P_{\perp}^2)}{\rmd^2 \bP}&\,=\,
   \frac{\alpha_s}{2\pi^2}\,\frac{1}{P_\perp^2}\,
\int_x^{z_{\rm max}}\rmd z\,P_{qq} (z)\,\frac{x}{z}
q_{\mathbb{P}}\left(\frac{x}{z}, x_{\mathbb{P}}, P_\perp^2\right),\nonumber\\	
\frac{\rmd xq_{\mathbb{P}}^{q\bar q (g)}
  (x, x_{\mathbb{P}}, P_{\perp}^2)}{\rmd^2 \bP}&\,=\,
   \frac{\alpha_s}{2\pi^2}\,\frac{1}{P_\perp^2}\,
\int_x^1\rmd z\,P_{qg} (z)\,\frac{x}{z}
G_{\mathbb{P}}\left(\frac{x}{z}, x_{\mathbb{P}}, P_\perp^2\right),
  \end{align}
  where we have changed the notations a bit as compared to the previous formulae.
The variable denoted as $x$ in Eqs.~\eqref{qDTMDNLO}--\eqref{qDTMDhard} --- the longitudinal fraction of the measured (anti)quark  w.r.t.~the Pomeron ---  was previously denoted as $\beta$ in 
Eqs.~\eqref{qqgSIDIS}--\eqref{qqgSIDISsoftg}. 
Also, the integration variable in \eqn{qDTMDhard} is the splitting fraction $z$ of the daughter (anti)quark at
the hard vertex (previously denoted as $\beta/x$). 
The upper limit $z_{\rm max}$ on $z$ corresponds to the lower limit $x_{\rm min}$ in the original equations
and has been obtained as $z_{\rm max}=\beta/x_{\rm min}=1/(1+4P_\perp^2/Q^2)\simeq 
1-4P_\perp^2/Q^2$. Finally, the virtuality argument of the  diffractive PDF  
has been approximated as $P_\perp^2$. 
  
 
At this level, it is intuitively clear that the analog of \eqn{qDTMDNLO} for the case of the gluon diffractive TMD reads
as follows
\begin{align}
  \label{GDTMDNLO}
  \frac{\rmd xG_{\mathbb{P}}
  (x, x_{\mathbb{P}}, P_{\perp}^2)}{\rmd^2 \bP}\,=\, \frac{\rmd xG_{\mathbb{P}}^{gg}
  (x, x_{\mathbb{P}}, P_{\perp}^2)}{\rmd^2 \bP}\,+\,2n_f\frac{\rmd xG_{\mathbb{P}}^{(q)\bar q g}
  (x, x_{\mathbb{P}}, P_{\perp}^2)}{\rmd^2 \bP}\,+\,\frac{\rmd xG_{\mathbb{P}}^{(g)gg}
  (x, x_{\mathbb{P}}, P_{\perp}^2)}{\rmd^2 \bP},
  \end{align}
  where the first piece is the tree-level contribution (the direct production of a semi-hard $gg$ dipole by the Pomeron), 
  as appearing in the TMD factorisation for $q\bar q g$ jets with a soft gluon \cite{Iancu:2022lcw}, cf.~Eqs.~\eqref{3jetsD1}--\eqref{Gscalarnew}, while the two other contributions involve convolutions
between the quark or gluon DPDF and  one hard vertex. Specifically,  
 the contribution with upper index $(q)\bar q g$ refers to the case where the Pomeron first produces a semi-hard
$q\bar q$ dipole and then the quark (or the antiquark) from this pair decays into a hard quark-gluon system; this is similar to \eqn{qqgSIDIS} except that, now, it is the hard gluon that is measured by the TMD:
\begin{align}
	\label{ghardTMDqqg}
	\frac{\rmd xG_{\mathbb{P}}^{(q)\bar q g}
  (x, x_{\mathbb{P}}, P_{\perp}^2)}{\rmd^2 \bP}\,=\, \frac{\alpha_s }{2\pi^2}\,\frac{1}{P_\perp^2}\,
\int_x^{1}\rmd z\,P_{gq}(z)\,\frac{x}{z}
q_{\mathbb{P}}\left(\frac{x}{z}, x_{\mathbb{P}}, P_\perp^2\right).
 \end{align}
Finally, the contribution with upper index ${(g)gg}$ refers to the case where the Pomeron first produces a semi-hard gluon-gluon pair
and then one of these gluons splits into a hard $gg$ pair:
\begin{align}
	\label{ghardTMDggg}
	\frac{\rmd xG_{\mathbb{P}}^{(g)gg}
  (x, x_{\mathbb{P}}, P_{\perp}^2)}{\rmd^2 \bP}\,=\, \frac{\alpha_s }{\pi^2}\,\frac{1}{P_\perp^2}\,
\int_x^{z_{\rm max}}\rmd z\,P_{gg}(z)\,\frac{x}{z}
G_{\mathbb{P}}\left(\frac{x}{z}, x_{\mathbb{P}},P_\perp^2\right),
 \end{align}
where the splitting function reads
\begin{align}
	\label{Pgg}
	P_{gg}(z) = 2N_c 
	\left[
	\frac{1-z}{z} + z(1-z) + \frac{z}{(1-z)}
	\right].
\end{align}
 These two contributions to the gluon DTMD could be probed via
 (3+1)--jet processes involving two steps of DGLAP evolution: the hard gluon produced either in \eqref{ghardTMDqqg},
 or in \eqref{ghardTMDggg}, splits into an even harder $q\bar q $ pair and one of the daughter fermions is absorbing 
the virtual photon.

The previous equations in this section have the right structure to signal the emergence  of the DGLAP evolution for 
the diffractive, quark and gluon, PDFs.
Roughly speaking, the quantities in the l.h.s.'s of Eqs.~\eqref{qDTMDNLO}  and \eqref{GDTMDNLO} can be seen as 
the momentum derivatives of the respective DPDFs, whereas the r.h.s.'s of these equations 
describe the change in the diffractive distributions due to one hard splitting. 
There are however some noticeable differences w.r.t.~the standard structure of the DGLAP equations 
in the literature:  \texttt{(i)} the absence of  the virtual contributions and, related to that, the appearance of the 
upper cutoff $z_{\rm max}< 1$ in the integration over the splitting fraction $z$, and
\texttt{(ii)} the presence of an additional, tree-level, piece, representing the respective
(quark or gluon) diffractive TMD as computed in the colour dipole picture,  which
acts as a {\it source} for the corresponding PDFs. 

The virtual DGLAP contributions were naturally absent in our previous
results, which focused on the explicit production of hard partons via the (hard) splitting of semi-hard ones. But clearly, they must
be added when such hard splittings get iterated; this in particular amounts to endowing the DGLAP splitting functions with
 the plus prescription. The source term represents the essential new ingredient provided by our analysis: it
 enables one to explicitly construct the quark and gluon diffractive PDFs at  small $x_{\mathbb{P}}$ via calculations in
 the dipole picture. The complete DGLAP equations emerging from our formalism will be presented and solved
 in the next section. Using their solutions, we shall in particular compute the diffractive structure function $F_2^D$ 
 according to (cf.~\eqn{xqD})
 \begin{align}
 \label{xqD3}
x_{\mathbb{P}} F_2^{D(3)}(\xbj, x_{\mathbb{P}}, Q^2)= 2
\sum_f e_f^2 \, xq_{\mathbb{P}}(x, x_{\mathbb{P}}, Q^2)\big |_{x=\beta}.
\end{align}

We conclude this section with a few remarks on the remaining LT contribution to the cross-section for diffractive
SIDIS, which as announced is associated with (2+1)--jet processes whose final states comprises  a soft quark 
($\vartheta_1\sim Q_s^2/Q^2\ll 1$) and a relatively symmetric antiquark-gluon dijet ($\vartheta_2\vartheta_3\sim 1/4$).
As explained in Appendix~\ref{sec:3SIDIS}, this contribution comes exclusively from the  the first 
process in Fig.~\ref{fig:3jets_TMD}, where the gluon is emitted by the antiquark in the final state. Besides
the LT contribution showing a logarithmic singularity  at $x=\beta$, that was included  in  \eqn{qqgSIDIS}, this process
also generates a contribution which exhibits stronger, power-like, singularities and which is therefore
controlled by the lower limit  $x_{\rm min}$ of the integral over $x$, that is, by values of $x$ close to $\beta$.
Recall that $x$ is the longitudinal momentum fraction  of a semi-hard antiquark in the wavefunction
of the Pomeron. Hence, the fact that $x\simeq \beta$ implies that the virtual photon is absorbed by this
semi-hard antiquark, prior to its hard splitting. Accordingly, this particular LT contribution, as explicitly
computed in \eqn{softqSIDIS1},  it is likely to correspond to {\it final-state} radiation also in the target picture. 
Still  in Appendix~\ref{sec:3SIDIS}, we show in
Fig.~\ref{fig:two-integrals} a numerical comparison between the LT contributions generated by symmetric and asymmetric
 $\bar q g$ dijets, Eqs.~\eqref{softqSIDIS1} and respectively \eqref{qqgSIDIS}. We see that the latter dominates,
 especially at small $4 \beta P_{\perp}^2/Q^2$, where the integral in Eq.~\eqref{qqgSIDIS} grows logarithmically.

\section{The parton distributions of the Pomeron}
\label{sec:qPomeron}


While in the previous sections (and notably in Sect.~\ref{sec:MS}) we have presented 
qualitative estimates for the quark DTMD and the quark DPDF, our goal here will be their more accurate determination, 
via both analytic and numerical studies. Many of the characteristic features that we shall find are similar to
 those of the diffractive gluon distributions, as thoroughly studied 
in \cite{Iancu:2022lcw}. Inevitably we will repeat some parts of that previous analysis in order to work out the quark case.
In the process, and for the sake of comparison, we will review some features of the gluon case. The basic ingredient in calculating the (quark and gluon) diffractive distributions is the scattering amplitude $\mcal{T}(R,\YP)$ in the appropriate
representation of the colour group. In what follows we shall give analytic expressions relying on the MV model \cite{McLerran:1993ni,McLerran:1994vd} valid at moderate values of $\alpha_s \YP \ll 1$. For higher values such that $\alpha_s \YP \gtrsim 1$, we will  resort to numerical solutions to the BK equation \cite{Balitsky:1995ub,Kovchegov:1999yj} and its collinear improvement \cite{Iancu:2015vea,Ducloue:2019ezk}. 
Finally we will study the (coupled)  DGLAP evolution of the quark and gluon DPDFs.

\subsection{The quark DTMD at tree-level}
\label{sec:qDTMD}

The quark DTMD at tree level is defined in Eq.~\eqref{qDTMD}, where the dimensionless scalar quantity $\QP(x, x_{\mathbb{P}}, K_{\perp}^2)$ is given in Eq.~\eqref{QPpm}, where we recall that $\mcal{M}^2=\frac{x}{1-x}K_\perp^2$.

\begin{figure}
	\begin{center}
		\includegraphics[width=0.9\textwidth]{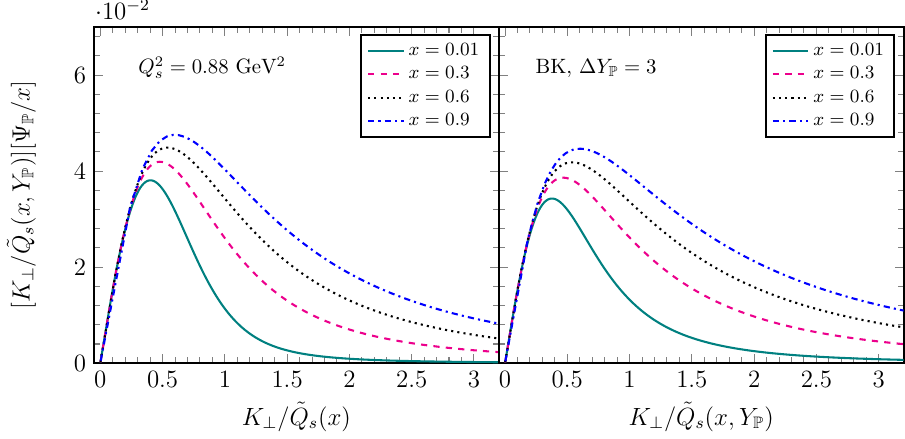}
	\end{center}
	\caption{\small The quark DTMD as obtained numerically from Eqs.~\eqref{qDTMD} and \eqref{QPpm}, scaled with $[K_{\perp}/\tilde{Q}_s(x,Y_{\mathbb P})]/x$ and plotted as a function of $K_\perp/\tilde Q_s(x,Y_{\mathbb P})$, for various values of $x$ and $Y_{\mathbb P}$. 
 	Left panel:  $\mathcal{T}({R})$ is given by the MV model with a saturation scale $Q_{s}^2=0.88\,\mathrm{GeV}^2$ (in the fundamental representation).
 	Right panel:  $\mathcal{T}({R,Y_{\mathbb P}})$ is obtained from the solution to the collinearly improved BK equation with the initial condition at 
	$Y_0$ given by the MV model. The respective saturation scale is $Q_{s}^2(\Delta\YP=3)=1.5\,\mathrm{GeV}^2$.}
\label{fig:PsiQq}
\end{figure}

We first present analytic estimates based on the MV model, cf.~\eqn{SMV}.
We start with the saturation regime at relatively
low transverse momenta $K_{\perp} \lesssim \tilde{Q}_s(x)$, with  $\tilde{Q}_s^2(x)=(1-x)Q_s^2$ (recall \eqn{qDTMDpw}). 
Then the integral in  Eq.~\eqref{QPpm}  is dominated by large dipoles of size $R \gtrsim 1/Q_s$, for which
the MV model is well approximated by a Gaussian (a.k.a. ``the GBW model'' \cite{GolecBiernat:1999qd}):
$\mcal{T}(R) \simeq 1- e^{-R^2 Q_s^2/4}$, with $Q_s$ the saturation momentum defined in \eqn{QsMV}.
We will  obtain fully analytical results for the two limiting cases $x \to 0$ and $x \to 1$ and then extrapolate
some of their features to generic values of $x$. 
When $x\ll 1$, or equivalently when $\mcal{M} \ll K_{\perp}$, Eq.~\eqref{QPpm} simplifies to 
\begin{align}
	\label{QPMVsatx0}
	\QP(x, K_{\perp}^2) 
	\simeq
	\mcal{M}
	\int_{0}^{\infty}\! \dif R\, 
	J_1(K_{\perp}R) 
	\left(1 - e^{-R^2 Q_s^2/4}\right)
	\simeq \sqrt{x}\, e^{-K_{\perp}^2/Q_s^2}
	\quad
	{\rm for}
	\quad
	K_{\perp} \lesssim Q_s,\,\, x\ll 1, 
\end{align}
which exhibits Gaussian momentum broadening. 
When $1-x \ll 1$, hence $\mcal{M} \gg K_{\perp}$, Eq.~\eqref{QPpm} becomes
\begin{align}
	\label{QPMVsatx1}
	\QP(x, K_{\perp}^2) 
	&\simeq
	\frac{\mcal{M}^2 K_{\perp}}{2}
	\int_{0}^{\infty} \dif R\, R^2
	K_1(\mcal{M} R) 
	\left(1 \!-\! e^{-R^2 Q_s^2/4}\right)
	\nn
	& \simeq 
	\sqrt{1-x}
	\left[
	1 - \frac{K_{\perp}^2}{\tilde{Q}_s^2}
	+ \frac{K_{\perp}^4}{\tilde{Q}_s^4}\,
	e^{K_{\perp}^2/\tilde{Q}_s^2}
	E_1 \big(K_{\perp}^2/\tilde{Q}_s^2\big)
	\right]
	\quad
	{\rm for}
	\quad
	K_{\perp} \lesssim \tilde{Q}_s,\,\, 1-x\ll 1,
\end{align}
where $E_1(z) = \int_{z}^{\infty} \dif t \, e^{-t}/t$ is the exponential integral function. While the above formulae reduce to the black disk limit expression \eqref{QPsmallK} when $K_{\perp} \ll \tilde{Q}_s$ as they should, it is important to notice that they are valid in a wide region which extends up to momenta of the order of $\tilde{Q}_s$. (In particular, when expanding for 
$K_{\perp}\ll \tilde{Q}_s$, Eqs.~\eqref{QPMVsatx0} and \eqref{QPMVsatx1} have the same dependence on $K_{\perp}$ up to quadratic order and they start to differ only at the fourth order.) 
These analytic approximations predict ``geometric scaling''
--- the function $\QP(x, K_{\perp}^2)/\sqrt{x (1-x)}$ appears to depend upon $K_\perp$ and $x$
only via the ratio  $K_{\perp}/\tilde{Q}_s(x)$ --- and we expect this property to be quite robust: a similar property should
be well satisfied by the exact result in the saturation regime. Recalling the definition \eqref{qDTMD} of the quark
occupation number in the Pomeron, we therefore expect the following scaling behaviour:
\begin{align}
	\label{scaleapprox}
	\frac{\Psi_{\mathbb P}(x,K_{\perp}^2)}{x}
	\simeq F(K_{\perp}/\tilde{Q}_s(x))
	\quad \textrm{for} \quad K_{\perp} \lesssim \tilde{Q}_s(x). 
\end{align}
Making use of Eqs.~\eqref{QPMVsatx0} and \eqref{QPMVsatx1}, we can also estimate the scaling violation in the transition regime around $\tilde{Q}_s$; for example at $K_{\perp} =\tilde{Q}_s/2$ we have
\begin{align}
	\label{scaleviol}
	\frac{\big[\Psi_{\mathbb P}(x, K_{\perp}= \tilde{Q}_s/2)/x\big]_{x\to 1}}
	{\big[ \Psi_{\mathbb P}(x,K_{\perp}= \tilde{Q}_s/2)/x \big]_{x\to 0}}
	=
	\left[\frac{12 e^{1/4}+ \sqrt{e} E_1(1/4)}{16}\right]^2 
	\simeq 1.15,
\end{align}
meaning  a small violation of 15\%. However, this violation rapidly increases when increasing  $K_{\perp}$ above
$\tilde{Q}_s/2$, to reach about 160\% at $K_{\perp} = \tilde{Q}_s$ (see also the left panel in Fig.~\ref{fig:PsiQq}).

Let us also consider large transverse momenta $K_{\perp} \gg \tilde{Q}_s$. Then the integration 
 in  Eq.~\eqref{QPpm}  is controlled by small dipoles of size $R \ll 1/Q_s$ and one can replace the MV model amplitude with its single scattering approximation, that is $\mcal{T}(R) \simeq (R^2 Q_A^2/4)\ln (4/R^2\Lambda^2)$. This calculation has been performed for arbitrary $x$ in Appendix \ref{sec:single}, but to our purposes here it suffices to present only the leading terms in the usual limiting cases. Using Eqs.~\eqref{app:QPKsmall} and \eqref{app:QPKlarge} we readily obtain
\begin{align}
\label{Psicases}
	\frac{\Psi_{\mathbb P}(x,K_{\perp}^2)}{x} 
	\simeq \frac{1}{2\pi}
	\begin{cases}
	{\displaystyle \frac{Q_A^4}{K_{\perp}^4}}
	&\quad \mathrm{for} \quad 
	K_{\perp} \gg Q_s,\,\,x\ll 1
	\\*[0.5cm]
	{\displaystyle \frac{4 (1-x)^2 Q_A^4}{K_{\perp}^4}\,
	\ln^2 \frac{K_{\perp}^2}{(1-x)\Lambda^2}}
	&\quad \mathrm{for} \quad K_{\perp} \gg \tilde{Q}_s,\,\,1-x\ll 1.
	\end{cases}
\end{align}
We note that when we express $Q_A^2$ in terms of the saturation momentum $Q_s^2$, cf.~\eqn{QsMV},
 the factor $(1-x)^2$ in the second case of the above can be absorbed 
 into the dependence on the effective saturation scale $\tilde{Q}_s(x)$. 
 Eventually this indicates that, even though Eq.~\eqref{Psicases} 
 does not scale with $K_{\perp}/\tilde{Q}_s(x)$, the scaling violation happens only through
  a logarithmic (and not a power) dependence on $x$. 

The MV model, on which the above analysis in based, is valid when the diffractive gap $\YP$ is not too large, such that
$\alpha_s\YP \ll 1$. High energy evolution becomes relevant when $\YP \gtrsim 1/\alpha_s$ and is taken into account by solving the BK equation \cite{Balitsky:1995ub,Kovchegov:1999yj}, or more precisely its collinearly improved version \cite{Beuf:2014uia,Iancu:2015vea,Ducloue:2019ezk}. Using the MV model amplitude $\mcal{T}(R)$ as an initial condition at $Y_0$ (typically $Y_0 \sim 4 $), one solves numerically this equation to obtain the amplitude $\mcal{T}(R,\YP)$
at $\YP = Y_0+\Delta \YP$.
 The solution $\mcal{T}(R,\YP)$ is characterised by a saturation momentum $Q_s(\YP)$ which increases with $\YP$.
 For large dipole sizes $R \gg 1/Q_s(\YP)$, the solution approaches the black disk limit $\mcal{T}(R,\YP) =1$.
For small dipoles with  $R \ll 1/Q_s(\YP)$, it exhibits the slightly softer tail (in comparison to the MV model) $\mcal{T}(R,\YP) \sim [R^2 Q_s^2(\YP)]^{\gamma}$ with $1/2<\gamma<1$ ($\gamma \simeq 0.63$ for the leading logarithmic evolution
and for asymptotically high $\YP$). Therefore, the high energy evolution will not alter the main qualitative 
features of the quark DTMD, except from modifying its hard momentum behaviour to 
$[\tilde{Q}_s^2(x,\YP)/K_{\perp}^2]^{2\gamma}$.

For the numerical implementation we regularize the infrared limit of the MV model according to
\begin{align}
	\label{MVreg}
	\mcal{T}(R) = 1 - \exp\left[
	-\frac{Q_A^2 R^2}{2} \ln \left(\frac{2}{\Lambda R} +e \right)
	\right].
\end{align}
Such a scattering amplitude is evolved in $\YP$ according to the collinearly improved BK evolution \cite{Ducloue:2019ezk} with the precise conventions given in Appendix D in \cite{Iancu:2020jch}. The corresponding amplitude in the adjoint representation at large-$N_c$ is determined by
\begin{align}
	\label{Tadjoint}
	\mcal{T}_g(R,\YP) = 1- [1- \mcal{T}(R,\YP)]^2.
\end{align} 
For the purposes of the numerical calculations we shall use a slightly modified definition of the saturation scale (than the one we have used so far for the analytical calculations and estimates, cf.~Eq.~\eqref{QsMV} in the MV model), namely $\mcal{T}(R=2/Q_s,\YP) = 1/2$ and similarly for $Q_{s,g}$. We choose the value of $Q_A^2$ in Eq.~\eqref{MVreg} in such a way that the initial adjoint saturation momentum at $Y_0$ is $Q_{s,g}^2=2\, \mathrm{GeV}^2$, while we also fix the QCD scale to $\Lambda=0.2\,  \mathrm{GeV}$. In turn, this leads to $Q_s^2=0.88\, \mathrm{GeV}^2$ for the saturation momentum in the fundamental representation.

\begin{figure}
	\begin{center}
		\includegraphics[width=0.9\textwidth]{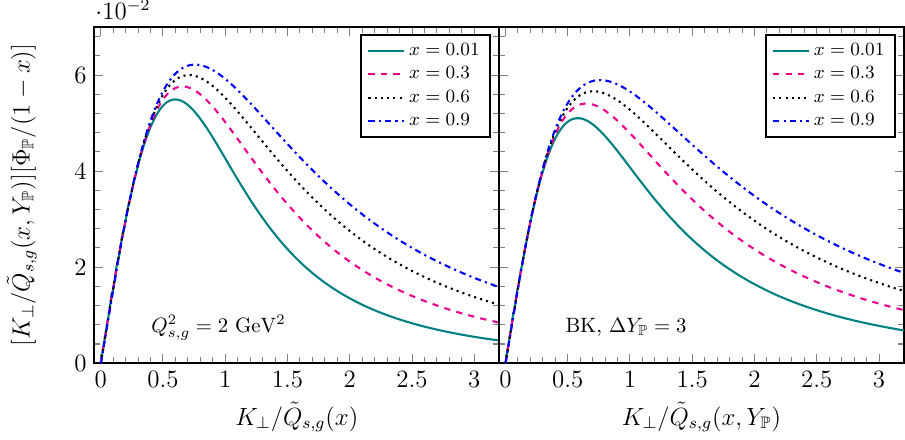}
	\end{center}
	\caption{\small The gluon DTMD as obtained numerically from Eqs.~\eqref{pomugddef} and \eqref{Gscalarnew}, scaled with $[K_{\perp}/\tilde{Q}_{s,g}(x,Y_{\mathbb P})]/(1-x)$ and plotted as a function of $K_\perp/\tilde Q_{s,g}(x,Y_{\mathbb P})$, for various values of $x$ and $Y_{\mathbb P}$. The amplitude in the adjoint representation is determined by Eq.~\eqref{Tadjoint}, where the amplitude $\mcal{T}(R,\YP)$ in the fundamental is obtained as in Fig.~\ref{fig:PsiQq}.
 	Left panel: MV model calculation with a corresponding adjoint saturation scale $Q_{s,g}^2=2\, \mathrm{GeV}^2$. Right panel: collinearly improved BK calculation with a corresponding adjoint saturation scale $Q_{s,g}^2(\Delta\YP=3)=3.9\, \mathrm{GeV}^2$.}
\label{fig:PhiPg}
\end{figure}

In Fig.~\ref{fig:PsiQq} we present the results of the numerical calculation of the quark DTMD, which confirm all the properties presented in this section. In this plot, we have scaled the distribution with a factor $K_{\perp}/\tilde{Q}_s(x)$,
in order to exhibit the radial distribution in $K_{\perp}$ and thus emphasise the fact that its bulk 
is concentrated at low momenta $K_{\perp} \lesssim \tilde{Q}_s$. The numerical results show that
 the scaling in Eq.~\eqref{scaleapprox} is rather good in the regime $K_{\perp} \lesssim \tilde{Q}_s/2$ and violated when $K_{\perp} \gg \tilde{Q}_s$, while we also observe the aforementioned softening of the high momentum behavior induced by the BK evolution. For comparison, in Fig.~\ref{fig:PhiPg} we also show the respective 
results for the gluon DTMD \cite{Iancu:2022lcw}. As anticipated, the qualitative behaviour (notably,  the approximate
geometric scaling and the pattern of its violation, the existence
of a pronounced peak around $K_{\perp} = \tilde{Q}_s$ and the power law decay at larger momenta) 
is very similar for quarks and gluons.

\subsection{The quark DPDF at tree level}
\label{sec:qDPDF}

\begin{figure}[t] 
\centerline{\includegraphics[width=0.8\columnwidth]{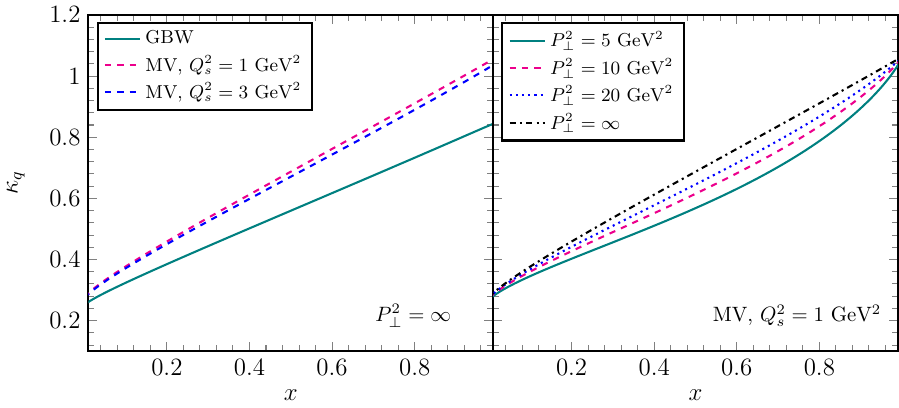}}
\caption{\small Left panel: The slowly varying function $\kappa_q(x)$ appearing in the quark DPDF in Eq.~\eqref{xqpinfty} for the GBW model (for which it is independent of $Q_s$) and for the MV model for two different saturation scales. Right panel: The function $\kappa_q(x,P_{\perp}^2)$ for various values of $P_{\perp}^2$ (with $P_{\perp}^2>Q_s^2$) as a function of $x$ in the MV model with $Q_s^2=1$ GeV$^2$.}
\label{fig:kappax}
\end{figure}

Now we turn our attention to the quark DPDF $xq_{\mathbb P}(x,x_{\mathbb P}, P_{\perp}^2)$ defined in Eq.~\eqref{xqP} for large momenta such that $P_{\perp}^2 \gg \tilde{Q}^2_s(x,\YP)$. As we have already seen, the unintegrated distribution, whether obtained in the MV model or from the solution to the BK equation, falls fast enough for momenta larger than the saturation momentum so that it is integrable. Hence, the dependence of the quark DPDF on $P_{\perp}^2$ is rather weak and to a first approximation we can take $P_{\perp}^2 \to \infty$. Eventually the integrand is controlled by momenta of the order of $\tilde{Q}_s(x,\YP)$, cf.~Eq.~\eqref{Psicases} and Fig.~\ref{fig:PsiQq}, and thus the quark DPDF inherits the approximate scaling properties of the quark DTMD. We readily obtain
\begin{align}
	\label{xqpinfty}
	x q_{\mathbb P}(x,x_{\mathbb P}, P_{\perp}^2=\infty)
	\simeq
	\frac{S_{\perp}N_c}{4\pi^3} \,
	\kappa_q(x) x (1-x) Q_s^2(\YP),
\end{align}
where $\kappa_q(x)$ is a slowly varying function of $x$ in the interval $[0,1]$. Its precise form depends on the details of the scattering amplitude $\mcal{T}(R,\YP)$, hence it also depends weakly on $\YP$, and can be easily determined numerically or even analytically in certain simple cases. For example, in the GBW model one can make use of Eqs.~\eqref{QPMVsatx0} and \eqref{QPMVsatx1} to find \cite{Buchmuller:1998jv,Hatta:2022lzj} $\kappa_q(0)=1/4$ and  $\kappa_q(1)=3\pi^2/16 -1 \simeq 0.85$. In the left panel of Fig.~\ref{fig:kappax} we show $\kappa_q(x)$ in the GBW and MV models.

One can generalize Eq.~\eqref{xqpinfty} to include the mild dependence on $P_{\perp}^2$ by writing
\begin{align}
	\label{xqpfinite}
	x q_{\mathbb P}(x,x_{\mathbb P}, P_{\perp}^2)
	\simeq
	\frac{S_{\perp}N_c}{4\pi^3} \,
	\kappa_q(x,x_{\mathbb P},P_{\perp}^2) x (1-x) Q_s^2(\YP),
\end{align}
and where it is a straightforward exercise to find that (by subtracting from Eq.~\eqref{xqpinfty} the contribution for momenta larger than $P_{\perp}$)
\begin{align}
	\label{kappafinite}
	\kappa_q(x,x_{\mathbb P},P_{\perp}^2) = \kappa_q(x) - 
	\frac{\pi \int_{P_{\perp}^2}^\infty \dif K_{\perp}^2\, \Psi_{\mathbb P}(x,x_{\mathbb P},K_{\perp}^2)}{x(1-x)Q_s^2(\YP)}.
\end{align}
Using the above and the large $K_{\perp}$ behavior of the MV model in Eq.~\eqref{Psicases}, one immediately finds that the correction due to the dependence on $P_{\perp}^2$ is of the order of $\tilde{Q}_s^2(x)/P_{\perp}^2$, and therefore very small as anticipated. In the right panel of Fig.~\ref{fig:kappax} we exhibit how $\kappa_q(x,x_{\mathbb P},P_{\perp}^2)$ approaches the asymptotic line $\kappa_q(x)$ in the MV model.

Finally, it is clear that BK evolution\footnote{Notice that we keep the ``tree-level'' denomination for the quark and gluon
DPDFs even after including the effects of the high-energy (BK/JIMWLK) evolution; indeed, the respective {\it evolved}
distributions will be defined as the solutions to the DGLAP equations to be discussed in the next section.}
 will lead to an overall strong increase of the quark DPDF, since the latter is proportional to $Q_s^2(\YP)$. This is the main dependence of the quark DPDF on $\YP$; indeed, as we have already pointed out $\kappa_q(x)$ depends very weakly on $\YP$ (so we have opted not to write it explicitly) while the $P_{\perp}^2$-dependent correction in Eq.~\eqref{kappafinite} now becomes of the order of $[\tilde{Q}_s^2(x,\YP)/P_{\perp}^2]^{2\gamma-1}$ and still remains very small.

\begin{figure}
	\begin{center}
		\includegraphics[width=0.9\textwidth]{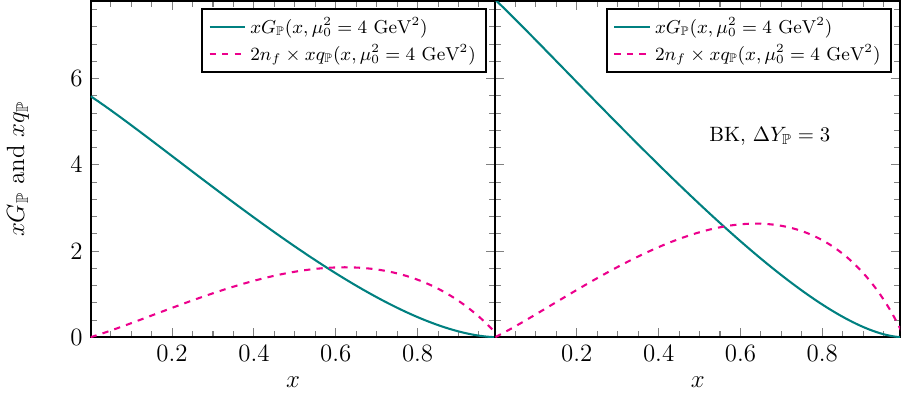}
	\end{center}
	\caption{\small The quark  DPDF (multiplied by the total number of active flavors including antiquarks) and 
	the gluon DPDF as functions of $x$ evaluated at the scale $\mu_0^2= 4\, \mathrm{GeV}^2$ and with the scattering amplitude given by the MV model (left), or by the solution to the collinearly improved BK equation (right). A factor $S_{\perp}/4\pi^3$ has been omitted.}
\label{fig:xdep_source}
\end{figure}

In Fig.~\ref{fig:xdep_source} we show how the quark and gluon DPDFs depend on $x$. We evaluate them at the scale $\mu_0^2= 4\, \mathrm{GeV}^2$ for reasons which will become clear in Sect.~\ref{sec:DGLAP} and with a scattering amplitude given, as usual, by the MV model or by the solution to the collinearly improved BK equation. Both panels verify the 
qualitative behaviour visible in Eqs.~\eqref{xqpfinite} (for quarks) and  \eqref{xGPhigh} (for gluons).
 For $x=0$ the quark DPDF vanishes while the gluon one is non-zero, while as $x\to 1$ the gluon vanishes faster than the quark. Moreover, confronting the right with the left panel, we also confirm the previously discussed dependence on $\YP$: the quark DPDF plot on the right panel is almost a perfect rescaling of the corresponding plot on the left with a factor $Q_s^2(\YP)/Q_s^2(Y_0)$, and similarly for the gluon.
 
\subsection{DGLAP evolution of the DPDFs}
\label{sec:DGLAP} 
When $P_{\perp}^2 \gg Q_s^2(\YP)$, there is a large phase space available for partonic emissions between the semi-hard jet and the hard dijet system (that is, outside the rapidity gap which remains unaltered). The dominant corrections come
from emissions  which are strongly ordered in transverse momenta and which express the DGLAP evolution of the parton DPDFs. As shown by our analysis of NLO corrections to diffractive SIDIS in  Sect.~\ref{sec:SIDIS}, this evolution
also emerges from the colour dipole picture, although our respective argument was  incomplete (we did not compute
the relevant virtual corrections). Assuming this to be the case, in this section we shall study the coupled DGLAP evolution
of the quark and gluon DPDFs. Yet another lesson from the discussion in  Sect.~\ref{sec:SIDIS} is the fact that
the tree-level results for the DTMDs naturally act as {\it source terms} in the DGLAP equations for the respective
DPDFs  (see also \cite{Iancu:2022lcw,Iancu:2023lel}).   Taking this into account, the relevant equations are
\begin{align}
	\label{dglapmatrix}
	\hspace{-0.4cm}
	\frac{\dif}{\dif \ln P_{\perp}^2} 
	\begin{bmatrix}
	x G_{\mathbb P}(x,x_{\mathbb P},P_{\perp}^2) \\*[0.1cm]
	x q_{\mathbb P}(x,x_{\mathbb P},P_{\perp}^2)
\end{bmatrix}
=\,\, &
\pi P_{\perp}^2
\begin{bmatrix}
	N_g \Phi_{\mathbb P} (x,x_{\mathbb P},P_{\perp}^2) \\*[0.1cm]
	N_c \Psi_{\mathbb P} (x,x_{\mathbb P},P_{\perp}^2)
\end{bmatrix}
\nn 
&+\frac{\alpha_s(P_{\perp}^2)}{2\pi}
\int_{x}^1 \dif z
\begin{bmatrix}
	P_{gg}(z) & 2 n_f P_{gq}(z) \\*[0.1cm]
	P_{qq}(z) & P_{qg}(z)
\end{bmatrix}
\begin{bmatrix}
	{\displaystyle (x/z) G_{\mathbb P}(x/z,x_{\mathbb P},P_{\perp}^2)} 
	\\*[0.1cm]
	{\displaystyle (x/z) q_{\mathbb P}(x/z,x_{\mathbb P},P_{\perp}^2)}
\end{bmatrix},
\end{align}
where we have omitted a factor $S_{\perp}/4\pi^3$, which is common to both gluon and quark distributions and can be easily reinserted in the final results. Still, we have kept the corresponding color factors which are different, namely $N_g=N_c^2-1$ and $N_c$. The factor $2 n_f=6$ in front of the $P_{gq}$ splitting function sums the quark and antiquark contributions from all light flavors. This simplification occurs, because in our framework all distributions for such different flavours are identical. The splitting functions in Eq.~\eqref{dglapmatrix} read as follows:
\begin{align}
	\label{splittingfs}
	&\hspace{-0.7cm}
	P_{gg}(z) = 2N_c 
	\left[
	\frac{1-z}{z} + z(1-z) + \frac{z}{(1-z)_+}
	\right]
	+\frac{11N_c - 2 n_f}{6}\, \delta(1-z),
	\nn
	&\hspace{-0.7cm}
	P_{gq}(z) = C_F\,\frac{1+(1-z)^2}{z},
	\quad
	P_{qq}(z) = C_F\, \frac{1+z^2}{(1-z)_+} + 2 \delta(1-z),
	\quad
	P_{qg}(z)=\frac{1}{2}\left[z^2 + (1-z)^2\right].
\end{align}
As compared to their previous expressions in Sect.~\ref{sec:SIDIS}, they now include additional pieces, notably
the  plus prescription defined as
\begin{align}
	\label{pluspres}
	\int_x^1
	\dif z\, \frac{f(z)}{(1-z)_+} \equiv
	\int_x^1 \dif z\,
	\frac{f(z) - f(1)}{1-z}
	+
	f(1) \ln(1-x),
\end{align}
that should be generated after adding the relevant virtual corrections.
The QCD running coupling  is evaluated at the one loop level, that is
\begin{align}
	\label{rcoupling}
	\alpha_s(P_{\perp}^2) = \frac{1}{b \ln P_{\perp}^2/\Lambda^2}
	\quad \mathrm{with} \quad
	b = \frac{11N_c - 2 n_f}{12 \pi}.
\end{align}

In these  equations, the rapidity gap $\YP$ plays the role of a parameter: the dependence upon $\YP$ only
enters via the source terms, which are obtained from solutions to the BK/JIMWLK equation (as explained in the
previous sections).

The solutions to \eqn{dglapmatrix} are physically meaningful only for values of the resolution scale $P_{\perp}^2$ 
which are significantly larger than the target saturation momentum $Q_s^2(\YP)$: indeed, as already mentioned,
the physical phase-space for DGLAP-like emissions is $\ln(P_\perp^2/Q_s^2)$. This means that the
DGLAP evolution must be ``turned on'' at some scale $\mu_0^2$, which is the lowest value of $P_{\perp}^2$ 
for which the effects of this evolution start to be important.  Specifically, this scale should be larger than
the {\it gluon} saturation momentum  $Q_{s,g}^2$ (which is the largest between the two saturation scales), 
but at the same time obey $\alpha_s \ln (\mu_0^2/Q_{s}^2) \ll 1$ (to avoid losing important
effects of the DGLAP evolution). Different choices for $\mu_0^2$ 
 which differ, say, by a factor of 2 should be seen as a source of scheme dependence.
Accordingly, the solutions to \eqn{dglapmatrix} can be fully trusted only for sufficiently large values $P_{\perp}^2\gg \mu_0^2$,
where this scheme dependence becomes less important. (Indeed, this dependence decreases 
with increasing $P_\perp$ above $\mu_0$; see e.g.~\cite{Iancu:2023lel} for some numerical tests.)

The most interesting values of $x$ are of the order of 1/2, i.e.~they are not very close to the endpoints of the allowed kinematic regime. Still, we can get a rather useful insight to the solution of the DGLAP equations by studying analytically the two limiting cases $x\to 0$ and $x\to 1$. 

(i) When $x\ll 1$, the gluon evolution is driven by the singular $ 2 N_c/z$ part of the $P_{gg}(z)$ splitting function at small $z$, which leads to logarithmically enhanced contributions, as coming from the lower limit $x$ of the integral over $z$.
Notice that the $P_{gq}(z)$ splitting function is singular too in that limit, but since $x q_{\mathbb P}(x=0,x_{\mathbb P}, P_{\perp}^2)=0$ at tree level (cf.~Eq.~\eqref{xqpfinite}), the corresponding contribution will be subleading. Indeed, in order to create a non-zero quark distribution at $x=0$, a gluon must split to a $q\bar{q}$ pair and such a process does not generate a large logarithm since $P_{qg}(z)$ is finite at small $z$. It is straightforward to resum the dominant double logarithms and obtain the leading term of the gluon DPDF as \cite{Iancu:2022lcw}
\begin{align}
	\label{gluonDGLAPx0}
	xG_{\mathbb{P}}(x, x_{\mathbb{P}}, P_{\perp}^2)
	\simeq
	N_g\kappa_g(0)
	Q_{s,g}^2(Y_{\mathbb P})\,
	I_0
	\left(\sqrt{\frac{4 N_c}{\pi b}
	\ln \frac{\alpha_s(\mu_0^2)}
	{\alpha_s(P^2_{\perp})}
	\ln\frac{1}{x}}
	\right)
	\quad 
	\mathrm{for} 
	\quad x \ll 1,
\end{align}
where $\mu_0^2$ is the scale at which we initiate the evolution. This gluon distribution acts also as a seed for the corresponding quark distribution. More precisely, the latter receives its dominant contribution from a series of sequential $g \to gg$ splittings followed by a single $g\to q\bar{q}$ splitting only in the very last step and we find 
\begin{align}
	\label{quarkDGLAPx0}
	\hspace*{-0.3cm}
	xq_{\mathbb{P}}(x, x_{\mathbb{P}}, P_{\perp}^2)
	\simeq
	\frac{N_g\kappa_g(0)
	Q_{s,g}^2(Y_{\mathbb P})}{3}\,
	\sqrt{\frac{
	\ln \frac{\alpha_s(\mu_0^2)}{\alpha_s(P^2_{\perp})}}
	{ 4 \pi b N_c \ln \frac{1}{x}}}\,
	I_1
	\left(\sqrt{\frac{4 N_c}{\pi b}
	\ln \frac{\alpha_s(\mu_0^2)}
	{\alpha_s(P^2_{\perp})}
	\ln\frac{1}{x}}
	\right)
	\quad 
	\mathrm{for} 
	\quad x \ll 1.
\end{align}
The aforementioned $g\to q\bar{q}$ transition, gives rise to the extra factor $\int_0^1 \dif z\, P_{qg}(z)=1/3$ in the above. To conclude, for asymptotically small values of $x$, both the gluon and the quark DPDFs increase with increasing $P_{\perp}^2$ and/or decreasing $x$, since they are determined by real emissions only. Needless to say, the DGLAP evolution generates a non-zero value of $xq_{\mathbb{P}}(x, x_{\mathbb{P}}, P_{\perp}^2)$ as $x\to 0$. 

(ii) When $1-x \ll 1$ the second term in Eq.~\eqref{pluspres}, which arises from virtual graphs and is proportional to a large $\ln(1-x)$ factor, controls the evolution as one can verify by successive iterations of the tree level term. Such a piece is present only in the diagonal splitting functions $P_{gg}(z)$ and $P_{qq}(z)$, hence it becomes clear that the two parton distributions decouple from each other. For the gluon DPDF we find \cite{Iancu:2022lcw}
\begin{align}
	\label{gluonDGLAPx1}
	xG_{\mathbb{P}}(x, x_{\mathbb{P}}, P_{\perp}^2)
	\simeq
	N_g\kappa_g(1)
	Q_{s,g}^2(Y_{\mathbb P})
	(1-x)^{\textstyle 2 + 
	\frac{N_c}{\pi b} 
	\ln \frac{\alpha_s(\mu_0^2)}
	{\alpha_s(P^2_{\perp})}}
	\quad 
	\mathrm{for} 
	\quad  1-x \ll 1
\end{align}
and similarly for the quark DPDF
\begin{align}
	\label{quarkDGLAPx1}
	xq_{\mathbb{P}}(x, x_{\mathbb{P}}, P_{\perp}^2)
	\simeq
	N_c \kappa_q(1)
	Q_s^2(Y_{\mathbb P})
	(1-x)^{\textstyle 1 + 
	\frac{C_F}{\pi b} 
	\ln \frac{\alpha_s(\mu_0^2)}
	{\alpha_s(P^2_{\perp})}}
	\quad 
	\mathrm{for} 
	\quad  1-x \ll 1.
\end{align}
The two above expressions exhibit an increased suppression for $x\to 1$ as compared to the respective tree level results. Furthermore, both DPDFs decrease with increasing $P_{\perp}^2$ (in sharp contrast to what happens when $x \ll 1$), a property which is a corollary of the fact that the virtual terms drive the evolution. All this is in agreement with a physical picture in which the modes with large $x$ get depleted due to the successive branchings, whereas the emerging partons populate the modes with smaller values of $x$.

Now we move on to present numerical solutions to the DGLAP equations.
 In \cite{Iancu:2023lel} it has been numerically confirmed that the scheme dependence related to the precise choice of $\mu_0^2$ is rather small. In what follows we shall take $\mu_0^2 =4\, \mathrm{GeV}^2$, the value for which we presented the DPDFs at tree level in Fig.~\ref{fig:xdep_source}.

\begin{figure}
	\begin{center}
		\includegraphics[width=0.9\textwidth]{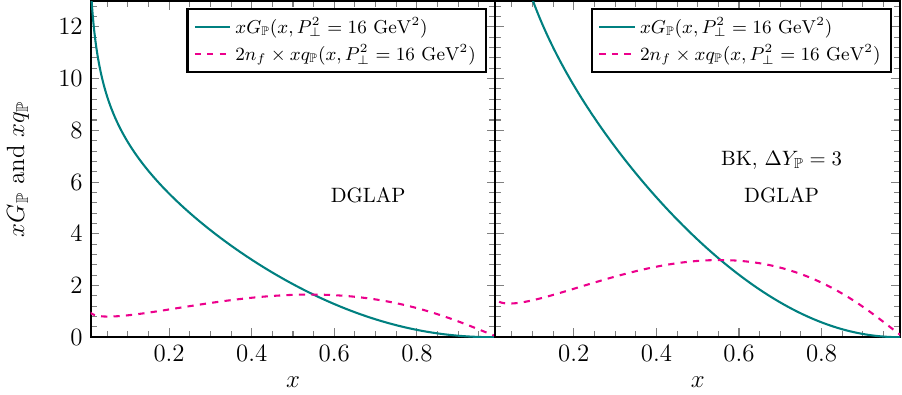}
	\end{center}
	\caption{\small The DGLAP evolved quark and gluon DPDFs as functions of $x$ evaluated at the scale $P_{\perp}^2 = 16\, \mathrm{GeV}^2$ and with the scattering amplitude given by the MV model (left) or by the solution to the collinearly improved BK equation (right). A factor $S_{\perp}/4 \pi^3$ has been omitted.}
\label{fig:xdep_DGLAP}
\end{figure}

\begin{figure}
	\begin{center}
		\includegraphics[width=0.9\textwidth]{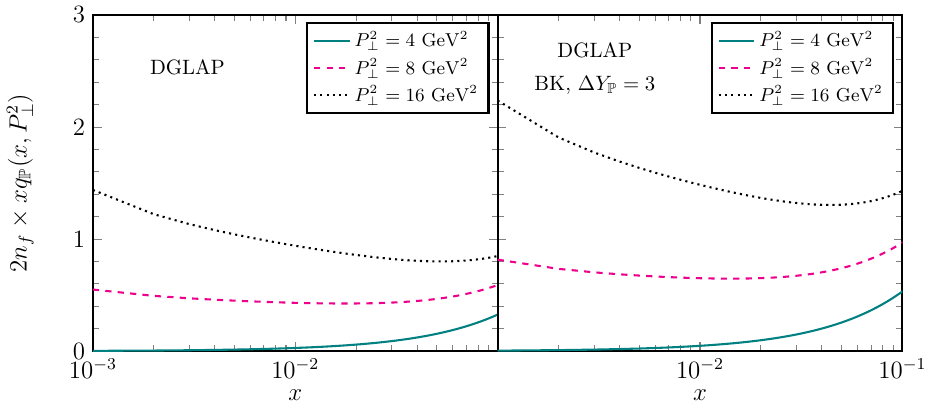}
	\end{center}
	\caption{\small The quark DPDF as a function of $x$ evaluated for various values of $P_{\perp}^2$ and with the scattering amplitude given by the MV model (left) or by the solution to the collinearly improved BK equation (right). A factor $S_{\perp}/4 \pi^3$ has been omitted.}
\label{fig:xqpsmallx}
\end{figure}

\begin{figure}
	\begin{center}
		\includegraphics[width=0.9\textwidth]{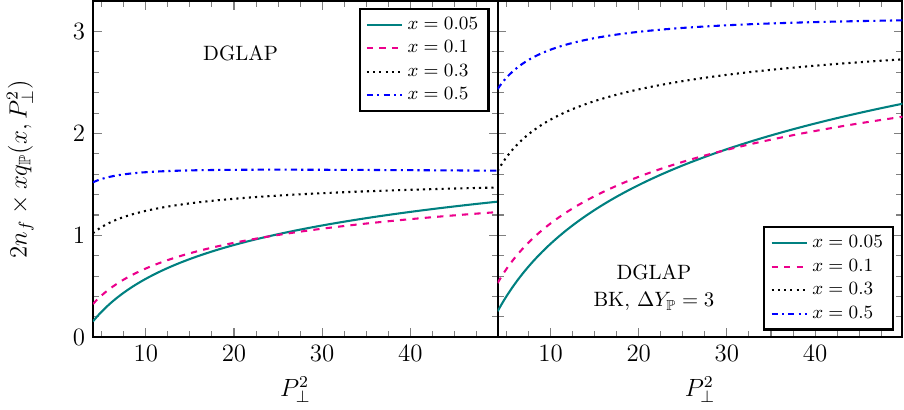}
		\\*[0.3cm]
		\includegraphics[width=0.9\textwidth]{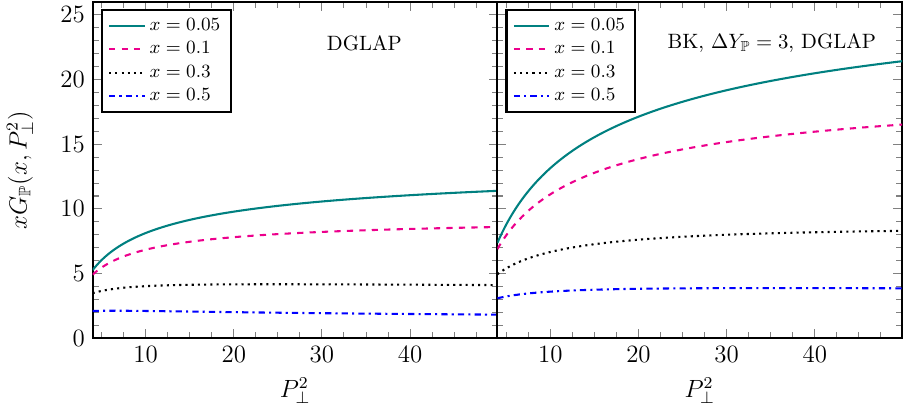}
	\end{center}
	\caption{\small The DGLAP evolved quark (top panel) and gluon (bottom panel) DPDFs as functions of $P_{\perp}^2$ for various values of $x$. The scattering amplitude is given by the MV model (left) or by the solution to the collinearly improved BK equation (right). A factor $S_{\perp}/4 \pi^3$ has been omitted.}
\label{fig:pdep_DGLAP}
\end{figure}

 In Fig.~\ref{fig:xdep_DGLAP} we show the $x$-dependence of the evolved DPDFs for a resolution scale 
 $P_{\perp}^2 = 16\, \mathrm{GeV}^2$, for both $\Delta \YP=0$ (where the source terms are constructed with the 
 MV model) and for $\Delta \YP=3$ (where they are obtained from the solution to the BK equation, as explained in
 Sect.~\ref{sec:qDTMD}).
As $x$ approaches zero, we observe that the evolution induces a sharp  increase in the gluon DPDF, 
which is much stronger than the one seen at tree level in Fig.~\ref{fig:xdep_source}. When $x\to 0$, 
the numerical solution appears to diverge, in agreement with the analytic expectation \eqref{gluonDGLAPx0}.  
Still for $x\to 0$, the quark DPDF is not vanishing any more, since its evolution is driven by the gluon distribution.
The asymptotic approximation \eqref{quarkDGLAPx0} predicts an increase in the quark DPDF as
 $x\to 0$ and the onset of this behaviour is marginally visible in Fig.~\ref{fig:xdep_DGLAP}.  To better
 visualise this behaviour, in Fig.~\ref{fig:xqpsmallx} we expand the scale at small $x$. (Clearly, the increase
 in the quark DPDF at small $x$ will become more and more pronounced with increasing  $P_{\perp}^2$; see also
 Fig.~\ref{fig:pdep_DGLAP}.)
  When $1-x \ll 1$, both distributions fall faster than the corresponding tree level ones in accordance with Eqs.~\eqref{gluonDGLAPx1} and \eqref{quarkDGLAPx1}. It goes without saying that the $\YP$-dependence is the same as in Fig.~\ref{fig:xdep_source}, that is, in going from the left to the right panel, the DPDFs scale with the respective saturation 
  scales, cf.~Eqs.~\eqref{xqpfinite} and  \eqref{xGPhigh}.


In Fig.~\ref{fig:pdep_DGLAP} we present the transverse momentum dependence of the DPDFs for various values of $x$. With increasing $P_{\perp}^2$ they increase when $x$ is small and they decrease when $x$ starts to get close to one (cf.~the discussions below Eqs.~\eqref{quarkDGLAPx0} and \eqref{quarkDGLAPx1}).

\begin{figure}
	\begin{center}
		\includegraphics[width=0.9\textwidth]{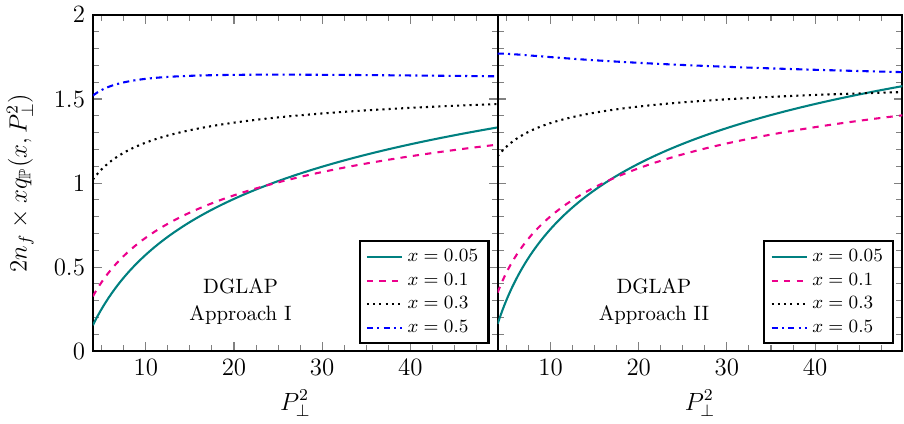}
		\\*[0.3cm]
		\includegraphics[width=0.9\textwidth]{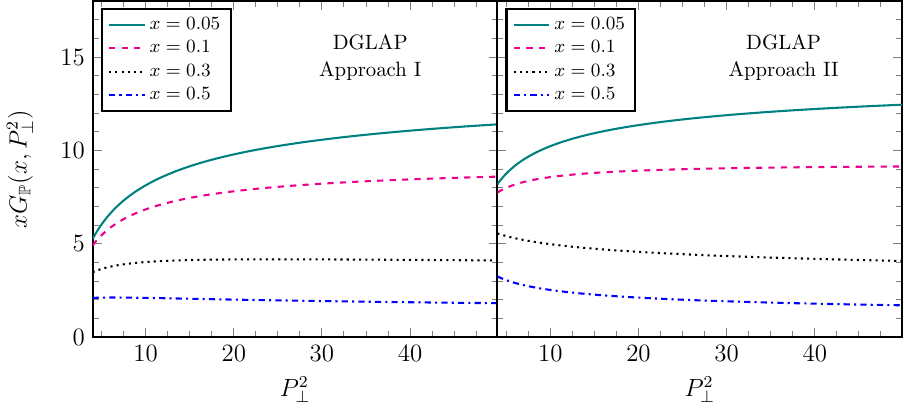}
	\end{center}
	\caption{\small The DGLAP evolved quark (top panel) and gluon (bottom panel) DPDFs as functions of $P_{\perp}^2$ for various values of $x$, obtained by employing two different ``schemes''. Left: DGLAP evolution with a source term (Approach I), as given in Eq.~\eqref{dglapmatrix}. DGLAP evolution as an initial value problem (Approach II), as discussed in Sect.~\eqref{sec:DGLAP}. The scattering amplitude is given by the MV model in all panels. A factor $S_{\perp}/4 \pi^3$ has been omitted.} 
\label{fig:schemedep}
\end{figure}

To conclude this section, we would like to point out that it is possible to reformulate Eq.~\eqref{dglapmatrix} as an
initial value problem, which is a more standard formulation of the DGLAP evolution. 
To that aim, one should replace the coupled evolution equations \eqref{dglapmatrix} with {\it sourceless} equations, 
which are endowed with initial conditions at
$P_\perp^2=\mu_0^2$, as obtained by integrating the tree-level quark and gluon DTMDs over $K_{\perp}^2$ either
up to $\mu_0^2$, or up to infinity. (The difference between the two choices for the initial conditions should be 
small, and part of our scheme dependence, due to the rapid fall off of the DTMDS when $K_{\perp}^2\gg \tilde Q_s^2(x)$.)
By construction, these two approaches to the DGLAP evolution --- that involving the source terms
and its more traditional version, which is sourceless ---
 should lead to similar results when we evolve up to a scale $P_{\perp}^2$ which is much larger than $\mu_0^2$. Their differences at finite $P_\perp^2$  should be seen as a form of scheme dependence. We have numerically tested this source of scheme dependence  in Fig.~\ref{fig:schemedep}, by  comparing 
the solutions numerically obtained in both approaches. (The initial conditions for the sourceless equations have
been inferred by integrating  the DTMDs over $K_{\perp}^2$ up to infinity.)  As expected, the difference becomes 
negligible at large momenta, particularly for values of $x$ which are not too small. In what follows we shall use only the source term approach, since it is physically better motivated for the problem under consideration (cf.~the discussion in Sect.~\ref{sec:SIDIS}).

\subsection{The diffractive structure function}
\label{sec:f2d3}

The diffractive structure function $x_{\mathbb{P}} F_2^{D(3)}$ has been defined in Eq.~\eqref{F2D} and to (strict) 
leading order and leading twist is determined by the quark (and antiquark) DPDFs as shown in Eq.~\eqref{xqD}. At the end of Sect.~\ref{sec:MS} we have already looked briefly at its behaviour and also emphasised the fact that,
in order to obtain a realistic behaviour near $x=1$, one must include a higher twist contribution coming from longitudinally polarised virtual photons. In Sect.~\ref{sec:SIDIS} we have shown the emergence of the DGLAP evolution 
 in the context of diffractive SIDIS and concluded by writing down a better approximation  for
$x_{\mathbb{P}} F_2^{D(3)}$, cf.~Eq.~\eqref{xqD3}, which includes the DGLAP evolution of the quark DPDF.
(Strictly speaking, it is this improved approximation that should be viewed as the {\it proper} leading-order and leading twist
result for the diffractive structure function.) In this section, we will revisit the tree level results presented in Sect.~\ref{sec:MS}, 
by including the effects of the DGLAP evolution. 
It would be clearly interesting to compare our subsequent results with the predictions of the recent 
next-to-leading order calculation
of the diffractive structure function in the colour dipole picture \cite{Beuf:2024msh}.
\begin{figure}
	\begin{center}
		\includegraphics[width=0.78\textwidth]{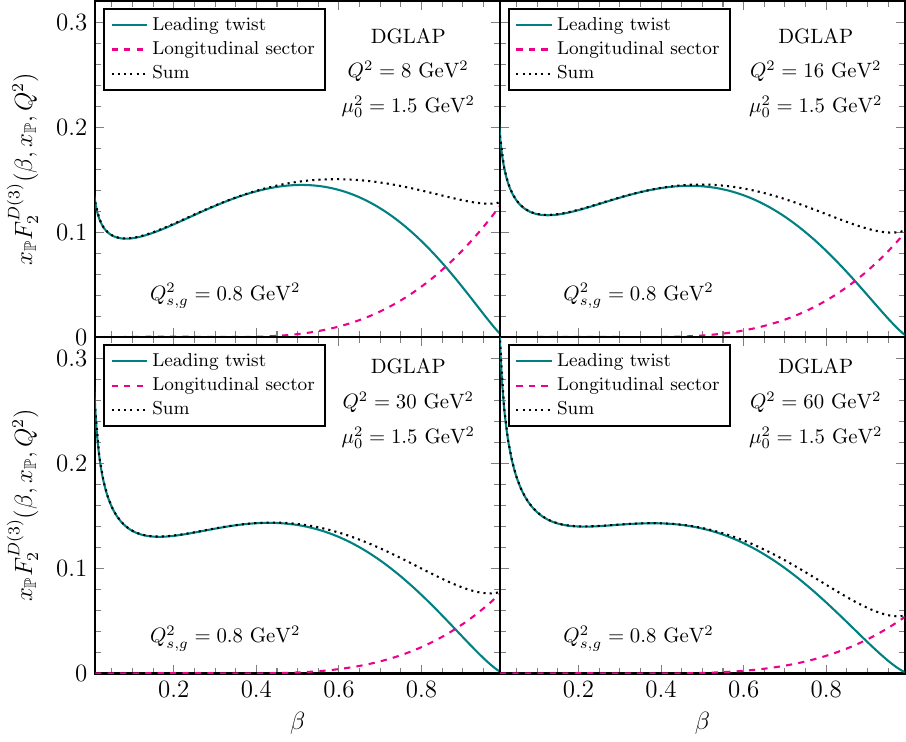}
	\end{center}
	\caption{\small The proton diffractive structure function $x_{\mathbb{P}} F_2^{D(3)}$ and the separate contributions from the transverse (leading twist) and longitudinal sector (higher twists) as functions of $\beta$ for various values of $Q^2$ and for two different initial scales $\mu_0^2$. The scattering amplitude is given by the MV model with
	 $Q_{s,g}^2=0.8\,\mathrm{GeV}^2$. A factor $S_{\perp}/4 \pi^3$ has been omitted.}
\label{fig:f2d3_p}
\end{figure}

The function $x_{\mathbb{P}} F_2^{D(3)}$ depends upon
three independent variables, that will be chosen as  $\beta, \, x_{\mathbb P}$ and $Q^2$ (like 
in the r.h.s.~of Eq.~\eqref{xqD3}).
Even though we do not aim at a realistic phenomenological study, we shall nevertheless show results for two
types of target: a large nucleus with tree-level (gluon) saturation momentum $Q_{s,g}^2=2\,\mathrm{GeV}^2$
 and a proton described as a ``small
nucleus'' --- that is, by the MV model with a lower value for the saturation momentum, 
chosen as\footnote{The corresponding scale in the fundamental representation is $Q_{s}^2=0.35\,\mathrm{GeV}^2$, which is consistent with the value extracted from the fit in \cite{Ducloue:2019jmy} at $x_{\mathbb P} \simeq 5 \cdot 10^{-3}$.}
$Q_{s,g}^2=0.8\,\mathrm{GeV}^2$. The above  ratio between the two saturation momenta squared
is significantly smaller than the mean-field value $A^{1/6}\simeq 6$ (for $A=200$) expected at asymptotically
large values of $A$. Such a smaller ratio indeed seems to be favoured by the phenomenology \cite{Ducloue:2019jmy}, 
yet its precise value  is not our main concern: clearly, the overall normalisations of the quark and gluon DPDFs are strongly
sensitive to the value of $Q_{s,g}^2$, cf.~Eqs.~\eqref{xqpfinite} and  \eqref{xGPhigh}, but here we
are merely interested in the functional dependences of $x_{\mathbb{P}} F_2^{D(3)}$, and not in its overall strength.
Rather, what we would like to demonstrate is the importance of the running coupling effects in conjunction
with a relatively low value for the  initial scale  $\mu_0^2$, as appropriate for a less dense target.

We recall that  $\mu_0^2$ is the scale for the onset of the DGLAP evolution and it must be low enough
to avoid losing a significant part of this evolution. For a large nucleus, we have chosen 
$\mu_0^2= 4\, \mathrm{GeV}^2$, which is twice as large as the respective saturation momentum squared.
Following the same argument, we will choose  $\mu_0^2=1.5\, \mathrm{GeV}^2$ for a proton target  with 
$Q_{s,g}^2=0.8\,\mathrm{GeV}^2$. (This value appears to be also
preferred by fits to the HERA data for $x_{\mathbb{P}} F_2^{D(3)}$ \cite{Buchmuller:1998jv,Golec-Biernat:2001gyl}.)

\begin{figure}
	\begin{center}
		\includegraphics[width=0.78\textwidth]{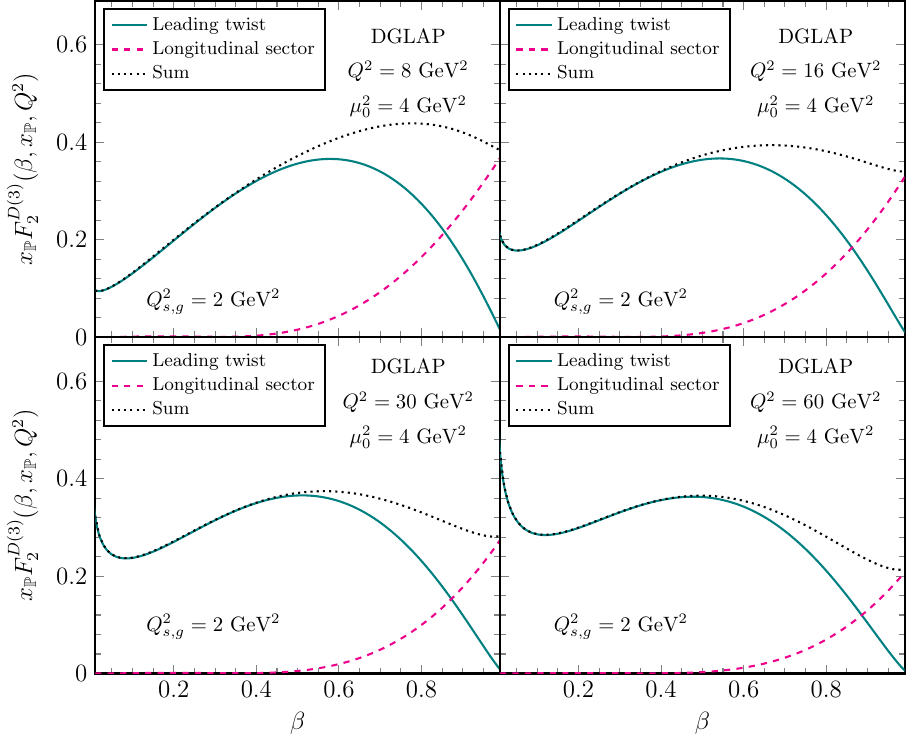}
	\end{center}
	\caption{\small The nucleus diffractive structure function $x_{\mathbb{P}} F_2^{D(3)}$ and the separate contributions from the transverse (leading twist) and longitudinal sector (higher twists) as functions of $\beta$ for various values of $Q^2$, in the MV model with $Q_{s,g}^2=2\,\mathrm{GeV}^2$. A factor $S_{\perp}/4 \pi^3$ has been omitted.}
\label{fig:f2d3_nuc}
\end{figure}

In Figs.~\ref{fig:f2d3_p} and~\ref{fig:f2d3_nuc} we show the numerical solutions to the DGLAP equations 
\eqref{dglapmatrix} with the source terms given by the MV model, for a proton target and for a nuclear target, 
respectively (as functions of $\beta$ for several values of $Q^2$). We emphasise that, at tree level, these curves
had exactly the same shapes for proton and nucleus, respectively (only their overall normalisations
were different). So, all the differences in shape that are visible between the proton and 
the nuclear curves in these figures are to be attributed to the DGLAP evolution and, more precisely,
to the different choices for the initial scale  $\mu_0^2$ in the two cases. We have previously advocated that such
differences should be seen as a form of scheme dependence. Yet, that was only the case when changing 
the value of $\mu_0^2$ for a {\it fixed} target. In the present case, the different choices for $\mu_0^2$  
are {\it physically} motivated: the phase-space for the DGLAP evolution is bounded from below by the
saturation momentum, hence it is truly different for dense and respectively dilute targets.
The results in Figs.~\ref{fig:f2d3_p} and~\ref{fig:f2d3_nuc} reflect these physical difference and,
in particular,  its interplay with the running of the coupling.

When comparing the corresponding proton and nuclear curves in Figs.~\ref{fig:f2d3_p} and~\ref{fig:f2d3_nuc},
one should distinguish the LT contributions (represented by full lines), which are sensitive to the DGLAP evolution,
from the higher-twist contribution of a longitudinal photon (shown in dashed lines), 
which is still computed at tree-level, like in Sect.~\ref{sec:MS}. The {\it relative} importance of the latter
is larger for the nucleus than for the proton, for a trivial reason: being higher-twist, it is proportional to a larger
power of $Q_s^2$ --- namely to $Q_s^4$. So, in order to compare the effects of the DGLAP evolution, we shall
focus on the curves represented by full lines.

The numerical results  in Figs.~\ref{fig:f2d3_p} and~\ref{fig:f2d3_nuc} 
show that  the effects of the evolution are considerably larger for a proton target than for a large
nucleus (compare also with the tree-level results shown in Fig.~\ref{fig:three-int}).
This is manifest both at low values of $\beta$ (where the proton diffractive structure function increases
faster than the nuclear one) and at  largish values of $\beta$ (where the proton curves decrease faster
with increasing $\beta$). As expected, the differences become more pronounced with increasing $Q^2$.
The difference $\ln(4/1.5)\simeq 1$ in the respective phase-spaces being small, one may wonder what is the true
reason for such a different behaviour.  A moment of thinking (confirmed by numerical tests) reveals that this 
should be mostly attributed to the running of the QCD coupling: the latter is not that small between 
$1.5\, \mathrm{GeV}^2$ and $4\, \mathrm{GeV}^2$, hence the contributions from this region are indeed substantial.
This argument is further corroborated by the fact that a similar change in the initial scale, but from 
$\mu_0^2= 4\, \mathrm{GeV}^2$ to $\mu_0^2 = 8\, \mathrm{GeV}^2$ (where the coupling is somewhat smaller),
introduced only minor differences in the results \cite{Iancu:2023lel}. 

Although it is not in our intentions to make predictions for the phenomenology, let us nevertheless observe that
the previous  considerations may have interesting consequences for the DIS data. Our results 
for a proton target in Fig.~\ref{fig:f2d3_p}, which exhibit a vivid DGLAP evolution, are qualitatively supported by the analysis
of the HERA data in Ref.~\cite{Golec-Biernat:2001gyl}. So, assuming one can also trust our results for a nuclear target,
cf.~Fig.~\ref{fig:f2d3_nuc}, we would conclude that the effects of the DGLAP evolution should be considerably 
less important for diffractive DIS off a large nucleus at the EIC than for diffractive DIS off a proton at HERA. 
For moderate values $Q^2\sim 10\,\mathrm{GeV}^2$,  one might even be able to recognise 
 the tree-level behaviour shown in Fig.~\ref{fig:three-int} in the actual data.


%

\subsection{(2+1)--jet contributions to the quark DTMD}

In Sect.~\ref{sec:SIDIS} in Eq.~\eqref{SIDISqqg} we wrote the diffractive SIDIS cross section in a factorized form in terms of the quark DTMD. The latter is a sum of three terms which are listed in  Eq.~\eqref{qDTMDNLO}. The first term, arises from the exclusive $q\bar{q}$ pair production and was studied in detail in Sect.~\ref{sec:qDTMD}. Here we shall take a closer look at the remaining two terms, i.e.~to the contributions which the quark DTMD receives from the production of 2+1 jets and are explicitly given in Eq.~\eqref{qDTMDhard}. 

It is important to keep in mind that the approximations underlying Eqs.~\eqref{qDTMDNLO}--\eqref{qDTMDhard} are
valid only for {\it hard} diffractive SIDIS, that is, for the production of a quark (or antiquark) with transverse momentum
$P_\perp$ much larger than the target saturation momentum $Q_s(\YP)$ (but much smaller than the photon virtuality
$Q$). Indeed, in  Sect.~\ref{sec:SIDIS} we have isolated the LT contributions in the regime  $Q^2\gg P_\perp^2\gg
Q_s^2(\YP)$. From Sect.~\ref{sec:qqbar} and also  Sect.~\ref{sec:qDTMD}, we know already that the $q\bar q $ contribution
to diffractive SIDIS is strongly suppressed at large transverse momenta, where it decays like $Q_s^4(\YP)/P_\perp^4$.
On the other hand, the respective contributions of (2+1)--jets decay much slower, like  $Q_s^2(\YP)/P_\perp^2$,
as visible in Eq.~\eqref{qDTMDhard}.  This argument seems to suggest that the (2+1)--jet contributions
to the r.h.s.~of Eq.~\eqref{qDTMDNLO} should largely dominate over the $q\bar q$ contribution at any $P_\perp^2\gg
Q_s^2(\YP)$. This is roughly true for generic values of $x$, which are not too close to 1. On the other hand, 
when $x\to 1$, the (2+1)--jet contributions vanish faster than the tree-level ($q\bar q$) piece.
This is due to the DGLAP evolution, which depletes the modes at large $x$, and also to kinematical constraints, 
like the condition $x <z_{\rm max}$ in the first integral in Eq.~\eqref{qDTMDhard}.  

\begin{figure}
	\begin{center}
		\includegraphics[align=c,width=0.47\textwidth]{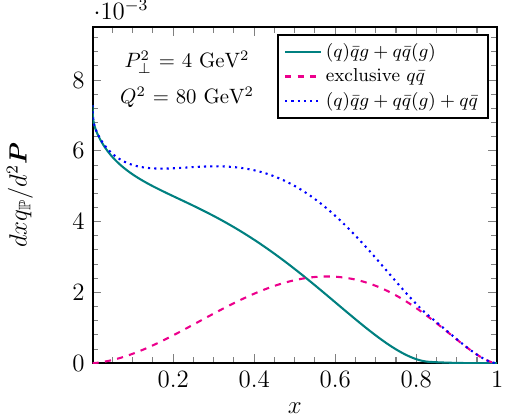}
		\hspace*{0.04\textwidth}
		\includegraphics[align=c,width=0.47\textwidth]{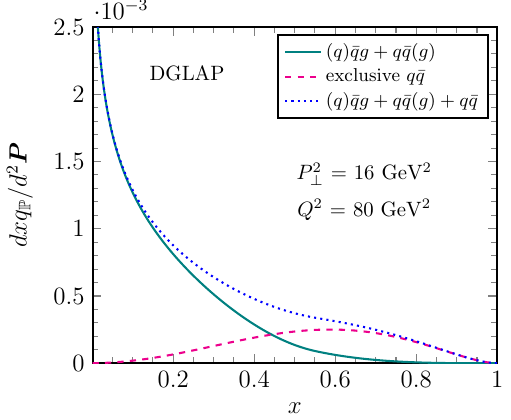}
	\end{center}
	\caption{\small The total quark DTMD and its two pieces, arising from the $q\bar{q}$ dijet (cf.~\eqn{qDTMD}) and,
	respectively, from (2+1)--jets (cf.~Eq.~\eqref{qDTMDhard}), as functions of $x$ for $P_{\perp}^2= 4\, \mathrm{GeV}^2$ (left panel) and  $P_{\perp}^2 = 16\, \mathrm{GeV}^2$ (right panel). The scattering amplitude is given by the MV model. The right panel includes the effects of the DGLAP evolution for the (2+1)--jet 
	 contributions in Eq.~\eqref{qDTMDhard}.	The  resolution scale $Q^2=80\, \mathrm{GeV}^2$ 
	 is necessary  to fix the upper limit $z_{\rm max}=1/(1+4P_\perp^2/Q^2)$ in the first integral \eqref{qDTMDhard}.  
	A factor $S_{\perp}/4 \pi^3$ has been omitted.}
\label{fig:dxqpall}
\end{figure}

The above discussion helps understanding the numerical
results exhibited in  Fig.~\ref{fig:dxqpall}, which compare the strength of the $q\bar q$ piece (computed as  
in Sect.~\ref{sec:qDTMD}) with the contributions of (2+1)--jets, as functions of $x$ and for two hard values
$P_\perp^2=4\, \mathrm{GeV}^2$ and $16\, \mathrm{GeV}^2$.
Namely, we immediately see that 2+1 jets dominate at relatively small $x\lesssim 0.5$, whereas the exclusive
dijets still prevail at larger values of  $x$ (and especially near $x=1$).
Interestingly, the two types of contributions are of the same order for $x$ around one half.

\begin{figure}
	\begin{center}
		\includegraphics[align=c,width=0.47\textwidth]{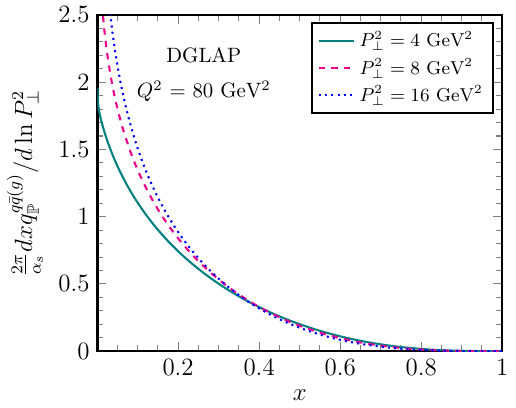}
		\hspace*{0.04\textwidth}
		\includegraphics[align=c,width=0.47\textwidth]{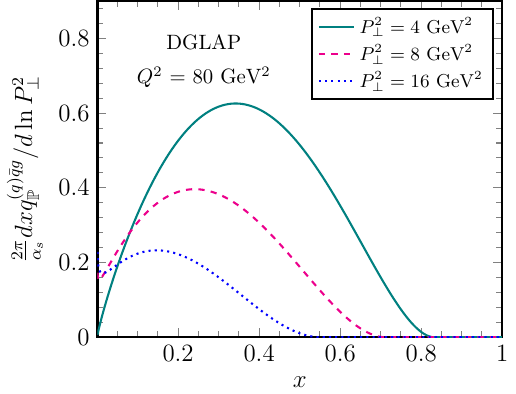}
	\end{center}
	\caption{\small  The soft gluon $q\bar{q}(g)$ (left panel) and  soft quark $(q)\bar{q}g$ (right panel) 
	 contributions to the quark DTMD, cf.~Eq.~\eqref{qDTMDhard}, as functions of $x$ for various values of $P_{\perp}^2$. 
	As compared to Fig.~\ref{fig:dxqpall}, the distributions are multiplied by a factor 
	$P^2_\perp /\alpha_s(P_\perp^2)$ in order to remove some strong overall dependences upon $P_\perp^2$,
	but keep only those introduced by the DGLAP evolution and by the upper limit 
	$z_{\rm max}=1/(1+4P_\perp^2/Q^2)$ in the first integral \eqref{qDTMDhard}.
	 In the case of a soft gluon, we see a rapid
	rise as $x\to 0$, due to the growth of the gluon DPDF.
	The contribution involving a soft quark vanishes faster when $x$ is near 1, due to the upper limit 
	$z_{\rm max}$. A factor $S_{\perp}/4 \pi^3$ has been omitted.}
\label{fig:dxqpfrom2plus1}
\end{figure}

Once again, it is instructive to support these numerical findings with analytical arguments.
Let us start from the $q\bar{q}(g)$ piece, that is, the one where the gluon is soft, and consider first the limiting case $x \ll 1$. Making use of Eq.~\eqref{gluonDGLAPx0} and performing the longitudinal fraction integration we obtain the leading behavior
\begin{align}
	\label{dxqpfromg0}
	\hspace*{-0.1cm}
	\frac{\rmd xq_{\mathbb{P}}^{q\bar q (g)}
  	(x, x_{\mathbb{P}}, P_{\perp}^2)}{\rmd \ln P_{\perp}^2}
	\simeq
	\frac{\alpha_s(P_{\perp}^2)}{2\pi}\,
	\frac{
	N_g\kappa_g(0)
	Q_{s,g}^2(Y_{\mathbb P})}{3}\,
	I_0
	\left(\sqrt{\frac{4 N_c}{\pi b}
	\ln \frac{\alpha_s(\mu_0^2)}
	{\alpha_s(P^2_{\perp})}
	\ln\frac{1}{x}}
	\right)
	\quad 
	\mathrm{for} 
	\quad x \ll 1.
\end{align}
It is important to notice that the above \emph{is equal} to the derivative w.r.t.~$\ln P_{\perp}^2$ of the quark DPDF in Eq.~\eqref{quarkDGLAPx0} as it should. Indeed, both equations have been obtained by taking into account $P_{qg}(z)$ (once) and the small-$z$ limit of $P_{gg}(z)$ (resummed). When $1-x \ll 1$ we employ Eq.~\eqref{gluonDGLAPx1} to get
\begin{align}
	\label{dxqpfromg1}
	\frac{\rmd xq_{\mathbb{P}}^{q\bar q (g)}
  	(x, x_{\mathbb{P}}, P_{\perp}^2)}{\rmd \ln P_{\perp}^2}
	\simeq
	\frac{\alpha_s(P_{\perp}^2)}{2\pi}\,
	\frac{N_g\kappa_g(1)
	Q_{s,g}^2(Y_{\mathbb P})}{3}
	(1-x)^{\textstyle 3 + 
	\frac{N_c}{\pi b} 
	\ln \frac{\alpha_s(\mu_0^2)}
	{\alpha_s(P^2_{\perp})}}
	\quad 
	\mathrm{for} 
	\quad  1-x \ll 1
\end{align}
and clearly this \emph{is not equal} to the derivative of Eq.~\eqref{quarkDGLAPx1}. Eq.~\eqref{dxqpfromg1} arises from a series of $g \to gg$ splitting followed by a  $g\to q\bar{q}$ one (which overall lead to a  subdominant contribution), but Eq.~\eqref{quarkDGLAPx1} is controlled by the virtual term of the $q \to qg$ splitting. 
The two above limiting behaviours are indeed consistent with the numerical results for the  $q\bar{q}(g)$ piece of the quark DTMD, as shown in the left plot of Fig.~\ref{fig:dxqpfrom2plus1}. 

The $(q)\bar{q}g$ piece for $x\ll 1$ can be computed by using the real part of the splitting function $P_{qq}(z)$ in Eq.~\eqref{splittingfs} and the asymptotic solution of the quark DPDF in Eq.~\eqref{quarkDGLAPx0} and we have
\begin{align}
	\label{dxqpfromq0}
	\hspace*{-0.1cm}
	\frac{\rmd xq_{\mathbb{P}}^{(q)\bar q g}
  	(x, x_{\mathbb{P}}, P_{\perp}^2)}{\rmd \ln P_{\perp}^2}
	\simeq
	& \,\frac{\alpha_s(P_{\perp}^2)}{2\pi}\,
	\frac{
	N_g\kappa_g(0)
	Q_{s,g}^2(Y_{\mathbb P})}{3}\,
	C_F\left[
	2 \ln \frac{1}{1\!-\!z_{\max}} - \frac{z_{\max}(z_{\max}+2)}{2}
	\right]
	\nn
	& 
	\sqrt{\frac{
	\ln \frac{\alpha_s(\mu_0^2)}{\alpha_s(P^2_{\perp})}}
	{ 4 \pi b N_c \ln \frac{1}{x}}}\,
	I_1 \left(\sqrt{\frac{4 N_c}{\pi b}
	\ln \frac{\alpha_s(\mu_0^2)}
	{\alpha_s(P^2_{\perp})}
	\ln\frac{1}{x}}
	\right)
	\quad 
	\mathrm{for} 
	\quad x \ll 1,
\end{align}
where the last factor in the first line is just the integral of the relevant splitting function. This contribution from the soft quark configuration is subleading when compared to the one from the soft gluon configuration in Eq.~\eqref{dxqpfromg0}. Finally, we trivially notice that the $(q)\bar{q}g$ piece vanishes due to kinematic constraints for $x \geq z_{\max}$. Once again,
these dependences are in agreement with the respective numerical results, as shown in the right plot of Fig.~\ref{fig:dxqpfrom2plus1}.  At this level, one can also observe the good agreement between the results for (2+1)--jets
in Fig.~\ref{fig:dxqpfrom2plus1} and, respectively,  Fig.~\ref{fig:dxqpall}.

\section{Summary and perspectives}
\label{sec:conc}

In this paper, we have studied the production of two and three jets via coherent diffraction
 in photon-nucleus interactions at high energy.  We have focused on the hard QCD regime 
 where the photon virtuality $Q$ and/or the relative transverse momentum $P_\perp$ of 
 two of the produced jets is much larger than the saturation momentum $Q_s$ of the nuclear target.
 The relevant value of $Q_s$ is evaluated at the rapidity gap $\YP$ and thus includes the high-energy
 evolution of the gluon distribution in the Pomeron.
 We have shown that, despite the presence of hard scales, the dominant contributions 
 at large $Q^2$ and/or  large  $P_\perp^2$ --- the {\it leading twist contributions} ---  
 come from relatively large partonic fluctuations of the virtual photon,
 with a typical transverse size $R\sim 1/Q_s$. Such large configurations avoid the colour transparency
 of small colour dipoles and thus allow for strong scattering in the vicinity of the black disk limit.
 In turn, strong scattering is favoured by the elastic nature of the diffractive processes.

 
 In the case of exclusive dijets, the partonic configurations which matter at leading twist (and to leading order) 
are quark-antiquark fluctuations of the virtual photon, which are very asymmetric: 
one of the quarks (the ``aligned jet'') carries most of the longitudinal momentum
 of the virtual photon, while the other quark is soft, with a longitudinal momentum fraction $\vartheta_1
 \sim Q_s^2/Q^2 \ll 1$. Such typical configurations scatter off the saturated gluons in the target wavefunction
 and thus acquire transverse momenta of order $Q_s$. They provide a leading-twist contribution to 
 {\it diffractive SIDIS}, in which the measured jet has transverse momentum of order $Q_s$.

 In the case of three-jet final states, one of the partons (quark, antiquark, or gluon) must be soft
 ($\vartheta_1\sim Q_s^2/Q^2 \ll 1$) and semi-hard ($K_\perp\sim Q_s$), for the overall system 
 to scatter strongly. Yet, the two other  jets can be hard ($P_\perp\gg Q_s$) and nearly back-to-back in the
 transverse plane. These asymmetric, ``(2+1)--jet'', configurations provide the dominant contribution
 to the diffractive production of a pair of hard jets, despite the fact that they are formally suppressed
 by a power of the QCD coupling: at large $P_\perp\gg Q_s$, this suppression is more than compensated
 by the fact that the respective cross-section is of leading-twist order, i.e. it only falls like $1/P_\perp^4$
 (to be compared with  $1/P_\perp^6$ for hard exclusive dijets).
 
 In all cases, that is, for both exclusive dijets and for diffractive (2+1)--jets, we found that the respective leading twist 
 contributions admit {\it diffractive TMD factorisation}. In practice, this means that the soft parton from the wavefunction
 of the virtual photon can be reinterpreted as a constituent of the target --- more precisely, of the Pomeron ---, with a 
 transverse-momentum dependent parton distribution which physically represents  the 
 unintegrated quark distribution of the Pomeron. In the case of exclusive dijets, this factorisation is most
 naturally formulated for diffractive SIDIS and involves the quark diffractive TMD, as originally pointed out in
 \cite{Hatta:2022lzj}. For diffractive (2+1)--jets, we meet both the quark 
 and the gluon diffractive TMDs, corresponding to the different possibilities for the soft parton.
 The case of (2+1)--jets with a soft gluon has been studied earlier \cite{Iancu:2022lcw,Iancu:2021rup} and we include
 here the corresponding results,  for completeness (see Eqs.~\eqref{3jetsD1}--\eqref{pomugddef}). Our main new
 result in this paper is the extension of this factorisation to the case of  diffractive (2+1)--jets with a soft quark,
 together with an explicit proof of the universality of the quark DTMD: indeed, we find exactly the same result
 for the latter  in the case of exclusive dijets  (see Eqs.~\eqref{2jetTMD}--\eqref{qDTMD})
 and  for the three
 contributions to the cross-section for diffractive (2+1)--jets with a soft quark
 (cf. Eqs.~\eqref{cross1new}--\eqref{hardtotal}).

 The colour-dipole approach to the virtual photon wavefunction supplemented with the CGC description of the dense
 nuclear target provides first principles estimates for the quark and gluon diffractive TMDs --- valid 
  at leading order in the QCD running coupling, of course. A main common characteristic of these LO results is the
  fact that both the  quark and the gluon DTMD have support at relatively low transverse momenta, $K_\perp\lesssim
  \tilde Q_s(x,\YP)$, with $ \tilde Q_s(x,\YP)$ the effective saturation momentum introduced in \eqn{Qsx}.
  This strongly suggests that, due to the fact that it is a colour-singlet, the Pomeron fluctutation of the target 
  is built with quarks
  and gluons {\it at saturation}. As discussed in Sect.~\ref{sec:MS}, parton saturation in the Pomeron wavefunction
  is dual to strong {\it elastic} scattering between an external dipole and the target.

Another distinctive  feature of our LO results for the quark and gluon TMDs is the fact that they apply for generic values of
$x$ (the longitudinal momentum fraction of the measured parton w.r.t. the Pomeron), and not only at small $x$,
as is the case for the gluon TMDs which emerge from  CGC studies of {\it inclusive} scattering \cite{Dominguez:2011wm}.
This makes it possible to use our tree-level results in order to initiate the DGLAP evolution at transverse momenta
of the order of $Q_s$. Moreover, despite the fact that we often refer to them as ``tree-level'', our
LO approximations  for the quark and gluon TMDs generally include the high-energy BK/JIMWLK evolution 
\cite{Balitsky:1995ub,Kovchegov:1999yj,JalilianMarian:1997jx,JalilianMarian:1997gr,Kovner:2000pt,Weigert:2000gi,Iancu:2000hn,Iancu:2001ad,Ferreiro:2001qy} with increasing $\YP$.
  These ``tree-level''  results are studied {\it in extenso} (via both analytic approximations and numerical solutions)
  in Sect.~\ref{sec:qDTMD}.
  
  By integrating the (tree-level) quark and gluon DTMDs over their transverse momentum argument $K_\perp^2$
up to the hard resolution scale ($Q^2$ or $P_\perp^2$), one deduces LO approximations
for the diffractive PDFs, which control more inclusive cross-sections, like the diffractive structure function.
In the context of collinear factorisation, the DPDFs are known to obey the DGLAP equations describing their
evolution with increasing the resolution scale. So, it is natural to ask: can one see a similar evolution emerging 
 from the dipole picture ? The calculation of (2+1)--jets gives us the possibility to answer this question:
 after integrating out two of the final jets (one soft and one hard), one finds a contribution to the cross-section for 
 measuring the other hard jet, which has the right kinematics to encode the DGLAP evolution. 
 And indeed, in Sect.~\ref{sec:SIDIS} we find that the leading-twist piece of this contribution
 takes the form of one step in the DGLAP evolution of the LO cross-section for SIDIS in Sect.~\ref{sec:qqbar}.
This is interesting from at least two perspectives. First, it demonstrates the emergence of the DGLAP evolution 
of the target wavefunction via quantum calculations at the level of the projectile. Second,
it suggests that the diffractive TMD factorisation that was established here and in
the previous papers \cite{Iancu:2022lcw,Iancu:2021rup,Hatta:2022lzj} at leading order 
may remain valid after adding quantum corrections.

Our study  of the emergence of the DGLAP evolution from the colour dipole picture is still
incomplete --- for instance, we have not computed  the virtual corrections --- and it 
would be interesting to complete this study and also extend it to the inclusive cross-sections  
which are known to admit TMD factorisation when computed in the dipole picture at leading order
\cite{Dominguez:2011wm}.

The  analysis in Sect.~\ref{sec:SIDIS}  also suggests that the natural matching between the tree-level results for the TMDs 
and the  DGLAP equations for the PDFs consists in using the results for  the former as {\it source terms} in the latter.
Using this strategy,  in Sect.~\ref{sec:DGLAP}  we have presented numerical solutions to the DGLAP equations
for the quark and gluon diffractive PDFs. By comparing with the respective tree-level approximations in 
Sect.~\ref{sec:qDPDF}, one can better appreciate both the effects of the DGLAP evolution and the influence
 of the tree-level results on the DGLAP solutions. As a simple application, in Sect.~\ref{sec:f2d3}
we have used the DGLAP solutions for the quark DPDF to compute the diffractive structure function.
Our results are qualitatively consistent with previous calculations for the case of a proton target
\cite{Golec-Biernat:2001gyl}, but they show an interesting twist --- a substantial reduction in the effects
of the DGLAP evolution  --- when the target is a large nucleus.  This effect can be attributed to the importance
of the running coupling corrections during the early stages of the DGLAP evolution.

 Another intriguing result which emerges from our calculations is the lack of angular correlations between
 the two transverse momenta which control  diffractive (2+1)--jet production: the relative momentum
 $\bP$ of the hard dijets and their imbalance $\bK$. Such correlations were systematically found in the literature,
in cases where the imbalance $\bK$ is caused by a net momentum transfer between
the produced jets and the target. This refers both to inclusive
dijets (where the angular correlations measure a tensorial component of the WW gluon TMD
~\cite{Metz:2011wb,Dominguez:2011br,Iancu:2013dta,Dumitru:2015gaa,Marquet:2017xwy,Dumitru:2018kuw})
and to exclusive dijets (where they probe the elliptic gluon Wigner distribution
\cite{Altinoluk:2015dpi,Hatta:2016dxp,Hagiwara:2017fye,Mantysaari:2019csc,Salazar:2019ncp,Mantysaari:2019hkq,Hatta:2021jcd}). In our calculations, we systematically neglected the net momentum transferred via the elastic scattering,
since this is tiny ($\Delta_\perp\sim 30$~MeV) and in any case much smaller than the recoil $K_\perp$ of the semi-hard jet. 
The (2+1)--jet cross-section depends upon both $\bP$ and $\bK$ and we see no {\it a priori} reason
to exclude tensorial components in the diffractive TMDs. Yet, we found no trace of them in our final results.
This is particularly surprising since our intermediate calculations  involve rather complex tensorial structures
built with these two vectors (notably, for the interference between the two channels for (2+1)--jets 
with a soft quark), yet they leave no imprint on the cross-section. 
It would be interesting to understand if the absence of angular correlations for diffractive
 jets is a generic feature and if it has a fundamental explanation.

Finally, it would be interesting to consider the phenomenological implications of our new results in more detail.
In a recent paper \cite{Iancu:2023lel}, some of us considered  2+1 diffractive dijets with a soft gluon in the context of
ultra-peripheral nucleus-nucleus collisions at the LHC. It would clearly be interesting to add the contributions
from the other channels, which involve a soft quark. More generally, one should systematically
study the consequences of our results for the future phenomenology at the Electron Ion Collider 
 \cite{Accardi:2012qut,Aschenauer:2017jsk,AbdulKhalek:2022hcn}. This also requires some theoretical refinements,
 like a better treatment of the final states (by constructing real jets, or including parton fragmentation into hadrons)
and a study of the impact of additional effects, like the Sudakov corrections \cite{Mueller:2013wwa} and the
soft gluon radiation in the initial and final states (which represents an additional source
of angular correlations) \cite{Hatta:2021jcd,Shao:2024nor}.

 \section*{Acknowledgements} 
The work of A.H.M.~is supported in part by the U.S.~Department of Energy Grant \# DE-FG02-92ER40699. S.Y.~Wei is supported by the Taishan fellowship of Shandong Province for junior scientists and the Shandong Province Natural Science Foundation under grant No.~2023HWYQ-011.

\appendix

\section{Longitudinal sector}
\label{sec:long}

In this Appendix we would like to study the case of a longitudinally polarised virtual photon $\gamma^*_L$. 
The steps to be followed are very similar to those in the transverse photon case which has been explicitly worked out in the main text, hence we shall only present the most relevant expressions and results. The only technical change, which one must implement in order to go from the transverse to the longitudinal sector, consists of the replacement
\begin{align}
	\label{TtoL}
	\varphi^{ij}(\vartheta,\lambda) k^j \longrightarrow 2 \vartheta (1- \vartheta) Q
\end{align}
in the $\gamma^* \to q\bar{q}$ vertex. Let us see how this modifies the various contributions to the cross sections of interest.

\subsection{ The $q\bar{q}$ component} 

The probability amplitude (the analog of Eqs.~\eqref{psiqq} and \eqref{psiqqnew}) reads
\begin{align}
	\label{psiqqL}
	\psi^{L}_{\lambda_1 \lambda_{2}} (\vartheta_1, \bk_1) =
	\frac{\delta_{\lambda_1\lambda_2}}{\sqrt{{2 q^+}}}
	\frac{ee_{f}}{(2\pi)^3}\,
	\frac{Q}
	{E_{q}+E_{\bar{q}}-E_{\gamma}}=
	\delta_{\lambda_1\lambda_2}\,
	\sqrt{\frac{q^+}{2}}\,
	\frac{ee_{f}}{(2\pi)^3}\,
	\frac{2 \sqrt{\vartheta_1 (1-\vartheta_1)}\, \bar{Q}}
	{k_{1\perp}^2 +\bar{Q}^2}.
\end{align} 
The transformation to coordinate space is achieved via the replacement (the analog of Eq.~\eqref{qqFT})
\begin{align}
	\label{qqFTL}
	\frac{\bar{Q}}{k_{1\perp}^2 +\bar{Q}^2}
	\to
	\frac{1}{2 \pi}\,
	\bar{Q} K_0(\bar{Q} R).
\end{align}
Including the diffractive scattering and performing the inverse Fourier transform leads to (the analog of Eq.~\eqref{psiqqdiff})
\begin{align}
	\label{psiqqdiffL}
	\psi^{L,D}_{\lambda_1 \lambda_{2}} (\vartheta_1, \bk_1) =
	- \delta_{\lambda_1\lambda_2}\,
	\sqrt{\frac{q^+}{2}}\,
	\frac{ee_{f}}{(2\pi)^3}\,
	2 \sqrt{\vartheta_1 (1-\vartheta_1)}\,
	\frac{\mcal{Q}_{L}(\bar{Q},k_{1\perp},\YP)}{\bar{Q}},
\end{align}
where we have defined the dimensionless scalar quantity (the analog of Eq.~\eqref{QP})
\begin{align}
\label{QL}
	\mcal{Q}_{L}(\bar{Q},K_{\perp}, Y_{\mathbb P} ) \equiv
	\bar{Q}^2 \int \dif R\, R\, J_0(K_{\perp} R)\,
	K_0(\bar{Q} R)\,
	\mcal{T}(R, Y_{\mathbb P}).
\end{align} 
We finally arrive at the differential cross section (the analog of Eq.~\eqref{crossqqnew}\footnote{In Eq.~\eqref{crossqq} there is an overall factor of 1/2 due to averaging over the two polarizations of the transverse virtual photon. Obviously such a factor is not present in the longitudinal sector.}
\begin{align}
	\label{crossqqnewL}
	\frac{\rmd\sigma_{\scriptscriptstyle}^{\gamma_{\scriptscriptstyle L}^* A\rightarrow q\bar q A}}
	{\rmd \vartheta_1\,
	\rmd^{2}\bK} = 
	\frac{2 S_{\perp} \alpha_{\rm em} N_c}{\pi^2} \left( \sum e_f^2\right)
	\left[ \vartheta_1 (1-\vartheta_1) \right]
	\frac{|\mcal{Q}_{L}(\bar{Q},K_{\perp}, Y_{\mathbb P})|^2}{\bar{Q}^2},
\end{align}
which, for a fixed value of the rapidity gap $\YP$ or, equivalently, of $\beta$, becomes (the analog of \eqn{2jetSIDIS})
\begin{align}
  \label{2jetSIDISL}
  \frac{\rmd \sigma^{\gamma_L^* A
  \rightarrow q\bar q A}}
  {\rmd \ln(1/\beta)\,\rmd^{2}\bm{K}} = 
  \frac{4 S_{\perp} \alpha_{\rm em} N_c}{\pi^2} \left( \sum e_f^2\right)\,\frac{1}{Q^2}\,\frac{\frac{\mcal{M}^2}{Q^2}}
 {\sqrt{1-\frac{4\mcal{M}^2}{Q^2}}}\,
	\frac{[\mcal{Q}_{L}(\mcal{M},K_{\perp}, Y_{\mathbb P})]^2}{1-\beta},
  \end{align}
where $\mcal{M}^2=[\beta/(1-\beta)]K_{\perp}^2$. We recall that the change of variables from $\vartheta_1$ to
$\beta$ has replaced $\vartheta_1(1-\vartheta_1) \to \mcal{M}^2/Q^2$ and hence $\bar Q \to \mcal{M}$, 
cf.~Eq.~\eqref{Q2M2}.
It is furthermore understood that $K_\perp\le K_{\max}$ with the upper limit $K_{\max}^2 = (1-\beta)Q^2/4\beta$ from Eq.~\eqref{K2max}.

By inspection of \eqn{2jetSIDISL}, it is apparent that this quantity is of higher-twist order --- it starts at twist-4 order, 
i.e.~it decays like $1/Q^4$ in the high $Q^2$ limit---, hence it brings no contribution to the quark diffractive TMD. 
Clearly, this suppression should be attributed to the factor $\vartheta_1(1-\vartheta_1)$ in the r.h.s.~of
\eqn{crossqqnewL}, which is the splitting function for
the decay $\gamma^*_L\to q\bar q$ of the longitudinal photon, cf.~\eqn{PgammaL}.
This factor is much smaller than one for the asymmetric $q\bar q$ configurations. Yet, as observed
in  Ref.~\cite{Golec-Biernat:2001gyl}, the longitudinal sector gives the dominant contribution to the diffractive structure function 
in the limit $\beta \to 1$, since it remains finite in that limit, unlike the leading-twist contribution of the transverse cross section, 
which vanishes like $1-x$, cf.~Eqs.~\eqref{DF2} and \eqref{xqPsat}.

%

In order to demonstrate this and to gain more insight in the physical content of \eqn{2jetSIDISL}, 
let us estimate this cross-section in two limiting regimes: for small and respectively large transverse momenta $K_\perp$.  
The discussion of $\mcal{Q}_L$ in these limits is similar to that of
the quantity $\mcal{Q}_{\mathbb{P}}$ in  Sect.~\ref{sec:MS}, except for the fact that the variable $x$ is now replaced by $\beta$.
In particular, $R_{\max}^2 \sim (1-\beta)/K_{\perp}^2$ and the effective saturation momentum is $\tilde Q_s(\beta, \YP)
= \sqrt{1-\beta}\,Q_s(\YP)$. For the purpose of parametric estimates, one can replace
$J_0(K_{\perp} R)\,K_0(\mcal{M} R)\sim\Theta(R_{\max}-R)$.

{\tt (i)} When $K_{\perp} \!\gg\! \sqrt{(1-\beta)}\,Q_s(\YP)$ the scattering is weak and we have (the analog of Eq.~\eqref{QPlargemom}) 
\begin{align}
	\label{QLhigh}
	\mcal{Q}_L
	\sim 
	\mcal{M}^2 Q_s^2 R_{\max}^4 
	\sim  
	\beta\,(1-\beta)\frac{Q_s^2(\YP)}{K_{\perp}^2}
	\qquad \mathrm{for} \qquad
	K_{\perp} \gg   \tilde{Q}_s(\beta,Y_{\mathbb{P}}),
\end{align}
where logarithmic factors have been neglected.

{\tt (ii)} When $K_{\perp} \!\ll\! \sqrt{(1-\beta)}\,Q_s(\YP)$, the integration over $R$ is dominated by large dipoles which scatter strongly off the target and gives (the analog of Eq.~\eqref{QPsmallmom})  
\begin{align}
	\label{QLlow}
	\mcal{Q}_L
	\sim 
	\mcal{M}^2 R_{\max}^2 
	\sim  
	\beta
	\qquad \mathrm{for} \qquad
	K_{\perp} \!\ll\! \  \tilde{Q}_s(\beta,Y_{\mathbb{P}}).
	\end{align}
For completeness, we point out that one can perform a precise calculation for case {\tt (i)} in the MV model (similar to the one in Appendix \ref{sec:single}), while the above estimate in case {\tt (ii)} is in fact equal to the exact result.

The above results can be conveniently summarised in the following, piece-wise approximation:
\begin{align}
	\label{QLpw}\,\frac{\mcal{M}^2}{Q^2}\,
	\frac{[\mcal{Q}_{L}]^2}{1-\beta}
	\,\simeq \,\beta^3
	\begin{cases}    
	\displaystyle{\,\frac{Q_s^4(\YP)}{Q^2 K_{\perp}^2} }
    &\quad {\rm for} \quad
    \tilde{Q}_s(\beta,Y_{\mathbb{P}})\ll K_\perp \le K_{\max}
    \\*[0.4cm]
    \displaystyle{\,\frac{1}{(1-\beta)^2}\,
    \frac{K_\perp^2}{Q^2}}
    &\quad {\rm for} \quad
    K_\perp \ll \tilde{Q}_s(\beta,Y_{\mathbb{P}})\,.
    \end{cases}
\end{align}
This shows that the cross-section \eqref{crossqqnewL} is peaked at $K_\perp \sim \tilde{Q}_s(\beta,Y_{\mathbb{P}})$,
but unlike in the transverse sector, the bulk of the distribution is located {\it above} this peak, in the slowly decaying tail at larger 
transverse momenta $K_\perp\gg  \tilde{Q}_s(\beta,Y_{\mathbb{P}})$. This becomes clear when integrating over $K_{\perp}$, to deduce
the contribution of the longitudinal sector to the diffractive structure function  $F_2^{D(3)}$ (cf.~\eqn{F2D}). Using the estimate
in \eqn{QLpw}, one can easily check that the dominant contribution comes from high-momentum tail $\propto 1/K_{\perp}^2$,
since the respective integral over $K_{\perp}^2$ yields a large transverse logarithm $\ln[Q^2/Q_s^2(Y_{\mathbb{P}})]$. To leading
logarithmic accuracy, one can assume that $K_\perp\ll   K_{\max}$ and thus approximate the square root in the denominator of
 \eqref{2jetSIDISL} by unity. This argument also shows that dominant contribution comes from aligned jet configurations:
 $\vartheta_1(1-\vartheta_1) = \mcal{M}^2/Q^2\ll 1$. One thus finds
\begin{align}
  \label{DF2L}
  \frac{\rmd \sigma^{\gamma_L^* A
  \rightarrow q\bar q A}}
 {\rmd \ln(1/\beta)} 
  \propto S_{\perp} \frac{4 \beta^3 \alpha_{em} N_c}{\pi} \left(\sum e_{f}^{2}\right)
  \frac{Q_s^4(\YP)}{Q^4}\,\ln\frac{Q^2}{4\beta Q_s^2(Y_{\mathbb{P}})}\,,
\end{align}
where the logarithm has been generated as $\ln\big[K_{\max}^2/ \tilde{Q}_s^2\big]=\ln\big[{Q^2}/{4\beta Q_s^2}\big]$.
Notice that the factor $1-\beta$ has cancelled in the ratio $K_{\max}^2/ \tilde{Q}_s^2$, that is, in the argument of the logarithm. The final result in \eqn{DF2L} does not vanish in the limit $\beta \to 1$, where it even dominates 
over the leading-twist contribution of the transverse sector, which vanishes there, cf.~\eqn{xqPsat}. A similar
observation was made in \cite{Golec-Biernat:2001gyl}. However, outside this limit, the longitudinal
contribution \eqref{DF2L} is of twist-4 order and thus is strongly suppressed at high $Q^2\gg Q_s^2(Y_{\mathbb{P}})$ compared to
the LT result in Eqs.~\eqref{DF2} and \eqref{xqPsat}. The derivation of Eq.~\eqref{DF2L} relies on the high momentum behavior in Eq.~\eqref{QLpw} which is valid for a scattering amplitude which is given by the GBW model. When using the MV model the precise high momentum behavior can be inferred from Eqs.~\eqref{QLKlarge} and \eqref{QLKlow} in Appendix \ref{sec:single}. Eventually, this results in an even stronger logarithmic dependence, namely the longitudinal cross section in Eq.~\eqref{DF2L} becomes proportional to $\ln^3(Q^2/4\beta^2\Lambda^2) - \ln^3(Q_s^2/\beta\Lambda^2)$. 

\subsection{The $q\bar{q}g$ component}

As explained in Sect.~\ref{sec:2plus1}, we consider $q\bar{q}g$ configurations which are asymmetric in the sense
that the (final) quark is both soft and semi-hard, whereas the (final) antiquark is hard and carries a large fraction of
the photon longitudinal momentum. Because of this asymmetry, gluon emissions from the quark and,
respectively, the antiquark lead to very different $q\bar{q}g$ states. This difference has also consequences
in the case where the incoming photon has longitudinal polarisation, as we now explain.

\paragraph{(i) Gluon emission from the antiquark.} 

The total amplitude after taking into account the gluon emission both before and after the scattering is (the analog of \eqref{Psitot})
\begin{align}
	\label{PsitotL}
	\Psi^{L, m}_{\lambda_1\lambda_2\,(\bar{q})}
	\simeq 
	-\,\delta _{\lambda_{1}\lambda_2}\,
	\frac{e e_f g q^+ }{(2\pi)^6}\,
	\frac{\vartheta_1}{\sqrt{\vartheta_3}}\,
	\frac{Q}
	{k_{1\perp}^2+\mcal{M}^2}\,
	\frac{\tau^{mn}(0,\vartheta_2,\lambda_1)P^n}{P_\perp^2},
\end{align}
valid for $\vartheta_1 \ll 1$ and $k_{1\perp} \ll P_{\perp}$. Taking the Fourier transform w.r.t.~the semi-hard momentum $\bk_1$, inserting the diffractive scattering and performing the inverse Fourier transform we get (the analog of \eqref{Psidiff})
\begin{align}
\label{PsidiffL}
	\Psi^{L,m,D}_{\lambda_1\lambda_2 (\bar{q})}
	\simeq 
	\,\delta _{\lambda_{1}\lambda_2}\,
	\frac{e e_f g q^+ }{(2\pi)^6}\,
	\frac{\vartheta_1}
	{\sqrt{\vartheta_3}}\,
	\frac{Q}{\mcal{M}}
	\frac{\tau^{mn}(0,\vartheta_2,\lambda_1)P^n}{P_\perp^2}\,
	\frac{\mcal{Q}_{L}(\mcal{M}, K_{\perp}, \YP)}{\mcal{M}}.
\end{align}  
Recalling the definition in Eq.~\eqref{taudef} we have\footnote{In the summation below, the coefficient of $\varepsilon^{nr}$ is proportional to $\lambda$ and therefore vanishes since $\sum\limits_\lambda \lambda =0$.}
\begin{align}
	\label{tausquare1}
	\sum_{m,\lambda}
	\tau^{mn}(0,\vartheta,\lambda)
	\tau^{mr*}(0,\vartheta,\lambda)
	= 4 (1+\vartheta^2) \delta^{nr},
\end{align}
which allows us to calculate the magnitude squared of Eq.~\eqref{PsidiffL} and arrive at the cross section
\begin{align}
\hspace{-0.5cm}
	\label{cross1L}
	\frac{\rmd\sigma_{(\bar{q}\bar{q})}^{\gamma_{\scriptscriptstyle L}^* A\rightarrow q\bar q g A}}
	{\rmd \vartheta_1 
	\rmd \vartheta_2
	\rmd \vartheta_3 \,
	\rmd^{2}\bP
  	\rmd^{2}\bK}
   = 
   \frac{S_{\perp} \alpha_{\rm em} N_c}{\pi^4} 
   \left( \sum e_f^2\right)
   \delta_{\vartheta}\,
   \frac{\alpha_s C_F \tilde{Q}^2}{P_{\perp}^2 (P_{\perp}^2 + \tilde{Q}^2)}\,
   \frac{\vartheta_1 (1+\vartheta_2^2)}{\vartheta_3}\,
   \frac{\big|\mcal{Q}_{L}(\mcal{M}, K_{\perp},Y_{\mathbb P})\big|^2}{\mcal{M}^2},
   \end{align}
where we have also used Eq.~\eqref{Mdef1}. We immediately see that this longitudinal cross section is power suppressed when compared to the corresponding transverse one in Eq.~\eqref{cross1} due to the presence of an extra factor $\vartheta_1 \sim K_\perp^2/Q^2\ll 1$. Physically, this suppression is related to the fact that the original $q\bar q$ pair created
by the decay of the longitudinal virtual photon is very asymmetric.

\paragraph{(ii) Gluon emission from the quark.} 

The respective amplitude is dominated by the diagram in which the gluon is emitted before the scattering and reads (the analog of \eqref{Psiqtot})
\begin{align}
	\label{PsiqtotL}
	\Psi^{Lm}_{\lambda_1\lambda_2(q)}
	\simeq  -\,
	\delta _{\lambda_{1}\lambda_2}\,
	\frac{e e_f g q^+ }{(2\pi)^6}\,
	\frac{\sqrt{\vartheta_2}}{\vartheta_3}\,
	\frac{\tilde{Q}}
	{P_{\perp}^2+ \tilde{Q}^2}\,
	\frac{\tau^{mn*}(\vartheta_2,0,\lambda_1)k_{1}^n}
	{k_{1\perp}^2+ \mcal{M}^2},
\end{align}
valid for $\vartheta_1 \ll 1$ and $k_{1\perp} \ll P_{\perp}$. We take the Fourier transform w.r.t.~the semi-hard momentum $\bk_1$, insert the diffractive scattering and perform the inverse Fourier transform to get (the analog of \eqref{Psiqdiff})
\begin{align}
	\label{PsiqdiffL}
	\Psi^{Lm,D}_{\lambda_1\lambda_2(q)}
	\simeq 
	-\delta _{\lambda_{1}\lambda_2}\,
	\frac{e e_f g q^+ }{(2\pi)^6}\,
	\frac{\sqrt{\vartheta_2}}{\vartheta_3}\,
	\frac{\tilde{Q}}
	{P_{\perp}^2+ \tilde{Q}^2}\,
	\frac{\tau^{mn*}(\vartheta_2,0,\lambda_1) K^n}
	{K_{\perp}}\,
	\frac{\mcal{Q}_{\mathbb{P}}(\mcal{M}, K_{\perp},\YP)}{\mcal{M}}.
\end{align}
Using
\begin{align}
	\label{tausquare2}
	\sum_{m,\lambda}
	\tau^{mn*}(\vartheta,0,\lambda)
	\tau^{mr}(\vartheta,0,\lambda)
	= 4 (1-\vartheta)^2 \delta^{nr},
\end{align}
we obtain the cross section
 \begin{align}
	\label{cross2L}
	\frac{\rmd\sigma_{(qq)}^{\gamma_{\scriptscriptstyle L}^* A\rightarrow q\bar q g A}}
	{\rmd \vartheta_1 
	\rmd \vartheta_2
	\rmd \vartheta_3 \,
	\rmd^{2}\bP
  	\rmd^{2}\bK}
   = 
   \frac{S_{\perp} \alpha_{\rm em} N_c}{\pi^4} 
   \left( \sum e_f^2\right)
   \delta_{\vartheta}\,
   \frac{\alpha_s C_F \tilde{Q}^2}{(P_{\perp}^2 + \tilde{Q}^2)^2}\,
   \vartheta_2\,
   \frac{\big|\mcal{Q}_{\mathbb P}(\mcal{M}, K_{\perp},Y_{\mathbb P})\big|^2}{\mcal{M}^2},
   \end{align}
which is of the same order with the corresponding transverse cross section in Eq.~\eqref{cross2}. In fact, we can also express 
this result in terms of the quark DTMD introduced in Eq.~\eqref{qDTMD}; that is, \eqn{cross2L} admits TMD factorisation at
the hard dijet level, similarly to Eq.~\eqref{cross2new}. This factorised form is exhibited in Eqs.~\eqref{cross2newL}--\eqref{hqqg2L}
in the main text.

Since one of the two amplitudes (that where the gluon is emitted from the antiquark) is power-suppressed by the factor
$\vartheta_1 \sim K_\perp^2/Q^2\ll 1$, the same is also true for the interference term, so we shall not attempt to compute the latter.

\section{Quark and gluon diffractive TMDs in the single scattering approximation}
\label{sec:single}

Here we shall present exact calculations of $\mcal{Q}_{\mathbb{P}}$ in Eq.~\eqref{QPpm}, 
$\mcal{Q}_L$ in Eq.~\eqref{QL}, and  $\mcal{G}_{\mathbb{P}}$ in Eq.~\eqref{Gscalarnew}, within the context
of the MV model and in the regime of weak scattering.  The results to be obtained here are referred at several
places in the main text. In particular, we shall demonstrate the power suppression of the cross-section for exclusive
dijet production at large $K_\perp\gg Q_s(\YP)$.

For the present purposes,  the dipole
scattering amplitude $\mcal{T}(R)$ is given by the single scattering approximation to the MV model,
meaning by the exponent of \eqn{SMV}. This is a legitimate approximation when at least one of the two scales $K_{\perp}$ and $\mcal{M}$ is much larger than the saturation momentum $Q_s$, i.e.~when $\max\{K_{\perp},\mcal{M}\} \gg Q_s$. The smaller of these two scales does not need to satisfy any constraint w.r.t.~$Q_s$. 
Due to the presence of the Bessel functions the integration has support only for dipoles with size such that $R \lesssim \min\{1/K_{\perp},1/\mcal{M}\}\ll 1/Q_s$, thus justifying the use of the single scattering approximation.  We shall not make any assumptions regarding the relation between the scales $\mcal{M}$ and $K_{\perp}$ and therefore no simplification can be done to any of the two Bessel functions.

\paragraph{(i) Quark DTMD.} We have (cf.~Eq.~\eqref{QPpm})
\begin{align}
	\label{app:QP}
	\mcal{Q}_{\mathbb{P}} = \frac{Q_A^2 \mcal{M}^2}{4}
	\int_0^{\infty} \dif R\, R^3 J_1(K_{\perp} R)
	K_1(\mcal{M} R) 
	\left(\ln \frac{K_{\perp}^2}{\Lambda^2}
	+ \ln \frac{4}{K_{\perp}^2R^2}\right),
\end{align}
where for our convenience we have split the logarithm of the MV model into two terms. The first logarithm is independent of the variable $R$ and the respective integration is standard, leading to the contribution
\begin{align}
	\label{app:QP1}
	\mcal{Q}_{\mathbb{P}}^{(1)} = 
	\frac{2 Q_A^2 K_{\perp}\mcal{M}^3}{(K_{\perp}^2 +\mcal{M}^2)^3}\,
	\ln \frac{K_{\perp}^2}{\Lambda^2}.
\end{align}
The second contribution is more involved because the logarithm depends on $R$ and we shall write it as
\begin{align}
	\label{app:QP2}
	\mcal{Q}_{\mathbb{P}}^{(2)} = 
	- \frac{Q_A^2 \mcal{M}^2}{4}
	\lim_{a\to 0} \frac{\dif}{\dif a}
	\int_0^{\infty} \dif R\, R^3 
	\left(\frac{K_{\perp}^2R^2}{4} \right)^a
	J_1(K_{\perp} R)
	K_1(\mcal{M} R). 
\end{align}
Next we expand the Bessel function $J_1(K_{\perp} R)$ in a power series to get
\begin{align}
	\label{app:QP22}
	\mcal{Q}_{\mathbb{P}}^{(2)} = 
	- \frac{Q_A^2 \mcal{M}^2 K_{\perp}}{8}
	\sum_{n=0}^{\infty}
	\frac{(-1)^n}{n! (n+1)!}\,
	\lim_{a\to 0} \frac{\dif}{\dif a}
	\int_0^{\infty} \dif R\, R^4 
	\left(\frac{K_{\perp}^2R^2}{4} \right)^{n+a}\!\!
	K_1(\mcal{M} R)
\end{align}
and a straightforward calculation of the last integral (which is clearly convergent for all the relevant values of $n$ and $a$) leads to
\begin{align}
	\label{app:QP23}
	\mcal{Q}_{\mathbb{P}}^{(2)} = 
	- \frac{Q_A^2 K_{\perp}}{\mcal{M}^3}
	\sum_{n=0}^{\infty}
	\frac{1}{n! (n+1)!}
	\left(-\frac{K_{\perp}^2}{\mcal{M}^2} \right)^n\!
	\lim_{a\to 0} \frac{\dif}{\dif a}
	\left[
	\left(\frac{K_{\perp}^2}{\mcal{M}^2} \right)^a\!
	\Gamma(n+2+a) \Gamma(n+3+a)
	\right].
\end{align}
Differentiating with respect to $a$ and subsequently setting $a=0$ gives
\begin{align}
	\label{app:QP24}
	\mcal{Q}_{\mathbb{P}}^{(2)} = 
	- \frac{Q_A^2 K_{\perp}}{\mcal{M}^3}
	\sum_{n=0}^{\infty}
	(n+1)(n+2)\!
	\left(-\frac{K_{\perp}^2}{\mcal{M}^2} \right)^n\!
	\left[\ln\frac{K_{\perp}^2}{\mcal{M}^2} 
	+\Psi(n+2) + \Psi(n+3)
	\right],
\end{align}
where $\Psi(n)$ is the logarithmic derivative of $\Gamma(n)$. For $K_{\perp} < \mcal{M}$ the series is convergent and we obtain
\begin{align}
	\label{app:QP25}
	\mcal{Q}_{\mathbb{P}}^{(2)} = 
	\frac{Q_A^2 K_{\perp} \mcal{M}^3}
	{(K_{\perp}^2 +\mcal{M}^2)^3}\,
	\left[
	4 \ln \frac{K_{\perp}^2 +\mcal{M}^2}{K_{\perp}\mcal{M}}
	+ \frac{K_{\perp}^2}{\mcal{M}^2} - (5-4 \gamma_{\scriptscriptstyle E}) 
	\right],
\end{align}
with $\gamma_{\scriptscriptstyle E} \simeq 0.577$ the Euler constant. By analytic continuation, the above gives the correct result for the second contribution to the integration in Eq.~\eqref{app:QP} for any value of $K_{\perp}$ and $\mcal{M}$. Putting together Eqs.~\eqref{app:QP1} and \eqref{app:QP25} we finally arrive at 
\begin{align}
	\label{app:QPtot}
	\mcal{Q}_{\mathbb{P}} = 
	\frac{Q_A^2 K_{\perp} \mcal{M}^3}
	{(K_{\perp}^2 +\mcal{M}^2)^3}\,
	\left[
	2 \ln \frac{(K_{\perp}^2 +\mcal{M}^2)^2}{\mcal{M}^2 \Lambda^2}
	+ \frac{K_{\perp}^2}{\mcal{M}^2} - (5-4 \gamma_{\scriptscriptstyle E}) 
	\right].
\end{align}
We also notice that the last, constant, term in the square bracket can be neglected, since it is always logarithmically suppressed w.r.t~the first term (or even power suppressed w.r.t.~the second term when $K_{\perp}$ is much larger than $\mcal{M}$).

Let us now examine how the above general expression simplifies in the various kinematic regimes as determined by the scales $K_{\perp}$ and $\mcal{M}$. We recall that at least one of the two must be much larger than $Q_s$ for $\eqref{app:QPtot}$ to be valid.

\begin{enumerate}[(1)]
	\item When $K_{\perp} \sim \mcal{M}$ (that is, when $x$ is neither close to 0 nor to 1), the logarithm in the square bracket dominates and to the order of accuracy can be approximated as $\ln K_{\perp}^2/\Lambda^2$, so that
		\begin{align}
			\label{app:QPsame}
			\mcal{Q}_{\mathbb{P}} \simeq 
	\frac{2 Q_A^2 K_{\perp} \mcal{M}^3}
	{(K_{\perp}^2 +\mcal{M}^2)^3}\,
	\ln \frac{K_{\perp}^2}{\Lambda^2} = \mcal{Q}_{\mathbb{P}}^{(1)}
	\quad {\rm for} \quad K_{\perp} \sim \mcal{M}.
		\end{align}
		We point out that the prefactor in the above is under control and that in this regime the answer is solely given by the first contribution in Eq.~\eqref{app:QP1}. Indeed, the argument of the logarithm in Eq.~\eqref{app:QP25} is of the order of one when $K_{\perp} \sim \mcal{M}$ and thus the second contribution can be neglected. The above result is in complete agreement with Eq.~(A.9) in \cite{Iancu:2022lcw}. 
		\item When $K_{\perp} \gg \mcal{M}$ (that is, when $x \ll 1$), the second term in the square bracket in Eq.~\eqref{app:QPtot} is by far the largest and hence
			\begin{align}
			\label{app:QPKlarge}
			\mcal{Q}_{\mathbb{P}} \simeq
			\frac{Q_A^2 \mcal{M}}
			{K_{\perp}^3}
			\simeq 
			\frac{\sqrt{x}\,Q_A^2}
			{K_{\perp}^2}
			\quad {\rm for} \quad K_{\perp} \gg \mcal{M}.
			\end{align}
		Clearly, this could have been obtained directly by keeping only the first term in the expansion of $K_1(\mcal{M} R)$. 
		\item When $K_{\perp} \ll \mcal{M}$ (that is, when $1-x \ll 1$), the logarithm in the square bracket in Eq.~\eqref{app:QPtot} is enhanced and we easily find
			\begin{align}
			\label{app:QPKsmall}
			\mcal{Q}_{\mathbb{P}} 
			\simeq 
			\frac{2 Q_A^2 K_{\perp}}
			{\mcal{M}^3}\,
			\ln \frac{\mcal{M}^2}{\Lambda^2}
			\simeq
			\frac{2 (1-x)^{3/2} Q_A^2}
			{K_{\perp}^2}\,
			\ln \frac{K_{\perp}^2}{(1-x)\Lambda^2}
			\quad {\rm for} \quad K_{\perp} \ll \mcal{M}.
			\end{align}
		In this case the result could have been obtained directly by keeping only the first term in the expansion of $J_1(K_{\perp} R)$.
\end{enumerate}

\paragraph{(ii) Hard exclusive dijets.} As a direct application of the above, let us consider the production of a pair
of hard dijets with $K_\perp\sim 
\bar Q \gg Q_s$ via elastic scattering. The cross-section is given by \eqn{crossqq} which,
after integrating out a couple of trivial $\delta$--functions, becomes \eqn{crossqqnew}. Notice that the relevant
kinematical variables are different w.r.t.~the previous discussion: on fixes the longitudinal fraction $\vartheta_1$,
rather than $\beta$ (which can be computed in terms of $\vartheta_1$ and $K_\perp$, cf.~\eqn{beta}). In particular,
the effective virtuality is now denoted as $\bar Q$, instead of $\mcal{M}$. Taking into account these minor changes,
one finds the following expression for the exclusive dijet cross-section in the hard regime  (cf.~\eqn{app:QPsame})
\begin{align}
	\label{crossqqhard}
	\frac{\rmd\sigma_{\scriptscriptstyle}^{\gamma_{\scriptscriptstyle T}^* A\rightarrow q\bar q A}}
	{\rmd \vartheta_1\,
	\rmd^{2}\bK} &\,\simeq
	\frac{S_{\perp} \alpha_{\rm em} N_c}{2\pi^2} \left( \sum e_f^2\right)
	\left[\vartheta_1^2 +(1-\vartheta_1)^2 \right] \frac{4  \bar Q^4 K_\perp^2}{(K_\perp^2+\bar Q^2)^6}
	\,\left[Q_A^2\ln \frac{K_\perp^2}{\Lambda^2}\right]^2\,.
\end{align}
For large and comparable values of $K_\perp$ and $\bar Q$, this cross-section decreases like $1/K_\perp^6$.

\paragraph{(iii) Longitudinal sector.} 
Following the exact same steps, we can calculate the corresponding quantity $\mcal{Q}_L$, introduced in Eq.~\eqref{QL}, in the longitudinal sector. In the single scattering approximation we have
\begin{align}
	\label{QLsingle}
	\mcal{Q}_L = \frac{Q_A^2 \mcal{M}^2}{4}
	\int_0^{\infty} \dif R\, R^3 
	J_0(K_{\perp} R)
	K_0(\mcal{M} R) 
	\ln \frac{4}{R^2 \Lambda^2}
\end{align}
and we obtain the exact result
\begin{align}
	\label{QLexact}
	\mcal{Q}_L= 
	\frac{Q_A^2 \mcal{M}^2}
	{(K_{\perp}^2 +\mcal{M}^2)^3}\,
	\left\{
	\big(\mcal{M}^2 - K_{\perp}^2\big)
	\left[
	\ln \frac{(K_{\perp}^2 +\mcal{M}^2)^2}{\mcal{M}^2 \Lambda^2}
	+ 2(\gamma_{\scriptscriptstyle E}-1)
	\right]
	+ 2 K_{\perp}^2
	\right\}.
\end{align}
To the order of accuracy the constant term $ 2(\gamma_{\scriptscriptstyle E}-1)$ can always be dropped.  When $K_{\perp} \gg \mcal{M}$ the above reduces to
\begin{align}
	\label{QLKlarge}
	\mcal{Q}_L 
	\simeq
	-\frac{Q_A^2 \mcal{M}^2}
	{K_{\perp}^4}\,
	\ln\frac{K_{\perp}^4}{\mcal{M}^2 \Lambda^2}
	\simeq 
	-\frac{x Q_A^2}
	{K_{\perp}^2}\,
	\ln\frac{K_{\perp}^2}{x\Lambda^2}
	\quad {\rm for} \quad K_{\perp} \gg \mcal{M},
\end{align}
whereas for $K_{\perp} \ll \mcal{M}$ one finds
\begin{align}
	\label{QLKlow}
	\mcal{Q}_L 
	\simeq
	\frac{Q_A^2}{\mcal{M}^2}\,
	\ln\frac{\mcal{M}^2}{\Lambda^2}
	\simeq 
	\frac{(1-x)Q_A^2}
	{K_{\perp}^2}\,
	\ln\frac{K_{\perp}^2}{(1-x) \Lambda^2}
	\quad {\rm for} \quad K_{\perp} \gg \mcal{M}.
\end{align}

\paragraph{(iv) Gluon  DTMD.} Finally, we can do the analogous calculation for the amplitude $\mcal{G}_{\mathbb{P}}$ in Eq.~\eqref{Gscalarnew} which determines the gluon DTMD. We start with 
\begin{align}
	\label{app:GP}
	\mcal{G}_{\mathbb{P}} = \frac{Q_{A,g}^2 \mkern1mu \mcal{M}^2}{4}
	\int_0^{\infty} \dif R\, R^3 J_2(K_{\perp} R)
	K_2(\mcal{M} R)
	\ln \frac{4}{R^2 \Lambda^2 },
\end{align}
where $Q_{A,g}^2$ is enhanced by a color factor $N_c/C_{\scriptscriptstyle F}$ w.r.t.~to $Q_A^2$ introduced in Eq.~\eqref{SMV}, to arrive at the final exact result
\begin{align}
	\label{app:GPtot}
	\mcal{G}_{\mathbb{P}} = 
	\frac{Q_{A,g}^2 K_{\perp}^2 (K_{\perp}^2+ 3\mcal{M}^2)}
	{(K_{\perp}^2 +\mcal{M}^2)^3}\,
	\bigg[
	&\ln \frac{K_{\perp}^2 +\mcal{M}^2}{\Lambda^2}
	+ \frac{(2 \gamma_{\scriptscriptstyle E}\!-\!1)  K_{\perp}^2 - 
	6(1 \!-\! \gamma_{\scriptscriptstyle E}) \mcal{M}^2}
	{K_{\perp}^2+ 3\mcal{M}^2}
	\nn 
	&+ \frac{\mcal{M}^4}
	{K_{\perp}^2(K_{\perp}^2+ 3\mcal{M}^2)}
	- \frac{\mcal{M}^4(3 K_{\perp}^2+ \mcal{M}^2)}
	{K_{\perp}^4(K_{\perp}^2+ 3\mcal{M}^2)}
	\ln \frac{K_{\perp}^2 +\mcal{M}^2}{\mcal{M}^2}
	\bigg].
\end{align}
Independently of the order between the scales $K_{\perp}$ and $\mcal{M}$, the first term dominates the sum in the square bracket (notice that when $\mcal{M} \gg K_{\perp}$ the last two terms give a combined contribution of the order of one although each of them is of the order of $\mcal{M}^2/K_{\perp}^2 \gg 1$) and therefore we can write to logarithmic accuracy
\begin{align}
	\label{app:GPapp}
	\mcal{G}_{\mathbb{P}} \simeq 
	\frac{Q_{A,g}^2 K_{\perp}^2 (K_{\perp}^2+ 3\mcal{M}^2)}
	{(K_{\perp}^2 +\mcal{M}^2)^3}\,
	\ln \frac{K_{\perp}^2 +\mcal{M}^2}{\Lambda^2}
	\quad {\rm for\,\, any} \quad K_{\perp},\mcal{M}.
\end{align}	 
The above is in exact agreement with (the second line in) Eq.~(5.8) in \cite{Iancu:2022lcw}.

\section{Instantaneous interaction terms}
\label{sec:inst}

In this Appendix, we will show that when studying 2+1 jet production, where the source of the semi-hard jet is a quark or an antiquark, the diagrams involving instantaneous interactions are subleading. This should be contrasted to the case of a semi-hard gluonic jet, in which the contribution of instantaneous terms is of the same order as the regular ones and is in fact crucial for obtaining the correct wavefunction of the gluon-gluon dipole.

\paragraph{(i) Gluon emission from the antiquark in the transverse sector.} 

From \cite{Iancu:2022gpw}, one can infer that the exact probability amplitude of interest reads
\begin{align}
	\label{Psiinst1} 
	\Psi^{ij,{\rm inst}}_{\lambda_1\lambda_2\,(\bar{q})}=
	\delta _{\lambda_{1}\lambda_2}\,
	\frac{e e_f g }{4(2\pi)^6}\,
	\frac{1}{\sqrt{\vartheta_3} (1-\vartheta_1)}\,
	\frac{\bar{\varphi}^{ij}(\lambda_1)}
	{E_q+E_{\bar q}+E_g -E_\gamma},
\end{align}
where
\begin{align}
	\label{barphi}
	\bar{\varphi}^{ij}(\lambda) \equiv
	\delta^{ij}+ 2 i \lambda \varepsilon^{ij}.
\end{align}
The photon decay and the gluon emission happen at the same time and this leads to various differences w.r.t.~to the regular terms. First, when comparing to Eq.~\eqref{Psi1}, we observe that now the intermediate energy denominator $E_q +E_{\bar{q}'}-E_{\gamma}$ is absent. Second, both the electromagnetic and the QCD vertex are independent of the momenta. Finally, it is clear that the scattering occurs after the gluon emission, which means that there is only diagram, the one in which the trijet state crosses the nuclear shockwave. 

One should be careful enough not to compare Eq.~\eqref{Psiinst1} with Eq.~\eqref{Psi1}, but with the total probability amplitude in Eq.~\eqref{Psitot}. Indeed, when one combines Eqs.~\eqref{Psi1} and \eqref{Psi2}, the aforementioned energy denominator $E_q +E_{\bar{q}'}-E_{\gamma}$ which involves semi-hard transverse momenta cancels, and the new denominator which appears is much harder, cf.~Eq.~\eqref{ED1}. Hence, the total amplitude is much smaller than the two individual contributions and it is not a priori clear that it will me much larger than the instantaneous term. Still, by making use of Eqs.~\eqref{ED2} and \eqref{Mdef1}, we can write the instantaneous term as
\begin{align}
	\label{Psiinst1new} 
	\Psi^{ij,{\rm inst}}_{\lambda_1\lambda_2\,(\bar{q})} \simeq
	\delta _{\lambda_{1}\lambda_2}\,
	\frac{e e_f g q^+}{2(2\pi)^6}\,
	\frac{\vartheta_1}{\sqrt{\vartheta_3}}\,
	\frac{\bar{\varphi}^{ij}(\lambda_1)}
	{k_{1\perp}^2 + \mcal{M}^2}
\end{align}
and it is straightforward to check that the above is power suppressed w.r.t.~to the regular contribution in Eq.~\eqref{Psitot} when $\vartheta_1 \ll k_{1\perp}/P_{\perp}$.

\paragraph{(ii) Gluon emission from the quark in the transverse sector.} As we have already seen in the analysis of the regular terms, the probability amplitude with gluon emission from the quark can be obtained from the one with gluon emission from the antiquark by changing the sign, taking the complex conjugate and letting $1 \leftrightarrow 2$. Then Eq.~\eqref{Psiinst1} changes to
\begin{align}
	\label{Psiinst2} 
	\Psi^{ij,{\rm inst}}_{\lambda_1\lambda_2\,(q)}=
	-\delta _{\lambda_{1}\lambda_2}\,
	\frac{e e_f g }{4(2\pi)^6}\,
	\frac{1}{\sqrt{\vartheta_3} (1-\vartheta_2)}\,
	\frac{\bar{\varphi}^{ij*}(\lambda_1)}
	{E_q+E_{\bar q}+E_g -E_\gamma},
\end{align}
and using Eq.~\eqref{ED2q} we obtain
\begin{align}
	\label{Psiinst2new} 
	\Psi^{ij,{\rm inst}}_{\lambda_1\lambda_2\,(\bar{q})} \simeq
	-\delta _{\lambda_{1}\lambda_2}\,
	\frac{e e_f g q^+}{2(2\pi)^6}\,
	\frac{\vartheta_1}{\vartheta_3^{3/2}}\,
	\frac{\bar{\varphi}^{ij*}(\lambda_1)}
	{k_{1\perp}^2 + \mcal{M}^2}.
\end{align}
Again, the above is power suppressed when compared to Eq.~\eqref{Psiqtot} for $\vartheta_1 \ll k_{1\perp}/P_{\perp}$.

\section{Details on 2+1 jet contributions to diffractive SIDIS}
\label{sec:3SIDIS}

In this Appendix, we shall compute in detail the contributions of 2+1 jets to diffractive SIDIS, for both the
case of a soft quark, and for that of a soft gluon. Our main goal is to identify all the leading twist contributions
in both cases.

\subsection{The case of a soft quark}

The starting point is \eqn{qqgDIFFR} where, for the purposes of the physical interpretation, it is preferable to separate
the three types of contributions to the cross-section for diffractive (2+1)-jets: gluon emission by the antiquark, 
gluon emission by the quark, and the interference term (recall the three graphs in Fig.~\ref{fig:3jets_TMD}).
Accordingly, the hard factor in the second line of \eqn{qqgDIFFR} is decomposed as follows:
\beq\label{smallh}
{H}_T(\vartheta_2,\vartheta_3,P_{\perp}^2,\tilde{Q}^2)\,=\,
\frac{4 \pi^2 \alpha_{\rm em}}{Q^2} \left( \sum e_f^2\right)
   	\delta_{\vartheta}\,
	\frac{\alpha_s C_F}{2\pi^2}\,\frac{1}{P^2_\perp}\,
	\frac{1}{\vartheta_3}
	\left(h_T^{(\bar q)} + 
	h_T^{(q)} + 
	h_T^{\rm int}\right)\,.\eeq
Note that we have defined the ``small'' hard factors  $h_T^{(\bar q)}$, $h_T^{(q)}$, and $h_T^{\rm int}$ in such a way
to be dimensionless and we have also extracted a factor $1/\vartheta_3$, which was indeed common to the
original hard factors, as obtained in Sect.~\ref{sec:tmd}. Specifically, one has
(cf.~Eqs.~\eqref{cross1new}--\eqref{cross3new})
\beq 
\label{shf}
h_T^{(\bar q)} =\,\frac{(1+\vartheta_2^2)\tilde{Q}^2}
	{P_{\perp}^2 + \tilde{Q}^2}\,,
	\qquad 
	h_T^{(q)} =\,\frac{[\vartheta_2^2 +(1-\vartheta_2)^2]\,\tilde{Q}^2 P^4_\perp}
	{(P_{\perp}^2 + \tilde{Q}^2)^3}\,,
	\qquad  
	h_T^{\rm int}=- \,\frac{2\vartheta_2^2\, \tilde{Q}^2P_{\perp}^2}
	{(P_{\perp}^2 + \tilde{Q}^2)^2}\,.
	\eeq
%
%
	
Furthermore,  the $\delta$--function occurring 
in the second line of \eqn{qqgDIFFR} is conveniently combined with the factor $1/\vartheta_3$ from ${H}_T$, to yield
\begin{align}\label{delta3}
\frac{1}{\vartheta_3}\delta\left(x-\beta\, \frac{P_\perp^2+\tilde{Q}^2}{\tilde{Q}^2}\right)
&\,=\,\frac{\vartheta_2}{x-\beta}\,\delta\left(\vartheta_2\vartheta_3-\frac{\beta}{x-\beta}\,\frac{P_\perp^2}{Q^2}\right)
\nonumber\\*[0.2cm]
&\,=\,\frac{\vartheta_2}{x-\beta}\,
 \frac{\delta\left(\vartheta_2-\vartheta_*\right) + \delta\left(\vartheta_2-1+\vartheta_*\right)}{1-2\vartheta_*},
   \end{align}
where $\vartheta_*$ denotes the solution smaller than 1/2:
\beq
	\label{xmin}
	\vartheta_*\,=\,\frac{1}{2}\left(1 -\sqrt{1-\frac{4\beta}{x-\beta}\frac{P_\perp^2}{Q^2}}\right)
	\qquad
	\mbox{with}\quad x-\beta \ge \,4\beta\,\frac{P_\perp^2}{Q^2}\,\,.
\eeq
Using this $\delta$--function together with $\delta_{\vartheta}= \delta(1-\vartheta_2-\vartheta_3)$ 
from the structure of ${H}_T$, we first compute
the integrations over $\vartheta_3$ and $\vartheta_2$ for a fixed value of $x$ and for each of the three contributions to the hard factor.
The direct gluon emission by the antiquark yields
\begin{align}\label{hbarq}
\int \rmd \vartheta_2
  	\rmd \vartheta_3\,\delta_{\vartheta}\,\delta\left(x-\beta\frac{P_\perp^2+\tilde{Q}^2}{\tilde{Q}^2}\right)\frac{h_T^{(\bar q)}}{\vartheta_3}
	&=\,\frac{\beta}{x(x-\beta)}\left[2-\frac{3\beta}{x-\beta}\frac{P_\perp^2}{Q^2}\right]\frac{1}{1-2\vartheta_*}\,,
\end{align}	
where we have used $\tilde{Q}^2=\frac{\beta}{x-\beta}P_\perp^2$ and
\beq
\int\rmd \vartheta_2\,\vartheta_2(1+\vartheta_2^2)
\left[ \delta\left(\vartheta_2-\vartheta_*\right) + \delta\left(\vartheta_2-1+\vartheta_*\right)
\right]\,=\,2-3\vartheta_*(1-\vartheta_*)=2-\frac{3\beta}{x-\beta}\frac{P_\perp^2}{Q^2}\,.
\eeq
For the direct gluon emission by the quark, we similarly find
\begin{align}\label{hq}
\int \rmd \vartheta_2
  	\rmd \vartheta_3\,\delta_{\vartheta}\,\delta\left(x-\beta\frac{P_\perp^2+\tilde{Q}^2}{\tilde{Q}^2}\right)\frac{h_T^{(q)}}{\vartheta_3}
	&=\,\frac{\beta(x-\beta)}{x^3}\left[1-\frac{2\beta}{x-\beta}\frac{P_\perp^2}{Q^2}\right]\frac{1}{1-2\vartheta_*}.
\end{align}	
Finally, the interference contribution reads
\begin{align}\label{hint}
\int \rmd \vartheta_2
  	\rmd \vartheta_3\,\delta_{\vartheta}\,\delta\left(x-\beta\frac{P_\perp^2+\tilde{Q}^2}{\tilde{Q}^2}\right)\frac{h_T^{\rm int}}{\vartheta_3}
	&=-\,\frac{2\beta}{x^2}\left[1-\frac{3\beta}{x-\beta}\frac{P_\perp^2}{Q^2}\right]\frac{1}{1-2\vartheta_*}.
\end{align}	

Recalling that (below, we set $t\equiv x-\beta$ and $ t_{\rm min} \equiv 4\beta {P_\perp^2}/{Q^2}$ to simplify writing)
\beq
1-2\vartheta_* = \sqrt{1-\frac{4\beta}{x-\beta}\frac{P_\perp^2}{Q^2}}\,=\,\sqrt{1-\frac{t_{\rm min}}{t}}\,=\,1-
\frac{t_{\rm min}}{2t}\,+\,\order{(t_{\rm min}/t)^2},
\eeq
it is clear that, so long as $x-\beta$ is a fixed quantity, it is straightforward to perform the twist expansion (the expansion in powers
of ${P_\perp^2}/{Q^2}$) of the above results. However, after integrating over $x$, this expansion gets complicated by the 
singularity of the above results at $x=\beta$ --- a point which lies in the vicinity of the lower limit $x_{\rm min}$ of the integration:
$x_{\rm min} -\beta = t_{\rm min} \ll 1$. This subtlety concerns only the contribution in \eqref{hbarq} (direct gluon emission by the 
antiquark, cf.~Fig.~\ref{fig:fromqbar} and the first graph in Fig.~\ref{fig:3jets_TMD}), since this exhibits the strongest singularity at $x=\beta$.

To see this, lets us focus on the singular structure of the integral over $x$ involving this contribution; to that purpose,
one can ignore non-singular factors, like $1/x$, or the quark DPDF. The essential integration is
\beq\label{integraltq}
\int_{x_{\rm min} }^1\frac{\rmd x}{x-\beta}\,\frac{2-\frac{3\beta}{x-\beta}\frac{P_\perp^2}{Q^2}}
{\sqrt{1-\frac{4\beta}{x-\beta}\frac{P_\perp^2}{Q^2}}}\,=
\int_{t_{\rm min} }^{1-\beta}\frac{\rmd t}{t}\frac{2-\frac{3t_{\rm min}}{4t}}{\sqrt{1-\frac{t_{\rm min}}{t}}}\,=
\int_{t_{\rm min} }^{1-\beta}\frac{\rmd t}{t}\left\{2-\frac{t_{\rm min}}{4t}+\,\order{(t_{\rm min}/t)^2}\right\},
\eeq
where in the last equality we expanded the integrand in powers of $t_{\rm min}/t$. The first term in this expansion
gives a logarithmic integration, which is controlled by values of $t$ in the bulk, $t_{\rm min} \ll t \ll 1-\beta$, meaning
 $x-\beta \gg  P_\perp^2/Q^2$. Via \eqn{xmin}, this implies that the typical values of $\vartheta_*$ that are 
 explored by this logarithmic integration are very small: $\vartheta_*\sim P_\perp^2/Q^2\ll 1$. Clearly, this corresponds
 to an {\it asymmetric} configuration for the hard antiquark-gluon dijet in Fig.~\ref{fig:fromqbar}, and more precisely to the case
 where the gluon is soft\footnote{Recall indeed the overall factor $\vartheta_2$ in \eqn{xmin}; when $\vartheta_*\ll 1$,
 this factor favours the solution $\vartheta_2=1-\vartheta_*\simeq 1$.} 
  w.r.t.~the antiquark parent: $\vartheta_3\ll \vartheta_2\simeq 1$.
The situation is however different for the higher-order terms in the expansion within the integrand of \eqn{integraltq}:
 for each of these terms, of the form $t_{\rm min}^n/t^{n+1}$ with $n\ge 1$,  the integral is 
controlled by the lower limit and yields a contribution of order one. To summarise, all the terms in the expansion of
the integrand contribute at leading twist after the integration over $x$. 
And all these contributions except for the first one (which is logarithmic) come
from $t\sim t_{\rm min}$, i.e.~from  $x-\beta \sim 4\beta P_\perp^2/Q^2$, 
or $\vartheta_*\sim 1/2$: they correspond to {\it symmetric}
$\bar q g$ hard dijets in Fig.~\ref{fig:fromqbar}. For physical considerations (notably, for the interpretation of results in terms of
target evolution), it is convenient to separate symmetric from asymmetric dijet contributions. Clearly, this can be done by
rewriting the integrand of \eqn{integraltq} as follows:
\beq\label{integraltqsep}
\int_{t_{\rm min} }^{1-\beta}\frac{\rmd t}{t}\left\{2+\left[\frac{2-\frac{3t_{\rm min}}{4t}}{\sqrt{1-\frac{t_{\rm min}}{t}}} -2 \right]
\right\}.
\eeq

 Consider now the remaining two contributions --- the direct gluon emission by the quark and the interference term.
 By inspection of Eqs.~\eqref{hq}--\eqref{hint}, one sees that they are less singular, or not at all, when $x\to \beta$,
 so the corresponding discussion is much simpler: the twist expansion can already be performed before integrating
 over $x$. To demonstrate this, let us consider the interference term \eqref{hint}, for which the analog of \eqn{integraltq} reads
 (up to  irrelevant  factors)
 \beq\label{integraltint}
\int_{x_{\rm min} }^1{\rmd x} \,\frac{1-\frac{3\beta}{x-\beta}\frac{P_\perp^2}{Q^2}}
{\sqrt{1-\frac{4\beta}{x-\beta}\frac{P_\perp^2}{Q^2}}}\,=
\int_{t_{\rm min} }^{1-\beta}{\rmd t}\frac{1-\frac{3t_{\rm min}}{4t}}{\sqrt{1-\frac{t_{\rm min}}{t}}}\,=
\int_{t_{\rm min} }^{1-\beta}{\rmd t}\left\{1-\frac{t_{\rm min}}{4t}+\,\order{(t_{\rm min}/t)^2}\right\}.
\eeq
 It is quite clear that only the first term in the above expansion\footnote{For instance, the second term yields
 a contribution  $\sim t_{\rm min}\ln\frac{1-\beta}{t_{\rm min}}$, which counts to next-to-leading twist,
 since it is suppressed by $t_{\rm min}\sim P_\perp^2/Q^2$.}, i.e., 
 the unity, contributes to LT order and that the respective
contribution comes from $x$ values in the bulk,  $x-\beta \gg  P_\perp^2/Q^2$, that is from asymmetric $\bar q g$ dijets
with a soft gluon: $\vartheta_3\ll \vartheta_2\simeq 1$. A similar conclusion applies for Eq.~\eqref{hq}.

To summarise, the process represented by the first graph in Fig.~\ref{fig:3jets_TMD}
(direct gluon emission by the antiquark) yields two types of LT contributions to diffractive SIDIS:
those coming from asymmetric $\bar qg$ dijets
with $\vartheta_3\ll \vartheta_2\simeq 1$ and, respectively, those from the symmetric dijets with
$\vartheta_2\vartheta_3\sim 1/4$. These two types of contributions can be separated by
rewriting the r.h.s.~of \eqref{hbarq} as (cf.~\eqn{integraltqsep})
\begin{align}\label{3hbarqLT}
h_T^{(\bar q)}	&\,\to\,\frac{2\beta}{x(x-\beta)}\,+\,
\frac{\beta}{x(x-\beta)}\left[4\vartheta_*-\frac{3\beta}{x-\beta}\frac{P_\perp^2}{Q^2}\right]\frac{1}{1-2\vartheta_*}\,.
\end{align}	
The two other processes in  Fig.~\ref{fig:3jets_TMD} (direct gluon emission by the quark and the
interference term) yield LT contributions solely via the asymmetric configurations with
$\vartheta_3\ll \vartheta_2\simeq 1$. These contributions can be computed by simplifying the r.h.s.'s of  Eqs.~\eqref{hbarq}--\eqref{hint} as follows
 \begin{align}\label{3hLT}
h_T^{(q)}&\,\to\,\frac{\beta(x-\beta)}{x^3}\,,\qquad h_T^{\rm int}\,\to\,-\,\frac{2\beta}{x^2}\,.
\end{align}	
The three contributions from asymmetric dijets ($\vartheta_3\ll \vartheta_2\simeq 1$) in
Eqs.~\eqref{3hbarqLT}--\eqref{3hLT} add up to a
result proportional to the  DGLAP splitting function $P_{qq}$:
\beq
\frac{2\beta}{x(x-\beta)}\,+\,\frac{\beta(x-\beta)}{x^3}\,-\,\frac{2\beta}{x^2}\,=\,
\frac{\beta}{x^3}\,\frac{x^2+\beta^2}{x-\beta}\,
  =\, \frac{\beta}{x^2}\,\frac{1}{C_F}\, P_{qq}\left(\frac{\beta}{x}\right).
\eeq
After including all the relevant factors and performing the integral over $x$, we finally recover the contribution
\eqref{qqgSIDIS} to the diffractive SIDIS cross-section, which is naturally interpreted as one step in the DGLAP evolution
of the quark DPDF.

\begin{figure}
	\begin{center}
		\includegraphics[width=0.5\textwidth]{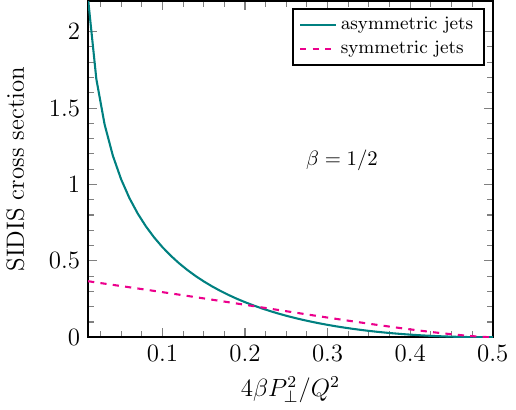}
	\end{center}
	\caption{\small Contributions to the SIDIS cross section of 2+1 jet configurations with a soft quark ($\vartheta_1 \ll 1$): the integrals appearing in Eqs.~\eqref{qqgSIDIS} (asymmetric $\bar{q}g$ dijet) and \eqref{softqSIDIS1} (symmetric $\bar{q}g$ dijet) as functions of the combined variable $4\beta P_{\perp}^2/Q^2 = x_{\rm min} - \beta$. The scattering amplitude is given by the MV model with a saturation scale $Q_{s}^2=0.88\,\mathrm{GeV}^2$ (in the fundamental representation).}
\label{fig:two-integrals}
\end{figure}

As for the remaining term in \eqn{3hLT} for $h_T^{(\bar q)}$, this yields a LT contribution associated with symmetric $\bar q g$ jets ($\vartheta_3\sim \vartheta_2\sim 1/2$), which is easily obtained as follows:
\begin{align}
	\label{softqSIDIS1}
	\frac{\rmd \sigma_{{\rm soft}\,q;\ {\rm sym}\,\bar q g}^{\gamma_{\scriptscriptstyle T}^* A\rightarrow q\bar q g A}}
	{ \rmd^{2}\bm{P} \,\rmd \ln(1/\beta)} 
	&\,= \frac{4 \pi^2 \alpha_{\rm em}}{Q^2} \left( \sum e_f^2\right) \, \frac{\alpha_s C_F }{2\pi^2}\frac{1}{P_\perp^2}
	\nonumber\\*[0.2cm] &\times
\int_{x_{\rm min} }^1\frac{\rmd x}{x}\,\frac{2\beta}{x-\beta}\, \frac{1-\sqrt{1-\frac{4\beta}{x-\beta}\frac{P_\perp^2}{Q^2}}
-\frac{3\beta}{2(x-\beta)}\frac{P_\perp^2}{Q^2}}
{\sqrt{1-\frac{4\beta}{x-\beta}\frac{P_\perp^2}{Q^2}}}\,
xq_{\mathbb{P}}\left(x, x_{\mathbb{P}}, {(1-x)P_\perp^2}\right).
 \end{align}	
 The integral is controlled by its lower limit and to the LT accuracy of interest it can be evaluated
 by replacing $x\to \beta$ in all the non-singular pieces of the integrand, including the quark DPDF.
 After this replacement, it is easy to check that (to LT accuracy, once again) the integral reduces to a pure number.
 With reference to the target picture, this LT contribution should be associated with final state radiation, that is,
 with a $\bar q \to \bar q g$ splitting which occurs after the absorption of the virtual photon by the antiquark. In Fig.~\ref{fig:two-integrals} we show the numerical results for such a part of the SIDIS cross section arising from symmetric jets, in comparison to the one in Eq.~\eqref{qqgSIDIS} which refers to asymmetric jets. We see that the latter dominates when $4 \beta P_{\perp}^2/Q^2$ is sufficiently small, since the integration in Eq.~\eqref{qqgSIDIS} becomes sensitive to its lower limit and as a consequence grows logarithmically.
 
Consider finally the case of a virtual photon with longitudinal polarisation. The respective contribution to diffractive
(2+1)--jets with a soft quark, as shown in Eqs.~\eqref{cross2newL}--\eqref{hqqg2L}, can be cast in the form of \eqn{smallh}
with 
\beq
\label{smallhL} 
 h_L =\,4\vartheta_2\vartheta_3
 \frac{\tilde{Q}^4 P^2_\perp}
	{(P_{\perp}^2 + \tilde{Q}^2)^3}\,=\,4\left(\frac{\beta}{x}\right)^3 \frac{P_{\perp}^2}{Q^2}\,,
	\eeq
where the second equality follows after also using the $\delta$--function \eqref{delta3}. This implies
\begin{align}\label{hqL}
\int \rmd \vartheta_2
  	\rmd \vartheta_3\,\delta_{\vartheta}\,\delta\left(x-\beta\frac{P_\perp^2+\tilde{Q}^2}{\tilde{Q}^2}\right)\frac{h_L}{\vartheta_3}
	&=\,\frac{\beta^2}{x^3}\, \frac{4\beta}{x-\beta}\frac{P_{\perp}^2}{Q^2}\,\frac{1}{1-2\vartheta_*}\,,
\end{align}	
which gives no LT contribution after integration over $x$. Indeed, the relevant integral reads
\beq\label{integralL}
\int_{x_{\rm min} }^1\frac{\rmd x}{x-\beta}\,\frac{4\beta P_\perp^2}{Q^2}\,\frac{1}
{\sqrt{1-\frac{4\beta}{x-\beta}\frac{P_\perp^2}{Q^2}}}\,=
\int_{t_{\rm min} }^{1-\beta}\frac{\rmd t}{t}\frac{t_{\rm min}}{\sqrt{1-\frac{t_{\rm min}}{t}}}\,,
\eeq
and is quite similar to that in \eqn{integraltq}, except that it involves an additional factor $t_{\rm min}\sim {P_{\perp}^2}/{Q^2}$.
Since the result of \eqref{integraltq} is of order one (up to logarithms),  it is clear that the result of \eqref{integralL} is 
of $\order{{P_{\perp}^2}/{Q^2}}$, so its contribution to the SIDIS cross-section  is of higher-twist order. From the above
manipulations, it should be clear that this suppression is ultimately due to the factor $\vartheta_2\vartheta_3$
introduced by the splitting of the longitudinal photon.

\subsection{The case of a soft gluon}

For (2+1)-jets with a soft gluon, the analog of \eqn{qqgDIFFR} reads (recall \eqn{gluondipColl})
\begin{align}
	\label{qqgDIFFRsoftg}
	\hspace{-0.4cm}
	\frac{\rmd \sigma^{\gamma_{T}^* A
	\rightarrow q\bar q (g)}}{ \rmd^{2}\!\bm{P}\, \rmd  \ln(1/\beta)} 
	&\,=\int \rmd \vartheta_1
  	\rmd \vartheta_2
	\frac{\rmd \sigma^{\gamma_{T}^* A
	\rightarrow q\bar q (g) XA}}
  	{\rmd \vartheta_1
  	\rmd \vartheta_2
  	\rmd^{2}\!\bm{P}
  	\,\rmd \ln(1/\beta)} \\*[0.2cm]
  	\nonumber
       &\,	=  \int\rmd \vartheta_2
		\int_\beta^1\rmd x\,\delta\left(x-\beta\, \frac{\bar Q^2+P_\perp^2}{\bar Q^2}\right)
  	 \tilde{H}_{T}(1-\vartheta_2,\vartheta_2, {Q}^2, P_{\perp}^2)\,
	 xG_{\mathbb{P}}\left(x, x_{\mathbb{P}}, {(1-x)P_\perp^2}\right).
 \end{align}
where the hard momentum $\bP$ is now the relative momentum of the quark-antiquark pair, whereas the semi-hard 
momentum $\bK$ refers to the gluon. Also, $\bar Q^2=\vartheta_1\vartheta_2 Q^2$, with 
$\vartheta_1+\vartheta_2 =1$. The hard factor corresponding to a virtual photon with transverse polarisation 
has been shown in \eqn{HardT}, 
 whereas the gluon diffractive TMD (or the UGD of the Pomeron) is exhibited
 in Eqs.~\eqref{pomugddef}--\eqref{Gscalarnew}.	
 
 This $\delta$--function is treated similarly to \eqn{delta3}, namely
 \begin{align}\label{delta4}
\delta\left(x-\beta\, \frac{P_\perp^2+\bar{Q}^2}{\bar{Q}^2}\right)
&\,=\,\frac{\vartheta_1\vartheta_2}{x-\beta}\,\delta\left(\vartheta_1\vartheta_2-\frac{\beta}{x-\beta}\,\frac{P_\perp^2}{Q^2}\right)
\nonumber\\*[0.2cm]
&\,=\,\frac{\beta}{(x-\beta)^2}\,\frac{P_\perp^2}{Q^2}\,
 \frac{\delta\left(\vartheta_2-\vartheta_*\right) + \delta\left(\vartheta_2-1+\vartheta_*\right)}{1-2\vartheta_*},
   \end{align}
with $\vartheta_*$ as shown \eqn{xmin}. This implies
  \beq\label{HFx}
\frac{P_{\perp}^4 + \bar{Q}^4}
	{(P_{\perp}^2 + \bar{Q}^2)^4}\,=\,\frac{1}{P_\perp^4}\left(\frac{x-\beta}{x}\right)^2\left[\left(1-\frac{\beta}{x}\right)^2+
	\left(\frac{\beta}{x}\right)^2\right]
	=\,\frac{2}{P_\perp^4}\left(\frac{x-\beta}{x}\right)^2 P_{qg}\left(\frac{\beta}{x}\right).
\eeq
After integrating over  $\vartheta_2$, one finds (notice that both solutions for  $\vartheta_2$ give identical results)
 \begin{align}\hspace*{-0.8cm}
	\label{qqgSIDISsoftgGEN}
	\frac{\rmd \sigma_{{\rm soft} \,g} ^{\gamma_{\scriptscriptstyle T}^* A\rightarrow q\bar q g A}}{ \rmd^{2}\bm{P} \,\rmd \ln(1/\beta)} 
	=  \frac{4 \pi^2 \alpha_{\rm em}}{Q^2} \left( \sum e_f^2\right)  \frac{\alpha_s}{\pi^2}\frac{1}{P_\perp^2}
\int_{x_{\rm min} }^1\frac{\rmd x}{x}\frac{\beta}{x}\, P_{qg}\left(\frac{\beta}{x}\right)\frac{1- \frac{2\beta}{x-\beta}\frac{P_\perp^2}{Q^2}}
{\sqrt{1-\frac{4\beta}{x-\beta}\frac{P_\perp^2}{Q^2}}}\,
xG_{\mathbb{P}}\left(x, x_{\mathbb{P}}, {(1-x)P_\perp^2}\right).
 \end{align}	
This differs from \eqn{qqgSIDISsoftg} in the main text via the lower limit, which is $x_{\rm min}>\beta$ (cf.~\eqn{xmin}),
and also via an additional factor inside the integrand, which for finite $x-\beta$ is formally of higher-twist
order, but which could bring in complications from the limit $x\to \beta$, as we have seen in the case of a soft quark.
Yet, in the present case, there are no such complications: the $g\to q\bar q$  splitting function $P_{qg}(\beta/x)$ is not
singular in that limit, hence the leading twist contribution to \eqn{qqgSIDISsoftgGEN} is correctly given by  \eqn{qqgSIDISsoftg}.
In fact, the structure of the above integral is very similar to that previously encountered in the case of a soft quark,
in relation with the interference contribution, cf.~\eqn{integraltint}.

 By the same argument, in this case there is no additional LT contribution from symmetric jets. 
 This can also be understood
 from a diagrammatic viewpoint: the gluon decay must occur inside the target wavefunction in order to produce
 the $q\bar q$ pair which absorbs the virtual photon. So, in this case, there is no place for final state radiation 
 within the context of collinear factorisation (and to the order of interest).

\providecommand{\href}[2]{#2}\begingroup\raggedright\endgroup


\end{document}